\documentclass[12pt,fleqn]{ucithesis}

\usepackage{amsmath}
\usepackage{amsthm}
\usepackage{array}
\usepackage{graphicx}
\usepackage[numbers]{natbib}
\usepackage{relsize}
\usepackage[titletoc]{appendix}

\usepackage{multirow}
\usepackage{tabularx}

\usepackage[linesnumbered,ruled]{algorithm2e}
\usepackage{wrapfig}

\usepackage{array}

\newcolumntype{L}[1]{>{\raggedright\let\newline\\\arraybackslash\hspace{0pt}}m{#1}}
\newcolumntype{C}[1]{>{\centering\let\newline\\\arraybackslash\hspace{0pt}}m{#1}}
\newcolumntype{R}[1]{>{\raggedleft\let\newline\\\arraybackslash\hspace{0pt}}m{#1}}

\usepackage{xcolor}

\definecolor{LightBlue}{HTML}{6EA6D5}
\definecolor{DarkBlue}{HTML}{0000A0}
\definecolor{MediumBlue}{HTML}{000000} 
\definecolor{MyColor}{HTML}{000000} 
\newcommand{\hl}[1]{\textcolor{MyColor}{#1}}

\newcommand{\mynoindent}{\vspace{0.3em} \noindent}
\newcommand{\boldstart}[1]{\mynoindent \textbf{#1}}

\definecolor{avinashcolor}{rgb}{0,0,0}
\newcommand{\avinash}[1]{{\color{avinashcolor} #1}}

\newcommand{\amberfrmname}{{\sf Amber}\xspace}
\newcommand{\frmname}{{\sf Reshape}\xspace}
\newcommand{\schfrmname}{{\sf Maestro}\xspace}

\mathchardef\mhyphen="2D

\usepackage[labelfont=bf]{caption}
\usepackage{subcaption}

\usepackage[plainpages=false,pdfborder={0 0 0}]{hyperref}

\newtheorem{definition}{\textsc{Definition}}[chapter]

\thesistitle{Towards Interactive, Adaptive and Result-aware Big Data Analytics}

\documenttitle{Dissertation}

\degreename{Doctor of Philosophy}

\degreefield{Computer Science}

\authorname{Avinash Kumar}

\committeechair{Professor Chen Li}
\othercommitteemembers
{
  Professor Michael J.~Carey\\
  Professor Sharad Mehrotra
}

\degreeyear{2022}

\copyrightdeclaration
{
  {\copyright} {\Degreeyear} \Authorname
}

\prepublishedcopyrightdeclaration
{
	Chapter 2 {\copyright} 2020 VLDB Endowment \\
	All other materials {\copyright} {\Degreeyear} \Authorname
}

\dedications
{
  To my parents Durga N.~Jha and Rita Jha.
}

\acknowledgments
{
  This thesis has been possible because of the contributions of a number of people.

First and foremost, I am extremely grateful to my advisor, Professor Chen Li, for his continuous support and guidance over the past several years. He has showed me how the process of system implementation can help in identifying and formulating research ideas. An important trait that I learnt from him is the ability to solve a big problem by breaking it down into manageable pieces. He trained me to write papers in a clear, concise and appealing way. He worked with me on my research tirelessly. His teachings have helped me grow and become an independent researcher.

I would like to thank Professor Michael Carey and Professor Sharad Mehrotra for joining my doctoral committee. Their knowledge and insightful comments have always helped me in understanding the topics better and making my work stronger. 

I would like to thank my colleagues Sadeem Alsudias and Zuozhi Wang for participating in numerous discussions and brain-storming sessions. Their comments have always helped me in seeing the bigger picture. 

I would like to thank the members of the Texera team - Zuozhi Wang, Sadeem Alsudais, Shengquan Ni, Yicong Huang and Xiaozhen Liu - for participating in various design and research discussions.

I would like to thank Qiushi Bai for being a patient neighbor of mine in the lab. The small discussions and walks helped me relax during stressful days.

I thank my friends Shubham, Shreya, Aman and Mukesh for being a patient listener and for their moral support. The conversation with them have always uplifted my mood and filled me with energy.

I would like to thank my parents for being a constant anchor in my life journey. Their support and sacrifices has given me the confidence to pursue bold paths and come this far.

The second chapter is adapted from my research paper publication in VLDB 2020 titled ``Amber: A Debuggable Dataflow System Based on the Actor Model''. The third chapter is adapted from my research paper available on Arxiv titled ``Reshape: Adaptive Result-aware Skew Handling for Exploratory Analysis on Big Data''. The co-authors listed in these papers are Sadeem Alsudais, Zuozhi Wang, Shengquan Ni, Yicong Huang and Chen Li. I thank the Microsoft Orleans team and Dr. Phil Bernstein for answering questions related to Orleans and Yuran Yan for helping in evaluating Apache Spark. This work has been supported in parts by NSF IIS-1745673 and IIS-2107150 awards and a grant from the Google Cloud Research Credits program. I have also been supported by cloud credits from the NSF Cloudbank program.
}

\newcommand{\mypubentry}[3]{
  \begin{tabular*}{1\textwidth}{@{\extracolsep{\fill}}p{4.5in}r}
    \textbf{#1} & \textbf{#2} \\ 
    \multicolumn{2}{@{\extracolsep{\fill}}p{.95\textwidth}}{#3}\vspace{6pt} \\
  \end{tabular*}
}

\curriculumvitae
{

\textbf{EDUCATION}
  
  \begin{tabular*}{1\textwidth}{@{\extracolsep{\fill}}lr}
    \textbf{Doctor of Philosophy in Computer Science} & \textbf{2022} \\
    \vspace{6pt}
    University of California, Irvine & \emph{Irvine, CA} \\
    \textbf{Masters in Computer Science} & \textbf{2022} \\
    \vspace{6pt}
    University of California, Irvine & \emph{Irvine, CA} \\
    \textbf{Masters in Information Technology} & \textbf{2015} \\
    \vspace{6pt}
    Indian Institute of Technology, Roorkee & \emph{Roorkee, Uttarakhand, India} \\
    \textbf{Bachelors in Computer Science and Engineering} & \textbf{2015} \\
    \vspace{6pt}
    Indian Institute of Technology, Roorkee & \emph{Roorkee, Uttarakhand, India} \\
  \end{tabular*}
  
\vspace{12pt}
\textbf{PUBLICATIONS}

  \mypubentry{Fries: Fast and Consistent Runtime Reconfiguration in Dataflow Systems with Transactional Guarantees}{ 2023}{Proceedings of the VLDB Endowment (PVLDB)}
  \mypubentry{Reshape: Adaptive Result-aware Skew Handling for Exploratory Analysis on Big Data}{ 2022}{Arxiv}
  \mypubentry{Demonstration of Collaborative and Interactive Workflow-Based Data Analytics in Texera}{ 2022}{Proceedings of the VLDB Endowment (PVLDB)}
  \mypubentry{Amber: A Debuggable Dataflow System Based on the Actor Model}{ 2020}{Proceedings of the VLDB Endowment (PVLDB)}
  \mypubentry{Demonstration of Interactive Runtime Debugging of Distributed Dataflows in Texera}{ 2020}{Proceedings of the VLDB Endowment (PVLDB)}

}

\thesisabstract
{
  As data volumes grow across applications, analytics of large amounts of data is becoming increasingly important. Big data processing frameworks such as Apache Hadoop, Apache AsterixDB, and Apache Spark have been built to meet this demand. A common objective pursued by these traditional cluster-based big data processing frameworks is high performance, which often means low end-to-end execution time or latency.

A typical user of these frameworks submits a job to the framework and waits for the results for minutes, hours or even days based on the size of input data and complexity of the job. There is often a need to interact with an executing job to check its states or modify parts of the job. Traditional big data processing frameworks offer little insight into an executing job. They provide simple statistics such as data size input into and processed by various operators of a job, which may not be enough information for the user.

The widespread adoption of data analytics has led to a call to improve the traditional ways of big data processing. There have been demands for making the analytics process more interactive and adaptive, especially for long running jobs. A typical data analytics workflow undergoes multiple iterations of refinement to become the final workflow that performs a task correctly. While performing these iterations, a data analyst is more interested in seeing the first few results quickly than the total execution time. If the results are undesirable, the analyst can terminate the workflow without waiting for it to execute completely. This underlines the importance of initial results in the iterative process of data wrangling and motivates a result-aware approach to big data analytics. 

This dissertation is motivated by these calls for improvement in data processing and the experiences over the past few years while working on the Texera project, which is a collaborative data analytics service being developed at UC Irvine. Texera is a GUI-based service that allows the users to drag-and-drop operators to create workflows that can be executed on computing clusters. This dissertation mainly consists of three parts. The first part is about the design of the Amber engine that serves as the backend data processing framework for the Texera service. Amber supports interactivity and adaptivity during data analysis. A key feature of Amber is the existence of fast control messages that allow the interaction and adaptation to happen with sub-second latency. The second part is about an adaptive and result-aware skew-handling framework called Reshape. Reshape uses fast control messages to implement iterative skew mitigation techniques for a wide variety of operators. The mitigation techniques in Reshape have also been analyzed from the perspective of their effects on the results shown to the user. Reshape is also capable of self-tuning its threshold parameter to lessen the technical burden on the users. The last part is about a result-aware workflow scheduling framework called Maestro. This part talks about how to schedule a workflow for execution on computing clusters and make result-aware decisions while doing so. This work improves the data analytics process by bringing interactivity, adaptivity and result-awareness into the process.
}

\hypersetup{
	pdftitle={\Thesistitle},
	pdfauthor={\Authorname},
	pdfsubject={\Degreefield},
}

\begin{document}

\preliminarypages

\chapter{Introduction}
\label{chap:introduction}

As information volumes in many applications continuously grow, analytics of large amounts of data is becoming increasingly important. Data-processing engines have been built to support this analytics demand. One of the earliest open-source cluster-based big data-processing frameworks is Apache Hadoop~\cite{misc/hadoopmapreduce}. It uses the MapReduce programming model~\cite{conf/osdi/DeanG04} to allow developers to easily write jobs to analyze bounded (finite) input data that can automatically be executed on large clusters. Apache Spark~\cite{misc/spark} improves on the MapReduce programming model by introducing resilient distributed datasets (RDDs) that do not need to be written to stable storage after every map or reduce stage of a job. This allows Spark to have better performance than MapReduce.  Apache AsterixDB~\cite{journals/corr/AlsubaieeAABBBCCCFGGHKLLOOPTVWW14} provides a scalable data management system for semi-structured data along with the support for a query language similar to SQL. Spark Streaming~\cite{misc/spark}, Apache Storm~\cite{apacheStorm}, and Apache Flink~\cite{misc/flink} are frameworks that have been built to support analytics over unbounded streams of data. There are also GUI-based workflow systems such as Alteryx~\cite{alteryx}, RapidMiner~\cite{rapidMiner}, Knime~\cite{knime}, and Einblick~\cite{Einblick} that provide a GUI interface where the users can drag-and-drop operators and create a workflow as a directed acyclic graph (DAG). These GUI-based frameworks lower the technical learning curve for its users and allow even non-technical users such as public health scientists to perform data analytics.

A common objective pursued by traditional big data-processing frameworks, especially cluster-based frameworks, is high performance, which often means low end-to-end execution time or latency. A typical user of these frameworks submits a job to the framework and waits for the results for minutes, hours, or even days based on the size of input data and complexity of the job. The frameworks aim to compile and optimize a submitted job to reduce the execution time or latency~\cite{conf/sigmod/BehmPAACDGHJKLL22}. They may try to decide how to parallelize the analytics jobs over a cluster of machines to achieve better performance~\cite{SparkAQE:website}. There have also been works that perform machine learning-based prediction to determine the resource allocation that optimizes the execution time of workloads~\cite{conf/sigmod/SabekUK22}.

The widespread adoption of data analytics has led to a call to improve the traditional ways of big data processing and introduced new requirements. For example, after starting the execution of a long-running workflow, the analyst may want to interact with the workflow to check the status of different operators and adapt parts of a running workflow according to changing circumstances. She may also want the initial results to be shown quickly so that she can identify any problems in the workflow early. Next we talk about these requirements in detail.

\boldstart{Interactivity.} As mentioned earlier, an analytics job may take a long time to complete. During the execution of such a long-running analytics job, there may arise a need to interact with the job to verify its correctness. Consider a task to collect tweets over multiple days about blunt smoking using tobacco-based wraps. The analyst uses ``blunt'' as one of the keywords to collect the tweets. After submitting the task, the analyst may want to periodically interact with the workflow to review samples of tweets being collected. After the workflow has run for a few hours, the analyst may observe that tweets related to the actress Emily Blunt are also being collected. When this happens, the analyst may want to make changes to the workflow. If such an interaction is not possible, the analyst will discover the erroneous tweets only when the task terminates after a few days. In general, the jobs analyzing large amounts of data are typically executed on a cluster of multiple machines. Popular cluster-based data-processing engines such as MapReduce, Spark, and Flink provide little feedback to users during the processing of data, leaving the analyst in the dark. They provide simple statistics such as data size input into and processed by various operators of a job, which may not be enough information for the analyst. This has led to the demand for engines that support interactivity in analytics jobs~\cite{journals/interactions/FisherDCD12}.

\begin{figure}[htbp]
 \begin{center}
	\includegraphics[width=5in]{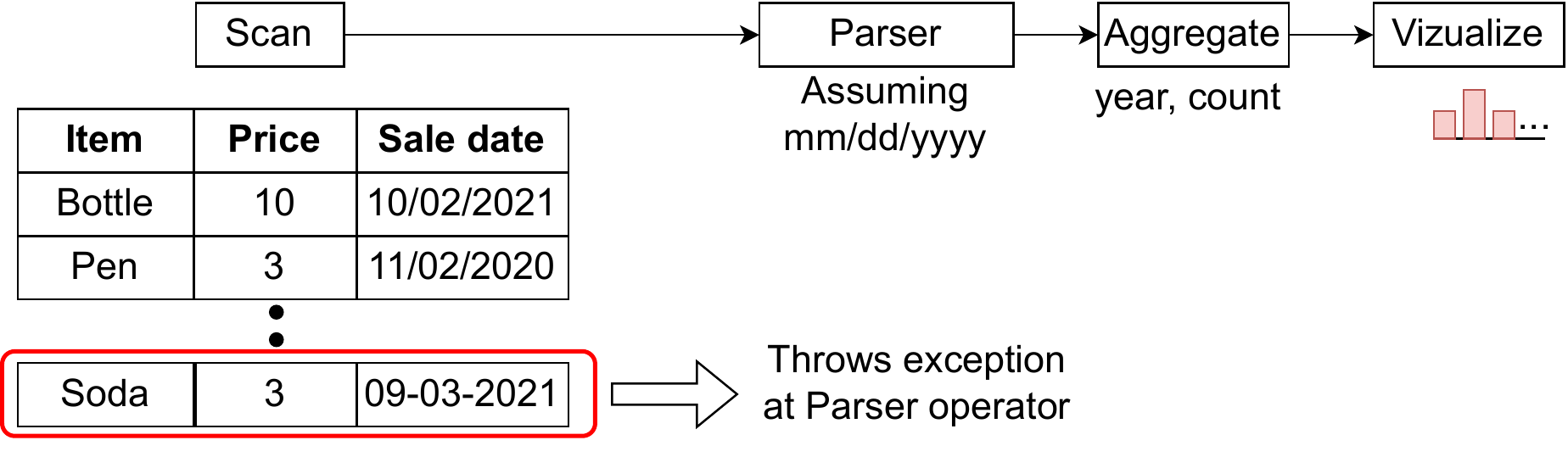}
	\caption{\label{fig:parse-error}
		\textbf{Workflow to plot the total sales by year. The sale date in the marked tuple is in a format that is not handled by the {\sf Parser} operator. If the workflow throws a parser exception and crashes, the results computed from earlier tuples are lost and the computational resources are wasted. Thus, the analyst may want to skip the problematic tuple or modify the parsing logic at runtime. }
	}
\end{center}
\end{figure}

\boldstart{Adaptivity.} During the execution of an analytics workflow, there may arise a need to adapt or modify parts of the workflow at runtime. A need for such runtime adaptation arises in tasks that are critical and their downtime should be reduced as much as possible. Consider a spam detection operator that needs to be constantly online in the cloud. When there is a sudden rise in spam emails directed to a network, the developer wants to modify the operator and set a stricter detection threshold without stopping and restarting the workflow. Another need for runtime adaptation arises in scenarios where a runtime adaptation is needed to preserve the already computed results of a task and avoid the wastage of computing resources. Consider the workflow shown in Figure~\ref{fig:parse-error}, containing a parser operator that parses the date column of the tuples to find the year. After a few thousand tuples have been processed, a tuple arrives that cannot be parsed by the operator because its date is in a different format. In this scenario, the analyst may prefer to have the ability to ignore this tuple and continue processing or the ability to modify the parser code to handle the tuple. The data-processing frameworks such as MapReduce, Spark and Flink cannot support such adaptation of workflows at runtime. In the first scenario containing the spam detection operator, the traditional data-processing frameworks usually require a complete stop and restart of the task. In the second scenario containing the parser operator, data-processing frameworks such as Spark crash when faced with exceptions leading to wasted earlier results and computational resources~\cite{conf/sigmod/GulzarICK17}. This has led to the demand for engines that support runtime adaptivity in analytics jobs~\cite{conf/sigmod/CarboneFKK20}.

\boldstart{Result Awareness.} The process of data analysis, especially in GUI-based analytics systems, is highly exploratory and iterative~\cite{journals/interactions/FisherDCD12, conf/sigmod/XuKAR22,conf/sigmod/VartakSLVHMZ16}. Often the analyst constructs an initial workflow and executes it to observe a few initial results. If they are not desirable, she terminates the current execution and revises the workflow. The analyst iteratively refines the workflow until finishing a final workflow to compute the results. While performing these iterations, a data analyst is more interested in seeing the first few results quickly than the total execution time. Also, it is valuable if the initial results are representative of the final results. Since data analysts spend almost $80$\% of their time performing such iterations during data wrangling~\cite{misc/nytimes-80percent-data-wrangling}, the data-processing frameworks need to also focus on result awareness, rather than just the total execution time.

In this dissertation, we focus on bringing interactivity, adaptivity, and result awareness into the data analytics process. This is motivated by the calls for improvement in the traditional ways of data processing and the experiences over the past few years while working on the Texera project~\cite{TexeraWebsite}, which is a  service being developed at UC Irvine to support collaborative data analytics. Texera is a GUI-based service that allows the users to drag-and-drop operators to create workflows that can be executed on computing clusters. In particular, we study three problems in this dissertation.

\boldstart{Designing an engine that supports interactivity and adaptivity.} We discuss the design of a data-processing engine called \amberfrmname that allows runtime interactivity and adaptivity while executing workflows. \amberfrmname is the backend engine of the Texera service~\cite{journals/pvldb/LiuWNAHKL22}. A main feature of \amberfrmname that allows it to be interactive and adaptive is the availability of fast control messages. We describe the implementation of these fast control messages in \amberfrmname and explain how to use these messages for features such as pausing the execution and changing the logic of operators with sub-second latency. We also discuss the concept of local and global conditional breakpoints in workflows and how to detect these in \amberfrmname. We discuss the challenges in supporting fault tolerance in Amber and present a technique to achieve it.

\boldstart{Adaptive result-aware skew handling.} We describe a framework called \frmname that uses the availability of fast control messages to handle partitioning skew in operators of a workflow during runtime and shows representative initial results. \frmname mitigates skew iteratively during the execution. We present different approaches of skew mitigation and analyze their impact on the results shown to the user. We also present a way to dynamically adjust the skew detection threshold to reduce the number of iterations of mitigation. We generalize \frmname to multiple operators such as HashJoin, Group-by, and Sort, and discuss challenges related to state migration. The \frmname framework has been implemented on Amber and Apache Flink.

\boldstart{Result-aware scheduler.} We discuss the design of a result-aware scheduler called \schfrmname. The scheduler takes a workflow as an input and breaks the workflow DAG into smaller sub-DAGs called regions that can be separately scheduled. We discuss the concept of a region graph that encapsulates the dependencies among the regions of a workflow and present an algorithm to obtain the region graph from a workflow DAG. In order to be scheduled, the region graph should not have any cycles. We show how to modify the workflow DAG to avoid the presence of cycles in the region graph. There are different ways to modify the workflow DAG and we present an algorithm to enumerate the various options. We then show how to choose an option in a result-aware manner.

The rest of the dissertation is organized as follows. Chapter~\ref{chap:amber} describes the design of the \amberfrmname engine. Chapter~\ref{chap:reshape} describes the skew-handling framework \frmname. Chapter~\ref{chap:maestro} describes the scheduling framework \schfrmname.  Finally, Chapter~\ref{chap:conclusion-future-work} concludes this dissertation and discusses future research directions.

\chapter{Amber: A Debuggable Dataflow System Based on the Actor Model}
\label{chap:amber}
\section{Introduction}
\label{sec:introduction}

As analytics of large amounts of data becomes increasingly important, many big data engines have been developed to support scalable analytics using computing clusters. In these systems, a main challenge faced by developers when running an analytic task on a large dataset is its long running time, which can take hours, days, or even weeks.  Such a long-running task often leaves the developer in the dark without providing valuable feedback about the status of the execution~\cite{journals/interactions/FisherDCD12}.  What is worse is that the job can fail due to various reasons, such as software bugs, unexpected data, or hardware issues.  In the case of failure, earlier computing resources and time are wasted, and a new job needs to be submitted from scratch.


Analysts have resorted to different techniques to identify errors in job execution. One could first run a job on a small dataset, with the hope of producing failures, and identifying and solving the problems. The analyst iteratively refines the workflow multiple times before arriving at the final workflow that executes on the small dataset without any errors~\cite{journals/interactions/FisherDCD12}. Then, the analyst runs the final workflow on a large dataset. Unfortunately, many runtime failures occur only on a big dataset. For instance, a software bug is triggered only by some rare, outlier data instances, which may not appear in a small dataset~\cite{conf/IEEEscc/HerathLMGSVMP12, conf/sigmod/GulzarICK17}. As another example, there can be an out-of-memory (OOM) exception that happens only when the data volume is large.

Another method is to instrument the software to generate log records to do post-execution analysis. This approach has several limitations. First, the developer has to add statements at many places in order to find bugs. These statements can produce an inordinate amount of log records to be analyzed offline, and most of them are irrelevant. Second, these log records may not reveal all the information about the runtime behavior of the job, making it hard to identify the errors. This situation is similar to the scenario of debugging a C program. Instead of using {\tt printf()} to produce a lot of output messages and do post-execution analysis, many developers prefer to use a debugger such as {\tt gdb} to investigate the runtime behavior of the program {\em during} its execution.


The aforementioned shortcomings of the debugging techniques have led data analysts to seek more powerful monitoring and debugging capabilities~\cite{conf/sigmod/OlstonR11, journals/cacm/BeschastnikhWBE16, journals/interactions/FisherDCD12}. There are several recent efforts to provide debugging capabilities to big data engines~\cite{conf/sigmod/GulzarICK17,conf/cloud/GulzarIHLCK17}. As an example, BigDebug~\cite{conf/sigmod/GulzarICK17} used a concept of {\em simulated breakpoint} during the execution of an Apache Spark job. Once the execution arrives at the breakpoint, the user can inspect the program state. More details about these approaches and their limitations are discussed in Section~\ref{ssec:related-work}.  A fundamental reason for their limitations is that they are developed on engines such as Spark that are not natively designed to support debugging capabilities, which limit their performance and usability. 

In this chapter, we consider the following question: 
\begin{quote}
{\em Can we develop a scalable data-processing engine that supports responsive debugging?}
\end{quote}
We answer the question by developing a parallel data-processing system called  Amber, which stands for ``actor-model-based debugger.''  A user of the system can interact with an analytic job during its execution.  For instance, she can pause the execution, investigate the states of operators in the job, and check statistics such as the number of processed records and average time to process each record in an operator.  Even if the execution is paused, she can still interact with the operators in the job.  The user can modify the job, e.g., by changing the threshold in a selection predicate, a regular expression in an entity extractor operator, or some parameters in a machine learning (ML) operator.  The user can also set conditional breakpoints, so that the execution can be paused automatically when a condition is satisfied. Examples of conditions are incorrect input formats, occurrences of exceptions etc. In this way, the user can skip many irrelevant iterations.  After doing some investigation, she can resume the execution to process the remaining data.  To our best knowledge,  Amber is the first system with these debugging capabilities.



 Amber is based on the {\em actor model}~\cite{conf/ijcai/HewittBS73,agha1985actors}, a distributed computing paradigm that provides concurrent units of computation called {\em actors}. The message-passing mechanism between actors in the actor model makes it easy to support both data messages and debugging requests, and allows low-latency control-message processing.  Also after the execution of a workflow is paused, the actor-based operators can still receive messages and respond to user requests. More details about the actor model and the motivation behind using it for  Amber are described in Section~\ref{ssec:actor-model}.  

The actor model has been around for decades and there are data-processing frameworks built on top of it~\cite{conf/icdm/NeumeyerRNK10, ludascher2006scientific}. A natural question is ``why do we develop  Amber now?''. The answer is twofold. First, as data is getting increasingly bigger, the need for a system that supports responsive debugging during big data processing is getting more important. Second, there are more mature and widely adopted actor model implementations on clusters recently, making it easy to develop our system without reinventing the wheel. 

There are several challenges in developing  Amber using the actor model. First, every actor has a single mailbox, which is a FIFO queue storing both data messages and control messages. (The actor model does not support priority messages natively.)  Large-scale data processing implies that data messages sent to an actor can be significantly more than its incoming control messages. Thus, the mailbox can already have many data messages when a control message arrives. Responsive debugging requires that control messages be processed quickly, but the control message can only be processed after  those data messages ahead of it. Second, a data message can take an arbitrarily long time to process (e.g., in an expensive ML operator). Real-time debugging necessitates that user requests should be taken care of in the middle of processing a data message instead of waiting for the entire message to be processed, which could take a long time depending on the complexity of the operator.

In this chapter, we tackle these challenges and make the following contributions. In Section~\ref{sec:debuggable-engines} we discuss important features related to debugging the execution of a data workflow, and analyze the requirements of an engine to support these features.  In Section~\ref{sec:system-overview} we present the overall architecture of  Amber and study how to construct an actor workflow for an operator workflow, how to allocate resources to actors, and how to transfer data between actors. In Section~\ref{sec:execution-lifecyle} we describe the lifecycle of executing a workflow, discuss how control messages are sent to the actors, how actors expedite the processing of these control messages, and how they save and load their states during pausing and resuming, respectively.  In Section~\ref{sec:ConditionalBreakpoints} we study how to support conditional breakpoints in  Amber, and present solutions for enforcing local conditional breakpoints (which can be checked by actors individually) and global conditional breakpoints (checked by the actors collaboratively in a distributed environment). In Section~\ref{sec:fault-tolerance}, we discuss challenges in supporting fault tolerance in Amber and present a technique to achieve it. In Section~\ref{sec:exps} we present the  Amber implementation on top of the Orleans system~\cite{orleans}, and report an experimental evaluation using real datasets on computing clusters to show its high performance and usability.  


\subsection{Related Work}
\label{ssec:related-work}





\noindent{\bf Spark-based debugging.}  Titian~\cite{journals/pvldb/InterlandiSTGYK15} is a library that enables high speed data provenance in Spark.  BigSift~\cite{conf/cloud/GulzarIHLCK17} is another provenance-based approach for finding input data responsible for producing erroneous results. It redefines provenance rules to prune input records irrelevant to given faulty output records before applying delta debugging~\cite{journals/tse/ZellerH02}.  BigDebug~\cite{conf/sigmod/GulzarICK17} uses the concept of {\em simulated breakpoint} in Spark execution.  A simulated breakpoint needs to be preset before the execution starts, and cannot be added or changed during the execution. Furthermore, after reaching a simulated breakpoint, the results computed till then are materialized, but the computation still continues. If the user makes changes to the workflow (such as modifying a filter condition) after the simulated breakpoint, the existing execution is cancelled, causing computing resources to be wasted.  In addition, the part of the workflow after the simulated breakpoint is executed again using the materialized intermediate results. Amber is different since the developer can set a breakpoint or explicitly pause the execution at any time, and the computation is truly paused.

Spark cannot support such features due to the following reason. In order for the driver (as ``controller'' in Amber) to send a Pause message to an executor (as ``actor'' in Amber) at an arbitrary user-specified time, the driver needs to send some state-change information to the executor.  Spark has two ways that might be possibly used to support communication from the driver to the executor, either through a broadcast variable or using an RDD.  Both are read-only to ensure deterministic computation, which is mandatory in the method used by Spark to support fault tolerance.  Any state change requires a modification of the content of a broadcast variable or an RDD, and such information cannot be sent to the executor from the driver.

\vspace{0.05in}

\noindent{\bf Workflow systems:}  Alteryx~\cite{alteryx}, Kepler~\cite{ludascher2006scientific}, Knime~\cite{knime}, RapidMiner~\cite{rapidMiner}, and Apache Taverna~\cite{missier2010taverna} allow users to formulate a computation workflow using a GUI interface.  They provide certain feedback to the user during data processing.  These systems do not run on a computing cluster, and do not support debugging either.  Texera~\cite{texera} is an open-source GUI-based workflow system we are actively developing in the past three years, and  Amber is a suitable backend engine.  Apache Airavata~\cite{conf/sc/MarruGHTPMSGCGSDPW11} is a scientific workflow system supporting pausing, resuming, and monitoring.  Its pause is coarse in nature since a user has to wait for an operator to completely finish processing all its data.  Apache Storm~\cite{apacheStorm} supports distributed computations over data streams, but does not support any low-level interactions with individual operators apart from starting and stopping the operators.  


\vspace{0.05in}


\noindent{\bf Debugging in distributed systems.} When debugging a program (e.g., in C, C++, or Java) in a distributed environment, developers often use pre-execution methods such as model-checking, running experiments on a small dataset, and post-execution methods such as log analysis to identify bugs in a distributed system~\cite{journals/cacm/BeschastnikhWBE16}, and their limitations are already discussed above. Although query-profiling tools such as Perfopticon~\cite{journals/cgf/MoritzHHH15} have simplified the process of analyzing distributed query execution, their application is limited to discovering runtime bottlenecks and problematic data imbalances. StreamTrace~\cite{conf/chi/BattleFDBCG16} is another tool that helps developers construct correct queries by producing visualization that illustrates the behavior of queries. Such pre-execution and post-execution analysis tools cannot be used to support debugging during the execution. On the other hand, breakpoints are an effective tool to debug the runtime behavior of a program. In prior studies, global conditional breakpoints in a distributed system are defined as a set of primitive predicates such as entering a procedure, which are local to individual processes (hence can be detected independently by a process), tied together using relations (e.g., conjunction, disjunction, etc.) to form a distributed global predicate~\cite{haban1988global,conf/icdcs/FowlerZ90,conf/icdcs/MillerC88,dao2009live}. Checking the satisfaction of a global condition given that all the primitive predicates have been detected was studied in~\cite{haban1988global, conf/icdcs/MillerC88, dao2009live}.  Our work is different given its focus on data-oriented conditions.



\vspace{0.05in}

\noindent{\bf Pausing/resuming in DBMS.}~\cite{conf/sigmod/ChandramouliBBY07} studied how to suspend and resume a query in a single-threaded pull-based engine. 
\cite{journals/pvldb/AntonopoulosKTU17} studied how to resume online index rebuilding after a system failure.  These existing approaches do not allow users to inspect the internal state after pausing.

\vspace{0.05in}

\noindent{\bf Actor model based data processing.} The use of the actor model for data processing has been explored before. For instance, S4~\cite{conf/icdm/NeumeyerRNK10} was a platform that aimed to provide scalable stream processing using the map-reduce paradigm and actor model.  Amber is different since it focuses on responsive debuggability during data processing, without compromising the scalability. Kepler~\cite{ludascher2006scientific} is a scientific workflow system using the Ptolemy II actor model implementation~\cite{ptolemyII}. It is limited to a single machine and treats a grid job as an outside resource included in the workflow as an operator.  Amber is different as it is a parallel runtime engine natively.




\section{Debuggable Dataflow Engines}
\label{sec:debuggable-engines}

In this section, we discuss important features related to debugging the execution of a data workflow, and analyze the requirements of an engine to support these features. We then give an overview of the actor model.


\subsection{Debugging Execution of Data Workflows}

A data workflow ({\em dataflow} for short) is a directed acyclic graph (DAG) of operators. An operator is {\em physical} (instead of {\em logical}) since it specifies how its computation is done exactly, such as a hash-join operator, which is different from a ripple-join operator.  We consider common relational operators as well as operators that implement user-defined functions. When running a workflow, data from sources is passed through the operators, and the results are produced from a final operator called {\sf Sink}.  For simplicity, we focus on the relational data model, in which data is modeled as bags of tuples, and the results generalize to other data models.

\begin{figure}[htbp]
\begin{center}
	\includegraphics[width=0.75\linewidth]{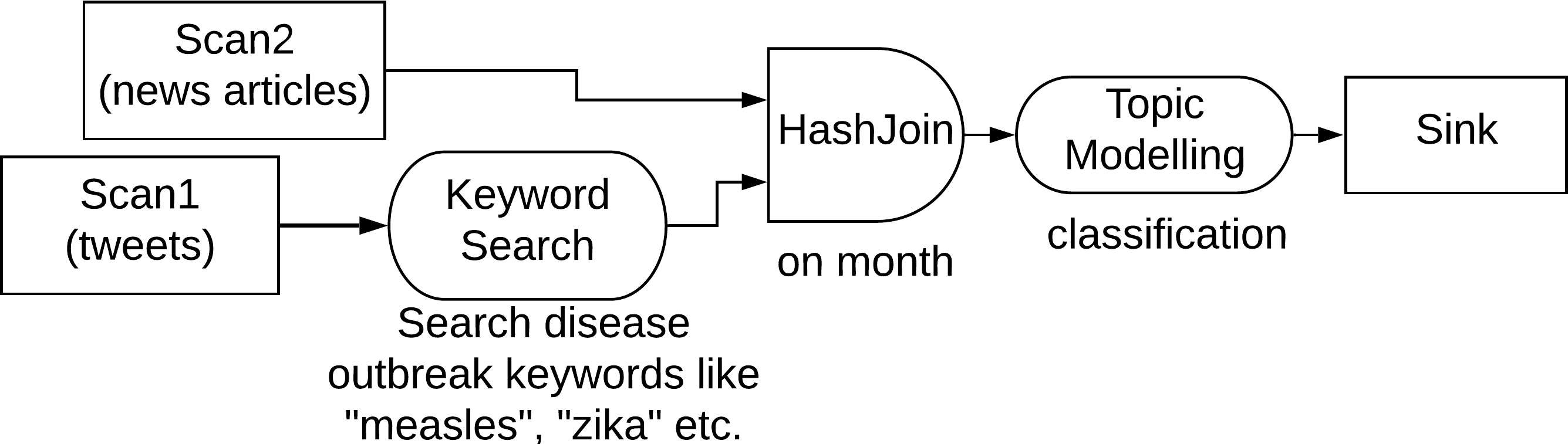} 
	\caption{\label{fig:sample-workflow}
		\textbf{A workflow to anlayze disease outbreaks from tweets and news.}
	}
    \vspace{-0.1in}
\end{center}
\end{figure}

Figure~\ref{fig:sample-workflow} shows an example workflow to identify news articles related to disease outbreaks using a table of news articles (timestamp, location, content, etc.) and a table of tweets (timestamp, location, text, etc.). The {\sf KeywordSearch} operator on the tweet table selects records related to disease outbreaks such as measles and zika. The next step is to find news articles published around the same time by joining them based on their timestamps (e.g., months). We then use topic modelling to classify the news articles that are indeed related to outbreaks.

During the execution of a workflow, we want to allow the developer to take any of the following actions. (1) {\em Pausing}: stop the execution so that all operators no longer process data. (2) {\em Investigating operators}: check the states of each operator, and collect statistics about its behaviors, such as the number of processed records and processing time. (3) {\em Setting conditional breakpoints}: stop the workflow once the condition of a breakpoint is satisfied, e.g., the number of records processed by an operator goes beyond a threshold.  Breakpoints can be set before or during the execution.  (4) {\em Modifying operators}: after pausing the execution, change the logic of an operator, e.g., by modifying the keywords in {\sf KeywordSearch}. (5) {\em Resuming}: continue the execution.


{\bf Engine requirements.} A dataflow engine supporting the abovementioned debugging capabilities needs to meet the following requirements. (1) {\em Parallelism}: To support analytics on large amounts of data, the engine needs to allow parallel computing on a cluster.  As a consequence, physically an operator can be deployed to multiple machines to run simultaneously. (2) {\em Supporting various messages between operators}: Developers control the execution by sending messages to operators, which should co-exist with data transferred between operators.  Even if the execution is paused, each operator should still be able to respond to requests.  (3) {\em Timely processing of control messages}: Debugging requests from the developers need to take effect quickly to improve the user experience and save computing resources. Thus control messages should be given a chance to be processed by the receiving operator without a long delay. Since processing data tuples can be time consuming, computation in an operator should be granulated, e.g., by dividing data into batches with a size parameter, so that it can handle control messages in midst of processing data. 



\subsection{The Actor Model}
\label{ssec:actor-model}

The {\em actor model}~\cite{conf/ijcai/HewittBS73,agha1985actors} is a computing paradigm that provides concurrent units of computation called ``actors.'' A task in this distributed paradigm is described as computation inside actors plus communication between them via messages. Every actor has a mailbox to store its received messages.  After receiving a message, the actor performs three basic actions: (i) sending messages to actors (including itself); (ii) creating new actors; and (iii) modifying its state. There are various open source implementations of the actor model such as Akka~\cite{Akka}, CAF~\cite{caf}, Orleans~\cite{orleans},  and ProtoActor~\cite{ProtoActor}, as well as large-scale applications using these systems such as YouScan~\cite{YouScan}, Halo 5~\cite{Halo}, and Tapad~\cite{Tapad}.  For instance, Halo 5 is an online video game based on Orleans that allows millions of users to play together, and each player's actions can be processed within milliseconds. There is a study to develop an actor-oriented database with support of indexing~\cite{conf/cidr/BernsteinDKM17}. These successful use cases demonstrate the scalability of these implementations.


We use the actor model due to its several advantages. First, it is intrinsically parallel, and many implementations support efficient computing on clusters. This strength makes our system capable of supporting big data analytics. Second, the actor model simplifies concurrency control by using message passing instead of distributed shared memory. Third, the message-passing mechanism in the actor model makes it easy to support both data computation via data messages and debugging requests via control messages. Streaming control messages in the same pipeline as data messages leads to high scalability~\cite{journals/pvldb/MaiZPXSVCKMKDR18}. As described in Section~\ref{sec:realtime-controlmessage-processing}, we can divide the logic of operators into a sequence of smaller actions using the actor model and thus support low-latency control-message processing.




\section{Amber System Overview}
\label{sec:system-overview}

In this section, we present the architecture of the {\sf Amber} system. We discuss how it translates an operator DAG to an actor DAG and delivers messages between actors.

\subsection{Architecture}
\label{ssec:architecture}

\begin{figure}[htbp]
\begin{center}
	\includegraphics[width=2.9in]{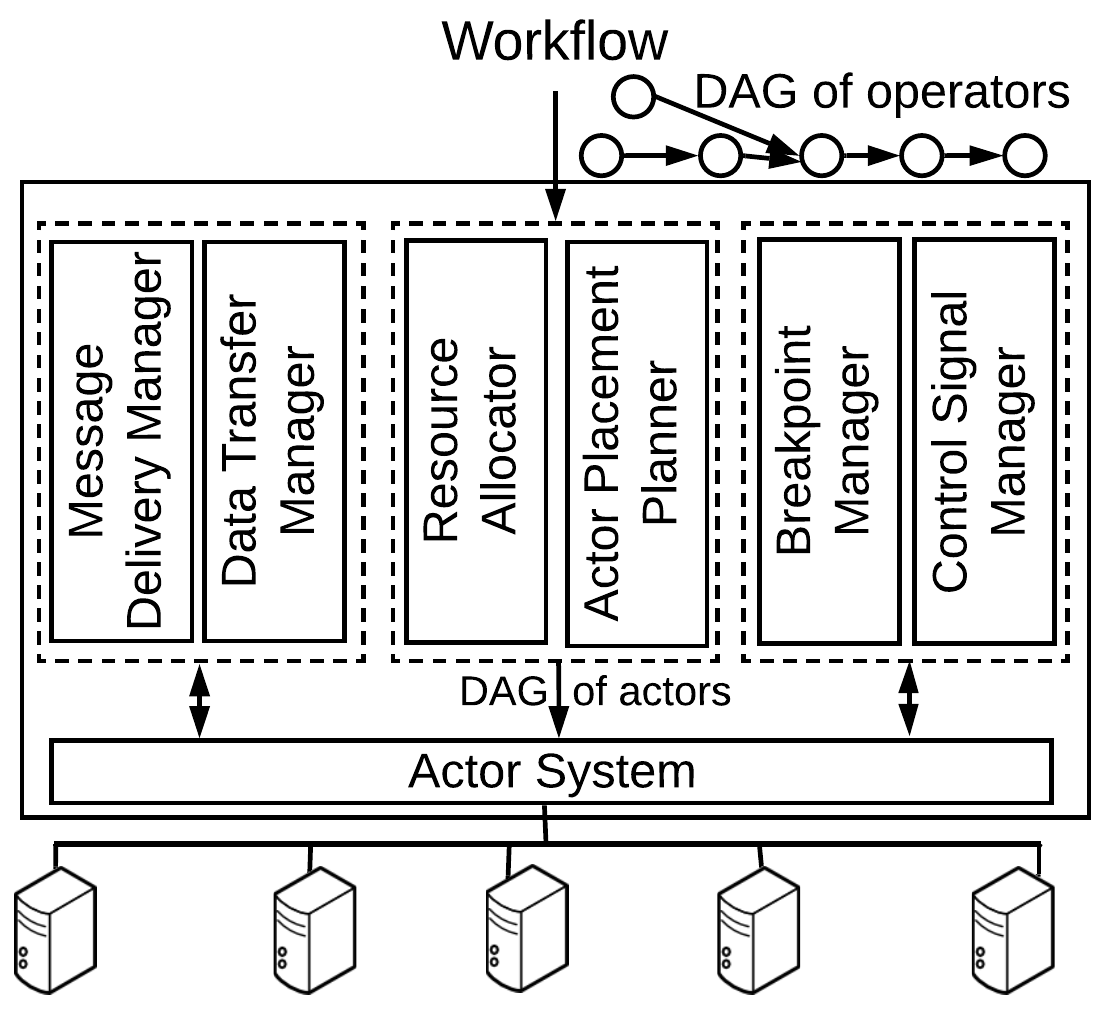} 
	\caption{\label{fig:architecture}
	 \textbf{Amber system architecture.}
	}
	\vspace{-0.2in}
\end{center}
\end{figure}

Figure~\ref{fig:architecture} shows the {\sf Amber} architecture.  The input to the system is a data workflow, i.e., a DAG of physical operators. This physical operator DAG is similar to the final optimized query plan in parallel DBMS~\cite{journals/csur/Kossmann00, journals/csur/Graefe93}. Based on the computational complexity of an operator, the {\em Resource Allocator} decides the number of actors allotted to each operator. The {\em Actor Placement Planner} decides the placement scheme of the actors across the machines of the cluster. An operator is translated to multiple actors, and the policy of how these actors send data to each other is managed by the {\em Data Transfer Manager}. These modules create a DAG of actors, allocate them to the machines, and determine how actors send data. The actor DAG is deployed to the underlying actor system, which is an implementation of the actor model, such as Orleans or Akka. The execution of the actor DAG takes place in the actor system, which places the actors on their respective machines, helps send messages between them, and executes the actions of an actor when a message is received. The actor system processes the data and returns the results to the client. The {\em Message Delivery Manager} ensures that the communication between any two actors is reliable and follows the FIFO semantics. More details about these modules are in Section~\ref{ssec:workflow-representation}.

During the execution, a user can send requests to the system, which are converted to control messages by the {\em Control Signal Manager}. The actor system sends control messages to the corresponding actors, and passes the responses back to the user.  The user can also specify conditional breakpoints, which are converted by the {\em Breakpoint Manager} to a form understandable by the engine. 



\subsection{Translating Operator DAG to Actor DAG}
\label{ssec:workflow-representation}

\begin{figure}[htbp]
\begin{center}
	\includegraphics[width=3.6in]{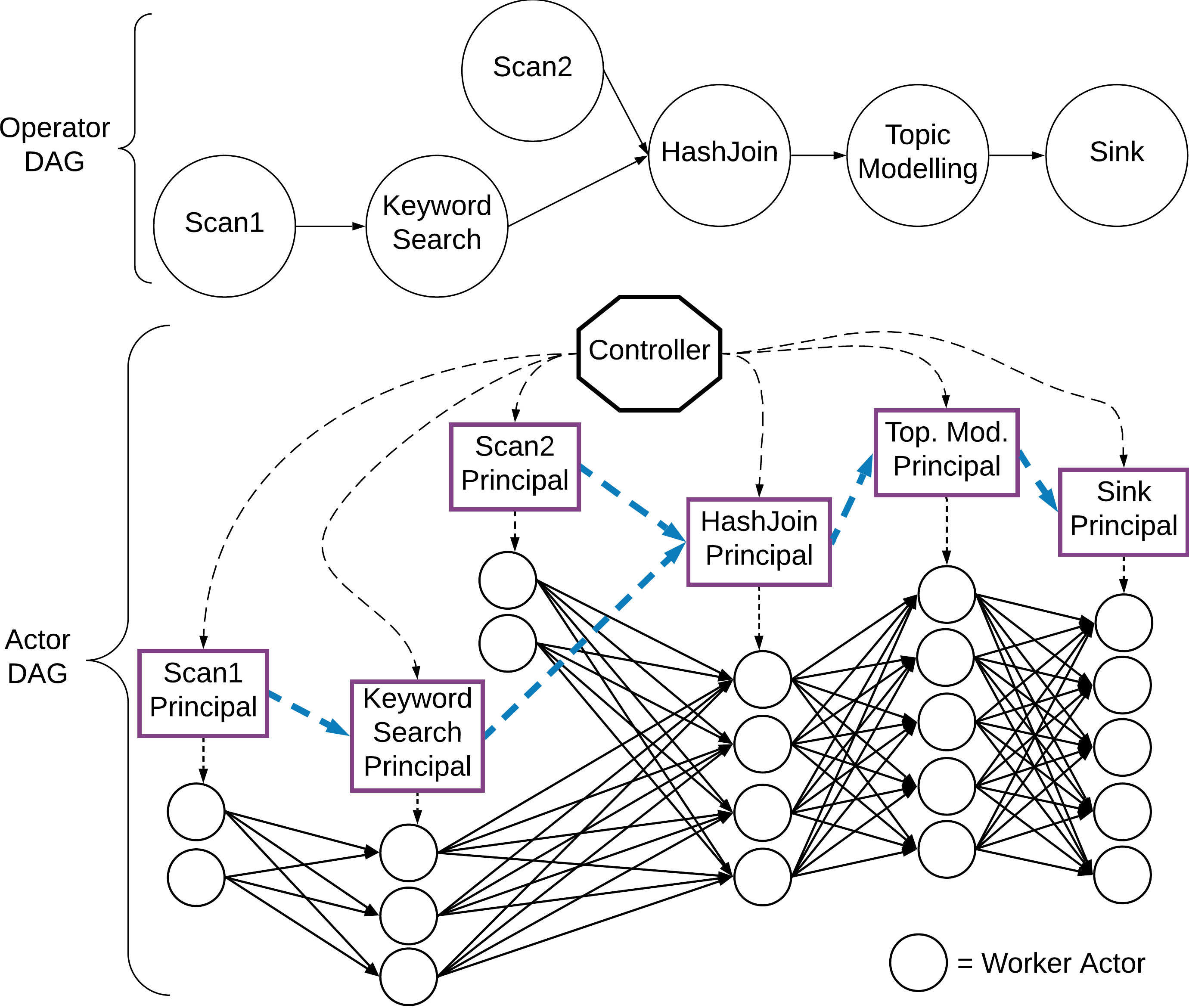} 
	\caption{\label{fig:actor-physical}
	 \textbf{Translating the disease-outbreak workflow to an actor DAG.} For clarity purpose, we show an edge from a principal actor to only one of its workers.}
	
	\vspace{-0.1in}
	\end{center}
\end{figure}

We use the example workflow of detecting disease outbreaks to show how {\sf Amber} translates the operator DAG to an actor DAG, as shown in Figure~\ref{fig:actor-physical}. 
A {\em controller} actor is the administrator of the entire workflow, as all control messages are first conveyed to this actor, which then routes them appropriately. The controller actor creates a {\em principal} actor for each operator and connects these principal actors based on the operator DAG. An edge $A \longrightarrow B$ between two actors $A$ and $B$ means that actor $A$ can send messages to $B$. The principal actor for an operator creates multiple {\em worker actors}, and each of them is connected to all the worker actors of the next operators. The worker actors conduct the data-processing computation and respond to control messages. The principal actor manages all the tasks within an operator, as well as collects runtime statistics, dispatches control signals, and aggregates control responses related to its operator. Placement of workers is planned to achieve load balancing and minimizing network communication overhead and the plan is included in the actor DAG. The workers of an operator are distributed uniformly across all machines. Workers do cross-machine communication only for shuffling data. Note that the structure of the final actor DAG is similar to the final task graph in Hyracks~\cite{conf/icde/BorkarCGOV11, phd/us/Borkar16} with a few differences such as existence of a principal actor and instantiation of workers as actors.




\subsection{Communication between Actors}
\label{comm-between-actors}


\noindent \textbf{Message-delivery guarantees.}  Data between actors is sent as data messages, where each message includes a batch of records to reduce the communication cost.  Control commands from the user are sent as control messages.  These two types of messages to an actor are queued into a {\em single} mailbox of the actor, and processed in their arrival order. Reliability is needed to avoid data loss during the communication and FIFO is needed for some operators such as {\sf Sort}. Thus, we made the communication channels between actors FIFO and exactly once. We use congestion control to regulate the rate of sending messages to avoid overwhelming a receiver actor and the network.

\noindent \textbf{Data-transfer policy on an incoming edge.}  For each edge $A \longrightarrow B$ from operator $A$ to operator $B$, the operator $B$ has a {\em data-transfer policy} on this incoming edge that specifies how  $A$ workers should send data messages to $B$ workers. If $B$ has multiple input edges, it has a data-transfer policy for each of them. The data-transfer policies used in Amber are similar to those in parallel DBMS~\cite{journals/cacm/DeWittG92, journals/tkde/DeWittGSBHR90}.  Following are a few example policies. (a) {\em One-to-one on the same machine:} An $A$ worker sends all its data messages to a particular $B$ worker on the same machine. (b) {\em Round-Robin on the same machine:} An  operator $A$ worker sends its messages to $B$ workers on the same machine in a round-robin order. (c) {\em Hash-based:}  Operators such as hash-based join require incoming data to be shuffled and put into specific buckets based on their hash value. An easy way to do so is to assign specific hash buckets to its workers.
 




\section{Lifecycle of Job Execution}
\label{sec:execution-lifecyle}



Each worker in Amber processes one data partition and forwards the results in batches to the workers of the downstream operators. Amber supports operator pipelining. In this section, we discuss the whole life cycle of the execution of a job, including how control messages are sent to the actors, how actors expedite the processing of control messages, and how each actor pauses and resumes its computation by saving and loading its states, respectively.



\subsection{Sending Control Messages to Actors}
\label{ssec:pausing-execution}


When the user requests to pause the execution, the controller actor sends a control message called ``{\sf Pause}'' to the actors. The message is sent in the following way, which is also applicable to other control messages.  The controller sends a {\sf Pause} message to all the principal actors, which forward the message to their workers. Due to the random delay in message delivery, the workers are paused in no particular order. For example, the source workers may be paused later than the downstream workers. Consequently, the workers may still receive data messages after being paused, and need to store them for later processing.

\subsection{Expedited Processing of Control Messages}
\label{sec:realtime-controlmessage-processing}

A critical requirement in  Amber is fast processing of control messages in order to support real-time response from the system during debugging.  Worker actors process a large number of data messages in addition to control messages. These two types of messages to a worker actor are enqueued in the same mailbox, which is a FIFO queue as specified in the actor model. Therefore, there could be a delay between the enqueuing of a control message and its processing.  This delay is affected mainly by two factors, the number of enqueued messages and the computation per batch.   For actor model implementations such as Akka that support priority messaging, we can expedite the processing of control messages by giving them a  priority higher than data messages. 

\begin{figure*}[htbp]
	\begin{center}
	    
	\includegraphics[width=\linewidth]{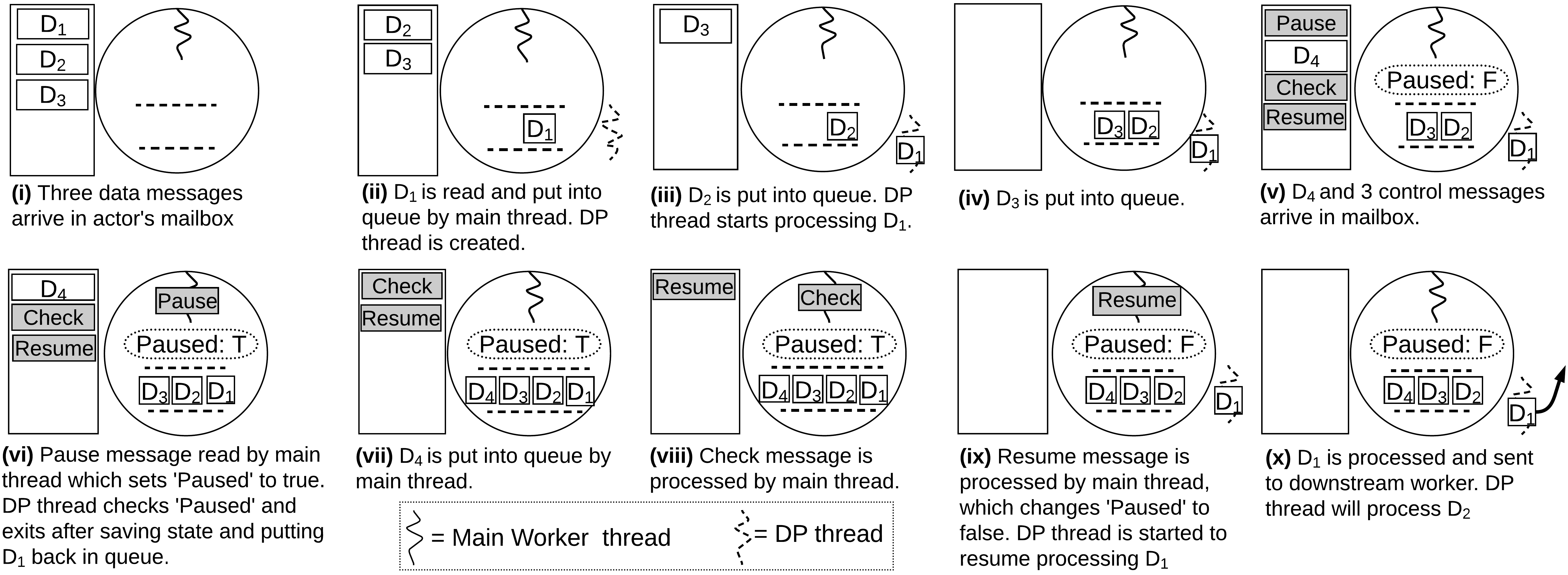} 
	\caption{\label{fig:solution2-dpthread}
		\textbf{Processing of control and data messages by a worker.} `Paused' variable and the queue are shared between the main thread and the data processing (DP) thread. The `Paused' shared variable is not shown in (i)-(iv) for simplicity.
	}
		\end{center}

	\vspace{-0.2in}
\end{figure*}

For actor model implementations that do not support priority such as Orleans,  Amber solves the problem by letting each actor delegate its data processing to an external thread, called {\em data-processing thread} or {\em DP thread} for short. This thread can be viewed as an external resource used by actors to do computation and send messages to other actors. The main thread shares a queue with the DP thread to pass data messages. After receiving a data message ($D_1$ in Figure~\ref{fig:solution2-dpthread}), the main thread enqueues it in the queue. The main thread offloads the data processing to the DP thread (steps (i) and (ii) in the figure). The DP thread dequeues data messages from the queue and processes them. After enqueuing a data message into the queue, the main thread is free to continue processing the next message in the mailbox.  The next data messages are also stored in the queue (messages $D_2$ and $D_3$ in steps (iii) and (iv)). If the next message is a {\sf Pause} message (step (v)), the main thread sets a shared variable {\em Paused} to true (step (vi)) to notify the DP thread. The DP thread, after seeing this new variable value, saves its states inside the worker, notifies its principal, and exits.  The worker then enters a {\em Paused} state. The details of these actions of the DP thread will be described in Section~\ref{ssec:pause-and-resume} shortly.

While in this {\em Paused} state, the main thread can still receive messages in its mailbox and take necessary actions. (More details are in Section~\ref{ssec:responding-in-pausing}.) A received data message is stored in the internal queue ($D_4$ in step (vii)) because no data processing should be done. After receiving a control message, the main thread can act and respond accordingly ($Check$ in step (viii)).  If the control message is a {\sf Resume} request, the main thread changes the {\em Paused} variable to false, and uses a new DP thread to resume the data processing (step (ix)). The DP thread continues processing the data messages in the internal queue, and sends produced data messages to the downstream worker (step (x)).

\subsection{Pausing Data Processing}
\label{ssec:pause-and-resume}


The DP thread associated with a worker actor needs to check the variable {\em Paused} to pause its computation so that the worker can enter the {\em Paused} state.  One way is to check the variable after processing every data message, but this method has a long delay, especially for a large batch size and expensive operators.  


Amber adopts a technique based on the observation that operators use an {\em iteration} model to process their tuples one by one and apply their computation logic on each tuple. Hence, the DP thread can check the variable after each iteration.  If the variable becomes true, the DP thread saves necessary states of data processing in the worker actor's internal state, then simply exits. When a {\sf Resume} message arrives, the main thread employs a new DP thread, which loads the saved states to resume the computation.  Thus, the worker actor can respond to a {\sf Pause} request quickly without introducing much overhead.  The delay of checking the shared variable is mainly decided by the time of each iteration, e.g., the time of processing one tuple.  Next we use several commonly used operators as examples to show how they are implemented in  Amber, and what state information is saved by a worker actor for pausing.

\vspace{0.05in}
\noindent \textbf{1. Tuple-at-a-time operators (e.g., selection, projection, and UDF operators)}:  In a non-blocking operator, its DP thread checks the {\em Paused} variable after processing each tuple.  When pausing, the thread needs to save the index (or offset) of the next to-be-processed tuple in the batch, called \emph{resumption-index}.


\vspace{0.05in}
\noindent \textbf{2. Sort}: Sort does not output results until receiving all its input data. A way to implement sort is using two layers of workers. The first layer sorts its data partition and the second layer contains a single worker that merges the sorted outputs of the first-layer workers. We use this method to show how to save and load states, and the solution works for other distributed sort implementations as well. A first-layer worker does its local sort using an algorithm such as {\sf InsertionSort}, {\sf MergeSort}, or {\sf QuickSort}. For online sorting algorithms such as {\sf InsertionSort}, the DP thread saves only the resumption-index. For offline sorting algorithm such as {\sf MergeSort}, the DP thread saves the state of merge sort to resume the operation later, such as the resumption index for each input chunk.  The worker in the second layer merges the sorted outputs of first-layer workers to produce the final sorted results.  When pausing the execution, the DP thread needs to store the resumption-index for each input batch.

\vspace{0.05in}
\noindent \textbf{3. Hash-based join.} This operator consists of two phases, namely hash-building for one input table (say table $R$) and probing the hash table using the tuples from the other table (say table $S$). When a worker receives a {\sf Pause} message, its DP thread can be in one of the two phases. If it is in the hash-building phase, it saves its states related to building the hash table and the resumption index for the interrupted batch of table $R$. If the DP thread is in the probing phase, it saves the corresponding state and the resumption index from the interrupted batch of table $S$. Note that Amber does not support spilling to disk currently.

\textbf{4. GroupBy.} We can implement this operator using two layers of workers. A first-layer worker does a local GroupBy for its input tuples. After that, it forwards its local aggregations to second-layer workers using a hash function on the GroupBy attribute. A second-layer worker produces final aggregated results for its own groups. When pausing the execution, the DP thread saves the resumption index of the interrupted batch and the current aggregate per group.

\subsection{Responding to Messages after Pausing}
\label{ssec:responding-in-pausing}

After pausing the execution, the user can investigate the states of the job.  For instance, she may want to know the number of tuples processed by each worker, or modify an operator, such as the constant in a selection predicate. Such requests can be implemented by sending control messages to the worker actors. Notice that even though an actor is in the {\em Paused} state, it can still receive and respond to messages, which is very important in debugging to support user interactions after pausing the execution (steps (vii) - (viii) in Figure~\ref{fig:solution2-dpthread}).  It is a unique capability of  Amber due to the adoption of the actor model. When the user wants to resume the computation, the controller actor sends a {\sf Resume} control message using the approach described above for the {\sf Pause} message. Each worker actor, after receiving {\sf Resume} message, uses a DP thread, to loads the saved states and continue the computation (steps (ix) and (x) in the figure).



\section{Conditional Breakpoints}
\label{sec:ConditionalBreakpoints}

The {\sf Amber} system allows users to set breakpoints before and during the execution of a workflow in order to detect bugs and data errors.  In this section, we present the semantics of conditional breakpoints, and discuss how to support two types of predicates in breakpoints.

\subsection{Semantics of Conditional Breakpoints}

We use an example to illustrate conditional breakpoints in {\sf Amber}.  Consider the workflow back in Figure~\ref{fig:sample-workflow} and assume the tweets are obtained from a tab-separated text file. In this scenario, an additional operator {\sf RegexParser} is required between {\sf Scan1} and {\sf KeywordSearch}. This operator reads the file line-by-line and uses tab as a delimiter to convert it to multiple attribute values. In this case, the data or the regex could have errors.  For example, the {\tt followerNum} value (i.e., number of followers of a twitter user) should always be a non-negative integer. Thus the user may want to do sanity checks on the output values of this operator. To do so, she puts a breakpoint on the output of this operator with a condition ``{\tt followerNum} $<$ 0'' When a tweet satisfies this condition, the system pauses the data processing and allows the user to investigate. 

Recall that an operator can have multiple inputs and outputs. A user specifies a breakpoint on a specific output of an operator with a conditional predicate.  When the predicate is satisfied, the data processing in the entire workflow should be paused.  There are two types of predicates.  A {\em local predicate} is a condition that can be satisfied by a single tuple, while a {\em global predicate} is a condition that should be satisfied by a set of tuples processed by multiple workers.  Next we will discuss how to check these two types of predicates.

\subsection{Evaluating Local Predicates}
\label{ssec:local-predicates}

Since a local predicate can be evaluated by a single actor, such a conditional breakpoint can be detected by a worker independently. An example use case of local predicate detection is validating the schema of input data into an ML operator. For instance, the values of the `ratings' column should always be from 1 to 5. Another use case is to pause the execution in case of an exception and show the culprit tuple to the user. Example local predicates for the workflow in Figure~\ref{fig:sample-workflow} are: 1) the {\tt followerNum} of a tweet is negative, 2) the maximum {\tt followerNum} among all the tweets is above 1,000.  Although the second predicate is a predicate over all the tuples, it is still a local predicate since it can be checked by an actor for its tuples independently.  Whenever the predicate is satisfied, the worker pauses its data processing and notifies its principal actor. Then, the principal pauses the workflow as described in Section~\ref{ssec:pausing-execution}.


\subsection{Evaluating Global Predicates}
\label{ssec:global-predicates}

A global conditional breakpoint relies on tuples processed by multiple workers of an operator, and cannot be detected by a single worker. The evaluation of such a predicate is done by the principal actor. Such predicates are valuable in scenarios where a performance metric of an operator has to be continuously monitored, e.g., the  number of emails marked as spam by a Spam-Detection operator within a time window. If this metric goes above a threshold, it indicates some problem that requires attention. A few possible causes can be cyber attack, input data corruption, or model degradation. It will be helpful if the system can detect such predicates. We use two example global predicates for the workflow in Figure~\ref{fig:sample-workflow} to show how to evaluate them. $G_1$: the total number of tweets output by {\sf KeywordSearch} is 15. $G_2$: the sum of {\tt followerNum} of all tweets produced by {\sf KeywordSearch} exceeds 90.

\begin{figure*}[htbp]
	\includegraphics[width=\linewidth]{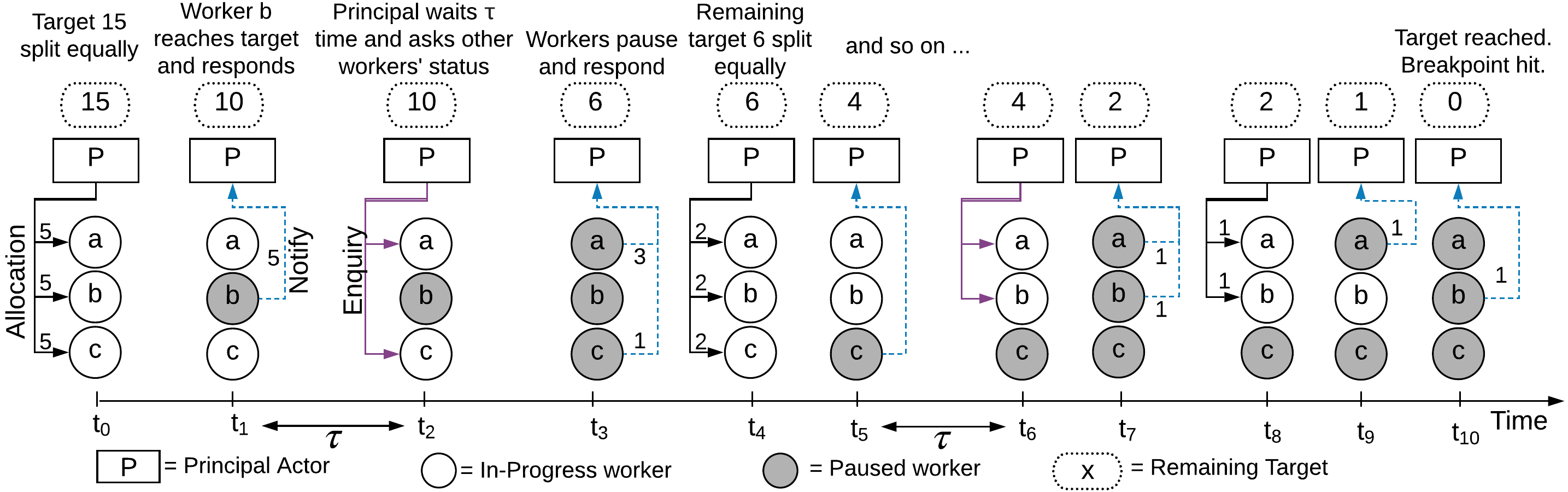} 
	\caption{\label{fig:breakpoint-impl}
		\textbf{Evaluating a global conditional breakpoint ``KeywordSearch operator producing 15 tuples.''} Solid lines are messages from the principal actor to workers, and dashed lines are responses from workers.
	}
\end{figure*}

\vspace{0.05in}
\noindent \textbf{Evaluating a global {\tt COUNT} predicate $G_1$.}  Suppose {\sf KeywordSearch} has three workers.  As illustrated in Figure~\ref{fig:breakpoint-impl}, the process of evaluating $G_1$ consists of several steps. At time $t_0$, the principal actor divides the target number 15 equally among the three workers, namely $a$, $b$, and $c$. Each worker, after producing a tuple, increments its counter by 1. Suppose worker $b$ is the first to produce 5 tuples. It pauses itself and notifies the principal actor (time $t_1$). The principal waits for a threshold time ($\tau$) with a timer. If the other two workers respond within the time limit $\tau$ (not shown in the figure), the conditional breakpoint is hit and the principal sends a message to the controller to pause the entire workflow. Otherwise, the principal inquires each worker that has not responded, about how many records it has produced (time $t_2$).  The figure shows the case where both workers $a$ and $c$ did not respond within $\tau$. These two workers pause themselves, and respond with their number, say, 3 and 1 (time $t_3$). Now the remaining target number becomes $15-5-(3+1) = 6$.

At time $t_4$, as before, the principal divides the new target number 6 equally, and sends a target number 2 to each worker to resume its data processing.  This reassignment is necessary so that all the three workers can be resumed and the operator processes data at the maximum parallelism. Assuming worker $c$ produces 2 tuples, it again pauses itself and notifies the principal (time $t_5$). The principal again waits for a threshold time after which it asks workers $a$ and $b$ (time $t_6$), who pause themselves and respond with their produced number of tuples, say 1 and 1 (time $t_7$).  For the new remaining target of 2, the principal gives a target of 1 to each of the first two workers $a$ and $b$ (time $t_8$). Suppose worker $a$ contacts the principal after producing one tuple (time $t_9$). After a while, worker $b$ also produces a tuple, reports the same after pausing itself (time $t_{10}$) and the conditional breakpoint is triggered.  

At the end, {\sf KeywordSearch}  has received 15 tuples and the conditional breakpoint is hit.  Notice that at time $t_9$, when worker $a$ contacts the principal after producing one tuple, the principal does not enquire the other workers for their tally. The reason is that there is only one tuple left to be computed, and reassigning this target to another worker will not increase parallelism. 


Before the conditional breakpoint is hit, the computation can be in one of the two states. A {\em normal processing state} starts when the workers have been assigned their targets by the principal and ends when one of the workers completes its target. A {\em synchronization state} starts when a worker completes its target and ends when the principal allocates new targets to the workers.  The amount of time spent in the synchronization state depends on the timeout threshold $\tau$ and the variance of the processing speeds of the workers.

\vspace{0.05in}
\noindent \textbf{Evaluating a global {\tt SUM} predicate $G_2$.} The evaluation of the second predicate $G_2$ follows a similar process. The principal starts by dividing the target into three parts of 30 each. A unique aspect of {\tt SUM} is that a tuple can bring the total value closer to the target by an arbitrary amount, unlike the {\tt COUNT} example where a tuple only causes a change of 1. For example, if the current sum of {\tt followerNum} at a worker is 24, and the next tuple has a {\tt followerNum} value of 15, then the target of 30 will be ``overshot'' by 9. Therefore, it is difficult to pause the execution at the exact target.

Our goal is to minimize the amount over the target.  To do so, the principal actor initially follows the same procedure as above till it gets close to the target and the overall needed value is below a threshold. The threshold can be decided based on the distribution of {\tt followerNum} values (obtainable online). Then the principal can give the target to only one worker in order to minimize the overshot amount. For example, say the sum of {\tt followerNum} values received by {\sf KeywordSearch} till now is 80 and it needs a total of 10 more to reach the target. If it gives the three workers a target of 3, 3, and 4, and the next tuples received by the three workers have a {\tt followerNum} value of 11, 14 and 13, respectively, the total {\tt followerNum} sum will be 118, which is 28 more than the target of 90. Instead, if the principal gives the target of 10 to only one worker and keeps the other two paused, then even if that worker receives a tuple with a {\tt followerNum} value such as 14, the excess is of just 4. Thus, the system pauses closer to the target.

The aforementioned methods are meant to allow the developer to pause the execution of the workflow when a conditional breakpoint is hit.  As in general debuggers, it is the developer's responsibility to decide what breakpoints to set and where in order to investigate the runtime behavior of the program and find bugs. As Amber is a distributed system, the execution of multiple actors to reach a global breakpoint could be non-deterministic.

\section{Fault Tolerance}
\label{sec:fault-tolerance}

Fault tolerance is critical due to failures in large clusters.  In traditional distributed data-processing systems such as Spark, recovery only ensures the correctness of final computed results.  As we will see below, the presence of control messages in Amber poses new challenges for fault tolerance, because Amber additionally needs to ensure the recovery of control messages and their resulting states.  In this section we first show why the Spark approach cannot be used, then present a solution to support fault tolerance in Amber.


\subsection{Why not the Spark Approach?}

Spark runs in a stage-by-stage fashion and allows checkpointing of the output of a stage. When failure happens, Spark reruns the computation of the lost data partitions from the last checkpoint using lineage information. This fault tolerance approach cannot be adopted in Amber for two reasons. Firstly, the computation of each data partition in Spark is fully independent, which allows Spark to recover only the failed partitions. In contrast, Amber has execution dependencies among workers of an operator. For example, the principal can split the global predicate in a breakpoint into multiple target numbers, which can be adjusted dynamically for each worker (see Section~\ref{ssec:global-predicates}). If Amber naively re-runs the computation of failed data partitions, the assigned intermediate target values are lost, which will lead to an incorrect detection of the global predicate. Secondly, a control message can alter the state of a worker, and in case of failure, Amber needs to recover the worker to the same consistent state.  For example, suppose before a failure happens, the worker is paused when processing the 10$^{th}$ record in the 1$^{st}$ data message, and the user has already seen the corresponding state of the operator of this worker, such as the number of  records processed so far.  After recovery, in order for the user to see the same operator state, we need to recover this worker to its state before the failure.  If we were to use the Spark fault tolerance approach, this worker will not pause at all, since this approach only reruns the computation without considering the control messages.  One way to support fault tolerance in Amber is using the Chandy-Lamport algorithm~\cite{journals/tocs/ChandyL85}, which records all the in-transit  data in a snapshot. This approach is not efficient since it can generate a large amount of checkpointed data.

\subsection{Supporting Fault Tolerance in Amber}



Next we develop a technique to support fault tolerance in Amber based on the following realistic assumptions. (A1) We treat the controller and the principal actors as a single unit (called ``coordinator''), which is placed on the same machine. (A2) Workers only exchange data messages, not control messages. (A3) For each worker, both its computation logic and response to a control message are deterministic, as assumed by many other data-processing systems.  
Our fault tolerance technique consists of two parts: 1) checkpointing, and 2) logging control messages and their arrival order relative to data messages. The approach used for checkpointing depends on the execution model used. If a stage-by-stage execution model (or batch execution model), like the one in Spark, is used, the data produced at the end of each stage needs to be checkpointed. If a pipelined execution model, like the one in Flink, is used, then the states of the workers need to be periodically checkpointed~\cite{journals/corr/CarboneFEHT15}. Recovery works by restarting the computation from the last checkpoint and replaying the control messages by injecting them in the original order relative to data messages. 


\begin{figure*}[htbp]
	\includegraphics[width=\linewidth]{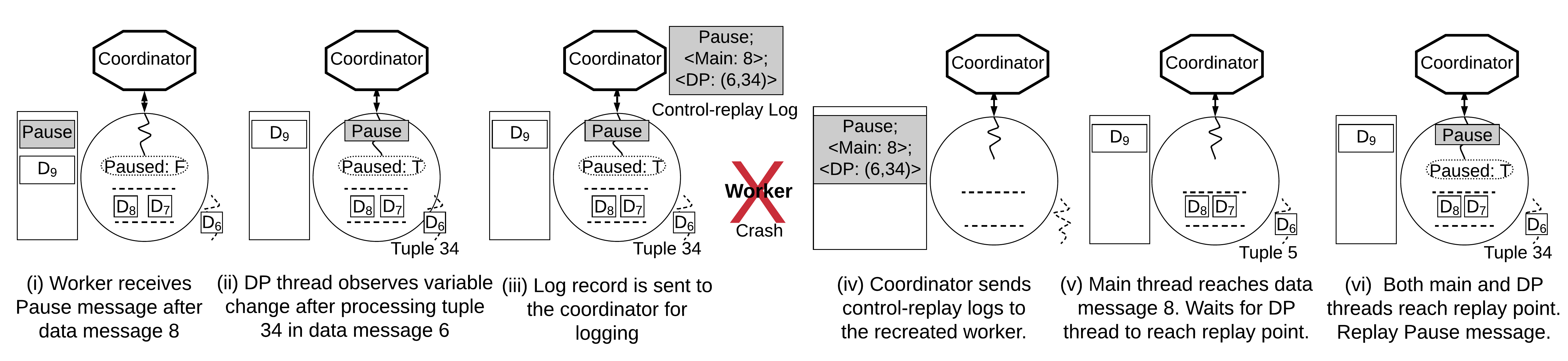} 
	\caption{\label{fig:fault-tolerance}
		\textbf{Fault Tolerance in Amber: logging control messages and recovery.}
	}
	\vspace{-0.1in}
\end{figure*}


We use an example to explain this technique.  Figure~\ref{fig:fault-tolerance} illustrates the logging process of control messages before a failure of a worker (steps (i)-(iii)). A {\sf Pause} message arrives at the worker after a data message with a sequence number 8 (step (i)). In step (ii), the main thread sees the {\sf Pause} message and saves the sequence number 8. The main thread  alters its internal state by setting the shared variable {\em Paused} to true. The DP thread observes the variable change after processing the 34$^{th}$ tuple in the 6$^{th}$ data message and notifies the main thread. In step (iii), the main thread sends the following log record to the coordinator:

\begin{center}
    \vspace{-0.06in}
    {\tt (Pause; <Main: 8>; <DP: (6, 34)>)}
    \vspace{-0.06in}
\end{center}

The record includes the content of the control message ({\sf Pause}), the sequence number (`8') of the last data message of the main thread when it received this control message, and the iteration status of the DP thread when it saw the shared-variable changed caused by this control message.  The iteration status includes the sequence number  (`6') of the currently processed data message and the index  (`34') of the last processed tuple in the message. After receiving this record, the coordinator stores it in a data structure called {\em control-replay log} for this worker. 

Suppose a machine failure happens, which causes the data partition of this worker to be lost. During recovery, the coordinator recreates all the workers of the failed partition and sends their control-replay log records respectively. During this period, the coordinator holds new control messages for each recreated worker until the worker has replayed all its control-replay log records.  Each recreated worker reruns the computation from the last checkpoint.  Since the computation is deterministic (assumption A3), a re-run of these recreated workers leads to the same content and sequence numbers of data messages received by each worker.  Now let us consider the recreated worker corresponding to the aforementioned worker. Steps (iv)-(vi) in the figure show the recovery process of this worker. In step (iv), it receives its control-replay log from the coordinator.  Intuitively, the main thread and DP thread of this worker continue processing their received data messages until both of them reach the control-replay point as specified in the log.  Specifically, after receiving a data message $D$, the main thread checks the sequence number of $D$, denoted $S(D)$.  If $S(D) < 8$, the main thread processes this message normally as before.  If $S(D) = 8$, it processes $D$, then waits to synchronize with the DP thread (step (v)).  Similarly, when the DP thread processes the tuples in a message, it handles those tuples ``before'' $(6, 34)$ (i.e., tuple 34 in message 6).  After processing this tuple, it will synchronize with the main thread.  After the synchronization, the control message is replayed by the main thread as if this message were just received (step (vi)).


There can be a case where a worker failed before being able to respond to a control message from the coordinator. For example, suppose the worker failed after step (ii) in Figure~\ref{fig:fault-tolerance}, before it sends the log record to the coordinator. The coordinator marks the processing of this {\sf Pause} message as {\em incomplete}. During recovery, the coordinator first allows the recreated worker to fully replay its existing control-replay log records. The coordinator then retries sending this {\sf Pause} message to the worker. Consequently, the worker can pause at a tuple different from the one before the failure. Fault tolerance is still valid because the processing of the {\sf Pause} message was incomplete, and the user never saw its effect. 

Amber's fault tolerance approach incurs little overhead on execution because it only saves control messages and the control-replay log, which have a much smaller size compared to data messages. To deal with the case of coordinator failures, we can use write-ahead logging or use backup coordinators to replicate the states of the coordinator. Notice that for an operator with multiple data inputs such as {\sf Join} or {\sf Union}, to satisfy assumption A3, they need to provide an ordering guarantee across their inputs, and the sequence numbers of each input will be maintained and recorded separately. If failure happens during recovery, the coordinator can simply restart the recovery procedure, which is idempotent.

\section{Experiments}
\label{sec:exps}

In this section, we present an experimental evaluation of the Amber system using real datasets on clusters.

\subsection{System Implementation and Setting}

\begin{figure}[htbp]
\begin{center}
	\includegraphics[width=3.6in]{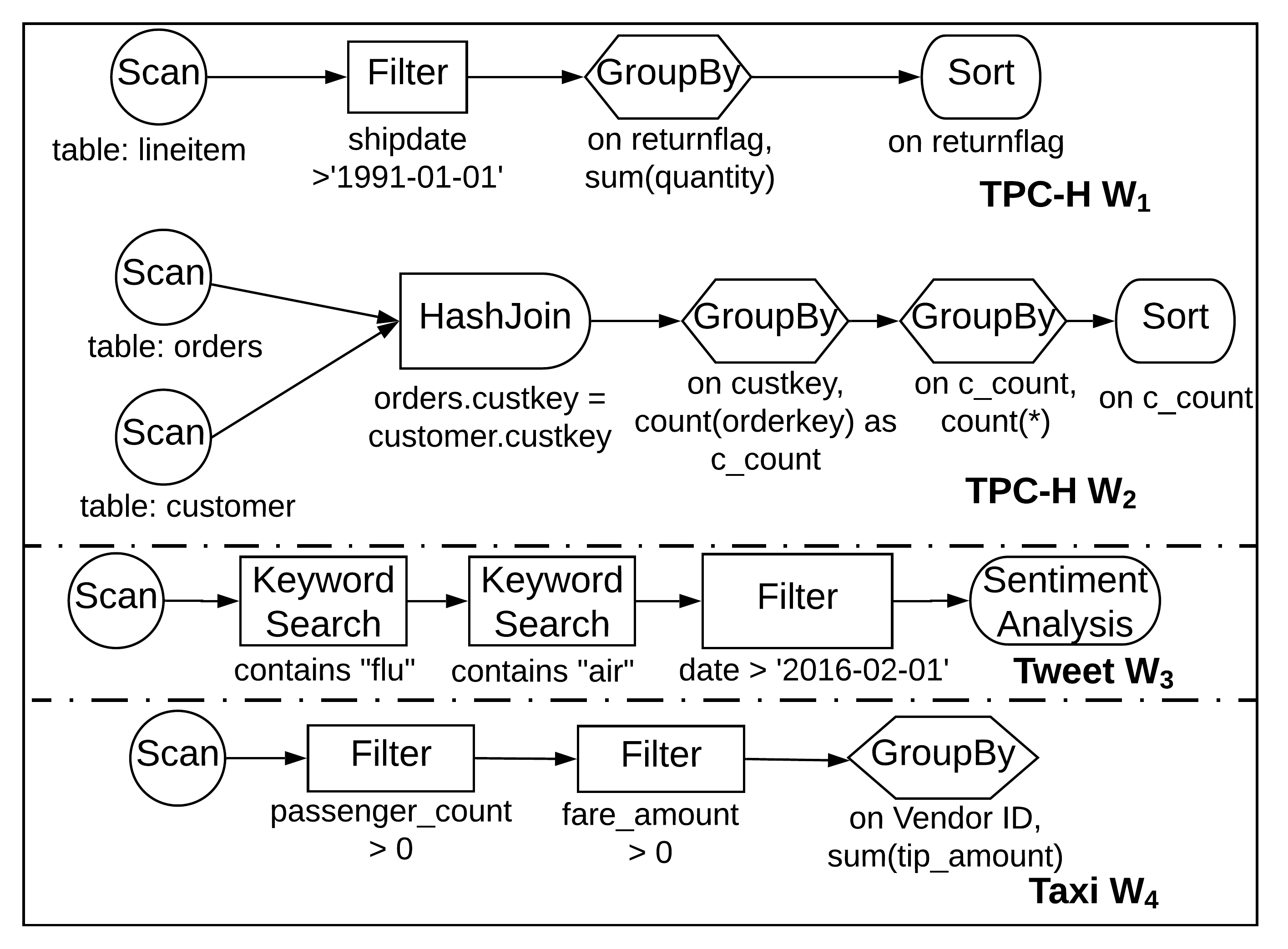} 
	\caption{\label{fig:tpch}
	\textbf{Workflows used in the experiments.}
	}
\end{center}
\end{figure}

\vspace{0.05in}
\noindent \textbf{Data and Workflows.} We used three real datasets, namely TPC-H, tweets, and New York taxi events. For the TPC-H benchmark~\cite{TPC-H}, we varied the scale factor to produce data of different sizes. Based on the TPC-H queries 1 and 13 we constructed two workflows, shown as $W_1$ and $W_2$ in Figure~\ref{fig:tpch}. Note that the {\sf Scan} operators of $W_1$ and $W_2$, had a built-in projection to read only the columns being used by the operators later. This improvement was used in the experiments for both Amber and Spark. The second dataset included 100M tweets in the US, on which we did sentiment analysis using an ML-based, computationally expensive operator. The third dataset included New York City Yellow taxi trips (about 210 GB), and each record had information about a trip, including its pick-up and drop-off geo-locations, times, trip distance, payment method, and fare~\cite{taxi-data}.  




\vspace{0.05in}
\noindent \textbf{Experiment Setting.} All the experiments were conducted on Google Cloud Platform (GCP).  The data was stored in an HDFS file system on a storage cluster of 51 {\tt n1-highmem-4} machines, each with 4 vCPU's, 26 GB memory, and 500GB standard persistent disk space (HDD's). The execution of a workflow was done on a {\em separate} processing cluster of 101 machines with the same type. The storage cluster and processing cluster were running Debian GNU/Linux 9.9 (stretch) and Debian GNU/Linux 9.11 (stretch) operating system respectively. The batch size used in data messages was 400 unless otherwise stated. Checkpointing was disabled by default in all experiments, except the experiment concerning fault tolerance (Section \ref{ssec:comp-with-spark-exp}). Out of the 101 machines, we used one just for the controller and principal actors of the operators, and the remaining 100 for data processing. When reporting the number of computing machines, we only included the number of data-processing machines.

For the experiments in this chapter, we implemented Amber in C\# on top of Orleans (version 2.4.2), running on the~.Net core runtime (version 3.0). The operators were implemented as discussed in Section~\ref{ssec:pause-and-resume}, and the workers of each operator were assigned uniformly across multiple machines. For example, if a {\sf Scan} operator had 10 workers and the processing cluster had 10 machines, then each machine had a single {\sf Scan} worker.

\subsection{Scaleup Evaluation}
\label{ssec:scaleup-exp}

We evaluated the scaleup of Amber using the TCP-H data. We started with a dataset of 10GB processed by 1 machine (4 cores), and gradually increased both the data size in the storage cluster and the machine number in the processing cluster linearly to 1TB processed by 100 machines (400 cores). For both workflows $W_1$ and $W_2$, the early operators did most of the work, leaving very few (less than 50) tuples for the final {\sf Sort} operator. Therefore, we allocated 2 workers for each operator on each machine,  except {\sf Sort} that was allocated 1 worker on each machine. The {\sf GroupBy} operator had two layers (Section~\ref{ssec:pause-and-resume}). The first layer was allocated 2 workers on each machine and the second layer was allocated 1 worker on each machine.





\begin{figure}[htbp]
\begin{center}
	\includegraphics[width=3.3in]{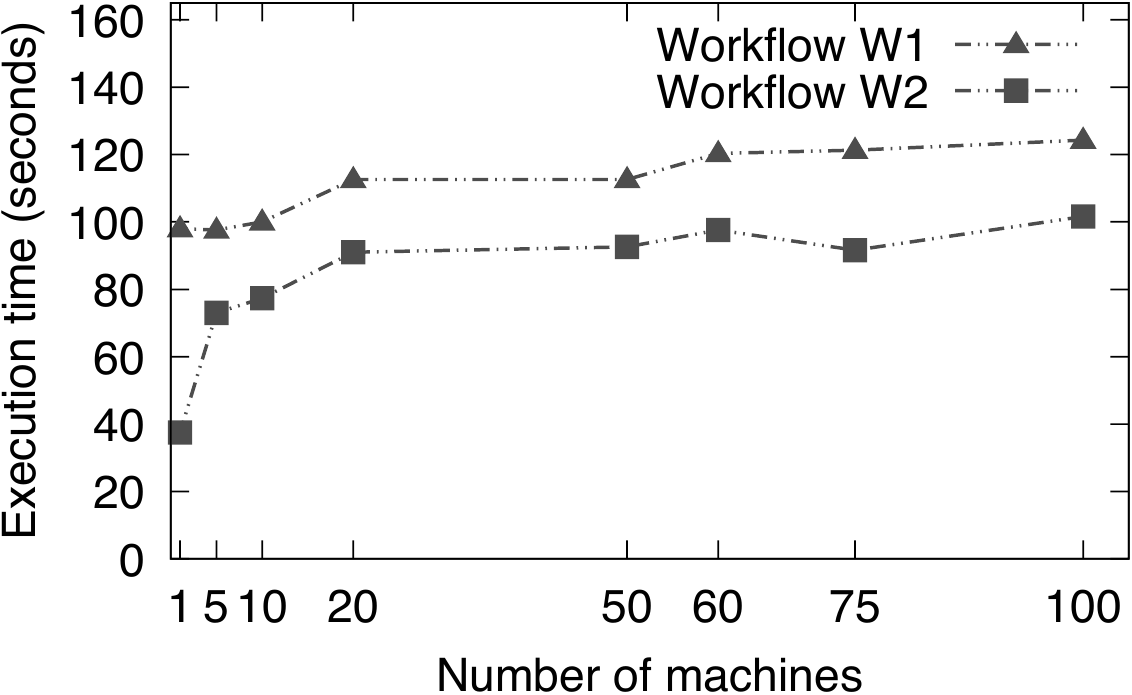} 
	\caption{\label{fig:scaleup-time-tpch}
		\textbf{Scaleup of TPC-H workflows $W_1$ and $W_2$}
	}
\end{center}
\end{figure}



Figure~\ref{fig:scaleup-time-tpch} shows the running time for workflow $W_1$. For the 10GB data processed by 1 machine, the total time was around 98s. When we increased the data size and the cluster size gradually, the time increased slightly. For the 1TB data processed by 100 machines, the total time was around 124.3s.  Figure~\ref{fig:scaleup-time-tpch} also shows the results for $W_2$. For the 10GB data processed by 1 machine, the total time was around 37.5s. When we increased the data size and the cluster size gradually, the time increased at a faster rate than $W_1$ because of the intrinsic quadratic complexity of {\sf Join}. It took 101.6s for 100 machines to process 1TB data.

\subsection{Speedup Evaluation}
\label{ssec:Speedup-exp}

To evaluate the speedup of Amber, we measured the time taken to execute workflows $W_1$ and $W_2$ on the 50GB data using 1 computing machine initially and gradually increased the number of computing machines to 100.



\begin{figure}[htbp]
\begin{center}
	\includegraphics[width=3.3in]{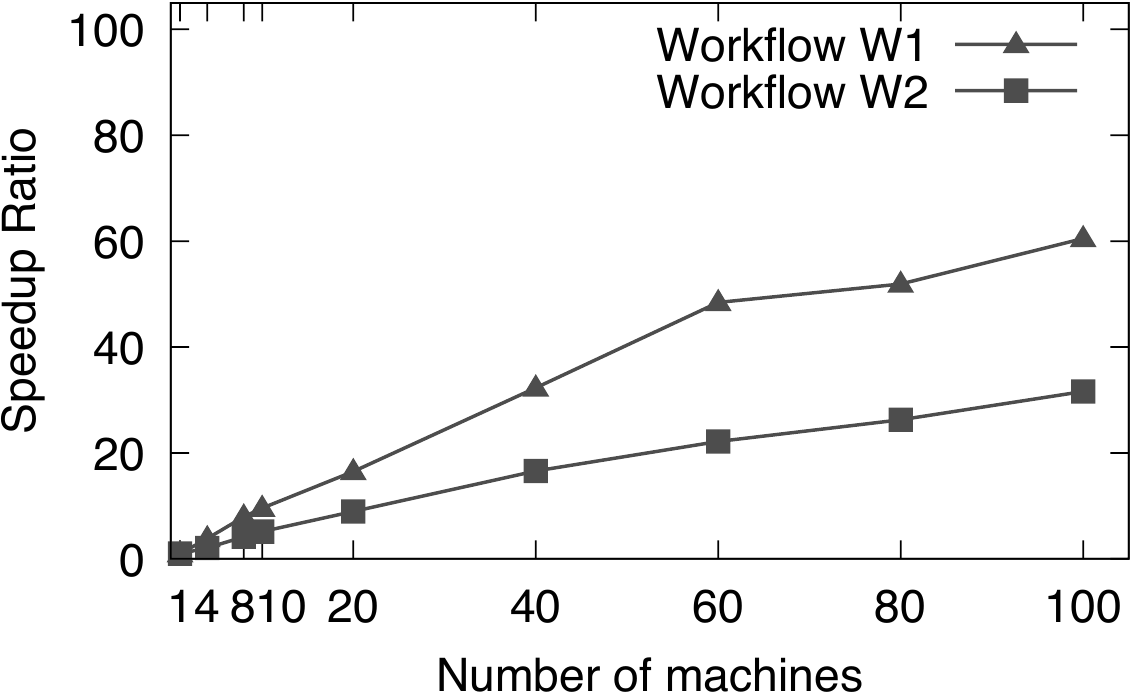} 
	\caption{\label{fig:speedup-time-tpch}
		\textbf{Speedup for TPC-H workflows $W_1$ and $W_2$}
	}
\end{center}
\end{figure}

Figure~\ref{fig:speedup-time-tpch} shows the speedup for workflow $W_1$. For the 50GB data processed by 1 machine, the total time was around 484.5s. When we increased the number of machines gradually, the time decreased. The time was about 10s when using 60 machines, with a speedup ratio of 48.4. When we increased the number of machines further to 80 and then 100, the total time taken did not decrease at the same rate. For 100 machines, the total time taken was 8s (with a speedup ratio of 60.5). This result was due to the fact that the total data to be processed was only 50GB and the machines were not fully utilized.
Figure~\ref{fig:speedup-time-tpch} also shows the results for $W_2$. 
Its speedup was sub-linear due to the intrinsic quadratic complexity of {\sf Join}. Increasing the number machines from 60 to 100 yielded little performance gain due to the increased communication cost.

\subsection{Time to Pause Execution}
\label{ssec:exp-pause}

We used {\sf Pause} and {\sf Resume} as examples to evaluate the time taken to process a control message while a workflow is running on a cluster. We did the experiment with the similar setting as the scaleup experiments.   Each execution was interrupted 8 times by sending a {\sf Pause} then a {\sf Resume} message, before its completion. 


\begin{figure}[htbp]
\begin{center}
	\includegraphics[width=3.3in]{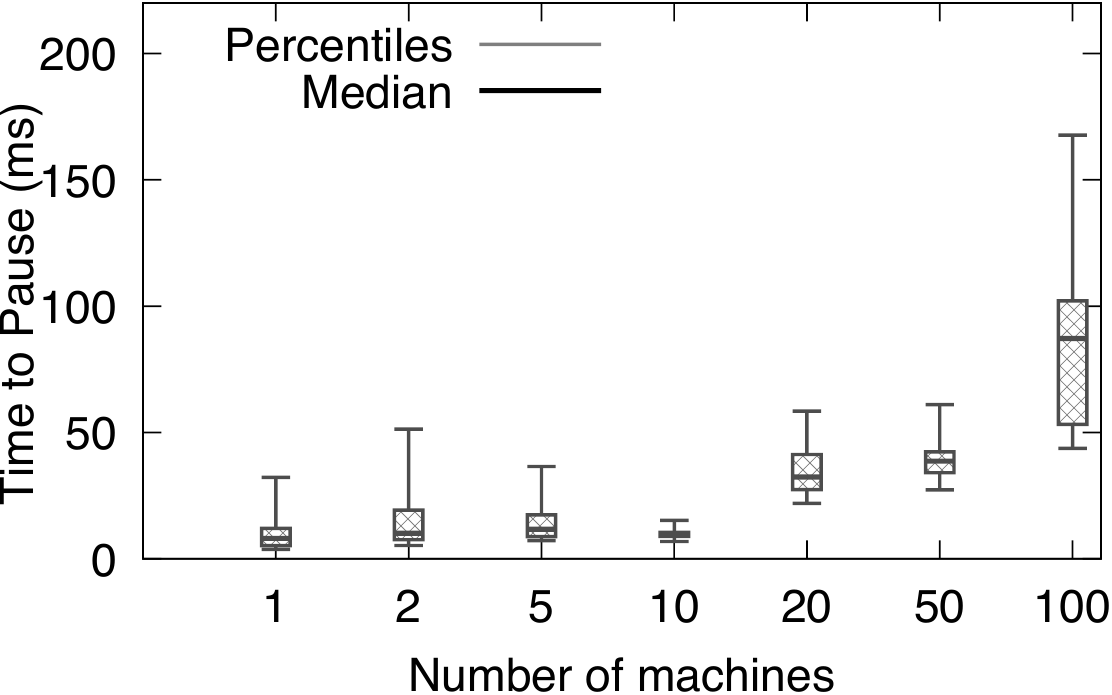} 
	\caption{\label{fig:pause-tpch1-scaleup}
		\textbf{Time taken to pause the execution while scaling up TPC-H workflow $W_1$.} 1$^{st}$ percentile,  1$^{st}$ quartile, median, 3$^{rd}$ quartile, and 99$^{th}$ percentile are shown.
	}
\end{center}
\end{figure}

\begin{figure}[htbp]
\begin{center}
	\includegraphics[width=3.3in]{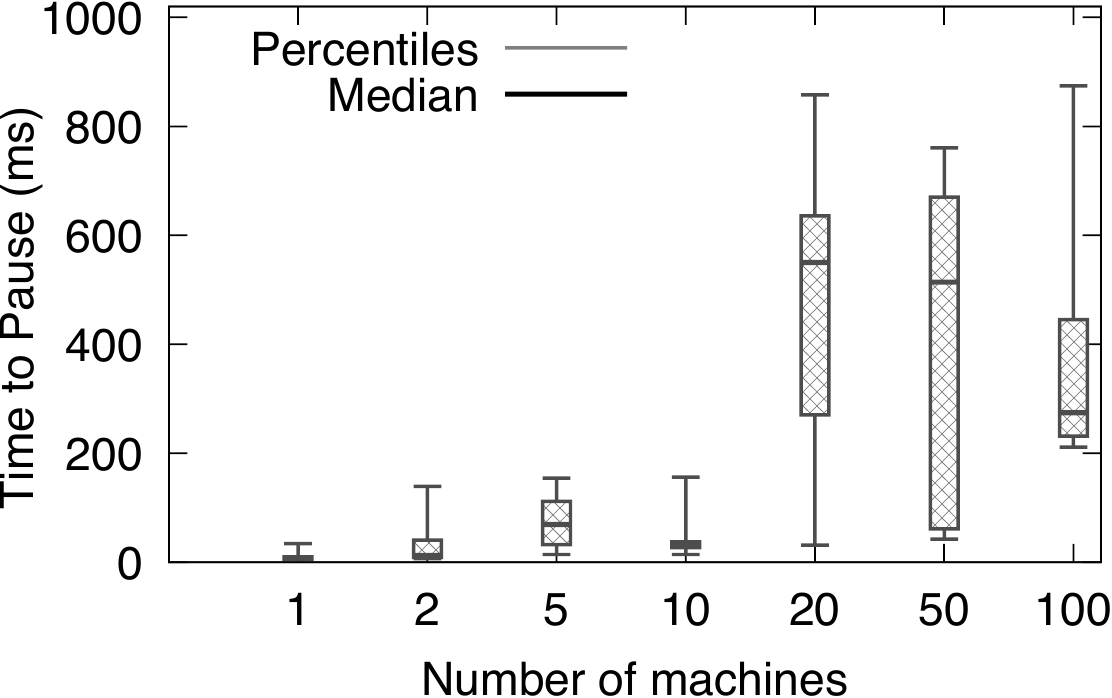} 
	\caption{\label{fig:pause-tpch13-scaleup}
		\textbf{Time taken to pause the execution while scaling up TPC-H workflows $W_2$.} 1$^{st}$ percentile,  1$^{st}$ quartile, median, 3$^{rd}$ quartile, and 99$^{th}$ percentile are shown.
	}
\end{center}
\end{figure}

Figure~\ref{fig:pause-tpch1-scaleup} and Figure~\ref{fig:pause-tpch13-scaleup} show the candlestick chart with 1$^{st}$ percentile,  1$^{st}$ quartile, median, 3$^{rd}$ quartile, and 99$^{th}$ percentile pause times for $W_1$ and $W_2$ respectively. All the times were less than 1 second. The time to pause $W_2$ was relatively more than that of $W_1$ due to the high number of data messages received by the {\sf Join} operator, resulting in more time to reach the {\sf Pause} message. The time to resume each workflow was also in milliseconds. The time to pause increased with the number of machines due to the inherent increase in the communication cost and higher number of data messages being received by the operators. These results show that control messages in Amber can be handled quickly during the processing of a large amount of data. Note that time to pause depends on the delay of checking the shared variable {\sf Pause} by the DP thread, which is approximately equal to the time required by the operator to process a single tuple. The delay in the relational operators was in milliseconds. For complex operators that need more time to process one tuple such as ML operators, the time to pause could be higher.



\subsection{Effect of Worker Number}
\label{ssec:workers-effect-exp}


A feature of Amber is that different operators can have different numbers of workers. We used workflow $W_3$ on the tweet dataset to evaluate the effect of the number of workers allocated to computationally expensive operators. It included a {\sf SentimentAnalysis} operator, which was based on the CognitiveRocket package~\cite{sentimentAnalysis} and needed about 4 seconds to process each tuple. We used it as an example of expensive ML operators.  The workflow took 100M tweets as the input and first applied {\sf KeywordSearch} and {\sf Filter} operators. The number of tweets going into the {\sf SentimentAnalysis} operator was 1,578. We varied the total number of workers allotted to the {\sf SentimentAnalysis} operator and measured its effect on the total running time. The total number of workers for all other operators was fixed at 10. The workflow was run on a cluster of 10 machines with a batch size of 25 for data from {\sf Filter} to {\sf SentimentAnalysis} operator. We used a smaller batch size because we only had 1,578 tuples to be distributed among sentiment analysis workers.

\begin{figure}[htbp]
\begin{center}
	\includegraphics[width=3.3in]{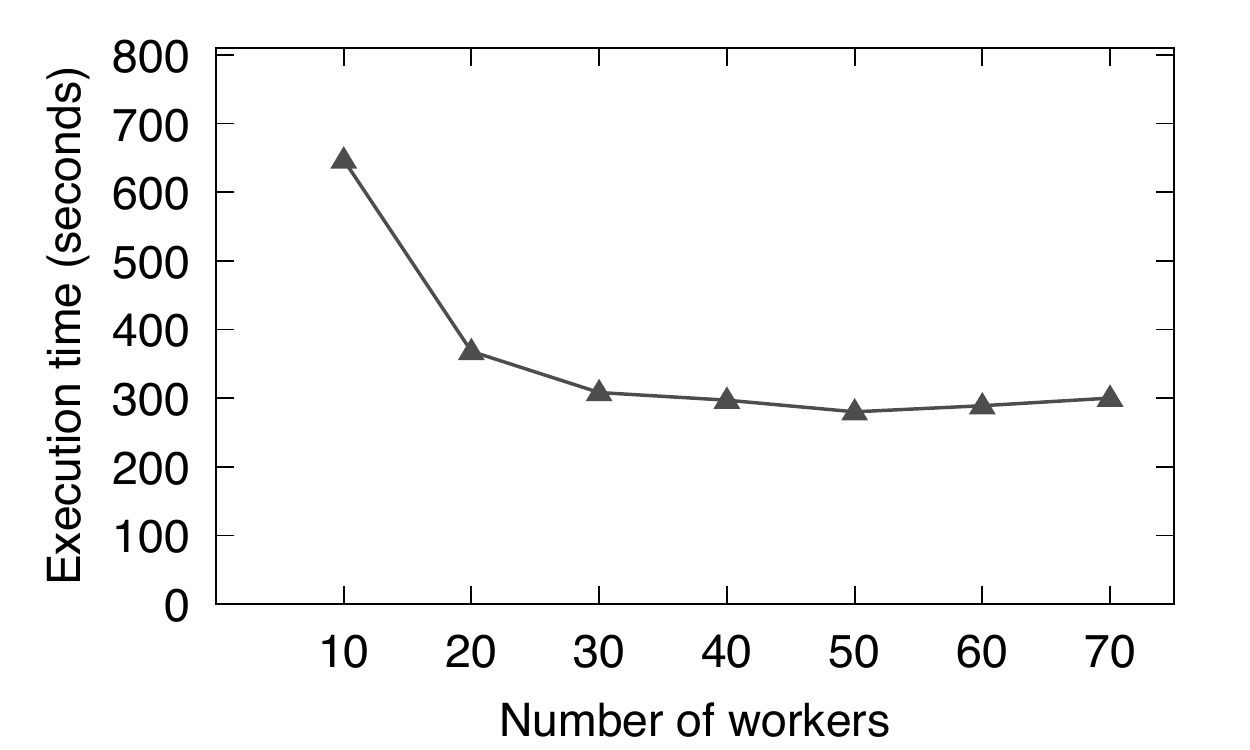} 
	\caption{\label{fig:workers-effect}
	\textbf{Changing the worker number of the SentimentAnalysis operator in workflow $W_3$ for tweets.}
	}
\end{center}
\end{figure}

Figure~\ref{fig:workers-effect} shows the results. The running time when the operator had 10 workers was 647s. When we doubled the number of workers, the execution time reduced to 368s.
The rate of decrease declined as the number of workers was further increased. When the number of workers was increased above 50, the execution time even started increasing, because data got distributed among many workers which competed for CPU (total number of CPU cores used in the experiment was $40$), thus increasing the overhead of context switching. Thus, workers took more time to finish their task. 

{\bf Dynamic resource allocation.} ML operators such as SentimentAnalysis can become a bottleneck in workflows due to their slow speed of computation and extra resources needed under peak load conditions. We implemented the technique of dynamic resource allocation as suggested in~\cite{journals/pvldb/MaiZPXSVCKMKDR18} to allocate extra machines to the SentimentAnalysis operator {\em during} the execution, and  evaluated the performance gain. First, we ran $W_3$ on a cluster of 6 machines, with each operator having 1 worker per machine, except SentimentAnalysis that had 5 workers per machine, and the total time to run $W_3$ was 422s. Then, we modified the setting by adding one more machine to the cluster every minute and allocating 5 SentimentAnalysis workers on the newly added machine. The total time to run $W_3$ reduced to 407s. This reduction of 15s was feasible because of Amber's capability to add more computing resources dynamically.





\subsection{Conditional Breakpoint Evaluation}
\label{ssec:cond-break}

\begin{figure}[htbp]
\begin{center}
	\includegraphics[width=3.3in]{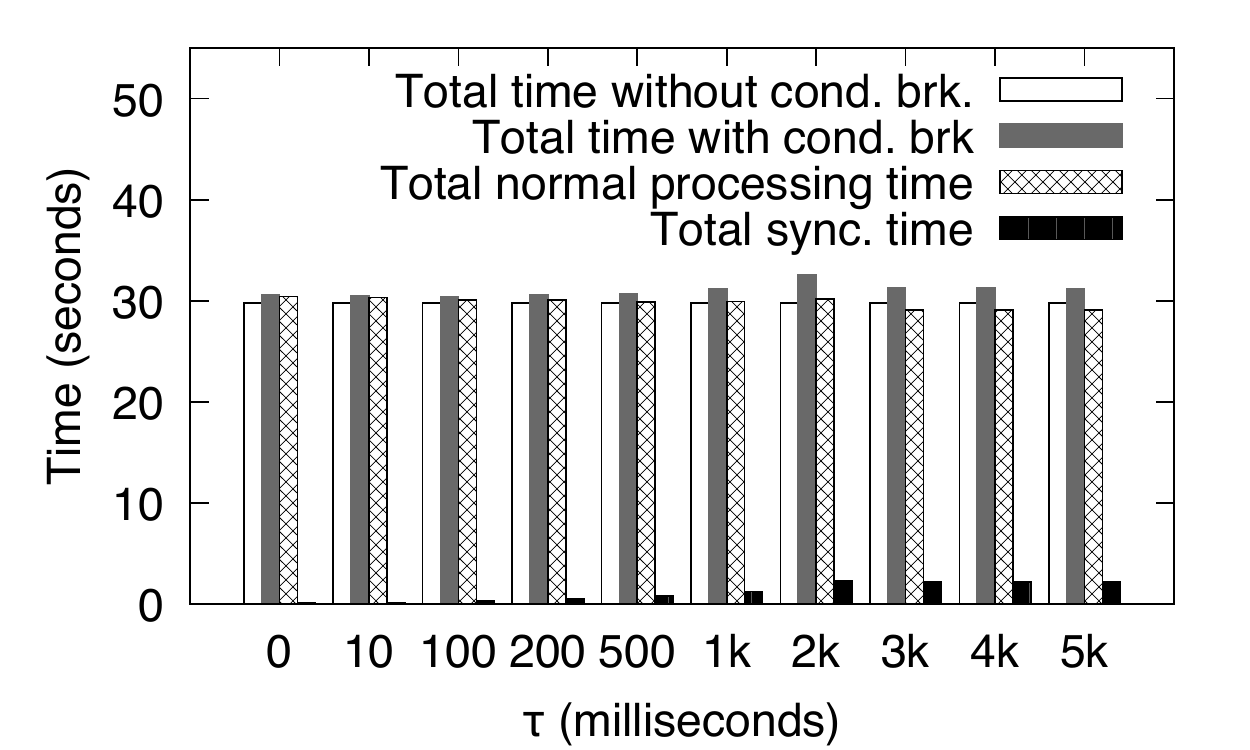}
	\caption{\label{fig:cond-brk}
			\textbf{Conditional breakpoint: running time versus principal's waiting threshold $\tau$.}
	}
\end{center}
\end{figure}

For the TPC-H workflow $W_1$ running on 10 machines, we used 119M tuples and set a conditional breakpoint on the output of the {\sf Filter} operator to pause the workflow after this operator produced 100M tuples.  We varied the timeout threshold ($\tau$) used by the principal from 0ms to 5s, and measured the time in the normal processing state and the time in the synchronization state, as discussed in Section~\ref{ssec:global-predicates}. Figure~\ref{fig:cond-brk} shows the results.  The normal processing time was about 30s.  The synchronization time was relatively small. When $\tau$ was high, the total synchronization time was around 2.15s. When we decreased $\tau$, the synchronization time decreased. The overall time decreased with decreasing $\tau$ since we had more data parallelism. The best setting was when $\tau$ was a few milliseconds.


To evaluate the overhead of conditional breakpoints, we measured the total time needed by {\sf Filter} operator to produce 100M tuples when the input had 119M tuples and no conditional breakpoint was set. Figure~\ref{fig:cond-brk} also shows that the time taken by the {\sf Filter} operator to produce the same 100 million tuples was about 29.8s, which was close to the overall time taken with the conditional breakpoint.

\subsection{Performance Comparison with Spark}
\label{ssec:comp-with-spark-exp}

We compared the performance of Amber with Apache Spark using TPC-H $W_1$ and $W_2$. Data checkpointing was disabled for Spark. The scaleup experiment settings were re-used for Spark. Similar to Amber, we put the Spark's driver on one dedicated machine and allowed its executors to run on the other 100 machines. We used two Spark API's, namely the DataFrame API on top of the Spark SQL engine, and the RDD API, which is its primary user-facing API~\cite{spark-rdd-vs-dataframe}. The RDD API is a more general API that supports user-defined data structures and transformations. The DataFrame API is a faster SQL-based API because of many optimizations, such as binary data formats, fast serialization, and code generation~\cite{spark-dataframe-optimization}. 


\begin{figure}[htbp]
\begin{center}
	\includegraphics[width=3.3in]{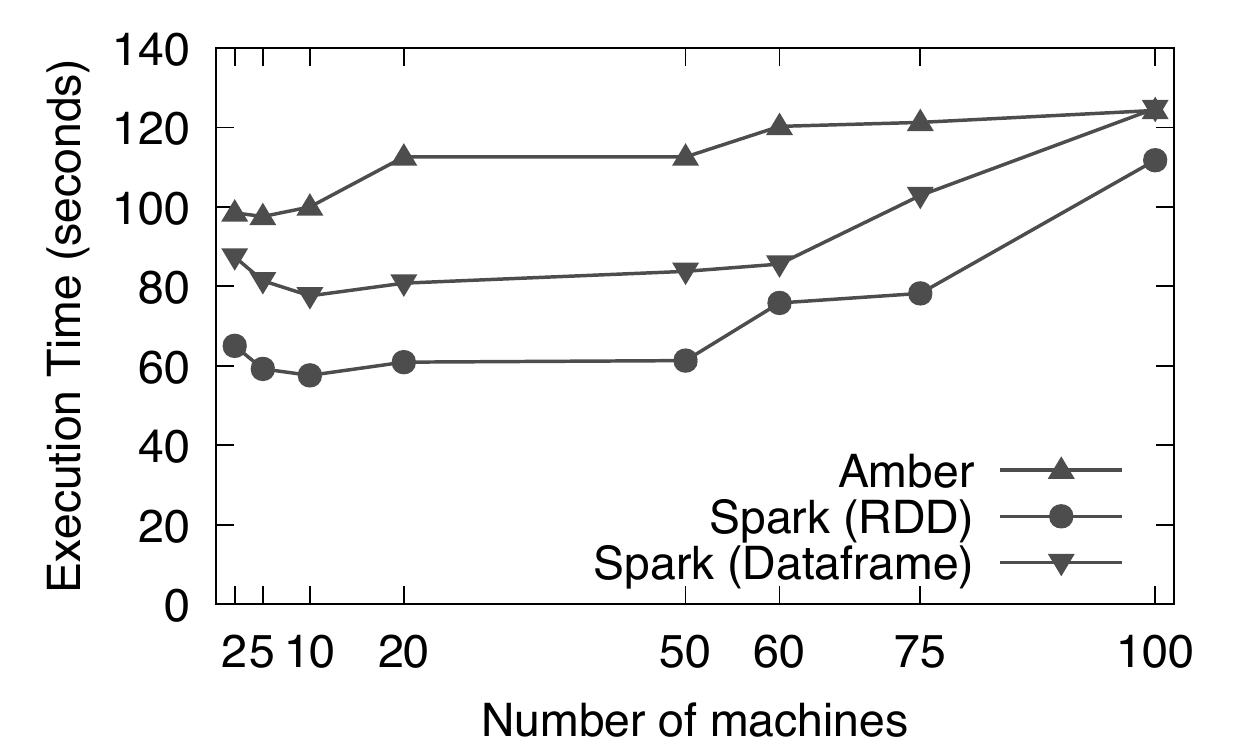}
	\caption{\label{fig:scaleup-time-tpch1-sparkVsAmber}
		\textbf{Scaleup for Amber and Spark for $W_1$.}
	}
\end{center}
\end{figure}

\begin{figure}[htbp]
\begin{center}
	\vspace{-0.1in}
	\includegraphics[width=3.3in]{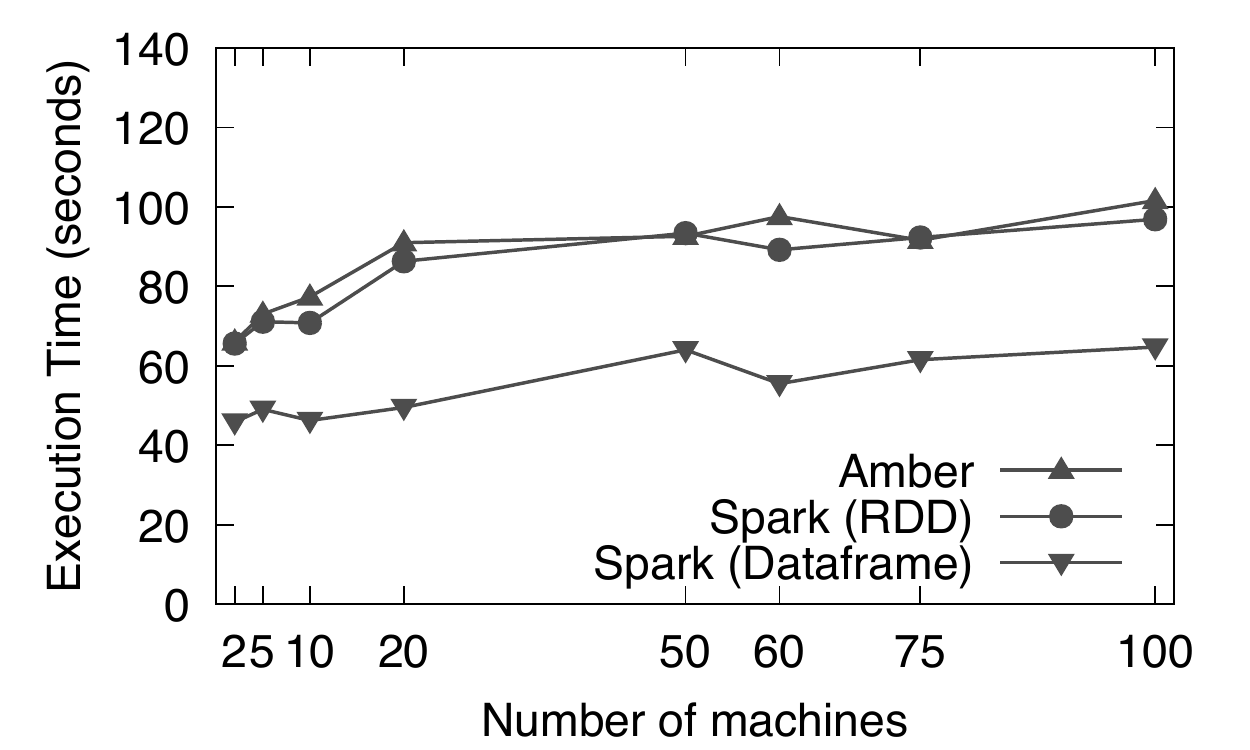}
	\caption{\label{fig:scaleup-time-tpch13-sparkVsAmber}
			\textbf{Scaleup for Amber and Spark for $W_2$.}
	}
\end{center}
\end{figure}

Figures~\ref{fig:scaleup-time-tpch1-sparkVsAmber} and~\ref{fig:scaleup-time-tpch13-sparkVsAmber} show the results for TPC-H $W_1$ and $W_2$, respectively. Amber achieved a performance comparable to Spark's DataFrame API and even more comparable to the RDD API for $W_2$. The performance gain of Spark's DataFrame API can be attributed to its optimizations discussed earlier. However, to our surprise, Spark's RDD API outperformed  Spark's DataFrame API for $W_1$. Amber performance remained quite comparable to both the API's for $W_1$ too.
We also compared Amber and Spark (DataFrame API) using the Taxi workflow $W_4$ on 10 machines. Spark took 442s, while Amber took 470s.






\subsection{Fault Tolerance in Amber and Spark}
\label{ssec:exp-fault-tolerance}

\begin{figure}[htbp]
\begin{center}
	\vspace{-0.1in}
	\includegraphics[width=3.3in]{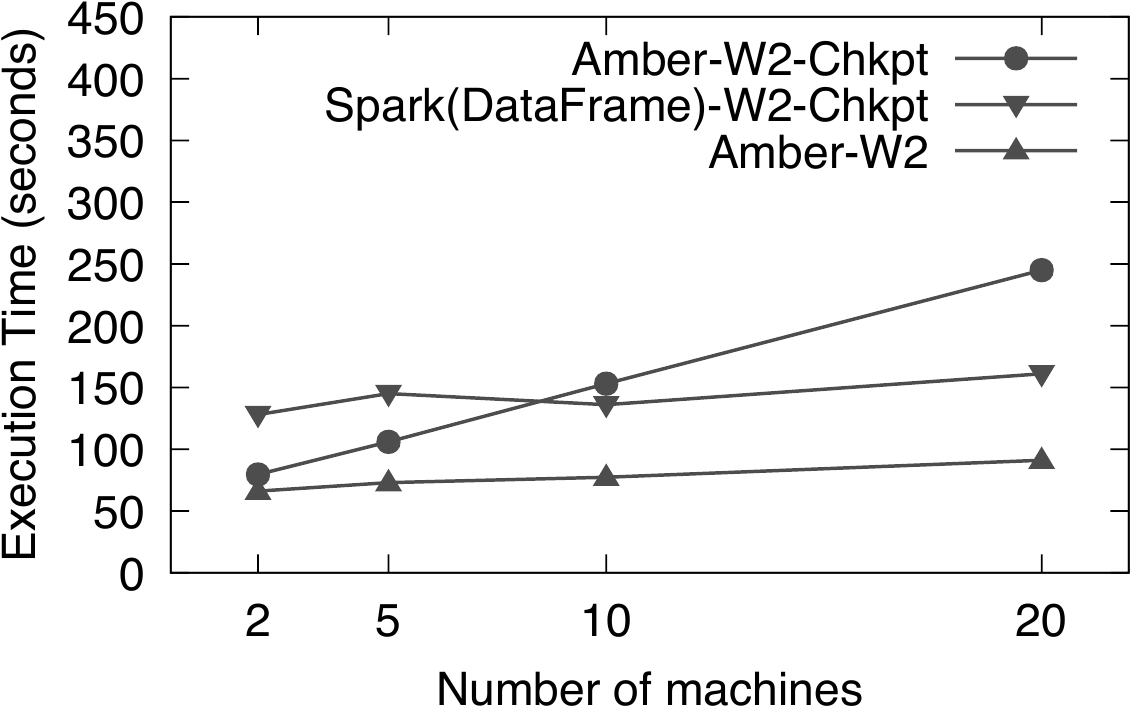}
	\caption{\label{fig:fault-tolerance-tpch13-sparkVsAmber}
			\textbf{Data-checkpointing overhead for Amber and Spark while executing $W_2$.} The time for Amber with data checkpointing disabled and time for Spark and Amber with data checkpointing enabled are shown.
	}
\end{center}
\end{figure}

We executed Amber in a stage-by-stage execution model (batch execution model) similar to Spark and evaluated the overhead of supporting fault tolerance. To evaluate the overhead of supporting fault tolerance, we turned on data checkpointing in Amber and Spark to write checkpointed data to a remote HDFS. In Amber, a worker created a separate file for each hash partition. In contrast, Spark consolidates its checkpoint data into HDFS block-sized files of 128MB. With data checkpointing on, we scaled the execution of $W_2$ on both systems from 2 machines (20GB data) to 20 machines (200GB data) and let the workflow run to completion. We chose $W_2$ because it has more number of stages than $W_1$.  In order to reduce the number of produced files, we used only 1 worker per operator per machine for Amber.  We used the DataFrame API of Spark as it was faster than the RDD API for $W_2$. Figure~\ref{fig:fault-tolerance-tpch13-sparkVsAmber} shows the execution times. Amber's data checkpointing initially performed better than Spark. For a higher number of machines, Amber took more time to complete the execution due to the quadratic increase in the number of files. For instance, for the 20-machine case, Amber produced 400 files (20 workers, each producing 20 partitions) at the end of each stage, while Spark produced only 66 files.

We evaluated crash recovery for Amber. When running $W_2$ on 10 machines (100GB data), after the Join operator ran for 5s, we simulated a crash by killing the workers of one data partition.  Amber took 176s in total (including crash and recovery) to run to completion, which was comparable to the case where there was no failure (153s). We also evaluated recovery of control messages in Amber. We paused the workflow after it entered the Join stage for 10s and then simulated a crash. Amber took 6s to recreate actors, and 10s of recomputation to recover to the same Paused state.

{\bf Summary}: The experiments showed that Amber can process control messages quickly and support conditional breakpoints with a low overhead.  It achieved a high performance (both scaleup and speedup) comparable to Spark. The capability of supporting dynamic resource allocation during the execution achieved a better performance.  

\section{Conclusions}
\label{sec:conclusions}


In this chapter we presented a system called Amber that supports powerful and responsive debugging during the execution of a dataflow.  We presented its overall system architecture based on the actor model, studied how to translate an operator DAG to an actor DAG that can run on a computing cluster. We described the whole lifecycle of the execution of a workflow, including how control messages are sent to actors, how to expedite the processing of these control messages, and how to pause and resume the computation of each actor. We studied how to support conditional breakpoints, and presented solutions for enforcing local conditional breakpoints and global conditional breakpoints. We developed a technique to support fault tolerance in Amber, which is more challenging due to the presence of control messages.  We implemented Amber on top of Orleans, and presented an extensive experimental evaluation to show its high usability and performance comparable to Spark. 
\chapter{Reshape: Adaptive Result-Aware Skew Handling for Exploratory Analysis on Big Data}
\label{chap:reshape}

\section{Introduction}
\label{sec:reshape-introduction}

We mentioned various existing systems for data analysis in Chapter~\ref{chap:introduction}. Data processing frameworks such as Hadoop~\cite{misc/hadoopmapreduce}, Spark~\cite{misc/spark}, and Flink~\cite{misc/flink} provide programming interfaces and GUI-based workflow systems such as Alteryx~\cite{alteryx}, RapidMiner~\cite{rapidMiner}, Knime~\cite{knime}, Einblick~\cite{Einblick}, and Texera~\cite{journals/pvldb/WangKNL20} provide a GUI interface that are used by analysts to formulate data processing jobs. Once the data processing job is created, it is submitted to an engine that executes the job. 

\begin{figure}[htbp]
\begin{center}
	\includegraphics[width=3.8in]{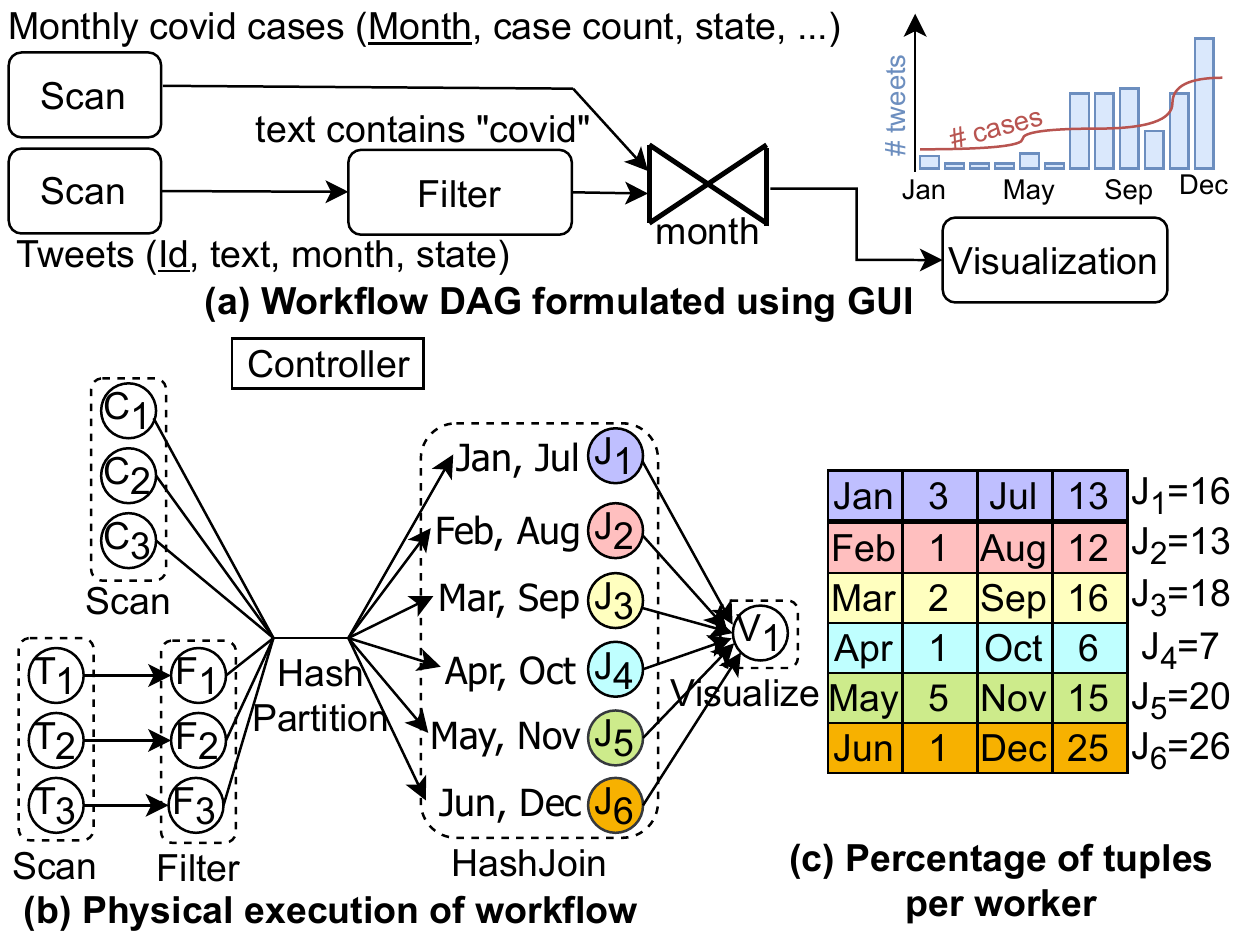}
	\caption{\label{fig:tweets-workflow}
		\textbf{Partitioning skew in a data science project of Covid tweet analysis.}
	}
    \end{center}
\end{figure}

The process of data analysis, especially in GUI-based analytics systems, has two important characteristics. \textbf{1) Highly exploratory:} The process of building a workflow can be very exploratory and iterative~\cite{journals/interactions/FisherDCD12, conf/sigmod/XuKAR22,conf/sigmod/VartakSLVHMZ16}. Often the user constructs an initial workflow and executes it to observe a few results. If they are not desirable, she terminates the current execution and revises the workflow. The user iteratively refines the workflow until finishing a final workflow to compute the results. As an example, Figure~\ref{fig:tweets-workflow} shows a workflow at an intermediate step during the task of covid data analysis. It examines the relationship between the number of tweets containing the keyword {\tt covid} and the number of Covid cases in 2020. The monthly details about the Covid cases are joined with tweets filtered on the {\tt covid} keyword on the month column. The result is plotted as a visualization operator that shows a bar chart about the total count of tweets about Covid per month and a line chart about the total Covid cases per month. The analyst may observe the visualization and choose to continue refining the workflow to do a deeper-dive analysis for specific US states. In the Texera system we are developing, we observe that the users refined a workflow about $80$ times on an average before reaching the final version. \textbf{2) Suitable for non-technical users:} GUI-based workflow execution systems significantly lower the technical learning curve for its users, thus enabling non-IT domain experts to do data science projects without writing code. Such systems also try to minimize the requirements on users to know the technical details of workflow execution, so that the user can focus solely on the analytics task at hand.


In exploratory data analytics, it is vital for a user to see results quickly to allow her to identify problems in the analysis early and take corrective actions without waiting for the entire workflow to finish executing. {\em Pipelined execution}~\cite{journals/csur/BenoitCRS13, books/sp/OzsuV20} is a popular workflow execution model that can produce results quickly. In pipelined execution, an operator does not wait for its entire input data to be produced before processing the input and sending results to its downstream operators. For example, when the workflow in Figure~\ref{fig:tweets-workflow} is executed using pipelined model, the {\sf HashJoin} operator starts executing and producing results as soon as the {\sf Filter} operator outputs initial results. The user can notice the initial results and make any changes, if needed.  Pipelined execution is adopted by data-processing engines such as Flink~\cite{misc/flink}, Samza~\cite{Samza:website}, and Storm~\cite{apacheStorm}.

As data volumes in these systems increase, it is indispensable to do parallel processing, in which data is partitioned and processed by multiple computation units in parallel. Data partitioning, either using hash partitioning or range partitioning, often results in skew.  As an example, the {\sf HashJoin} operator in Figure~\ref{fig:tweets-workflow} receives hash partitioned inputs from the two upstream operators.  Although the hash function allots the same number of months to each join worker, load imbalance still exists because of different numbers of tweets for those months. It is well known that partitioning skew adversely affects the efficiency of engines as it increases the processing time and reduces the throughput~\cite{conf/vldb/DeWittNSS92,conf/sigmod/KwonBHR12}.

The problem of partitioning skew has been extensively studied in the literature, mainly from the perspective of increasing the end-to-end performance. However, there is little research on the following important problem:

\begin{quote}
{\em In exploratory data analytics, how to consider the results shown to the user when mitigating skew?}
\end{quote} 

In exploratory data analysis, it is valuable to the analyst if the initial results are representative of the final results because they allow her to identify issues early and make necessary changes. Partitioning skew may lead to the production of misleading results during the execution. Let us consider the production rate of October and December tuples from the {\sf HashJoin} operator in the running example. Assume that the {\sf HashJoin} operator is the bottleneck of the execution, and its workers receive input at an equal or higher rate than they can process. Although there are more December tuples than October, their production rates are similar because the total amounts of data received by $J_4$ and $J_6$ are different (details in Section~\ref{sec:load-transfer-mechanism}). Thus, the bar chart shows similar heights for October and December bars till $J_4$ completes processing, whereas the December bar is almost four times taller than the October bar in the final result. In this chapter, we explore partitioning skew mitigation in the setting of exploratory data analysis and analyze the effect of mitigation strategies on the results shown to the user.

A common solution to handle partitioning skew at an operator is blocking the partitioning of its input data till the entire input data is produced by its upstream operator~\cite{SparkAQE:website,conf/vldb/DeWittNSS92,conf/icde/VitorovicE016,journals/tpds/ChenYX15,conf/sigmod/AbdelhamidMDA20} and then sampling the input data to create an optimal partitioning function. For example, in Figure~\ref{fig:tweets-workflow}, the {\sf HashJoin} operator waits for the {\sf Filter} operator to completely finish. Then, the output of the {\sf Filter} operator is sampled to create an optimal partitioning function to send data to the {\sf HashJoin} operator. Such blocking is not allowed in pipelined execution which makes these solutions infeasible in pipelined execution setting. Even temporarily blocking the partitioning till a small percentage of input (e.g., 1\%~\cite{conf/icde/RodigerIK016}) is collected for sampling can result in a long delay if there is an expensive upstream operator.


A different solution applicable to the pipelined execution setting is to detect the overloaded workers of an operator at runtime and transfer the processing of a few keys of the overloaded worker to a more available worker. For example, $J_6$ is detected to be overloaded at runtime and the processing of June tuples is transferred to $J_4$. However, this transfer has little effect on the results shown to the user (details in Section~\ref{sec:load-transfer-mechanism}). In order to show representative initial results, the data of December has to be split between $J_4$ and $J_6$. Thus, these two approaches of transferring load from $J_6$ to $J_4$ have different impacts on the initial results shown to the user.




In this chapter, we analyze the effect of different skew mitigation strategies on the results shown to the user and present a novel skew handling framework called \frmname that adaptively handles skew in a pipelined execution setting. Reshape monitors the workload metrics of the workers and adapts the partitioning logic to transfer load whenever it observes a skew.  These modifications can be done multiple times as the input distribution changes~\cite{conf/sigir/BeitzelJCGF04,conf/wsdm/KulkarniTSD11} or if earlier modifications did not fully mitigate the skew. The command to adapt the partitioning logic is sent from the controller to the workers using low latency control messages that are supported in various engines such as Flink, Chi~\cite{journals/pvldb/MaiZPXSVCKMKDR18}, and Amber~\cite{journals/pvldb/KumarWNL20}. 

We make the following contributions. (1) Analysis of the impact of mitigation on the shown results: We present different approaches of skew mitigation and analyze their impact on the results shown to the user. (Section~\ref{sec:load-transfer-mechanism}). (2) Automatic adjustment of the skew detection threshold: We present a way to dynamically adjust the skew detection threshold to reduce the number of iterations of mitigation to minimize the technical burden on the user
(Section~\ref{sec:adaptive-handling}). (3) Applicability to multiple operators: Since a data analysis workflow can contain many operators that are susceptible to partitioning skew, we generalize \frmname to multiple operators such as {\sf HashJoin}, {\sf Group-by}, and {\sf Sort}, and discuss challenges related to state migration (Section~\ref{sec:other-operators}). (4) Generalization to broader settings: We consider settings such as high state-migration time and multiple helper workers for an overloaded worker and discuss how \frmname can be extended in these settings (Section~\ref{sec:broader-settings}).  (5) Experimental evaluation: We present the implementation of \frmname on top of two big-data engines, namely Amber and Flink, to show the generality of this approach. We report an experimental evaluation using real and synthetic datasets on large computing clusters (Section~\ref{sec:experiments}).

\subsection{Related work}
\label{ssec:reshape-related-work}

There have been extensive studies about skew handling in two major execution paradigms in big data engines -- batch execution and pipelined execution. Batch-execution systems such as MapReduce~\cite{misc/hadoopmapreduce} and Spark~\cite{misc/spark} materialize complete input data before partitioning it across workers. Pipelined-execution systems such as Flink~\cite{misc/flink}, Storm~\cite{apacheStorm} and Amber~\cite{journals/pvldb/KumarWNL20} send input tuples to a receiving worker immediately after they are available. The complete input is not known to the operator in pipelined execution, which makes skew handling more challenging.

\boldstart{Skew handling in batch execution.}  A static technique is to sample and obtain the distribution of complete input data and use it to partition data in a way that avoids skew~\cite{SparkAQE:website, conf/vldb/DeWittNSS92, conf/icde/VitorovicE016,journals/tpds/ChenYX15,conf/sigmod/AbdelhamidMDA20}. Adaptive skew-handling techniques adapt their decisions to changing runtime conditions and mitigate skew in multiple iterations. For instance, SkewTune~\cite{conf/sigmod/KwonBHR12} and Hurricane~\cite{conf/eurosys/BindschaedlerMS18} handle skew adaptively. That is, upon detecting skew, SkewTune stops the executing workers, re-partitions the materialized input, and starts new workers to process the partitions. Hurricane clones overburdened workers and uses a special storage that allows fine-grained data access to the original and cloned workers in parallel. Hurricane can split the processing of a key over multiple workers and thus has a fine load-transfer granularity. SkewTune cannot split the processing of a key.




\boldstart{Static skew handling in pipelined execution.} Flow-Join~\cite{conf/icde/RodigerIK016} avoids skew in a {\sf HashJoin} operator. It samples the first 1\% of input data of the operator to decide the overloaded keys and does a broadcast join for the overloaded keys. Since it makes the decision based on an initial portion of the input, it cannot handle skew if the input distribution changes multiple times during the execution. Partial key grouping (PKG)~\cite{conf/icde/NasirMGKS15,conf/icde/NasirMKS16} uses multiple pre-defined partitioning functions. It results in multiple candidate workers sharing the processing of the same key. Since the partitioning logic is static, a worker may process multiple skewed keys, which makes it more burdened than other workers. PKG cannot be used to handle skew in operators such as {\sf Sort} and {\sf Median}.

\boldstart{Adaptive skew handling in pipelined execution.} Flux~\cite{conf/icde/ShahHCF03} divides the input into many pre-defined mini-partitions that can be transferred between workers to mitigate skew. Thus, the load-transfer granularity is fixed and pre-determined. Also, it cannot split the load of a single overloaded key to multiple workers. Another adaptive technique minimizes the input load on workers that compute theta joins by dynamically changing the replication factor for data partitions~\cite{journals/pvldb/ElseidyEVK14}. This approach uses random partitioning schemes such as round-robin and hence is not prone to partitioning skew. \frmname handles skew adaptively over multiple iterations. It determines the keys to be transferred dynamically and allows an overloaded key to be split over multiple workers for mitigation.


\section{\frmname: Overview}
\label{sec:overview}

\begin{figure}[htbp]
\begin{center}
	\includegraphics[width=3.6in]{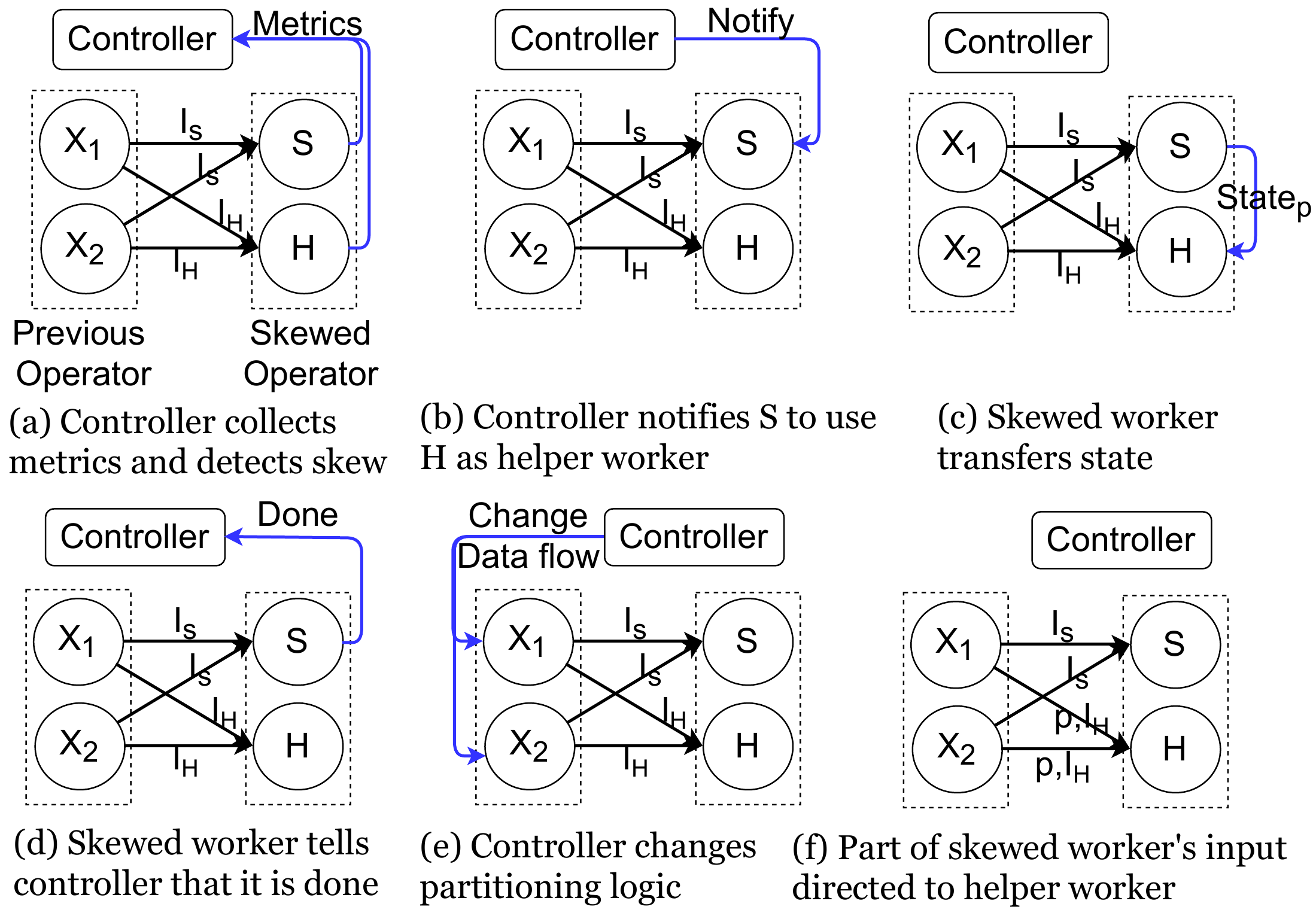} 
	\caption{\label{fig:overall-approach}
		\textbf{Steps of skew-handling in \frmname. Skew detected in (a) and mitigated in (b)-(f).}
	}
    \end{center}
\end{figure}

We use Figure~\ref{fig:overall-approach} to give an overview of \frmname.


\subsection{Skew detection} 
\label{ssec:skew-detection}
During the execution of an operator, the controller periodically collects workload metrics from the workers of the operator to detect skew (Figure~\ref{fig:overall-approach}(a)). There are different metrics that can represent the workload on a worker such as CPU usage, memory usage and unprocessed data queue size~\cite{conf/eurosys/BindschaedlerMS18, conf/sigmod/FernandezMKP13}.  Skew handling in \frmname is independent of the choice of workload metric, and we choose unprocessed queue size as a metric in this chapter. We choose this metric because the results seen by the analyst depend on the future results produced by a worker, which in turn depend on the content of its unprocessed data queue. We refer to a computationally overburdened worker as a {\em skewed worker} and workers that share the load as its {\em helper workers}. 

\boldstart{Skew test.} Given two workers of the same operator, say $C$ and $L$, the controller performs a {\em skew test} to determine whether $C$ is a helper candidate for $L$. The skew test uses the following inequalities to check if $L$ is computationally burdened and the workload gap between $L$ and $C$ is big enough:
\begin{equation}\label{eq:greater-than-eta}
    \phi^{}_L \geq \eta,
\end{equation}
\vspace{-0.22in}
\begin{equation}\label{eq:tau-diff}
    \phi^{}_L - \phi^{}_C \geq \tau,
\end{equation}
where $\eta$ and $\tau$ are threshold parameters and $\phi^{}_w$ is the workload on a worker $w$. 

\boldstart{Helper workers selection.} The skew tests may yield multiple helper candidates for $L$. For simplicity, we assume till Section~\ref{sec:other-operators} that one helper worker is assigned per skewed worker. In Section~\ref{sec:broader-settings}, we generalize the discussions by considering multiple helpers per skewed worker. The controller chooses the helper candidate with the lowest workload that has not been assigned to any other overloaded worker as the helper of $L$. In our discussions forward, we use $S$ and $H$ to refer to a skewed worker and its chosen helper worker respectively.

\subsection{Skew mitigation} 
\label{ssec:skew-mitigation}
Suppose the skewed worker $S$ and its helper $H$ have been handling input partitions $I^{}_S$ and $I^{}_H$, respectively (Figure~\ref{fig:overall-approach}(a)). \frmname transfers a fraction of the future input of $S$ to $H$ to reduce the load on $S$. Here future input refers to the data input that is supposed to be received by a worker but has not yet been sent by the previous operator. The controller notifies $S$ about the part {\em p} of partition $I^{}_S$ that will be shared with $H$ to reduce the load of $S$ (Figure~\ref{fig:overall-approach}(b)). The downstream results shown to the user have a role to play in deciding {\em p}, which will be discussed in Section~\ref{sec:load-transfer-mechanism}. Worker $S$ sends to $H$ its state information $State_p$ corresponding to the partition {\em p}  (Figure~\ref{fig:overall-approach}(c)). Details about state-migration strategies are in Section~\ref{sec:other-operators}. We assume the state migration time to be small till Section~\ref{sec:other-operators}. In Section~\ref{sec:broader-settings}, we consider the general case where the state-migration time can be significant. Worker $H$ saves the state information and sends an {\em ack} message to $S$, which then notifies the controller (Figure~\ref{fig:overall-approach}(d)). The controller changes the partitioning logic at the previous operator (Figure~\ref{fig:overall-approach}(e,f)).

\boldstart{Fault Tolerance.} The \frmname framework supports the fault tolerance mechanism of the Flink engine~\cite{journals/corr/CarboneFEHT15} that checkpoints the states of the workers periodically. During checkpointing, a checkpoint marker is propagated downstream from the source operators. When an operator receives the marker from all its upstream operators, it takes a checkpoint which saves the current states of the workers of the operator. Every checkpoint has the information about the current partitioning logic at the workers. If checkpointing occurs during state migration, then the skewed worker additionally forwards the checkpoint marker to each of its helper workers. A helper worker needs to wait for the checkpoint marker from its corresponding skewed worker. Since the skewed workers and the helper workers are two disjoint sets of workers, there is no cyclic dependency in marker propagation and the checkpointing process successfully terminates. During recovery, the workers restore their states from the most recent checkpoint and then continue the execution.
\section{Result-aware load transfer}
\label{sec:load-transfer-mechanism}

After helper workers are selected for the skewed workers, the load needs to be transferred from the skewed workers to the corresponding helper workers. \avinash{In this section, we assume that the state information in the workers is fixed (immutable) and the fixed state belonging to a partition has already been transferred from the skewed worker to the helper worker before the load of that partition is transferred. More details about the state-migration process will be discussed in Section~\ref{sec:other-operators}.} In Section~\ref{ssec:two-load-transfer-approaches}, we consider the different approaches of load transfer between workers and analyze their impact on the results shown to the user. Unlike other skew handling approaches that focus on evenly dividing the future incoming load among the workers, \frmname has an extra phase of load transfer at the beginning that removes the existing load imbalance between the workers. In Section~\ref{ssec:two-phases}, we discuss these two phases of load transfer and the significance of the first phase. 



\subsection{Mitigation impact on user results}
\label{ssec:two-load-transfer-approaches}

There are broadly two approaches to transfer the load from a skewed worker to its helper worker. We use the probe input of the {\sf HashJoin} operator in Figure~\ref{fig:tweets-workflow} (from the {\sf Filter} operator) as an example to explain the concepts in this section. It is assumed that the build phase of the join has finished. Suppose \frmname detects $J_6$ and $J_5$ as the skewed workers in the running example and $J_4$ and $J_2$ are their corresponding helpers, respectively.  The load-transfer approaches are implemented by changing the partitioning logic at the {\sf Filter} operator and affects the future tuples going into the {\sf HashJoin} operator.

{\em 1. Split by keys (SBK).} In this approach, the keys in the partition of the skewed worker are split into two disjoint sets, say $p_1$ and $p_2$. The future tuples belonging to $p_2$ are redirected to the helper worker, while tuples belonging to $p_1$ continue to be sent to the skewed worker. For example, the partition of the skewed worker $J_6$ is divided into $p_1$ = \{December\} and $p_2$ = \{June\}, and the future June tuples are sent to $J_4$, while December tuples continue to go to $J_6$.


{\em 2. Split by records (SBR).} In this approach, the records of the keys in the partition of the skewed worker are split between the skewed and the helper worker. The ratio of the split decides the amount of load transferred to the helper worker. For example, if the {\sf Filter} operator needs to redirect $\frac{9}{26}$ of the input $J_6$ to $J_4$, then it redirects $9$ tuples out of every $26$ tuples in $J_6$'s partition to $J_4$.

\boldstart{Impact of the two approaches on user results.} The two load transfer approaches have their own advantages and limitations. For example, {\sf SBK} incurs an extra overhead compared to {\sf SBR} because {\sf SBK} requires the workers to store the distribution of workload per key. On the other hand, {\sf SBR} may require transfer of a larger state size compared to {\sf SBK}, if all the keys of a skewed worker are shared with the helper. There are existing works in literature that address these concerns~\cite{conf/icdt/MetwallyAA05, conf/icde/RodigerIK016, conf/bigdataconf/YanXM13, conf/icde/GuflerARK12, conf/sigmod/MonteZRM20, journals/pvldb/HoffmannLMKLR19}. In the remainder of this subsection, we compare these two approaches from the perspective of their effects on the results shown to the user.

\begin{figure}[htbp]
\begin{center}
	\includegraphics[width=3.5in]{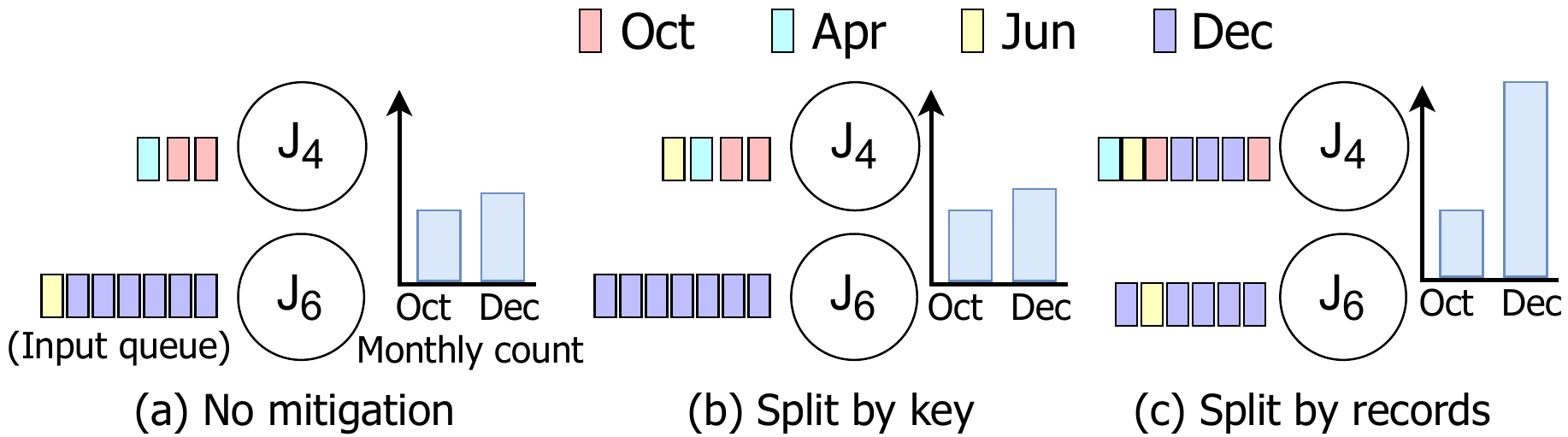}
	\caption{\label{fig:split-by-record-better}
		\textbf{SBR splits December tuples on both workers and shows representative bar charts.}
	}
    \end{center}
\end{figure}

\boldstart{a) Representative initial results.} As discussed before, it is valuable to the user if the initial results are representative of the final results. Partitioning skew may lead to the production of misleading results during the execution as shown next. Let us consider the bar chart visualization for October and December in the running example. The total count of December tweets, according to Figure~\ref{fig:tweets-workflow}(c), is about four times that of October tweets, i.e., the December bar is about four times longer than the October bar in the final visualization. Assume that the join operator is the bottleneck of the execution, and its workers receive input at an equal or higher rate than what they can process. Also assume that the processing speeds of the workers of {\sf HashJoin} are the same, say $t$ per second. $J_4$ produces $\frac{6}{7}*t$ October tuples and $J_6$ produces $\frac{25}{26}*t$ December tuples per second in the unmitigated case (Figure~\ref{fig:split-by-record-better}(a)). The rate of production of October and December tuples are similar because the total amount of data received by $J_4$ and $J_6$ are different. The bar chart shows similar heights for October and December bars in the unmitigated case till $J_4$ completes its processing.

When {\sf SBK} is used to mitigate the skew, the processing of June tuples is transferred to $J_4$ (Figure~\ref{fig:split-by-record-better}(b)). However, this transfer has little effect on the results shown to the user. The production rates of October and December after the transfer are $\frac{6}{8}*t$ and $t$ respectively.  That is, the heights of the December and October bars are still about the same, which is not representative of the final results.

{\sf SBR} has more flexibility for transferring load than {\sf SBK} because {\sf SBR} can split the tuples of a key over multiple workers. It leads to more representative initial results than {\sf SBK} as shown next. The processing of December and June tuples can be split between $J_6$ and $J_4$. For simplicity of calculation, we assume that only December tuples are shared with $J_4$. Since December tuples are now processed by two workers instead of one, the speed of production of these tuples increases. In order to make the future workloads of $J_4$ and $J_6$ similar, {\sf SBR} redirects $\frac{9}{26}$ of the input of $J_6$ to $J_4$, which increases the total percentage load on $J_4$ to $16$ and decreases that on $J_6$ to $17$. This is implemented by redirecting $9$ December tuples out of every $26$ tuples in $J_6$'s partition to $J_4$. The production rates of October tuples after the transfer is $\frac{6}{16}*t$. The December tuples are produced by $J_4$ and $J_6$. The production rate of December by $J_4$ is $\frac{9}{16}*t$ and by $J_6$ is $\frac{16}{17}*t$, which results in a total of approximately $\frac{24}{16}*t$. Thus, using {\sf SBR} leads to a more representative production ratio of December to October tuples of about $24{\,:\,}6$, which is similar to the actual ratio of $25{\,:\,}6$.

\begin{figure}[htbp]
\begin{center}
	\includegraphics[width=3.5in]{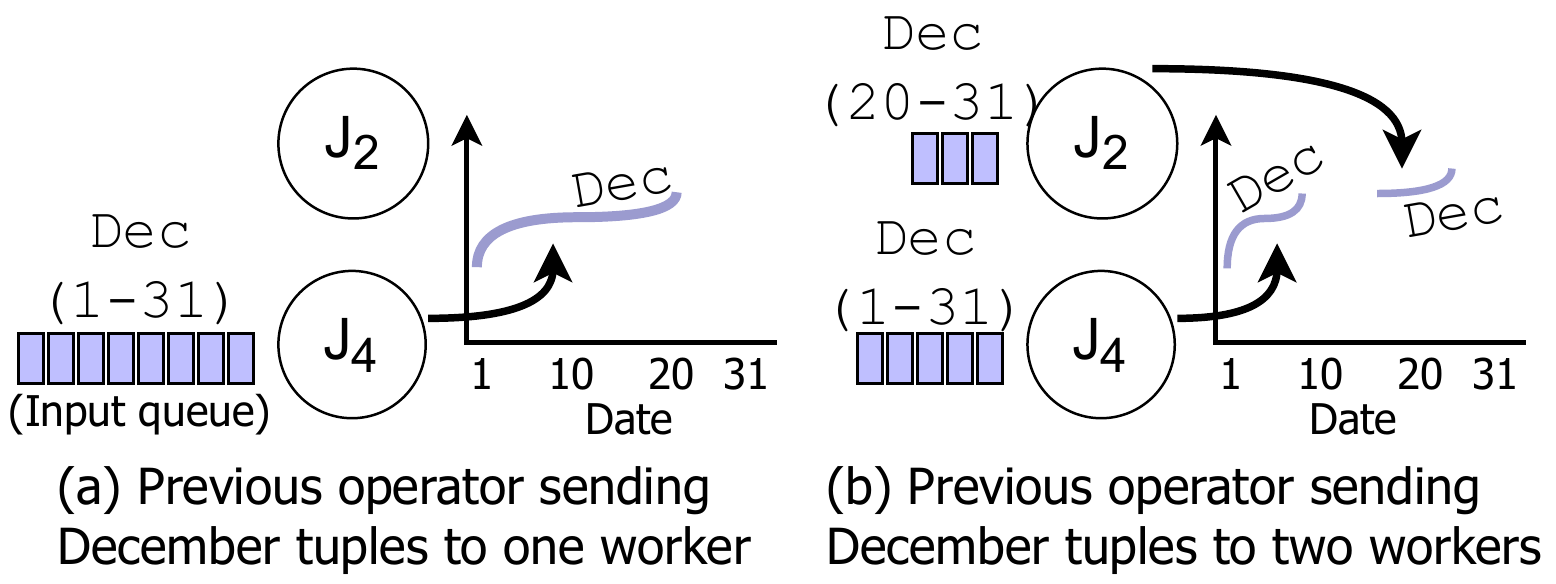}
	\caption{\label{fig:split-by-key-better}
		\textbf{Processing a key at multiple workers by {\sf SBR} leads to a broken line chart. Only December tuples have been shown for simplicity.}
	}
    \end{center}
\end{figure}

\boldstart{b) Preserving order of tuples.} If the tuples of a key being input into an operator are in a particular order and they need to be processed in that order, then {\sf SBK} is the suitable approach because it enforces a processing order by restricting the processing of the tuples of a key to a single worker at a time. If the processing of a key needs to be transferred to another worker, the migration can be synchronized using techniques such as pause and resume~\cite{conf/sigmod/ArmbrustDTYZX0S18,journals/pvldb/CarboneEFHRT17,conf/icde/ShahHCF03} or markers~\cite{journals/pvldb/ElseidyEVK14} (details in Section~\ref{sec:other-operators}) so that the tuples are processed in order. In contrast, {\sf SBR} distributes the tuples of a key over multiple workers to be processed simultaneously, which may cause them to be processed out of order. Consider the following example where an out-of-order processing of the tuples of a key is not desirable. Let us slightly modify the visualization operator in the running example to plot a line chart that shows daily count of covid related tweets. The daily count for each month is plotted as a separate line in the line chart. Figure~\ref{fig:split-by-key-better} shows the plot for  December in the line chart. Applications may want to show such plots as a continuous line with no breaks, starting from day 1 and extending towards increasing dates as execution progresses, for user experience purposes~\cite{Cloudberry}. In order to achieve this, the tuples of a month input into the {\sf HashJoin} operator are sorted in the increasing order of date. It is expected that {\sf HashJoin} produces tuples sorted by date, which can be consumed by the visualization operator to create a continuous plot.

{\sf SBK} assures that the December key is processed by only one worker at a time. Thus, it preserves the order of December tuples in the output sent to the visualization operator (Figure~\ref{fig:split-by-key-better}(a)). When {\sf SBR} is used, the December tuples are split between $J_4$ and $J_2$. In the example shown in Figure~\ref{fig:split-by-key-better}(b), the {\sf Filter} operator starts partitioning December tuples by {\sf SBR} when the tuples around the $20^{th}$ of December are being produced by the {\sf Filter} operator. Consequently, $J_2$ starts receiving the tuples from the date of the $20^{th}$ December and above. As $J_2$ and $J_4$ concurrently process data, the visualization operator receives the tuples out of order, resulting in broken line chart plots as shown in the figure.

In conclusion, {\sf SBR} allows more flexibility and enables the production of representative initial results than {\sf SBK}, but {\sf SBR} does not preserve the order of tuples. Thus, {\sf SBR} can be chosen unless there exists a downstream operator that imposes some requirement over the input order of the tuples. Such operators can be found at the workflow compilation stage. The operators before such an operator in the workflow can adopt {\sf SBK}.


\subsection{Extra phase in load transfer}
\label{ssec:two-phases}

The goal of skew mitigation is to use one of the two  approaches to transfer the load from the skewed worker to the helper worker in such a way that both workers have a similar workload for the rest of the execution. The skew handling works in literature usually have a single phase of load transfer that focuses on splitting the incoming input such that the workers receive similar load in future. \frmname has an extra phase of load transfer at the beginning that removes the existing load imbalance between the workers. We first give an overview of the two phases in \frmname, and explain the significance of the first phase.

\begin{figure}[htbp]
\begin{center}
	\includegraphics[width=4.4in]{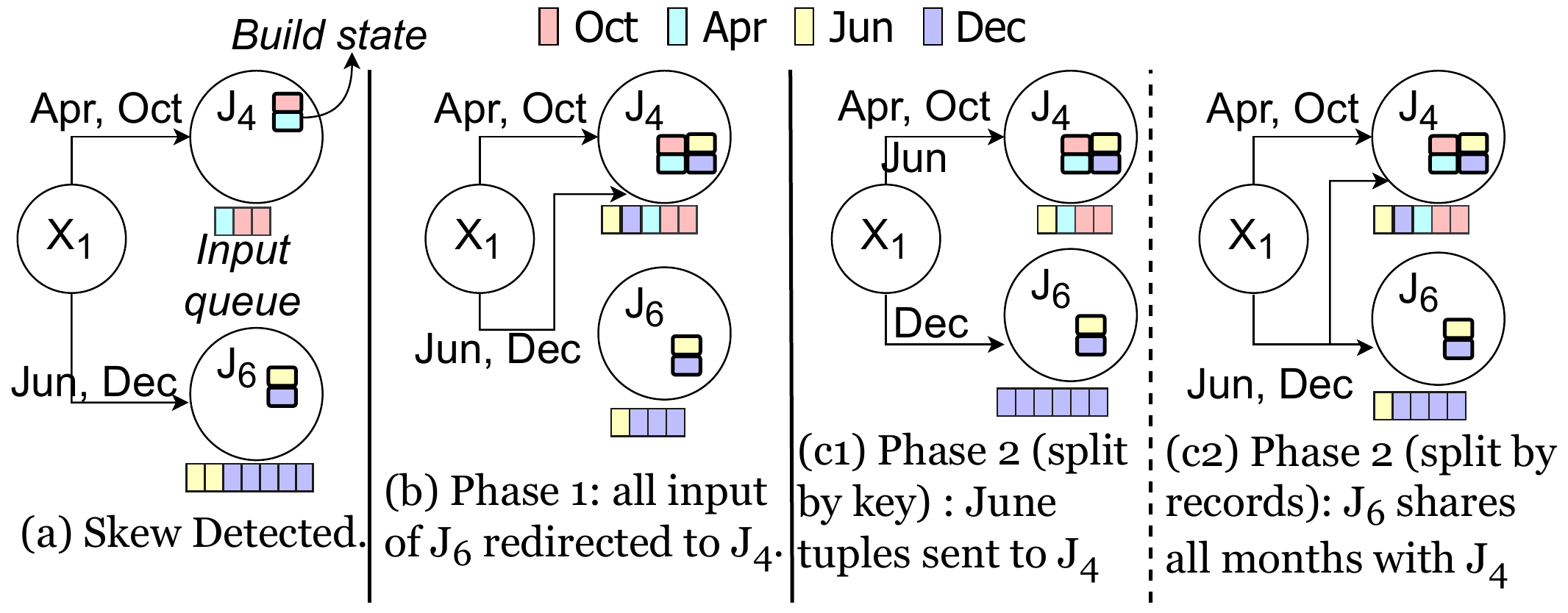}
	\caption{\label{fig:phases-of-mitigation}
		\textbf{An implementation of the two phases using the ``SBK'' and ``SBR'' approach. $X_1$ is a previous operator worker.}
	}
    \end{center}
\end{figure}

\boldstart{First Phase.} After the detection of skew (Figure~\ref{fig:phases-of-mitigation}(a)), the controller starts the first phase of load transfer. The first phase lets the helper ``catch up'' quickly with the skewed worker. One implementation of the partitioning logic in the first phase at the {\sf Filter} operator is that it sends all future tuples of $J_6$ to $J_4$ (Figure~\ref{fig:phases-of-mitigation}(b)). Note that $J_6$ will continue to process the data in its queue.  An alternative implementation is to send only a portion of $J_6$'s partition, such as the December data, to $J_4$. This alternative reduces the amount of state transfer, but it will take longer time for $J_4$ to catch up with $J_6$.

\boldstart{Second Phase.} Once the queue sizes of the two workers become similar, the controller starts the second phase. Its goal is to modify the partitioning logic at the {\sf Filter} operator to redirect part of the future input of $J_6$ in such a way that both the workers receive a comparable workload. In order to do this, first the incoming workload of the workers needs to be estimated. A sample of workloads needs to be collected to estimate the future workload of the workers~\cite{conf/cloud/RamakrishnanSU12, journals/tpds/ChenYX15, conf/bigdataconf/YanXM13, conf/icde/GuflerARK12} using a prediction function~$\psi$. \frmname can use the sample from the recent history collected during the current execution~\cite{conf/IEEEcloud/Kim0QH16, conf/cloud/ShenSGW11}.  If historical data is available, it can complement the recent data and improve the prediction accuracy~\cite{conf/isorc/GarraghanOTX15, conf/icde/PopescuEBBA12}.

To simplify the discussion of the second phase, we make the following assumptions: 
\begin{itemize}
 \item The two workers receive data at constant rates. 
 \item We have a perfect estimator to accurately predict the incoming data workload on the workers. 
\end{itemize}

In Section~\ref{sec:adaptive-handling} we will relax these two assumptions. In Figure~\ref{fig:tweets-workflow}(c), the original load ratio of $J_6$ to $J_4$ is $26{\,:\,}7$. {\sf SBK} cannot handle the skew between $J_6$ and $J_4$. The approach transfers the June month to $J_4$ (Figure~\ref{fig:phases-of-mitigation}(c1)), which does not mitigate the skew. However, {\sf SBR} can redirect $\frac{9}{26}$ of the input of $J_6$ to $J_4$, which mitigates the skew by increasing the percentage load on $J_4$ to $16$ and decreasing the percentage load on $J_6$ to $17$. An example where {\sf SBK} can mitigate the skew is the case of skew between the skewed worker $J_5$ and its helper $J_2$. {\sf SBK} can transfer the processing of May to $J_2$, which brings the two workers to a similar workload. Specifically, the percentage load on $J_2$ increases to $18$ and that on $J_5$ decreases to $15$.

It should be noted that two phases do not mean that the state transfer has to be done twice necessarily. There are implementations where the state transfer during the first phase is enough and the second phase does not require another state transfer. If the keys whose workloads are being transferred from the skewed worker to the helper worker in the second phase are the same as the keys whose workloads were transferred in the first phase, then there is no need to transfer state in the second phase. For example, in {\sf SBR}, the state of all keys are sent to $J_4$ in the first phase, and there is no state migration needed for the second phase. In other words, if the workload of a key that was not transferred to the helper in the first phase, is being transferred to the helper in the second phase, then its state needs to be transferred.

\begin{figure}[htbp]
\begin{center}
	\includegraphics[width=3.9in]{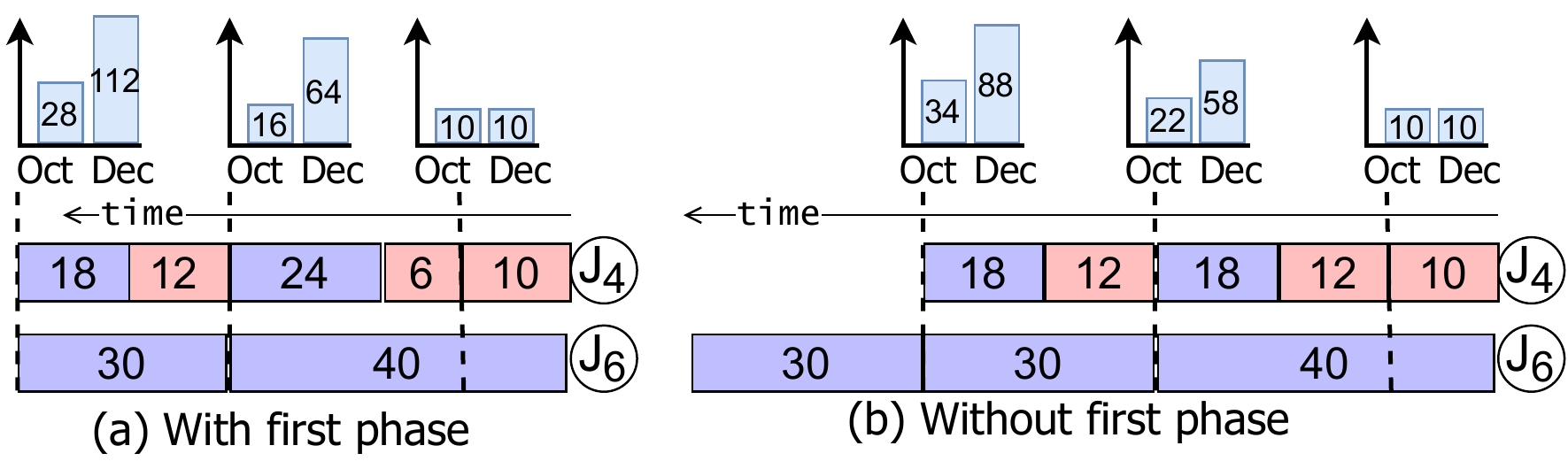}
	\caption{\label{fig:first-phase-significance}
		\textbf{First phase helps to reflect the actual ratio of December and October tuples early. The bar charts show the progression of results as the workers process tuples}
	}
    \end{center}
\end{figure}

\boldstart{Significance of the first phase.} \frmname has an extra phase for two reasons. First, it gives some immediate respite to the skewed worker and avoids imminent risks of the skewed worker going out of computing resources, invoking back-pressure~\cite{misc/backpressure} etc. Second, it may allow the user to see the representative results earlier compared to the case where there is only one phase. Figure~\ref{fig:first-phase-significance} illustrates this idea. For simplicity of calculation, we assume that $J_4$ processes October and $J_6$ processes December only. Notice that December tuples are almost four times the tuples of October (Figure~\ref{fig:tweets-workflow}(c)). Suppose the {\sf HashJoin} operator receives $2$ October and $8$ December tuples every second and the skew is detected when the unprocessed queue sizes of $J_4$ and $J_6$ are $10$ and $40$, respectively. Figure~\ref{fig:first-phase-significance}(a) shows the case where there exists a first phase. Suppose the first phase redirects all December tuples to $J_4$. In $3$ seconds, $J_4$ receives $24$ December and $6$ October tuples and catches up with the queue of $J_6$. After this, the second phase starts and redirects $3$ out of every $8$ December tuples to $J_4$. Assuming the workers process tuples at similar rates, the bar charts show the October and December tuples count shown to the user as the workers process more data. When the workers have processed $10$ tuples each, the bar chart shows $10$ tuples for both months. After that the effect of first phase starts. When both workers have processed $40$ tuples each, the bar chart shows $16$ tuples for October and $64$ tuples for December, which is representative of the ratio of October to December tuples in the input data. Figure~\ref{fig:first-phase-significance}(b) shows the case where there is no first phase. After detection of skew, the second phase starts and redirects $3$ out of every $8$ December tuples to $J_4$. In this case, even after both the workers have processed $40$ tuples each, the bar chart shows $22$ tuples for October and $58$ tuples for December. The ratio gradually moves towards the actual ratio of $1{\,:\,}4$ between October to December tuples. 

\section{Adaptive Skew Handling}
\label{sec:adaptive-handling}

In the previous section, we assumed that data arrives at constant rates to the workers and the second phase has a perfect estimator. In this section, we study the case when these assumptions are not true. In particular, variable patterns in incoming data rates and an imperfect estimator can result in erroneous workload predictions. Consequently, the second phase may not be able to keep the workload of the skewed and helper workers at a similar level. Thus, the controller may start another iteration of mitigation. Since, each iteration may incur an overhead, such as state transfer, we should try to make better workload predictions so that the number of iterations is reduced. We show that the workflow prediction accuracy depends on the skew detection threshold $\tau$ (Sections~\ref{ssec:benefit} and~\ref{ssec:tau-impact}). In order to reduce the technical burden on the user to fix an appropriate $\tau$, we develop a method to adaptively adjust $\tau$ to make better workload predictions (Section~\ref{ssec:adjust-tau-over-iterations}).




\subsection{Load reduction from mitigation}
\label{ssec:benefit}


We measure the {\em load reduction} ($LR$) from mitigation as the difference in the maximum input size received by a skewed worker and its helper without and with mitigation. Formally, let $S$ and $H$ represent the skewed worker and the helper worker, respectively. The load reduction is defined as:
\begin{equation}\label{eq:benefit-as-diff}
  LR = \left[max(\sigma^{}_S, \sigma^{}_H) \right]_{unmitigated} - \left[max(\sigma^{}_S, \sigma^{}_H) \right]_{mitigated},
\end{equation}
where $\sigma_w$ is the size of the total input received by a worker $w$ during the entire execution. 

\begin{figure}[htbp]
\begin{center}
	\includegraphics[width=3.3in]{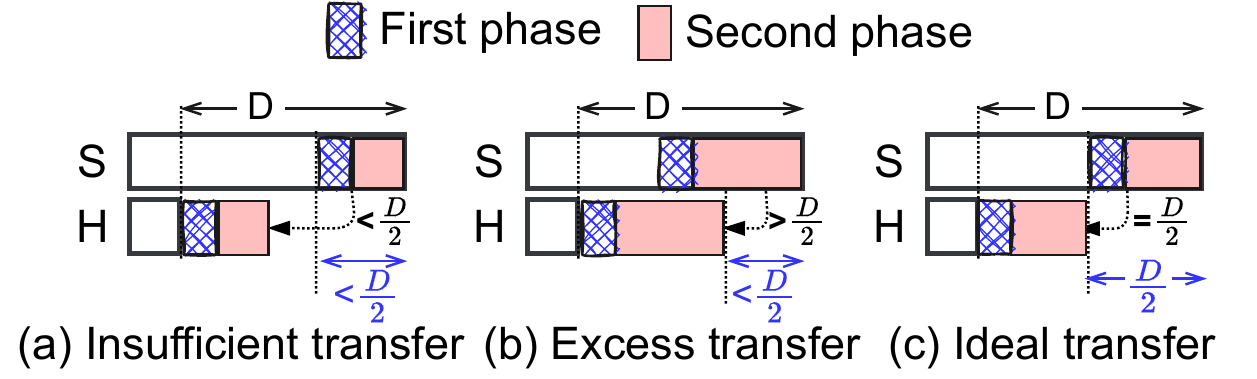} 
	\caption{\label{fig:ideal-mitigation}
		\textbf{Effect of the amount of transferred data on the load reduction. The shaded boxes represent the input of $S$ redirected to $H$ in the two phases.}
	}
    \end{center}
\end{figure}

In Figure~\ref{fig:ideal-mitigation}, $D$ represents the difference in the total input sizes of $S$ and $H$ in the unmitigated case. When mitigation is done, due to workload estimation errors, the second phase may not be able to redirect the precise amount of data to keep the workloads of $S$ and $H$ at a similar level. In Figure~\ref{fig:ideal-mitigation}(a), less than $\frac{D}{2}$ tuples of $S$ are redirected to $H$. Thus, $S$ receives more total input than $H$ and the load reduction is less than $\frac{D}{2}$. Similarly, in Figure~\ref{fig:ideal-mitigation}(b), more than $\frac{D}{2}$ tuples of $S$ are redirected to $H$. As a result, the load reduction is again less than $\frac{D}{2}$. The ideal mitigation, shown in Figure~\ref{fig:ideal-mitigation}(c), makes the total input of the two workers equal so that they finish around the same time. In particular, $\frac{D}{2}$ tuples of $S$ are sent to $H$, which is the maximum load reduction ($LR_{max}$) that can be achieved.


\vspace{-0.08in}
\subsection{Impact of $\tau$ on load reduction}
\label{ssec:tau-impact}


In this subsection, we discuss how the load reduction is affected by the value of $\tau$ at which the mitigation starts. Assume that the operator can have only one iteration of mitigation consisting of two phases. If the second phase uses a perfect estimator and the incoming data rates are constant, as assumed in Section~\ref{sec:load-transfer-mechanism}, then the maximum load reduction of $\frac{D}{2}$ can be achieved. That is:
\begin{equation}
\label{eq:ideal-total-benefit}
  LR_1 + LR_2 = \frac{D}{2},
\end{equation}
where $LR_1$ and $LR_2$ are the load reduction resulting from the first phase and second phase, respectively.



In general, the workloads estimations have errors~\cite{conf/sigmod/ChaudhuriMN98, conf/cloud/RamakrishnanSU12, journals/tpds/ChenYX15, conf/bigdataconf/YanXM13, conf/icde/GuflerARK12}. These errors can cause the second phase to redirect less or more than the ideal amount of $S$ tuples (Figure~\ref{fig:ideal-mitigation}(a,b)). In other words, the load reduction from the second phase depends on the accuracy of workload estimation.
The workload estimation accuracy depends on $\tau$ as shown next. If $\tau$ increases, then it takes a longer time for the workload difference of $S$ and $H$ to reach $\tau$, resulting in a higher sample size. Suppose the estimation accuracy increases as the sample size increases. Then a higher $\tau$ means that the system makes a more accurate workload estimation. Thus, the total load reduction can be computed as the following:
\begin{equation}\label{eq:actual-total-benefit}
  LR = LR_1 + (1-f(\tau))LR_2,
\end{equation}
where $f(\tau)$ is a function representing the error in the estimation of the future workloads. As $\tau$ increases, $f(\tau)$ decreases. 

The above analysis shows that a higher $\tau$ results in a higher load reduction. However, setting $\tau$ to an arbitrarily high value means that the system waits a long time before starting the mitigation.  Consequently, there may not be enough future input left to mitigate the skew completely. Thus the value of $\tau$ should be chosen properly to achieve a balance between a high estimation accuracy and waiting so long that the opportunity to mitigate skew is lost. This is a classic exploration-exploitation dilemma~\cite{journals/ml/AuerCF02}.

\begin{figure}[htbp]
\begin{center}
	\includegraphics[width=3.5in]{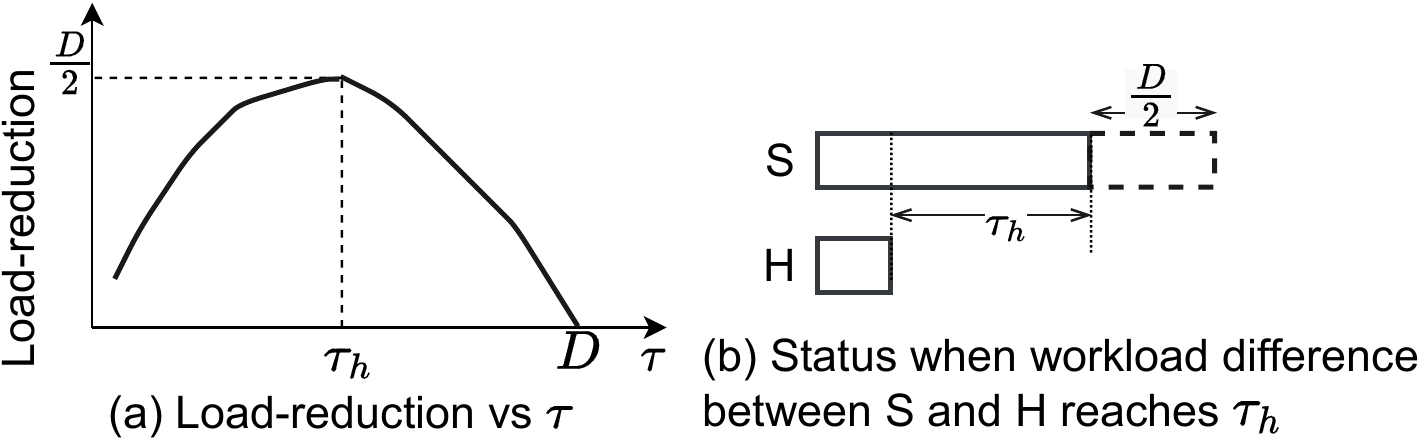} 
	\caption{\label{fig:analysis-return}
		\textbf{Dependence of load reduction on the $\tau$.}
	}
    \end{center}
\end{figure}

Figure~\ref{fig:analysis-return}(a) shows the relationship between $\tau$ and load reduction. A small $\tau$ results in a small load reduction because of a high estimation error. As $\tau$ increases, $f(\tau)$ decreases and load reduction increases. The load reduction cannot exceed $LR_{max}=\frac{D}{2}$. As $\tau$ further increases, the future input of $S$ is not enough to mitigate the skew completely. Thus, load reduction starts to decrease. Figure~\ref{fig:analysis-return}(b) shows the time when $S$ has $\frac{D}{2}$ future tuples left. The difference in the workloads of the workers at this time is denoted by $\tau_h$.


\subsection{Adaptive mitigation iterations}
\label{ssec:adjust-tau-over-iterations}

When the workloads of $S$ and $H$ diverge due to workload estimation errors, the controller may start another mitigation iteration. Section~\ref{sssec:multiple-iterations} discusses how multiple iterations of mitigation are performed. In the previous subsection, we saw that $\tau$ should be chosen appropriately to maintain a balance between workload estimation accuracy and a long delay in the start of mitigation. Section~\ref{sssec:choose-tau} shows how to autotune $\tau$ adaptively to make better workload estimations, rather than asking the user to supply an appropriate value of $\tau$.

\subsubsection{Multiple iterations of mitigation}
\label{sssec:multiple-iterations}

\begin{figure}[htbp]
\begin{center}
	\includegraphics[width=3.5in]{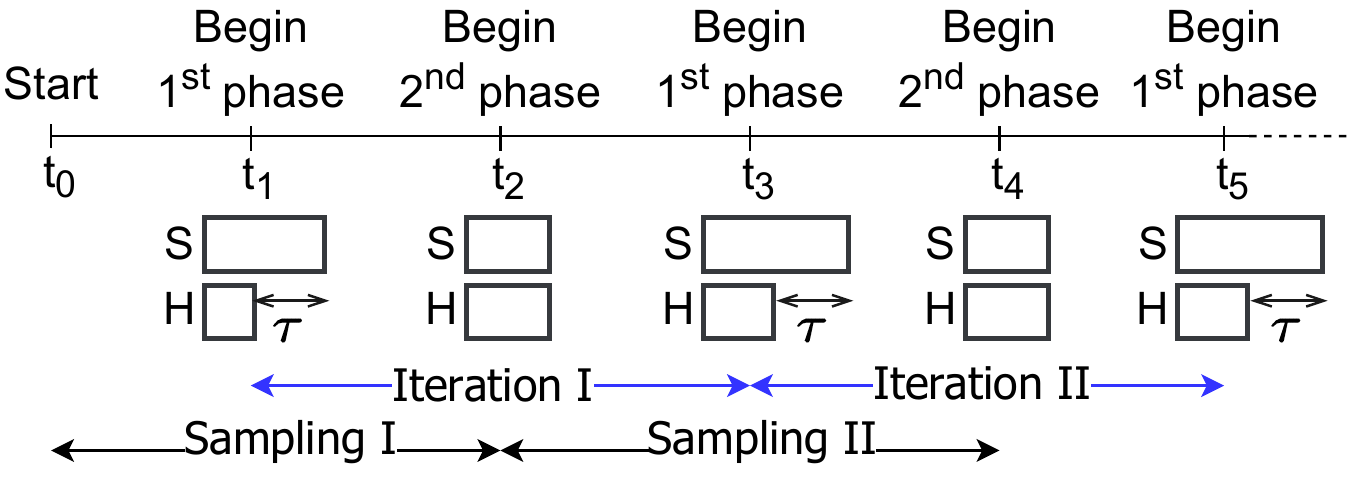}
	\caption{\label{fig:multiple-iterations}
		\textbf{Multiple mitigation iterations}
	}
    \end{center}
\end{figure}

Figure~\ref{fig:multiple-iterations} shows an example timeline of two successive iterations of mitigation. The first iteration starts at $t_1$ when the difference of the workloads of $S$ and $H$ exceeds $\tau$. Their workloads are brought to a similar level at $t_2$. Then, the second phase starts. Due to workload estimation errors, the second phase redirects less than the ideal amount of tuples. Thus, the workload of $S$ gradually becomes greater than $H$. At $t_3$, their workload difference exceeds $\tau$ and the second iteration starts.

A question is how to decide the time interval from which the sample is used to do prediction~\cite{journals/tpds/XiaoSC13, conf/sc/DiKC12}. Figure~\ref{fig:multiple-iterations} shows an example that uses the sample collected since the last time when $S$ and $H$ had a similar load. Specifically, at $t_2$, the second phase of the first iteration uses the sample collected since $t_0$. The second phase of the second iteration uses the sample collected since $t_2$.

\subsubsection{Dynamically adjusting $\tau$}
\label{sssec:choose-tau}

A low value of $\tau$ causes high errors in workload estimation due to a small sample size, which in turn results in more mitigation iterations. On the other hand, a high $\tau$ may start the mitigation too late when there are not enough future tuples to mitigate the skew. Rather than using a fixed user-provided value of $\tau$, which may be too low or too high, we adaptively adjust $\tau$'s value during execution to make better workload predictions, reduce the number of iterations, and achieve higher load reduction.



In Section~\ref{sec:load-transfer-mechanism}, we introduced an estimation function $\psi$ that uses a workload sample to estimate future workloads. Let $\varepsilon$ denote the standard error of estimation~\cite{misc/prediction-interval}, which is a measure of predicted error in workload estimation. For example, the standard error for mean-model~\cite{misc/statistical-forecasting, misc/prediction-interval} estimator is $\varepsilon = d \sqrt{1 + \frac{1}{n}}$, where $d$ is the sample standard deviation and $n$ is the sample size. As mentioned in Section~\ref{ssec:tau-impact}, $\varepsilon$ decreases as $\tau$ increases.  We want $\varepsilon$ to be in a user-defined range $[\varepsilon^{}_l,\varepsilon^{}_u]$, where $\varepsilon^{}_l$ and $\varepsilon^{}_u$ are the lower and upper limits, respectively. In particular, when $\varepsilon > \varepsilon^{}_u$, we assume the error is too high and will lead to a low load reduction. Similarly, when $\varepsilon < \varepsilon^{}_l$, the error is low enough to make a good estimation.

The controller keeps track of $\varepsilon$ and adaptively adjusts $\tau$ in order to move $\varepsilon$ towards the $[\varepsilon^{}_l,\varepsilon^{}_u]$ range. Algorithm~\ref{alg:adjusting-tau} describes the process of adjusting $\tau$. For a worker $w$, let $\phi^{}_w$ represent the current workload and $\hat{\phi}^{}_w$ represent the workload predicted by $\psi$. The controller periodically collects the current workload metrics from the workers (line~\ref{alg:collect-workload}) and adds them to the existing sample (line~\ref{alg:add-to-sample}). The function $\psi$ uses the workload sample to predict future workloads and outputs $\varepsilon$ in the prediction (line~\ref{alg:estimate}). Once $\varepsilon$ is obtained, $\tau$ can be adjusted.

\begin{algorithm}[ht]
\caption{\hl{Dynamic $\tau$ adjustment by the controller.} \label{alg:adjusting-tau}}
        \KwIn{$[\varepsilon^{}_l, \varepsilon^{}_u]  \gets$ Standard error acceptable range}
        \KwIn{$\texttt{W}$: collected workloads sample}
        \KwIn{$\tau$: current threshold}
        \KwOut{Adjusted threshold}
       \SetKwFunction{AdjustTau}{adjust-threshold}

    \SetKwProg{myproc}{Procedure}{}{}
    $\phi^{}_S, \phi^{}_H \gets$ Collect current workloads of $S$ and $H$ \\ \label{alg:collect-workload}
    Add $\langle \phi^{}_S, \phi^{}_H \rangle$ to \texttt{W} \\ \label{alg:add-to-sample}
    $\hat{\phi}^{}_S, \hat{\phi}^{}_H, \varepsilon \gets $ Estimate future workloads of $S$ and $H$ using $\psi$   \\ \label{alg:estimate}
    \vspace{0.05in}
    {\tt // adjust threshold} \\
    \uIf{$\phi^{}_S - \phi^{}_H >= \tau$ \textbf{and} $\varepsilon > \varepsilon^{}_u$}{
        \tcp{Higher sample size needed to lower $\varepsilon$}
        \textbf{return} increase-threshold($\tau$)
    }
    \uElseIf{$\phi^{}_S - \phi^{}_H < \tau$ \textbf{and} $\varepsilon < \varepsilon^{}_l$}{
        \tcp{$\varepsilon$ has become quite low }
        \textbf{return} decrease-threshold($\tau$)
    }
    \uElse{
        \textbf{return} $\tau$
    }
\end{algorithm}

\boldstart{Increasing $\tau$.} The need to increase $\tau$ arises when the workers $S$ and $H$ pass the {\em skew-test} (Section~\ref{ssec:skew-detection}), but $\varepsilon > \varepsilon^{}_u$. This means that a higher sample size is needed to lower $\varepsilon$. At this point, the mitigation is started and an increased $\tau$ is chosen for the next iteration to achieve a smaller $\varepsilon$. The threshold $\tau$ should be cautiously increased so as to not set it to a very high value (Section~\ref{ssec:tau-impact}). 


\boldstart{Decreasing $\tau$.} Now consider the case where $S$ and $H$ do not pass the {\em skew-test} because their workload difference is less than $\tau$, but $\varepsilon < \varepsilon^{}_l$. This means that $\varepsilon$ is low and the sample size is big enough to yield a good accuracy. If we wait for the workload difference to reach $\tau$, there may not be enough data left to mitigate the skew. Thus, $\tau$ is decreased to the current workload difference ($\phi^{}_S - \phi^{}_H$) and mitigation starts right away, thus yielding a higher load reduction. 

\section{\frmname on more operators}
\label{sec:other-operators}

Till now we used the running example of skew in the probe input of {\sf HashJoin}. A data analysis workflow can contain many
operators that are susceptible to partitioning skew such as {\sf sort} and {\sf group by}. In this section, we generalize \frmname to a broader set of operators. Specifically, we formalize the concept of ``operator state mutability'' in Section~\ref{ssec:formalize-state}. In Section~\ref{ssec:immutable-state-operators}, we discuss the impact of state mutability on state migration. In Sections~\ref{ssec:approach-1-mutable} and \ref{ssec:approach-2-mutable}, we use the load-transfer approaches described in Section~\ref{sec:load-transfer-mechanism} to handle skew in mutable-state operators. We discuss a state migration challenge when using the ``split by records'' approach and explain how to handle it.

\subsection{Mutability of operator states }
\label{ssec:formalize-state}

In this subsection, we define two types of operator states, namely {\em immutable state} and {\em mutable state}. When an operator receives input partitioned by keys, the state information of keys is stored in the operator as {\em keyed states}~\cite{journals/pvldb/CarboneEFHRT17}. Each keyed state is a mapping of type 
\vspace{-0.06in}
$$scope\rightarrow val,$$

\vspace{-0.06in}
\noindent
where $scope$ is a single key or a set or range of keys, and $val$ is information associated with the $scope$. For example, in {\sf HashJoin}, each join key is a $scope$, and the list of build tuples with the key is the corresponding $val$. Similarly, in a hash-based implementation of {\sf group-by}, each individual group is a $scope$, and the aggregated value for the group is the corresponding $val$. In a range-partitioned {\sf sort} operator, a range of keys is a $scope$, and the sorted list of tuples in the range is the corresponding $val$. In the rest of this section, for simplicity, we use the term ``state'' to refer to ``keyed state.''

An input tuple uses the state associated with the $scope$ of the key of the tuple. If the $val$ of this $scope$ cannot change, we say the state is {\em immutable}; otherwise, it is called {\em mutable}.
For example, the processing of a probe tuple in {\sf HashJoin} does not modify the list of build tuples for its key. Such operators whose states are immutable are called {\em immutable-state operators}. On the other hand, an input tuple to {\sf sort} is added to the sorted list associated with its $scope$ (range of keys), thus it modifies the state. Such operators that have a mutable state are called {\em mutable-state operators}. 

Notice that the execution of an operator can have more than one phase.  For instance, a {\sf HashJoin} operator has two phases, namely the build phase and the probe phase. The concept of mutability is with respect to a specific phase of the operator.  In {\sf HashJoin}, the states in the build phase are mutable, while the states in the probe phase are immutable.  \frmname is applicable to a specific phase, and its state migration depends on the mutability of the phase. Table~\ref{table:operator-types} shows a few examples of immutable-state and mutable-state operators.

\begin{table}[htbp]
\small{
\begin{tabular}{|p{3.3cm}|p{11.5cm}|}
\hline
\textbf{Immutable-state operator} & {\sf HashJoin}~(Probe phase), {\sf HB Set Difference}~(Probe phase), {\sf HB Set Intersection}~(Probe phase) \\ \hline
\textbf{Mutable-state operator} & {\sf HashJoin}~(Build phase), {\sf HB Group-by}, {\sf RB Sort}, {\sf HB Set Difference}~(Build phase), {\sf HB Set Intersection}~(Build phase), {\sf HB Set Union} \\ \hline 
\end{tabular}
}
\textit{}
\caption{Examples of physical operators based on state mutability. $HB$ means hash-based and $RB$ means range-based.}
\label{table:operator-types}
\end{table}

\subsection{Impact of mutability on state migration}
\label{ssec:immutable-state-operators}

\begin{figure}[htbp]
\begin{center}
	\includegraphics[width=4.0in]{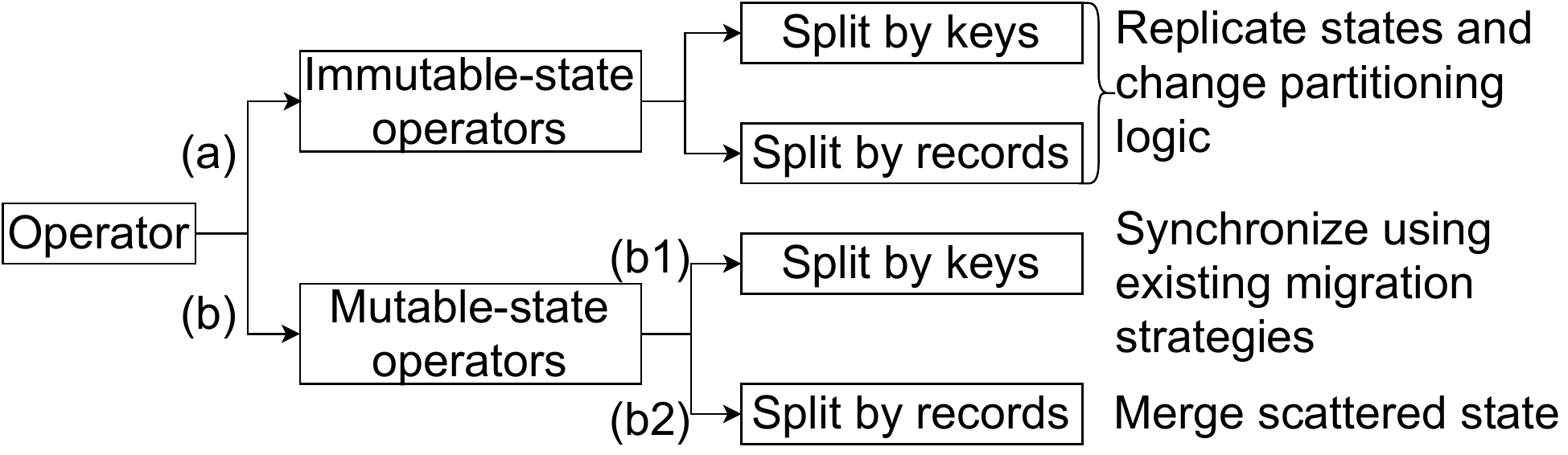} 
	\caption{\label{fig:operator-tree}
		\textbf{Operator state mutability and state migration.}
	}
    \end{center}
\end{figure}

Figure~\ref{fig:operator-tree} shows how to handle state migration for operators when using the two load-transfer approaches discussed in Section~\ref{sec:load-transfer-mechanism}. 

The state-migration process for immutable-state operators, as shown in branch~(a) in Figure~\ref{fig:operator-tree}, involves replicating the skewed worker's states at the helper, followed by a change in the partitioning logic. Thus, the tuples redirected from the skewed worker to the helper can use the state of their $scope$ at the latter. In contrast, the state-migration process is more challenging for mutable-state operators (branch~(b)) because it is difficult to synchronize the state transfer and change of partitioning logic for a mutable state~\cite{journals/pvldb/MaiZPXSVCKMKDR18}. State-migration strategies that focus on such synchronization exist in the literature and will be briefly discussed in Section~\ref{ssec:approach-1-mutable}. As we show in Section~\ref{ssec:approach-2-mutable}, such a synchronization is not always possible. Next, we discuss how to do state migration when using the two load-transfer approaches in mutable-state operators.



\subsection{Mutable-state operators: split by keys}
\label{ssec:approach-1-mutable}

The {\sf SBK} approach offloads the processing of certain keys in the skewed worker partition to the helper. Consider a {\sf group-by} operator that receives covid related tweets and aggregates the count of tweets per month. The skewed worker offloads the processing of a month (say, June) to the helper. There needs to be a synchronization between state transfer and change of partitioning logic so that the redirected June tuples arriving at the helper use the state formed from all June tuples received till then. In the case of {\sf group-by}, this state is the count of all June tuples received by the operator.  Existing work on state-migration strategies focuses on this synchronization. A simple way to do this synchronization is to pause the execution, migrate the state, and then resume the execution~\cite{conf/sigmod/ArmbrustDTYZX0S18,journals/pvldb/CarboneEFHRT17,conf/icde/ShahHCF03}. A drawback of this approach is that pausing multiple times for each iteration may be a significant overhead. Another strategy is to use markers~\cite{journals/pvldb/ElseidyEVK14}. The workers of the previous operator emit markers when they change the partitioning logic. When the markers from all the previous workers are received by the skewed and helper workers, the state can be safely migrated. Thus, skew handling in mutable-state operators using the ``split by keys'' approach can be safely done by using one of these state-migration strategies (branch~(b1) in Figure~\ref{fig:operator-tree}).

\subsection{Mutable-state operators: split by records}
\label{ssec:approach-2-mutable}

\begin{figure}[htbp]
\begin{center}
	\includegraphics[width=4.5in]{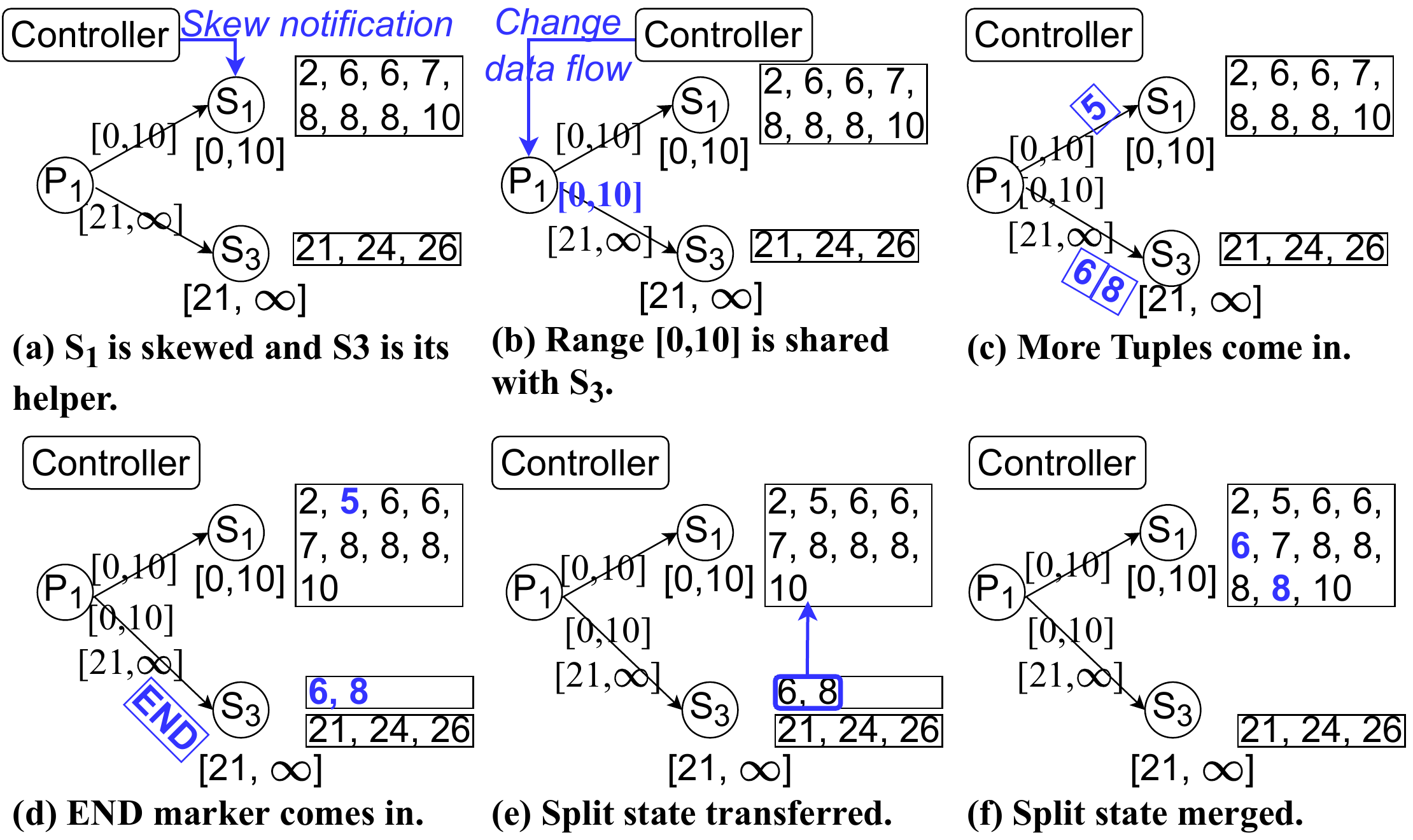} 
	\caption{\label{fig:reshape-on-sort}
		\textbf{Skew handling using the ``split by records'' approach in the sort operator. $S_2$ is omitted for simplicity.}
	}
    \end{center}
\end{figure}

In this subsection, we use the {\sf SBR} approach in mutable-state operators (branch~(b2) in Figure~\ref{fig:operator-tree}). We show that the synchronization between state transfer and change of partitioning logic is not possible when using this approach and discuss its effects. Consider a sort operator with three workers, namely $S_1$, $S_2$, and $S_3$, which receive range-partitioned inputs. The ranges assigned to the three workers are $[0,10]$, $[11,20]$, and $[21,\infty]$. As shown in Figure~\ref{fig:reshape-on-sort}(a), $S_1$ is skewed and $S_3$ is its helper. The controller asks the previous operator to change its partitioning logic and send the tuples in $[0,10]$ to both $S_1$ and $S_3$ (Figures~\ref{fig:reshape-on-sort}(b,c)). The synchronization of state migration and change of partitioning logic by the aforementioned state-migration strategies relies on an assumption that, at any given time, the partitioning logic sends tuples of a particular $scope$ to a single worker only. When the tuples of $[0,10]$ are sent to both $S_1$ and $S_3$, this assumption is no longer valid. Worker $S_3$ saves the tuples from the range $[0,10]$ in a separate sorted list (Figure~\ref{fig:reshape-on-sort}(d)). Such a scenario where the $val$ of a $scope$ is split between workers is referred to as a {\em scattered state}.

This scattered state needs to be merged before outputting the results to the next operator. Now we explain a way to resolve the scattered state problem. When a worker of the previous operator finishes sending all its data, it notifies the sort workers by sending an {\sf END} marker (Figure~\ref{fig:reshape-on-sort}(d)). When $S_3$ receives {\sf END} markers from all the previous workers, it transfers its tuples in the range $[0,10]$ to the correct destination of those tuples, i.e., $S_1$ (Figure~\ref{fig:reshape-on-sort}(e,f)), thus merging the scattered states for the $[0,10]$ range. 

We specify sufficient conditions for a mutable-state operator to be able to resolve the scattered state issue. The above approach of merging the scattered parts is suited for blocking operators such as group-by and sort, which produce output only after processing all the input data. Thus, the above approach can be used by mutable-state operators if they can 1) combine the scattered parts of the state to create the final state, and 2) block outputting the results till the scattered parts of the state have been combined. 
\vspace{-0.05in}
\section{\frmname in Broader Settings}
\label{sec:broader-settings}

Our discussion about \frmname so far is based on several assumptions in Section~\ref{sec:overview} for simplification. Next we relax these assumptions.

\vspace{-0.05in}
\subsection{High state-migration time}

The state-migration time is assumed to be small till now.  In this subsection, we study the case where this time could be significant. 

\boldstart{Precondition for skew mitigation.} In the discussion in Section~\ref{sec:overview}, state migration is started immediately after skew detection. If the time to migrate state is more than the time left in the execution, the state migration is futile. Thus, the controller checks if the estimated state-migration time is less than the estimated time left in the execution and only then proceeds with state migration. The state-migration time can be estimated based on factors such as state-size and serialization cost~\cite{conf/networking/YunLWRK20,journals/corr/DingFMWYZC15}. The time left in the execution can be estimated by monitoring the input data remaining to be processed and the processing speed~\cite{conf/sigmod/KwonBHR12} or by using the historical data~\cite{conf/icac/GuptaMD08}. 

\begin{figure}[htbp]
\begin{center}
	\includegraphics[width=3.3in]{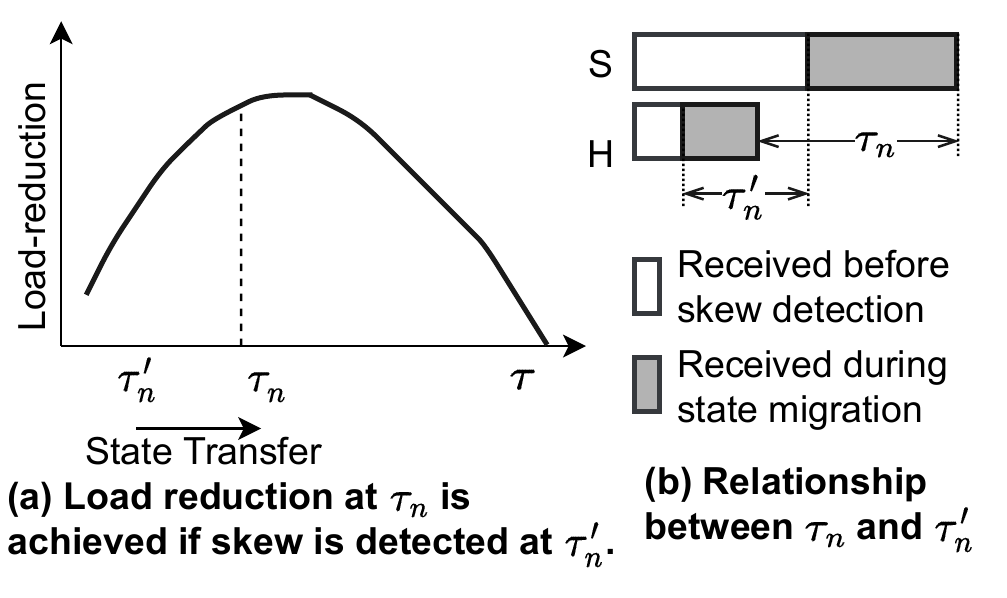} 
	\caption{\label{fig:state-transfer-effect}
		\textbf{Adapt $\tau$ by considering the state-transfer time.}
	}
    \end{center}
\end{figure}

\boldstart{Dynamic adaptation of $\tau$.} Suppose the adapted value of $\tau$ output by Algorithm~\ref{alg:adjusting-tau} to be used in the next iteration is $\tau_{n}$. The discussion in Section~\ref{sssec:choose-tau} assumes that the load transfer begins when the workload difference is around $\tau_{n}$. This is possible only when the state-migration time is small. When the time is significant, the load transfer will start when the workload difference becomes considerably greater than $\tau_{n}$. In order to start the load transfer at $\tau_{n}$ (as assumed by Section~\ref{sssec:choose-tau}), the skew has to be detected earlier. Thus, we adjust the skew detection threshold to $\tau_n'$, which is less than $\tau_n$, such that the state migration starts when the workload difference is $\tau_n'$ and ends when the workload difference is $\tau_n$ (Figure~\ref{fig:state-transfer-effect}). 

Formally, suppose $t$ is the number of tuples processed by the operator per unit time, $M$ is the estimated state-migration time, and $\hat{f}_S$ and $\hat{f}_{H}$ are the predicted workload percentages of $S$ and $H$, respectively. The estimated difference in the tuples received by $S$ and $H$ during the state migration is $(\hat{f}_S - \hat{f}_H)*t*M$. Therefore, given $\tau_{n}$, the value of $\tau_{n}'$ can be calculated as follows:
\vspace{-0.03in}
$$ \tau_{n}' = \tau_{n} - (\hat{f}_S - \hat{f}_H)*t*M.$$



\subsection{Multiple helper workers}
\label{ssec:multiple-helper-workers}

Till now we have assumed a single helper per skewer worker. Next, we extend \frmname to the case of multiple helpers.



\boldstart{Load reduction.} The load reduction definition (Section~\ref{ssec:benefit}) can be extended for $S$ and its helpers $h_1,\ldots,h_n$ as follows:
$$
  LR = \max_{w \in \{S,h_1,h_2,\ldots,h_n\}}(\sigma_w) - \max_{w \in \{S,h_1,h_2,\ldots,h_n\}}(\sigma'_w).
$$
In the equation, $\sigma_w$ and $\sigma'_w$ are the sizes of the total input received by worker $w$ during the entire execution in the unmitigated case and mitigated case, respectively. Suppose $T$ is the total number of tuples received by the operator and $f_w$ is the actual workload percentage of a worker $w$. In the unmitigated case, $S$ receives the maximum total input among $S$ and its helpers, which is $f_S*T$ tuples. In the ideal mitigation case, $S$ and its helpers have the same workload, which is the average of the workloads that they would have received in the unmitigated case. As discussed in Section~\ref{ssec:benefit}, the ideal mitigation results in maximum load reduction denoted as:
$$
  LR_{max} = \big(f_S - \dfrac{\sum_{w\in \{S,h_1,h_2,\ldots,h_n\}}{f_w}}{n+1}\big) * T.
$$

\begin{figure}[htbp]
\begin{center}
	\includegraphics[width=3.5in]{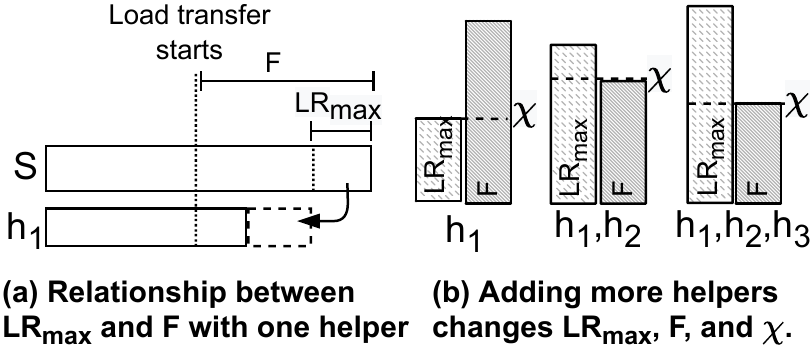} 
	\caption{\label{fig:multiple-helpers}
		\textbf{Choosing appropriate helpers.}
	}
    \end{center}
\end{figure}

\boldstart{Choosing appropriate helpers.} We examine the trade-off between the load reduction and the state-migration overhead to determine an appropriate set of helpers for $S$. Let $h_1,\ldots,h_c$ be $c$ helper candidates for $S$ in the increasing order of their workloads. From the definition above, increasing the number of helpers results in a higher $LR_{max}$, provided the average workload percentage reduces. However, increasing the number of helpers may result in higher state-migration time since more data needs to be transferred. Suppose $L$ is the number of future tuples to be processed by the operator at the time of skew detection. The estimated number of future tuples left to be processed by $S$ after state migration is $F = (L-M*t)*\hat{f}_S$.
Increasing the number of helpers may increase the state-migration time ($M$) and thus decrease $F$, which means that there are fewer future tuples of $S$ to do load transfer. Thus, given a set of helpers, the highest possible load reduction after state migration is $\chi = min(LR_{max}, F)$. As we add more helpers, $\chi$ initially increases and then starts decreasing. The set of helpers chosen right before $\chi$ starts decreasing are appropriate. Figure~\ref{fig:multiple-helpers} illustrates an example. Let $W$ be the set of helper workers, which is initially empty. After adding $h_1$ to $W$, we have $LR_{max} < F$, thus $\chi = LR_{max}$. Then, we add $h_2$ to $W$, which decreases $F$, and $\chi = F$. Then, we add $h_3$ to $W$, which decreases $F$ further and causes $\chi$ to start decreasing. Hence, the final set of helpers for $S$ is $\{h_1,h_2\}$.


\subsection{Unbounded data}
\label{ssec:infinte-data}

The input has been assumed to be bounded till now. Next, we discuss a few considerations when the input is unbounded.

\boldstart{Load reduction and impact of $\tau$.} In Section~\ref{ssec:benefit}, the load reduction was calculated based on the total input received by the workers. For the unbounded case, the load reduction can be calculated based on the input received by the workers in a fixed period of time. The impact of $\tau$ on the load reduction holds for unbounded case too. A small value of $\tau$ results in high errors in workload estimation, which leads to a small load reduction. A large value of $\tau$ that takes too long to reach is not preferred in the unbounded case either. If a large $\tau$ delays mitigation, it can lead to back pressure, loss of throughput, and even crashing of data-processing pipelines. The latency of processing can increase, causing adverse effects on time-sensitive applications such as image classification in surveillance~\cite{conf/osdi/HsiehABVBPGM18}.

\boldstart{Merging scattered states.} For bounded data, the scattered states in mutable-state operators were merged after the operator processed all the input. For unbounded data, the scattered states can be merged when the operator has to output results, e.g., when a watermark is received~\cite{journals/pvldb/BegoliACHKKMS21}.

\section{Experiments}
\label{sec:experiments}

In this section, we present an experimental evaluation of \frmname using real and synthetic data sets on clusters.

\subsection{Setting}
\begin{figure*}[htbp]
	\includegraphics[width=\linewidth]{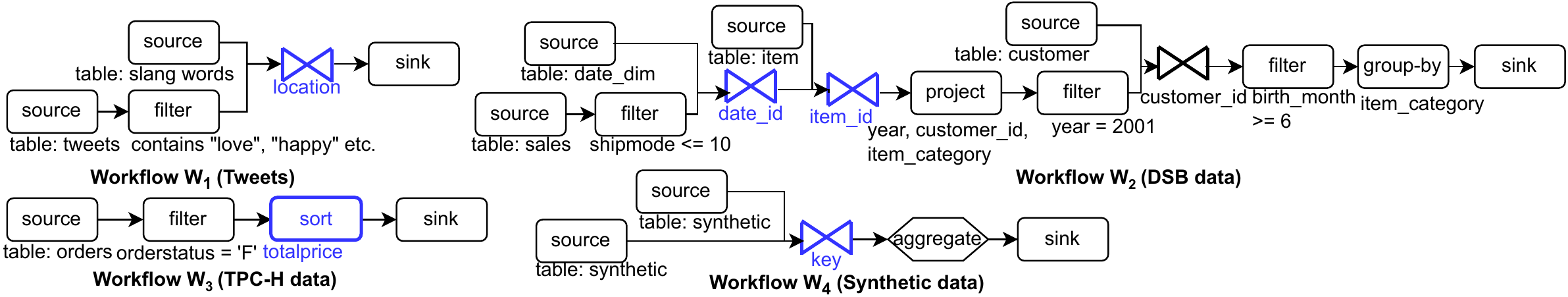} 
	\caption{\label{fig:workflows-experiments}
		\textbf{Workflows used in the experiments. The operators with skew are the join operator on location in $W_1$, the join operators on date\_id and item\_id in $W_2$, the sort operator in $W_3$ and the join operator on key in $W_4$.}
	}
\end{figure*}
\noindent \textbf{Datasets and workflows.} 
We used four datasets in the experiments. The first one included $180$M tweets in the US between 2015 and 2021 collected from Twitter. The second dataset was generated using the {\sf DSB} benchmark~\cite{journals/pvldb/DingCGN21}, which is an enhanced version of {\sf TPC-DS} containing more skewed attributes, to produce record sets of different sizes ranging from $100$GB to $200$GB by varying the scaling factor. The third dataset was generated using the {\sf TPC-H} benchmark~\cite{misc/TPC-H} to produce record sets ranging from $50$GB to $200$GB. The fourth dataset was generated to simulate a changing key distribution during the execution. It included a synthetic table of $80$M tuples and another table of $4,200$ tuples, and each table had two numerical attributes representing keys and values. 

\begin{figure}[htbp]
     \centering
     \begin{subfigure}[t]{0.32\columnwidth}
          \includegraphics[width=0.9\linewidth]{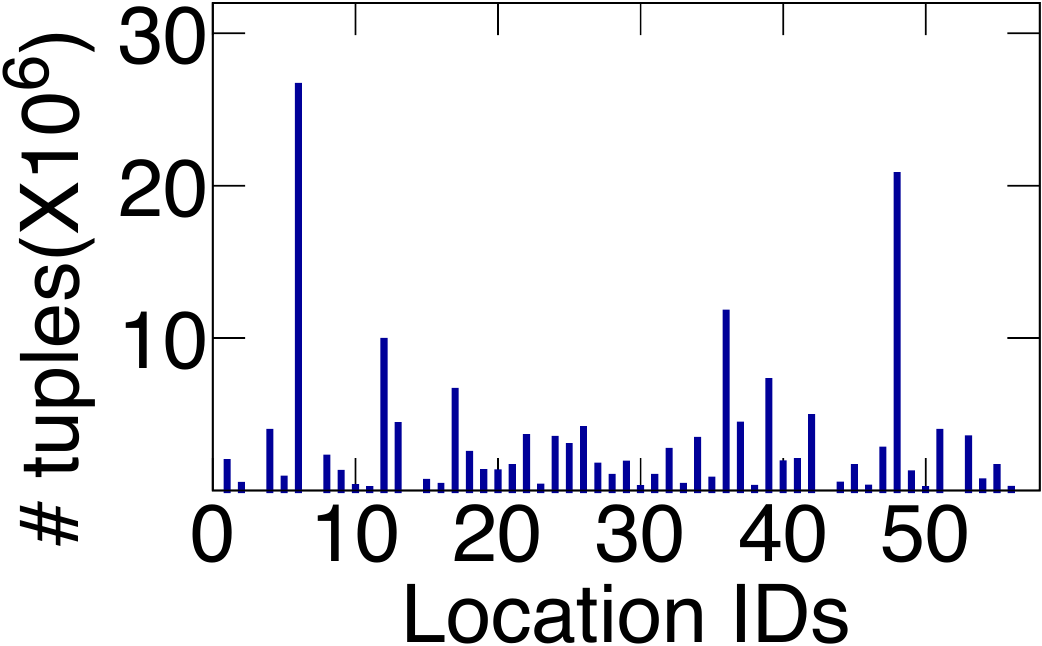}
          \caption{Tweet data.}
          \label{fig:tweet}
     \end{subfigure}
     \hfill
     \begin{subfigure}[t]{0.32\columnwidth}
          \includegraphics[width=0.9\linewidth]{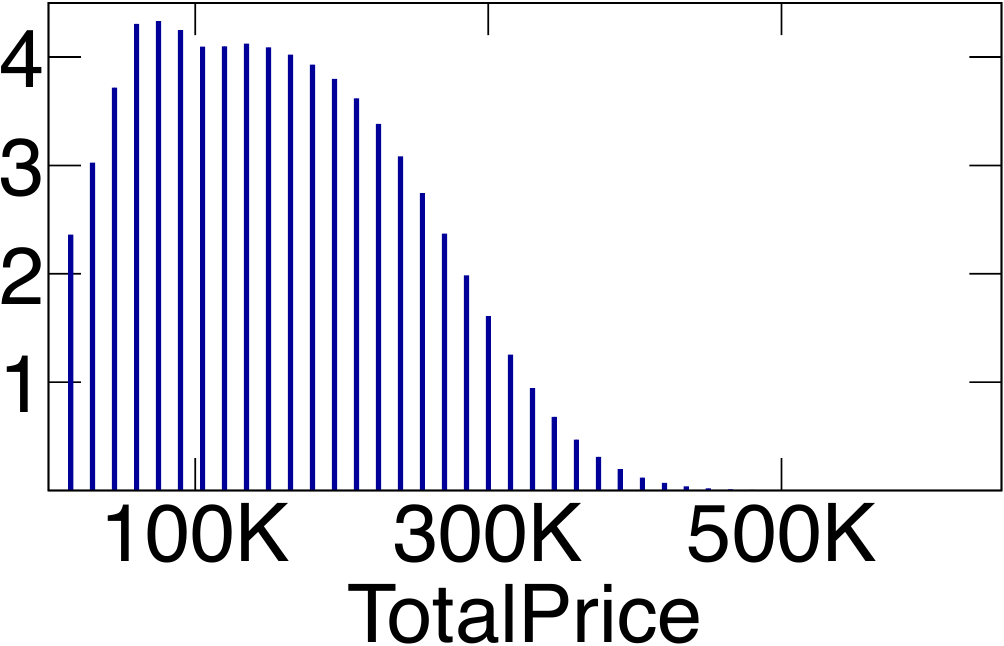}
          \caption{TPC-H data.}
          \label{fig:reshape-tpch}
     \end{subfigure}
     \begin{subfigure}[t]{0.32\columnwidth}
          \includegraphics[width=0.9\linewidth]{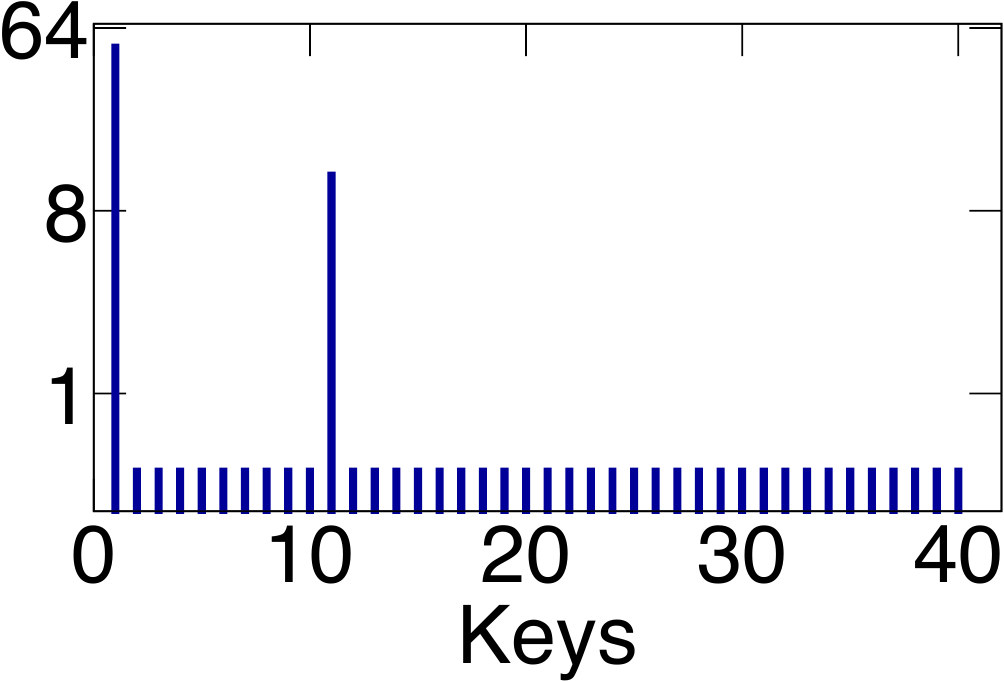}
          \caption{Synthetic data.}
          \label{fig:synthetic}
     \end{subfigure}
     \begin{subfigure}[t]{0.32\columnwidth}
          \includegraphics[width=0.9\linewidth]{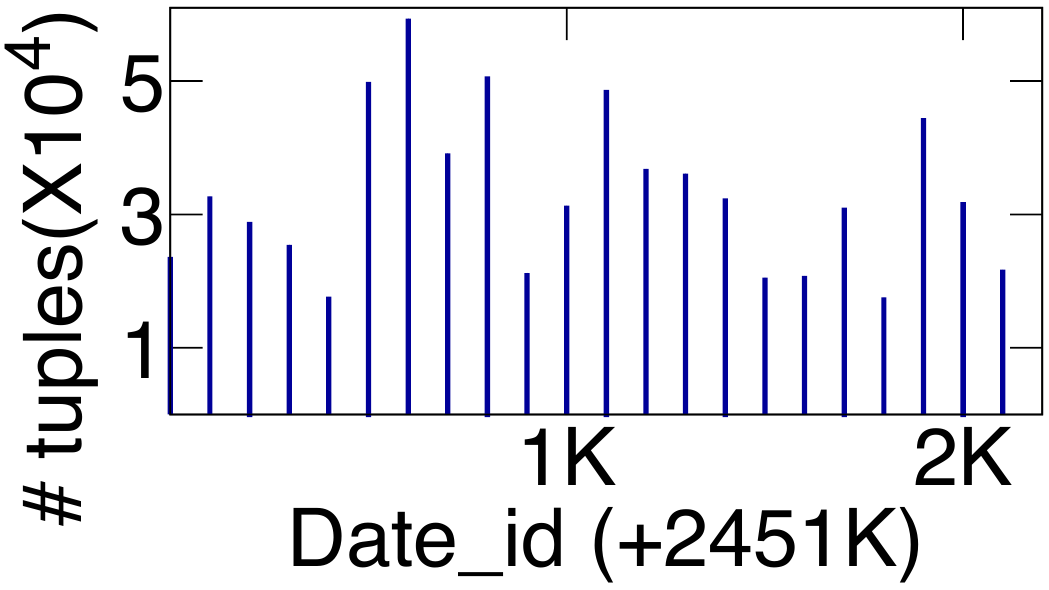}
          \caption{DSB sales data (date column).}
          \label{fig:dsb-date}
     \end{subfigure}
     \begin{subfigure}[t]{0.32\columnwidth}
          \includegraphics[width=0.9\linewidth]{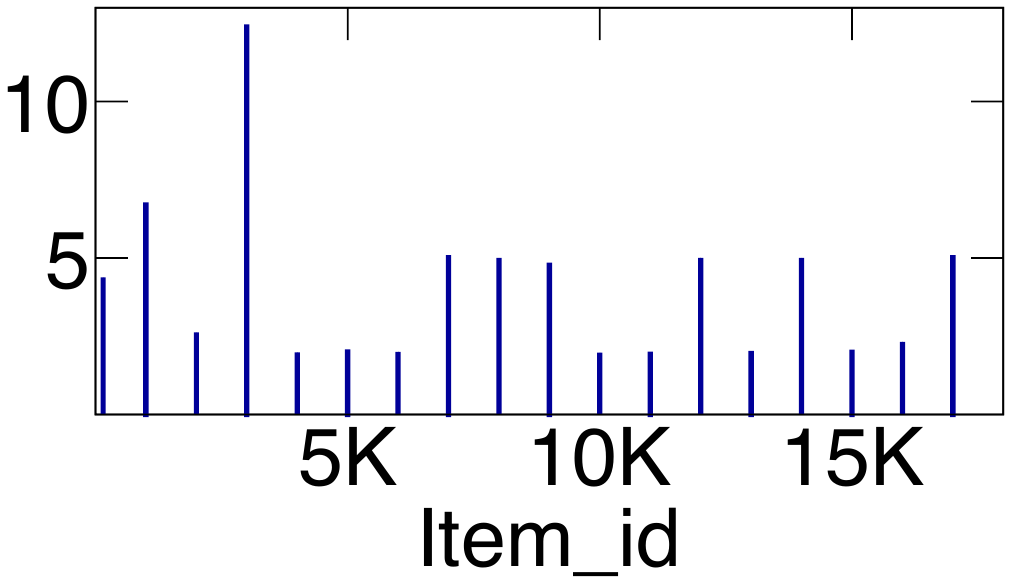}
          \caption{DSB sales data (item column).}
          \label{fig:dsb-item}
     \end{subfigure}
     \begin{subfigure}[t]{0.32\columnwidth}
          \includegraphics[width=0.9\linewidth]{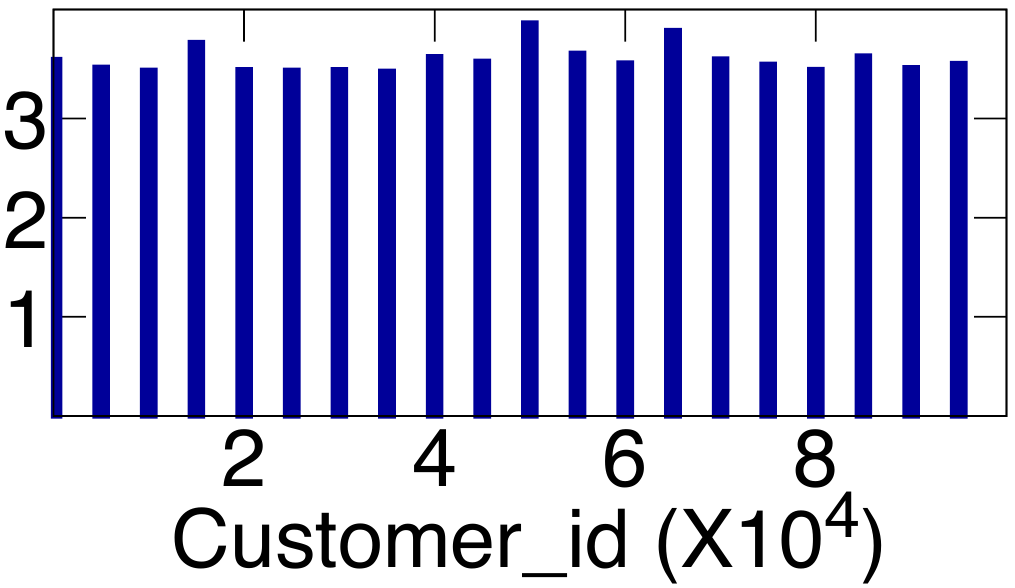}
          \caption{DSB sales data (customer column).}
          \label{fig:dsb-customer}
     \end{subfigure}
    \caption{\textbf{Partitioning-key distributions for the datasets.}}
    \label{fig:key-distribution}
\end{figure}

We constructed workflows of varying complexities as shown in Figure~\ref{fig:workflows-experiments}. 
Workflow $W_1$ analyzed tweets by joining them with a table of the top slang words from the location of the tweet. This workflow is used for social media analysis to find how often people use local slang in their tweets.
The tweets were filtered on certain keywords to get tweets of a particular category. Workflow $W_2$ was constructed based on {\sf TPC-DS} query $18$, and it calculated the total count per item category for the web sales in the year $2001$ by customers whose $birth\_month>=6$. Workflow $W_3$ read the {\tt Orders} table from the {\sf TPC-H} dataset and filtered it on the {\tt orderstatus} attribute before sorting the tuples on the {\tt totalprice} attribute.  Workflow $W_4$ joined the two synthetic tables on the key attribute. Figure~\ref{fig:key-distribution} shows the distribution of the datasets that may cause skew in the workflows. Figure~\ref{fig:tweet} shows the frequency of tweets, used in $W_1$, based on the location attribute. Figure~\ref{fig:reshape-tpch} shows the distribution of the {\tt Orders} table on its {\tt totalprice} attribute, used in $W_3$, for a $100$GB TPC-H dataset. Figure~\ref{fig:synthetic} shows the distribution of the larger synthetic table in $W_4$ on the key attribute. Figures~\ref{fig:dsb-date}-\ref{fig:dsb-customer} show the distribution of the three attributes of the sales table in $W_2$ used in the three join operations for a $1$GB dataset.

\boldstart{Reshape implementation.} We implemented \frmname on top of two open source engines, namely {\sf Amber}~\cite{journals/pvldb/KumarWNL20} and Apache {\sf Flink} (release 1.13).  In Amber, we used its native API to implement the control messages used in \frmname. Unless otherwise stated, we set both $\tau$ and $\eta$ to $100$. We used the mean model~\cite{misc/statistical-forecasting} to predict the workload of workers. In Flink, we used the {\em busyTimeMsPerSecond} metric of each task, which is the time ratio for a task to be busy, to determine the load on a task. We leveraged the mailbox of tasks (workers) to enable the control messages to change partitioning logic.  The control messages are sent to the mailbox of a task, and these messages are processed with a higher priority than data messages in a different channel. Using these control messages, we implemented the two phases of the {\sf SBR} load transfer approach on Flink as discussed in Section~\ref{sec:load-transfer-mechanism}.


\boldstart{Baselines.} For comparison purposes, we also implemented {\sf Flow-Join}~\cite{conf/icde/RodigerIK016} and {\sf Flux}~\cite{conf/icde/ShahHCF03} on Amber with a few adaptations. For {\sf Flow-Join}, we used a fixed time duration at the start to find the overloaded keys. The workload on a worker was measured by its input queue size. For {\sf Flow-Join}, after skew is detected, the tuples of the overloaded keys are shared with the helper worker in a round-robin manner. For {\sf Flux}, after skew is detected, the processing of an appropriate set of keys is transferred from the skewed worker to its helper. For both \frmname and the baselines,  one helper worker was assigned per skewed worker, unless otherwise stated. Also, unless otherwise stated, {\sf Flux} used a $2$ second initial duration to detect overloaded keys. To be fair, when running \frmname, we also had an initial delay of $2$ seconds to start gathering metrics and subsequent skew handling by \frmname.

All experiments were conducted on the Google Cloud Platform (GCP). The data was stored in an HDFS file system on a GCP dataproc storage cluster of 6 e2-highmem-4 machines, each with 4 vCPU's, 32 GB memory, and a 500GB HDD. 
The workflow execution was on a separate processing cluster of e2-highmem-4 machines with a 100GB HDD, running the Ubuntu 18.04.5 LTS operating system. In all the experiments, one machine was used to only run the controller. We only report the number of data-processing machines. The number of workers per operator was equal to the total number of cores in the data-processing machines and the workers were equally distributed among the machines.

\subsection{Effect on results shown to the user}
\label{ssec:results-produced}
\begin{figure}[htbp]
\begin{center}
	\includegraphics[width=3.3in]{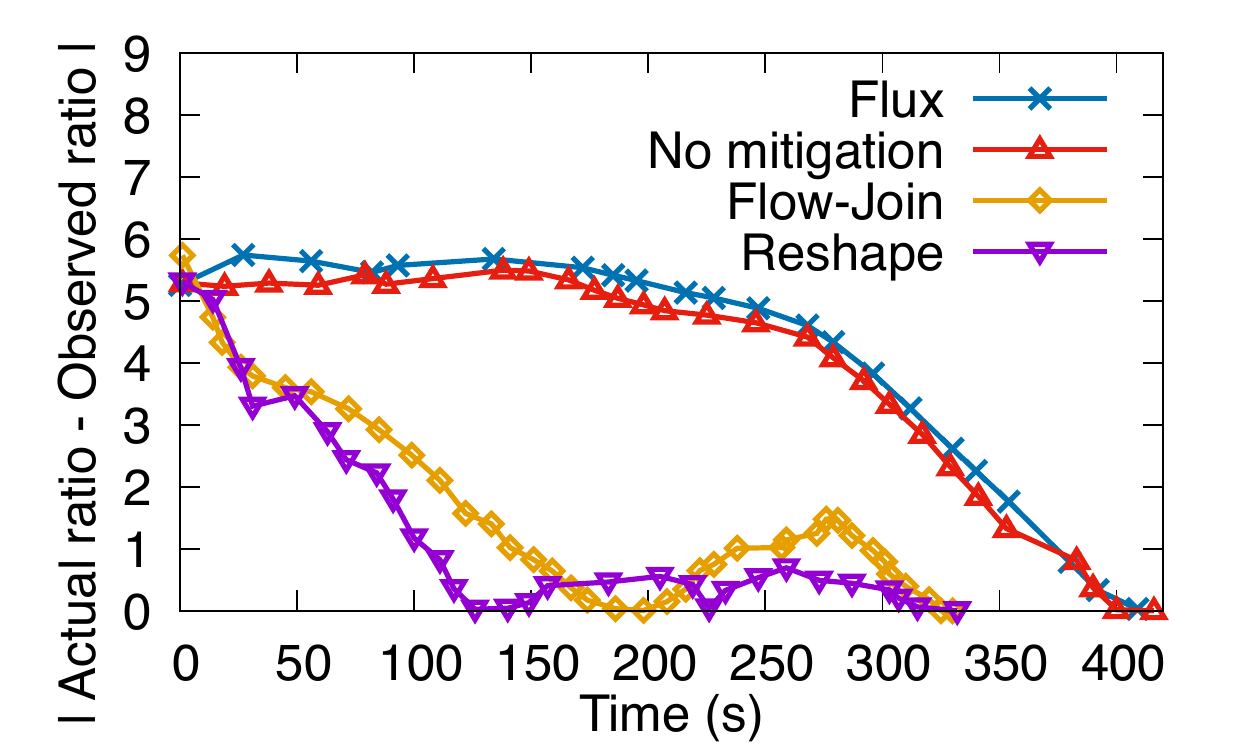} 
	\caption{\label{fig:tweet-ratio-az}
	\textbf{Effect of the mitigation strategies on the ratio of CA to AZ tweets. The ideal curve is a straight line at $y=0$.}}
\end{center}
\end{figure}

\begin{figure}[htbp]
\begin{center}
	\includegraphics[width=3.3in]{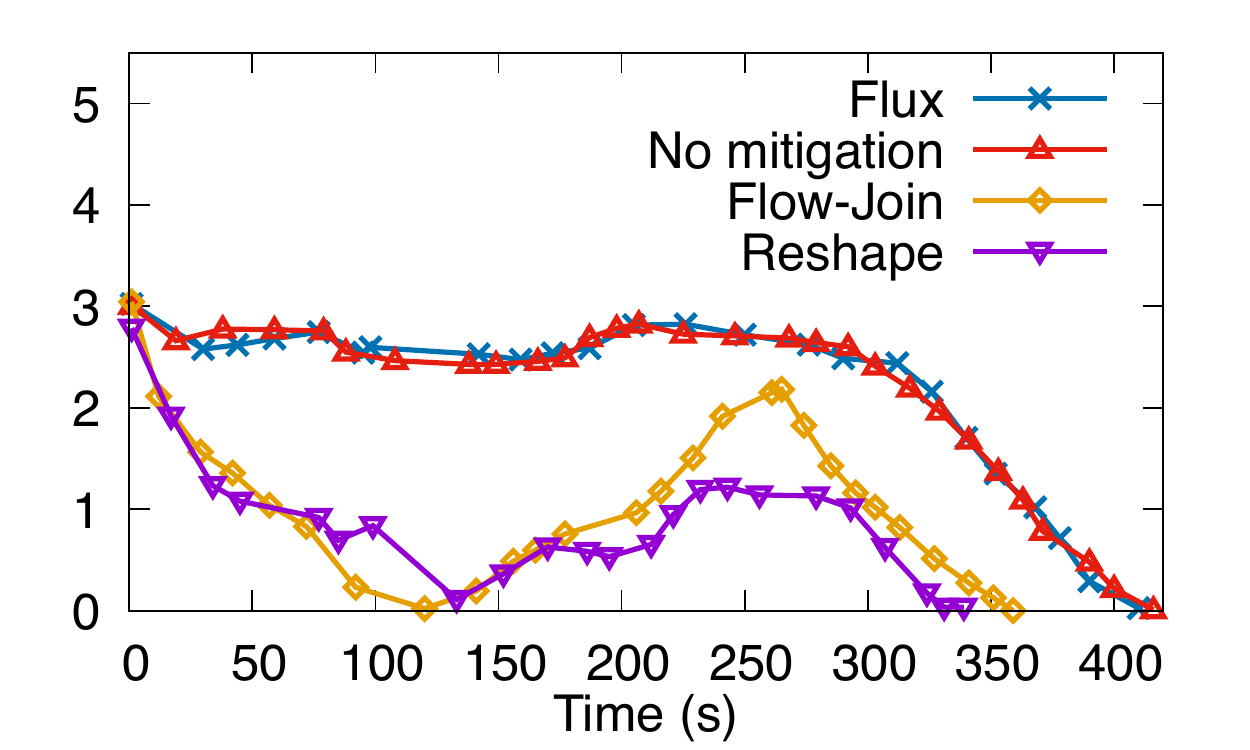} 
	\caption{\label{fig:tweet-ratio-il}
	\textbf{Effect of the mitigation strategies on the ratio of CA to IL tweets. The ideal curve is a straight line at $y=0$.}}
\end{center}
\end{figure}

We evaluated the effect of skew and the different mitigation strategies on the results shown to the user. We ran the experiment on $48$ cores ($12$ machines). California (location $6$) produced the highest number of tweets ($26$M) in the tweet dataset. Arizona (location $4$) and Illinois (location $17$) produced $3.8$M and $6.5$M tweets, respectively. In the unmitigated case, the tuples of California (CA), Arizona (AZ), and Illinois (IL) were processed by workers $6$, $4$, and $17$, respectively. We performed two sets of experiments, in which we mitigated the load on worker $6$ processing CA tweets by using different helper workers. In the first set of experiments, we used worker $4$ as the helper and monitored the ratio of CA to AZ tweets processed by the join operator. In the second set, we used worker $17$ as the helper and monitored the ratio of CA to IL tweets processed by the join operator. The line charts in Figure~\ref{fig:tweet-ratio-az} and \ref{fig:tweet-ratio-il} show the absolute difference of the observed ratio from the actual ratio as execution progressed. In the tweet dataset, the actual ratio of CA to AZ tweets was $6.85$ and CA to IL tweets was $4.05$. 


{\bf No mitigation}: When there was no mitigation, the CA, AZ, and IL tweets were processed at a similar rate as explained in Section~\ref{ssec:two-load-transfer-approaches}. The observed ratio remained close to $1$ till worker $4$ was about to finish processing AZ tweets in Figure~\ref{fig:tweet-ratio-az} and worker $17$ was about to finish IL tweets in Figure~\ref{fig:tweet-ratio-il}. The observed ratio started to increase (absolute difference of observed ratio with actual ratio started to decrease) after that because worker $6$ continued to process CA tweets. The actual ratio was observed near the end of execution (about $416$ seconds) in the unmitigated case.

{\bf Flux}: It used the {\sf SBK} load-transfer approach. It had the limitation of not being able to split the processing of a single key over multiple workers. The skewed worker $6$, apart from CA, was also processing the tweets from West Virginia. The processing of the tweets from West Virginia (about $600$K) was moved to the helper worker by Flux. However, this did not affect the observed ratio of tweets much.

{\bf Flow-Join}: It used the {\sf SBR} approach. The execution finished earlier because the approach mitigated the skew in worker $6$. {\sf Flow-Join} had two drawbacks. First, it did not perform mitigation iteratively. It changed its partitioning logic only once based on the heavy hitters detected initially. Second, it did not consider the loads on the helper and the skewed worker while deciding the portion of the skewed worker's load to be transferred to the helper. It always transferred $50$\% of the load of the skewed worker to the helper. The observed ratio of tweets started increasing once skew mitigation started. It reached the actual ratio $198$ seconds in Figure~\ref{fig:tweet-ratio-az} and around $120$ seconds in Figure~\ref{fig:tweet-ratio-il}. Due to the aforementioned drawbacks, the observed ratio of tweets continued to increase even after reaching the actual ratio because the skewed worker continued to transfer 50\% of its load to the helper. The observed ratio continued to increase till it reached about $8.3$ in Figure~\ref{fig:tweet-ratio-az} (absolute difference = $1.5$) and $6.2$ in Figure~\ref{fig:tweet-ratio-il} (absolute difference = $2.1$). At this point, the execution was near its end and the ratio started to decrease to the actual final ratio. 

{\bf Reshape}: It used the {\sf SBR} approach and could split the processing of the CA key with a helper worker. \frmname had the advantage of iteratively adapting its partitioning logic and considered the current loads on the helper and the skewed worker while deciding the portion of load to be transferred in the second phase (Section~\ref{ssec:two-phases}). Thus, \frmname kept the workload of the skewed worker and the helper at similar levels. In Figure~\ref{fig:tweet-ratio-az} and \ref{fig:tweet-ratio-il}, after  the observed ratio reaches the actual ratio at about $120$ seconds and $130$ seconds, respectively, \frmname kept the observed ratio near the actual ratio.

\subsection{Benefits of the first phase}
\label{ssec:first-phase-benefit}

\begin{figure}[htbp]
\begin{center}
	\includegraphics[width=3.3in]{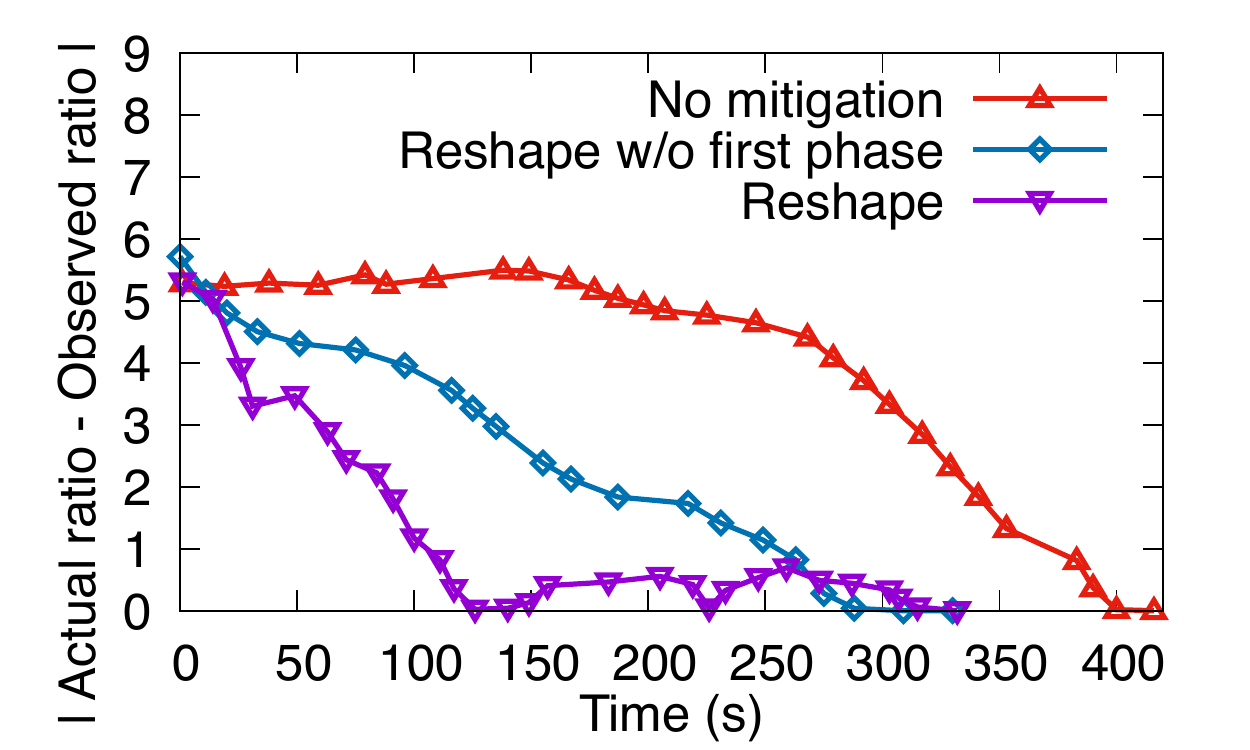}
	\caption{\label{fig:tweet-ratio-az-first-phase-benefit}
	\textbf{Effect of first phase on the ratio of CA to AZ tweets.}}
\end{center}
\end{figure}

\begin{figure}[htbp]
\begin{center}
	\includegraphics[width=3.3in]{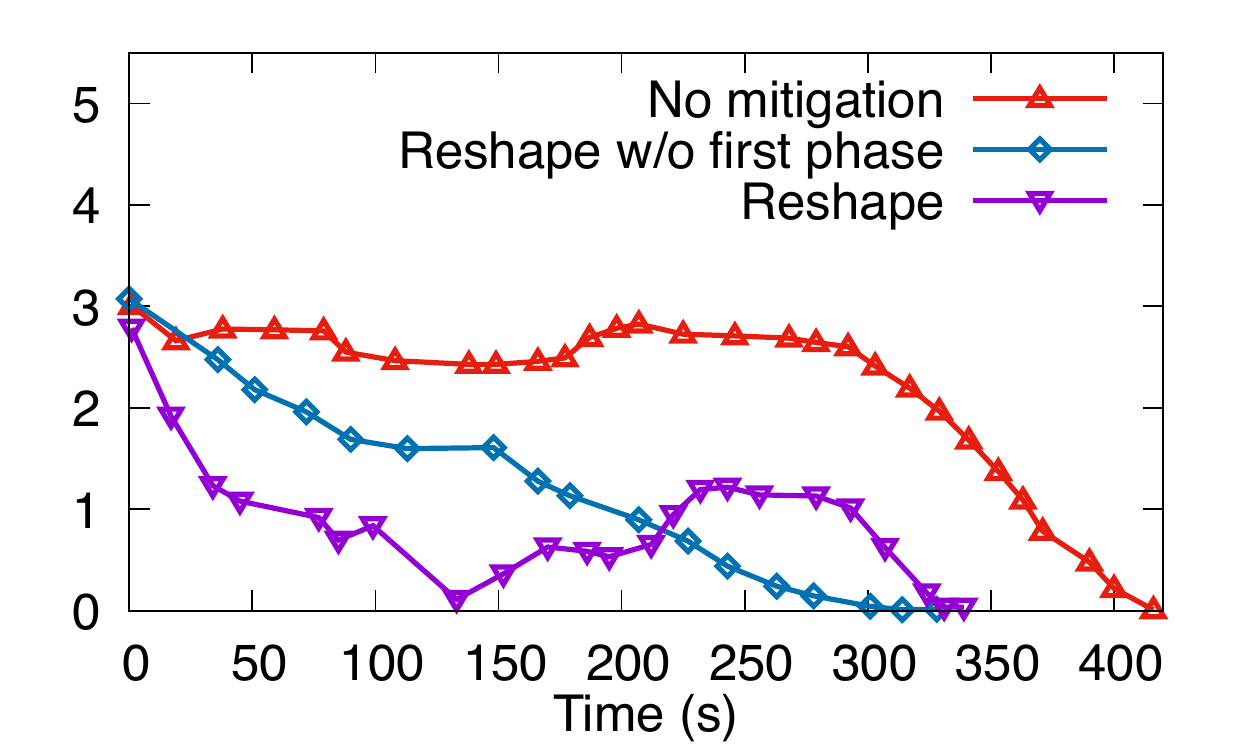}
	\caption{\label{fig:tweet-ratio-il-first-phase-benefit}
	\textbf{Effect of first phase on the ratio of CA to IL tweets.}}
\end{center}
\end{figure}

We evaluated the benefits of the first phase in Reshape as discussed in Section~\ref{ssec:two-phases}. We followed a similar setting as in the experiment in Section~\ref{ssec:results-produced} to monitor the ratio of processed tweets. There were two mitigation strategies used in this experiment. The first one was normal Reshape, with the two phases of load transfer. In the second strategy, we disabled the first phase in Reshape and just did load transfer using the second phase. The results are plotted in Figure~\ref{fig:tweet-ratio-az-first-phase-benefit} and \ref{fig:tweet-ratio-il-first-phase-benefit}.

The first phase quickly removed the existing imbalance of load between the skewed and the helper worker when skew was detected. When the first phase was present, \frmname reached the actual ratio around $120$ and $130$ seconds in Figures~\ref{fig:tweet-ratio-az-first-phase-benefit} and \ref{fig:tweet-ratio-il-first-phase-benefit}, respectively. When the first phase was disabled, \frmname reached the actual ratio around $288$ and $310$ seconds in Figures~\ref{fig:tweet-ratio-az-first-phase-benefit} and \ref{fig:tweet-ratio-il-first-phase-benefit}, respectively. Thus, the first phase allowed \frmname to show representative results earlier. Both strategies showed more representative results than the unmitigated case.

\subsection{Effect of heavy-hitter keys}
\label{ssec:heavy-hitter}

California (location $6$) produced the highest number of tweets ($26$M) and was a heavy-hitter key in the tweet dataset. We present the results for the mitigation of the skewed worker that processed the California key.

\textbf{Load balancing ratio.} The load balancing ratio at a moment during the execution is calculated by obtaining the total counts of tuples allotted to the skewed worker and its helper till that moment, and dividing the smaller value by the larger value. We periodically recorded multiple load balancing ratios during an execution and calculated their average to get the average load balancing ratio for an execution. A higher ratio is better because it represents a more balanced workload between the skewed worker and its helper.

\begin{figure}[htbp]
\begin{center}
	\includegraphics[width=3.3in]{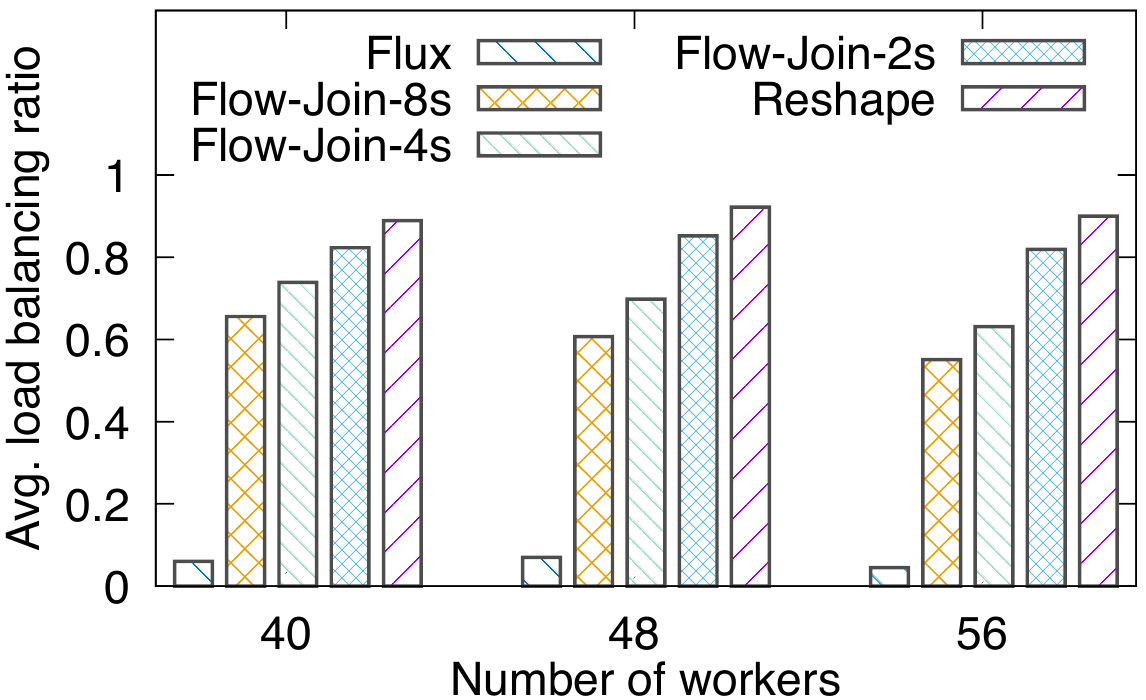} 
	\caption{\label{fig:baseline}
	\textbf{Evaluating different methods of handling heavy-hitter keys in $W_1$ using tweets. The three {\sf Flow-Join} bars correspond to the initial delay of $2$, $4$, and $8$ seconds.}}
\end{center}
\end{figure}

The average load balancing ratio for the skewed worker that processed the California key and its helpers is plotted in Figure~\ref{fig:baseline}. A higher ratio is better because it represents a more balanced workload between the skewed worker and its helper worker. We ran the experiments on three settings by varying the number of cores up to $56$ (on $14$ machines), which was the total number of distinct locations.

{\bf Flux}: It had the limitation of not being able to split the processing of a single key over multiple workers. Thus, the skewed worker processed the entire California input. The skewed worker was also processing another key with only a few hundred thousand tuples, which was moved to the helper when skew was detected. Flux had a low average load balancing ratio of about $0.06$.

{\bf Flow-Join}:  Its main drawback was the inability to do mitigation iteratively. It changed its partitioning logic once based on the heavy-hitters detected initially. The longer it spent to detect heavy-hitters with a higher confidence, the less was the amount of future tuples left to be mitigated for finite datasets. We varied the initial duration used by {\sf Flow-Join} to detect heavy-hitters from $2$ seconds to $8$ seconds. When the initial time spent was $2$ seconds, the average load balancing ratio was about $0.85$ and the final counts of tuples processed by the skewed and helper workers were approximately $14$M and $12$M, respectively. On the other hand, when the duration was $8$ seconds, the ratio was about $0.6$ and the final counts were approximately $17$M and $9$M, respectively. {\sf Flow-Join} was able to reduce the execution time of $W_1$ on $48$ cores from $416$ seconds to $302$ seconds, when the initial detection duration was $2$ seconds.

\textbf{Reshape}: It split the processing of the California key with a helper worker. \frmname had the advantage of iteratively changing its partitioning logic according to input distribution using fast control messages. Thus, the skewed and helper workers ended up processing almost similar amounts of data and the average load balancing ratio was about $0.92$. The execution time was reduced by 27\%. In particular, \frmname was able to reduce the execution time from $416$ seconds to $302$ seconds, by mitigating the skew in $W_1$ running on $48$ cores.


\subsection{Effect of latency of control messages}
\label{ssec:control-latency}

\begin{figure}[htbp]
     \centering
     \begin{subfigure}[t]{0.35\columnwidth}
  \includegraphics[width=\linewidth]{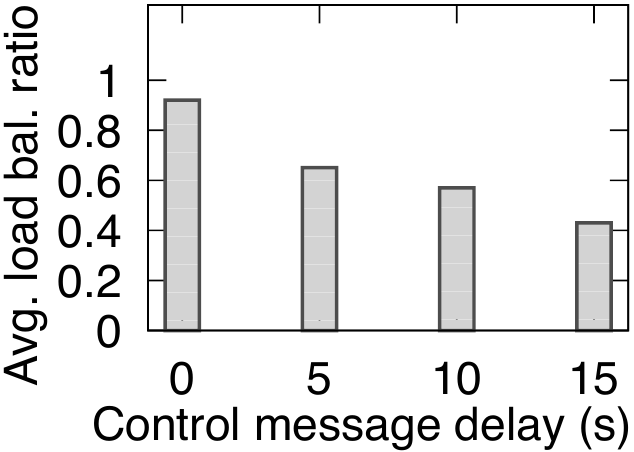}
  \caption{California data.}
  \label{fig:control-message-ca}
     \end{subfigure}
     \hspace{0.1\columnwidth}
     \begin{subfigure}[t]{0.35\columnwidth}
  \includegraphics[width=\linewidth]{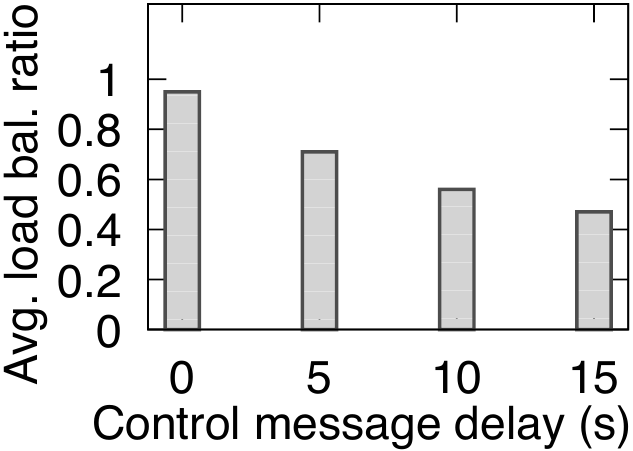}
  \caption{Texas data.}
  \label{fig:control-message-tx}
     \end{subfigure}
        \caption{\textbf{Effect of control message delay ($W_1$ on tweets).}}
        \label{fig:control-message}
\end{figure}

To evaluate the effect of the latency of control messages on skew handling by \frmname, we purposely added a delay between the time a worker receives a control message and the time it processes the message. Figure~\ref{fig:control-message} shows the result of varying the simulated delay from $0$ second (i.e., the message is processed immediately) to $15$ seconds on the mitigation of $W_1$ on $48$ cores. The figure shows the average load balancing ratio for the two pairs of skewed and helper workers processing the locations of California (location $6$) and Texas (location $48$), which had the highest counts of tweets.

{\bf Impact on responsiveness of mitigation:} As the control message delivery became slower, the delay between the controller sending a message and the resulting change in partitioning logic increased. Consider the example where the controller detected a workload difference of $350$ between the skewed worker and the helper worker and sent a message to start the first phase. In the case of no delay in control message delivery, the helper worker reached a similar workload as the skewed worker within $10$ seconds. In case of a delayed delivery, the workload difference continued to increase and got larger than $350$ before the first phase was started. For example, when there was a $5$-second delay, the workload difference was at $300$ after $10$ seconds of sending the message.


{\bf Impact on load balancing.} The latency in control messages affected the load sharing between skewed and helper workers. In the case of no delay, the two workers had almost similar loads and the average load balancing ratio was about $0.94$ as shown in Figures~\ref{fig:control-message-ca} and \ref{fig:control-message-tx}. As the delay increased, the framework was slow to react to the skew between workers, which resulted in imbalanced load-sharing. In the case of a $15$-second delay, the average load balancing ratio reduced to about $0.45$. Thus, low-latency control messages facilitated load balancing between a skewed worker and its helper.

\subsection{Benefit of dynamically adjusting $\tau$}
\label{ssec:dynamic-tau-exp}

\begin{figure}[htbp]
\begin{center}
	\includegraphics[width=3.3in]{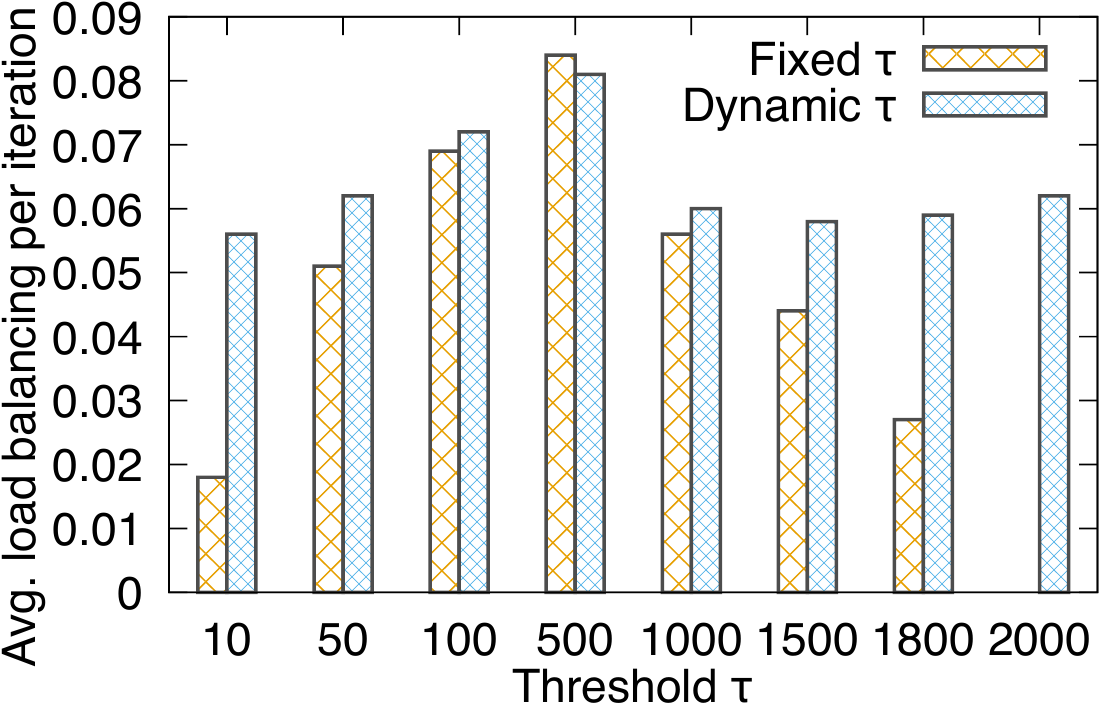} 
	\caption{\label{fig:dynamic-tau}
	\textbf{Benefit of dynamically adjusting $\tau$ ($W_1$ on tweets).}}
\end{center}
\end{figure}

We evaluated the effect of the dynamic adjustment of $\tau$ on skew mitigation in $W_1$ by \frmname on $48$ cores. We chose different values of $\tau$ ranging from $10$ to $2,000$, and did experiments for two settings. In the first setting, $\tau$ was fixed for the entire execution. In the second setting, $\tau$ was dynamically adjusted during the execution. The mean model estimated the workload of a worker as its expected number of tuples in the next $2,000$ tuples and the preferred range of standard error (Section~\ref{sssec:choose-tau}) was set to $98$ to $110$ tuples. We allowed up to three adjustments during an execution. Whenever $\tau$ had to be increased, it was increased by a fixed value of $50$. 
We calculated the average load balancing ratios for the workers processing the California and Texas keys and divided them by the total number of mitigation iterations during the execution. This resulted in the metric of average load balancing per iteration, shown in Figure~\ref{fig:dynamic-tau}. A higher value of this metric is better because it represents a more balanced workload of skewed and helper workers in fewer iterations. 

Let us first consider the cases where $\tau$ was dynamically adjusted to an increased value. Setting $\tau$ to a small value of $10$ resulted in a large number of iterations, i.e, $41$, in the fixed $\tau$ setting. In the dynamic $\tau$ setting, the controller observed that the standard error at the beginning of the second phase was greater than $110$ and increased $\tau$. Consequently, the number of iterations decreased to $14$, which resulted in a substantial increase in the metric of average load balancing per iteration. For the cases of $\tau = 50$ and $100$ in the fixed setting, the average load balancing per iteration increased with $\tau$ because the number of iterations decreased. The dynamic setting slightly decreased the iteration count in these cases.

Now let us consider the case where $\tau$ remained unchanged or decreased as a result of dynamic adjustment. When $\tau = 500$, the standard error was in the range $[98,110]$. Thus, the dynamic adjustment did not change $\tau$. When $\tau = 1000$ in the fixed setting, the mitigation started late and the workload of skewed and helper workers were not balanced. The mitigation was delayed even more for $\tau = 1500$ and $1800$ in the fixed setting and the mitigation did not happen for $\tau=2000$. In the dynamic setting for the cases of $\tau = 1000$, $1500$, $1800$, and $2000$, the controller observed that the standard error went below $98$ when the workload difference was about $700$. Thus, the controller reduced $\tau$ to $700$. The advantage of dynamically reducing $\tau$ was that it automatically started mitigation at an appropriate $\tau$, even if the initial $\tau$ was very high.

\subsection{Effect of different levels of skew}
\label{ssec:complex-workflow}

We evaluated the load balancing achieved by \frmname for different levels of skew. We used $W_2$ for this purpose. The data distributions in Figures~\ref{fig:dsb-date}-\ref{fig:dsb-item} show that the join on {\sf item\_id} was highly skewed and the join on {\sf date\_id} was moderately skewed. We evaluated the load balancing achieved for these two join operators. We scaled the data size from $100$GB to $200$GB. Meanwhile, we scaled the number of cores from $40$ to $80$ and did the experiments in each configuration.

\begin{figure}[htbp]
\begin{center}
	\includegraphics[width=3.3in]{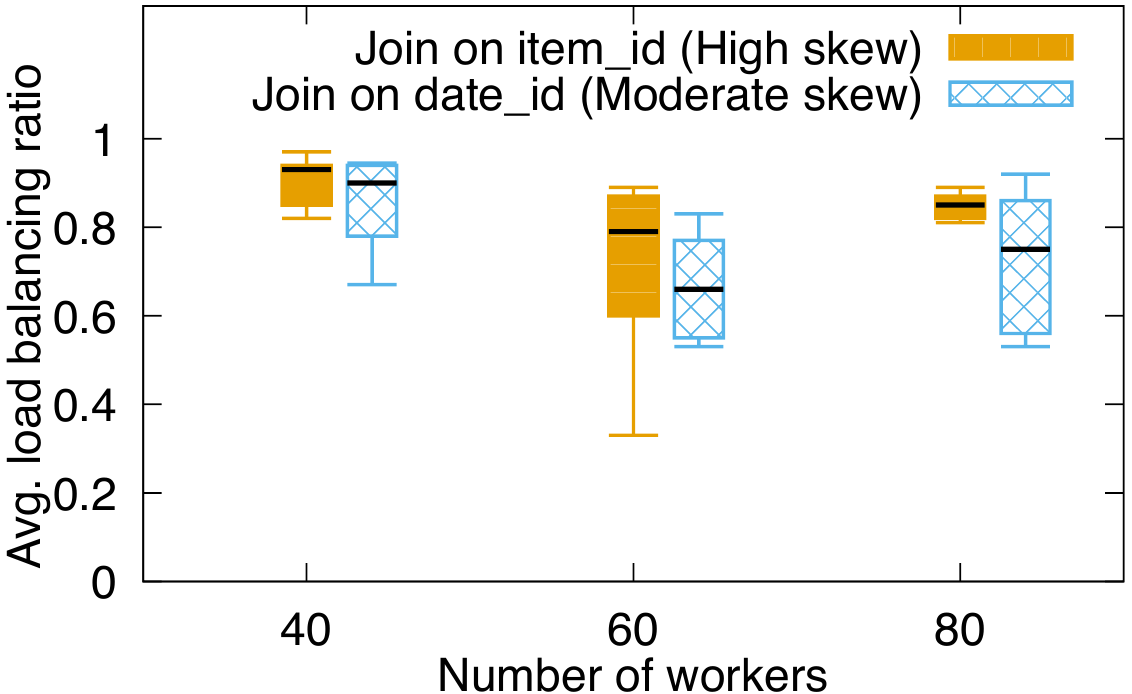} 
	\caption{\label{fig:different-skew-levels}
	\textbf{Effect of different levels of skew ($W_2$ on DSB data). Each candlestick body represents the $25^{th}$ to $75^{th}$ percentile.}}
\end{center}
\end{figure}

Figure~\ref{fig:different-skew-levels} shows the candlestick charts of the average load balancing ratios for the top five skewed workers from each of the two joins. For the highly skewed join on {\sf item\_id}, the skew was detected early, and there was enough time to transfer the load of the skewed workers to the helper workers. The $25^{th}$ and $75^{th}$ percentiles of the average load balancing ratios remained above $0.6$ for all the configurations. The median of the ratios was more than $0.77$. This result shows that \frmname was able to mitigate the skew and maintain comparable workloads on the skewed and helper workers when both the input and processing power were scaled up. The join on {\sf date\_id} had only a moderate skew, which resulted in a delayed detection of a few of its skewed workers. Due to the delayed detection, there were fewer future tuples of skewed workers to be transferred to the helpers. Thus the ratios for the join on {\sf date\_id} were lower than that for the join on {\sf item\_id}. The performance of \frmname was also shown by the reduction in the execution time. Specifically, in the case of $40$ cores, the mitigation reduced the execution time of $W_2$ from $267$ seconds to $243$ seconds. In the case of $80$ cores, the mitigation reduced the time from $335$ seconds to $269$ seconds.

\subsection{Effect of changes in input distribution}
\label{ssec:changing-data-distribution}

\begin{figure}[htbp]
\begin{center}
	\includegraphics[width=3.3in]{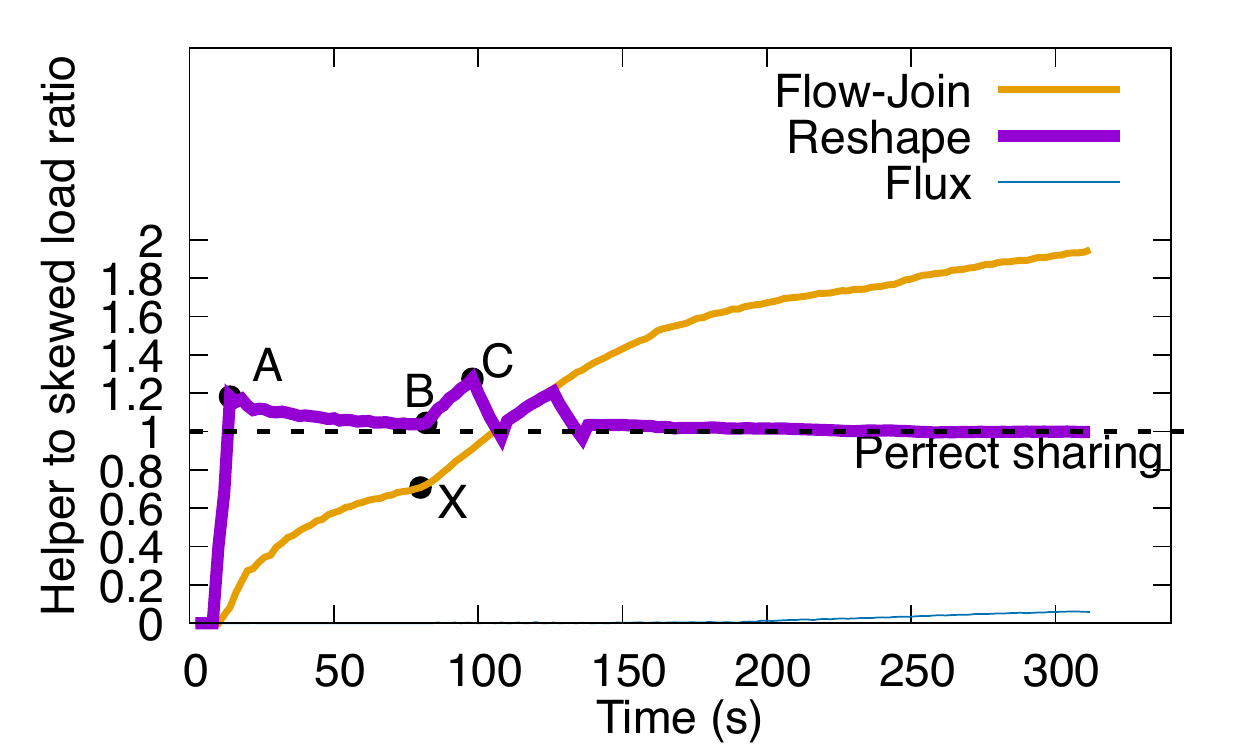} 
	\caption{\label{fig:dynamic-distribution}
	\textbf{Effect of changes in input data distribution on load sharing ($W_4$ on the synthetic dataset).}}
\end{center}
\end{figure}

We evaluated how load sharing was affected when the input distribution changed during the execution. We used the synthetic dataset and workflow $W_4$ running on $40$ cores. Both tables in the dataset had $42$ keys.  The first table contained $4,200$ tuples uniformally distributed across the keys. The second table contained $80$M tuples and was produced by the {\sf source} operator at runtime. We fixed worker $0$ and worker $10$ as the skewed and helper worker, respectively. We altered the load on key $0$ and $10$, which were processed by worker $0$ and $10$ respectively. Specifically, for the first $20$M tuples, $80$\% was allotted to the key $0$ and the rest $20$\% was uniformally distributed among the remaining keys. For the next $60$M tuples, $60$\% was allotted to the key $0$, $20$\% to key $10$, and the rest was uniformally distributed. Figure~\ref{fig:dynamic-distribution} shows the ratio of the workloads of the helper worker $10$ to the skewed worker $0$ as time progressed. We used $\tau=2,000$ to clearly show the effects of changing distributions.

\textbf{Flux.} The skewed worker was processing keys $0$ and $40$. {\sf Flux} had the limitation of not being able to split the processing of a single key over multiple workers. Upon detecting skew, {\sf Flux} can only move the key with smaller load (key $40$) to the helper. Thus, the workload ratio of helper to skewed worker remained close to $0$.

\textbf{Flow-Join:} We used a $2$-second initial duration to detect the overloaded keys. {\sf Flow-Join} identified key $0$ as overloaded and started to transfer half of its future tuples to the helper. Thus, the workload of the helper began to rise. At $80$ seconds (point {\tt X}), the input distribution changed. Since {\sf Flow-Join} cannot do mitigation iteratively, half of the tuples of key $0$ continued to be sent to the helper. The helper worker started receiving $50$\% ($=60\%*0.5 + 20\%$) and the skewed worker started receiving $30$\% ($=60\%*0.5$) of the input. Thus, the load on the helper rose and became more than the skewed worker.

\textbf{Reshape:} It started the first phase to let the helper worker quickly catch up with the skewed worker. Thus, the load of the helper sharply increased initially. After that the second phase started and the workload ratio got closer to $1$. At $80$ seconds (point {\tt B}), the input distribution changed. At point (point {\tt C}), \frmname started another iteration of mitigation and adjusted the partitioning logic according to the new input distribution. As a result, the ratio of the workloads of the workers remained close to $1$.

\subsection{Metric-collection overhead}
\label{ssec:metric-collection-overhead}

\begin{figure}[htbp]
\begin{center}
	\includegraphics[width=3.3in]{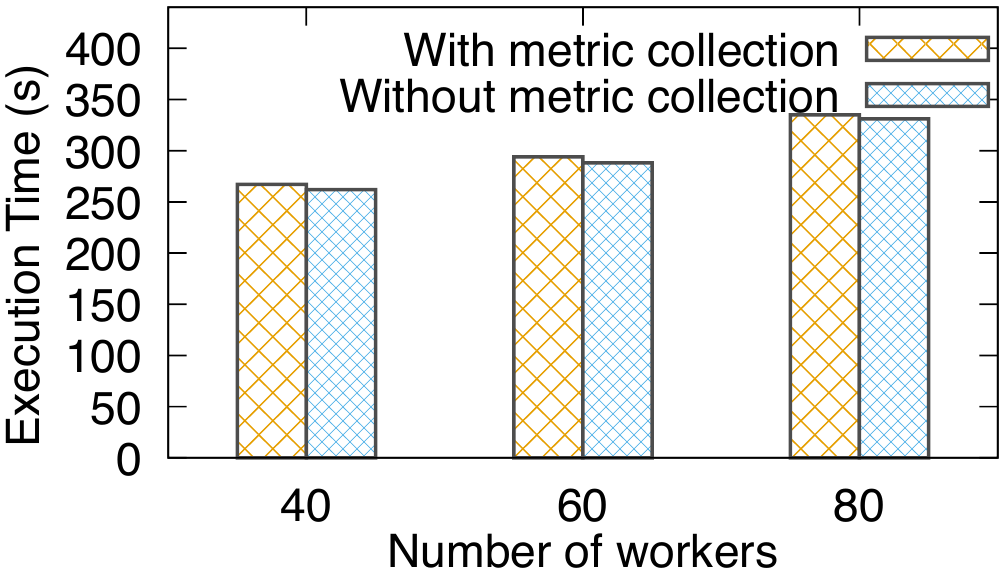} 
	\caption{\label{fig:exp-metric-collection}
	\textbf{Metric-collection overhead ($W_2$ on DSB data).}}
\end{center}
\end{figure}

We evaluated the metric-collection overhead of \frmname on the workflow $W_2$. We scaled the data size from $100$GB to $200$GB. Meanwhile, we scaled the number of cores from $40$ (on $10$ machines) to $80$ (on $20$ machines) and did the experiments in each configuration. We disabled skew mitigation and executed $W_2$ with and without metric collection to record the metric-collection overhead. As shown in Figure~\ref{fig:exp-metric-collection}, the overhead was around $1$-$2$\% for all the configurations.

\subsection{Performance of \frmname on {\sf sort}}
\label{ssec:exp-other-operators}

To evaluate its generality to other operators, we implemented \frmname for the sort operator. We used the workflow $W_3$ for this experiment. The {\em Orders} table was range-partitioned on its {\tt totalPrice} attribute. Table~\ref{table:exp-sort-reshape} lists the various percentile values of the average load balancing ratio for the skewed workers that received more than $3.5$M tuples in the unmitigated case (Figure~\ref{fig:reshape-tpch}). We scaled the data size and number of cores simultaneously from $50$GB on $20$ cores to $200$GB on $80$ cores, and did the experiment in each configuration.


\begin{table}[htbp]
\centering
\small{
\begin{tabular}{|@{ }p{2.5cm}|@{ }p{1cm}|@{ }p{1cm}|@{ }p{1cm}|@{ }p{1cm}|@{ }p{1cm}|}
\hline
{\bf \# workers} & {\bf $P_1$} & \textbf{$P_{25}$} & \textbf{$P_{50}$} & \textbf{$P_{75}$} & \textbf{$P_{99}$} \\ \hline
20 & 0.90 & 0.92 & 0.93 & 0.935 & 0.95 \\ \hline
40 & 0.84 & 0.87 & 0.89 & 0.90 & 0.91 \\ \hline
60 & 0.83 & 0.85 & 0.90 & 0.91 & 0.92 \\ \hline
80 & 0.83 & 0.84 & 0.86 & 0.87 & 0.90 \\ \hline
\end{tabular}
}
\textit{}
\caption{Average load balancing ratios when \frmname is applied on {\sf sort} ($W_3$ using the TPC-H data).}
\label{table:exp-sort-reshape}
\end{table}

As the number of cores increased, the $25^{th}$ and $75^{th}$ percentiles of the average load balancing ratios remained close to $0.9$. This result shows that the skewed and helper workers had balanced workloads when both the input and processing power were scaled up. The consistent performance of \frmname was also shown by about $20$\% reduction in the execution time. Specifically, in the case of $20$ cores, the time reduced from $789$ seconds to $643$ seconds. In the case of $80$ cores, the time reduced from $809$ seconds to $667$ seconds. 


\subsection{Effect of multiple helper workers}
\label{ssec:exp-multile-helpers}

We evaluated the load reduction achieved when multiple helper workers are assigned to a skewed worker. The experiment was done on $W_1$ running on $48$ cores. The most skewed worker among the $48$ workers received about $27$M tuples in the unmitigated case. We allotted different numbers of helpers to the skewed worker and calculated the load reduction. We set the build hash-table in each worker to have $10,000$ keys, so that the state size became significant and the state-migration time was noticeable.

\begin{figure}[htbp]
\begin{center}
	\includegraphics[width=3.3in]{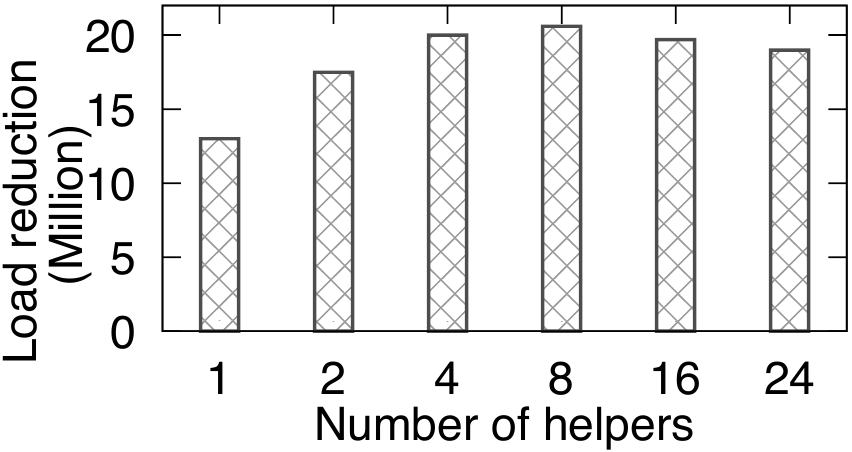} 
	\caption{\label{fig:exp-multiple-helpers}
	\textbf{Effect of multiple helper workers ($W_1$ on tweets).}}
\end{center}
\end{figure}

The results are plotted in Figure~\ref{fig:exp-multiple-helpers}. When a single helper was used, the state migration happened in $17$ seconds. The skewed worker transferred about half of its total workload to the helper, resulting in a load reduction of $13$M tuples. When $2$ helpers were used, the skewed worker transferred about two thirds of its tuples to the two helpers (about $9$M each). With more helpers, the state-migration time also increased. For $8$ helpers, the state-migration time was about $26$ seconds. Thus, there were fewer future tuples left, which resulted in a small increase in the load reduction. For $16$ helpers, the state-migration time became $32$ seconds and the load reduction decreased to $19.7$M. For $24$ helpers, the state-migration time was $39$ seconds and the load reduction decreased to $19$M.

\subsection{Performance of \frmname on Flink}
\label{ssec:flink-exp}

\begin{figure}[htbp]
     \centering
     \begin{subfigure}[t]{0.35\columnwidth}
  \includegraphics[width=\linewidth]{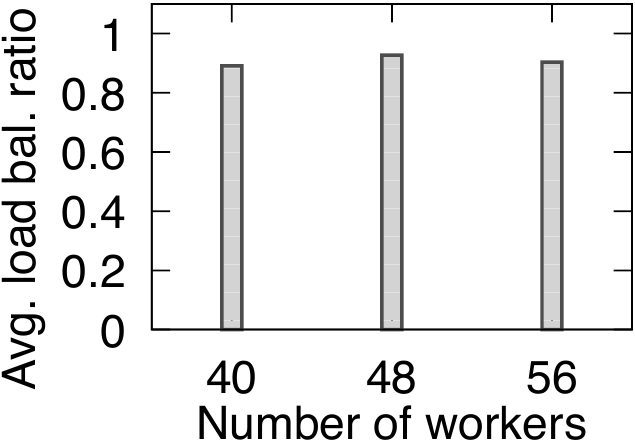}
  \caption{California data.}
  \label{fig:flink-ca}
     \end{subfigure}
     \hspace{0.1\columnwidth}
     \begin{subfigure}[t]{0.35\columnwidth}
  \includegraphics[width=\linewidth]{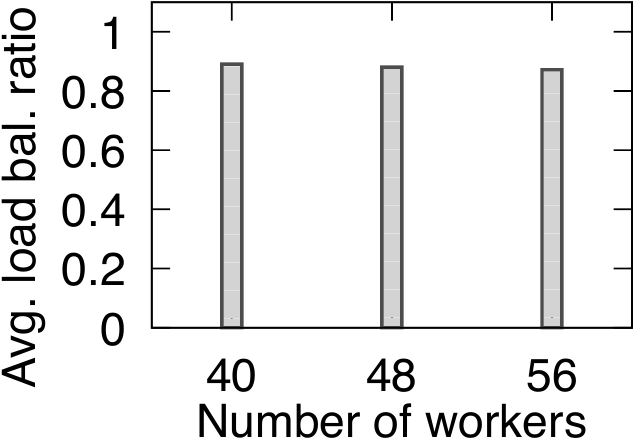}
  \caption{Texas data.}
  \label{fig:flink-tx}
     \end{subfigure}
        \caption{\textbf{Mitigation by \frmname on Flink ($W_1$ on tweets)}}
        \label{fig:flink}
\end{figure}

We implemented \frmname on Apache Flink and executed $W_1$ on $40$, $48$, and $56$ cores. A worker was classified as skewed if its {\em busyTimeMsPerSecond} metric was greater than $80$\%. Figure~\ref{fig:flink} shows the average load balancing ratio for the workers processing the California and Texas tweets. The ratio was about $0.9$, which means that the skewed and helper workers had similar workloads throughout the execution. For the $48$-core case, the final counts of tuples processed by the skewed and helper workers for California were $13$M and $14$M, respectively. The final counts of tuples processed by the two workers for Texas on $48$ cores were $10$M each. The execution time decreased as a result of the mitigation. For example, for the $48$-core case, the execution time decreased from $407$ seconds to $320$ seconds. 

\section{Conclusions}
\label{sec:reshape-conclusions}

In this chapter, we presented a framework called \frmname that adaptively handles partitioning skew in the exploratory data analysis setting. We presented different approaches for load transfer and analyzed their impact on the results shown to the user. We presented an analysis about the effect of the skew-detection threshold on mitigation and used it to adaptively adjust the threshold. We generalized \frmname to multiple operators and broader execution settings. We implemented \frmname on top of two big data engines and presented the results of an experimental evaluation.

\chapter{Maestro: Result-aware Scheduling for Exploratory Data Analysis on Big Data}
\label{chap:maestro}

\section{Introduction}
\label{sec:maestro-introduction}

Big data processing systems have become quite popular in the last two decades. We have mentioned various examples of such systems in the previous chapters. We also discussed that these systems accept workflows broadly through two interfaces - programming APIs and GUIs. The input workflow is compiled and executed by the engine.

The task to execute an input workflow to completion is performed by a scheduler. Scheduling a workflow involves decisions such as how to execute the various operators in the workflow and the resources to execute them on. A scheduler analyzes a workflow and comes up with an order to execute the operators. It also decides how many resources (such as CPU cores and memory) will be needed to execute these operators. Frameworks such as LSched~\cite{conf/sigmod/SabekUK22} and Decima~\cite{conf/sigcomm/MaoSVMA19} can help the scheduler decide the appropriate amount of resources to be assigned. Once the resource needs are determined, the scheduler may need to apply for a lease of the resources on shared clusters. Cluster managers such as YARN~\cite{conf/cloud/VavilapalliMDAKEGLSSSCORRB13} and Borg~\cite{conf/eurosys/VermaPKOTW15} have been developed to provide such leases.

The division of a workflow into multiple parts, each containing one or more operators, for scheduling depends on the type of execution. In batch-based execution adopted by systems such as Apache Hadoop~\cite{misc/hadoopmapreduce} and Apache Spark~\cite{misc/spark}, the workflow is broken down at data-shuffle boundaries~\cite{spark-stage} into a sequence of map and reduce phases. All the tasks of a phase should finish before the next phase starts. There are some optimizations to allow the reduce phase to partially overlap with the map phase using a slow-start configuration~\cite{misc/hadoop-mapreduce-configs}. On the other hand, in pipelined execution adopted by systems such as Apache Flink~\cite{misc/flink} and Hyracks~\cite{conf/icde/BorkarCGOV11}, the workflow is broken down at blocking links into a sequence of pipelined regions~\cite{misc/flink-pipelined-region-scheduling}. Blocking links are edges in a workflow DAG whose destination operator does not produce any output till it processes all the input data on that edge. Examples of blocking links are the build input of a two-phase {\sf HashJoin} operator and the input to {\sf GroupBy} operator.

We discussed the benefits of pipelined execution in exploratory data analysis in Chapter~\ref{chap:reshape}. Pipelined execution allows the results to be produced quickly because an operator does not wait for its entire input data to be produced before processing the input and sending results to its downstream operators. For example, when a workflow containing a two-phase {\sf HashJoin} operator is executed in a pipelined mode, the {\sf HashJoin} operator starts executing and producing results as soon as the upstream operator producing probe input outputs initial results. In contrast, when the workflow is executed in a batch mode, the {\sf HashJoin} operator waits for the entire probe input to be produced before joining the two inputs. Note that in both the execution models, the {\sf HashJoin} operator does not produce any output while processing the build input. In this chapter, we consider the pipelined execution setting to enable the users to see results quickly.

The problem of scheduling has been extensively studied in the literature, mainly from the perspective of increasing the end-to-end performance~\cite{conf/middleware/PengHHFC15, conf/debs/AnielloBQ13, conf/infocom/ChengCZGM17}. However, there is little research on the following important problem:

\begin{quote}
{\em In exploratory data analytics, how to consider the timing of results shown to the user when scheduling a workflow?}
\end{quote}

A few works are related to this problem. A work ranks the partitions of input data in the order of their likelihood to produce results and higher ranked partitions are processed first~\cite{conf/icde/EngOST03}. Other works focus on a particular operator such as join and customize the implementation of the operator so that it produces results quicker~\cite{conf/vldb/UrhanF01, conf/vldb/Lawrence05}. 

In this chapter, we take a different perspective. Considering a workflow DAG as an input in a pipelined execution setting, we explore how to divide the workflow into smaller parts that can be separately scheduled and how to schedule the different parts so that results are shown to the user quickly.

\begin{figure}[htbp]
 \begin{center}
	\includegraphics[width=3.5in]{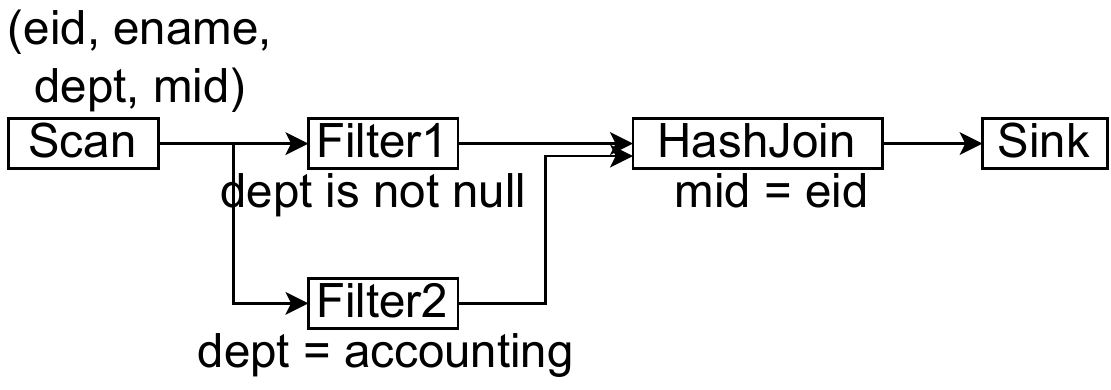}
	\caption{\label{fig:simple-self-join}
		\textbf{A workflow where probe and build inputs may arrive at join simultaneously. This leads to an exception if the join operator expects build input to be processed completely before the probe input arrives.}
	}
\end{center}
\end{figure}

A simple way to divide a workflow into parts for scheduling is based on the blocking links as discussed above. However, there may exist operators that expect the input data to arrive on their input links in a particular order and cause exceptions if that expectation is not met. For example, a two-phase {\sf HashJoin} operator requires the complete input to arrive on its build input link before input data arrives on its probe input link. To cater to such a join operator, the part of the workflow producing build input should execute completely before the part of the workflow producing probe input executes. However, in many cases, the build and probe input are produced by the same part of the workflow. Figure~\ref{fig:simple-self-join} shows such a workflow. Consider the execution of this workflow in the pipelined mode. When the {\sf Scan} operator starts producing data, both {\sf Filter1} and {\sf Filter2} process data and output the results to {\sf HashJoin} simultaneously, which causes an exception. Engines typically take a heuristic-based approach to handle such workflows. For example, Apache AsterixDB~\cite{asterix:website} adds materialization at an output of the operator that replicates the data, i.e., materialization is added on the link between the {\sf Scan} and the {\sf Filter2} operators. However, it can be seen in the figure that materialization can also be added to the link between the {\sf Filter2} and the {\sf HashJoin} operator. In this chapter, we identify the different places where materialization can be added in the workflow and examine their effect on the results shown to the user.

In this chapter, we present a novel result-aware scheduling framework called \schfrmname that analyzes a workflow DAG to create regions in a result-aware manner that are executed in a pipelined execution setting. We make the following contributions. (1) Dividing the workflow into regions and creating a region graph: We present a technique to divide a workflow into regions and an algorithm to create a region graph that encapsulates the dependencies among the regions of a workflow (Section~\ref{sec:schedule-a-workflow}). (2) Enumeration of materialization choices: When there is a need for materialization, there may be multiple places in the workflow where the materialization can take place. We present an algorithm to enumerate all the options of materialization. (3) Result-aware materialization choice: We perform an analysis to compare the different choices of materialization from the perspective of result awareness.

\subsection{Related work}
\label{ssec:maestro-related-work}

\boldstart{Cluster resource managers.} When a cluster is shared by different applications, the applications need to interact with the resource managers to execute their tasks. Apache Hadoop YARN~\cite{conf/cloud/VavilapalliMDAKEGLSSSCORRB13} is a resource manager that publishes APIs that can be used by applications to send resource requests (e.g., CPU cores, memory) and receive resource leases to execute their tasks. Once the lease has been obtained the application takes care of placing the tasks to be executed on those resources. YARN allocates resources based on different policies such as FIFO, FAIR, and capacity-based scheduling. Mesos~\cite{conf/nsdi/HindmanKZGJKSS10} works in a fashion similar to YARN where it offers resources to application level schedulers. Borg~\cite{conf/eurosys/VermaPKOTW15} is a resource manager that accepts job requests and, unlike YARN and Mesos, also manages placing the tasks of a job on different machines. In contrast, \schfrmname focuses on the scheduling of a particular workflow and can contact the cluster resource managers to ask for the computational resources needed for execution.

\boldstart{Early result production.} Symmetric hash join~\cite{conf/pdis/WilschutA91,journals/debu/UrhanF00} aims towards early production of join results by keeping hash tables for both inputs in its state. When a tuple arrives from one input, it is used to probe the hash table of the other input. Early hash join~\cite{conf/vldb/Lawrence05} is a join algorithm that allows early production of results without much overhead on the total execution time. It runs in two phases where the first phase focuses on quickly producing results and the second phase focuses on reducing the execution time. Another way to produce data early is to rank the partitions of input data in the order of their likelihood to produce results and process the higher ranked partitions first~\cite{conf/icde/EngOST03}. In contrast, \schfrmname focuses on the execution of a workflow as a whole and is not specific to a particular operator. The techniques in \cite{conf/vldb/Lawrence05, conf/icde/EngOST03} can complement \schfrmname to get even better results.

\boldstart{Workflow schedulers.} The schedulers in Apache Hadoop and Apache Spark divide the workflow into a sequence of stages~\cite{spark-stage}. Each stage has a predefined parallelism that is equal to the number of tasks in that stage. Tasks are allocated to the computation nodes when the nodes become available~\cite{misc/spark-job-scheduling}. The scheduler in Apache Flink and Hyracks divides a workflow into a sequence of pipelined regions based on blocking links~\cite{misc/flink-pipelined-region-scheduling}. The pipelined regions, similar to stages, have predefined parallelism. In systems such as Apache Storm~\cite{apacheStorm}, the traffic of data during the execution of the workflow can be monitored to change the placement of operators and improve performance. R-Storm~\cite{conf/middleware/PengHHFC15} is a scheduler for Apache Storm that requires each task of a workflow to provide its resource needs and allocates them accordingly.  If a task receives data from another task in the workflow, then it tries to put them on the same node. A-scheduler~\cite{conf/infocom/ChengCZGM17} is a scheduler for Spark Streaming that follows a scheduling policy that does FIFO scheduling for dependent tasks in a workflow and FAIR scheduling for independent tasks. LSched~\cite{conf/sigmod/SabekUK22} and Decima~\cite{conf/sigcomm/MaoSVMA19} are query schedulers that use machine learning to predict how to schedule individual queries so as to reduce the overall execution time of the workload. Davos~\cite{journals/pvldb/ShangZBEKMRK21} uses a publish-subscribe pattern between operators for data transmission and schedules operators one at a time depending on which operator has pending input. It uses a priority-based scheduling strategy to execute the operator in a workflow that has its input ready and is nearest to the output. \schfrmname is different from the Davos scheduler because Davos assumes a publish-subscribe based execution, whereas \schfrmname assumes a pipelined-based execution. Most workflow schedulers focus on improving the performance of a workflow, i.e., reducing the total execution time. In contrast, \schfrmname approaches the scheduling problem from the perspective of result timing awareness.

\section{Preliminaries}
\label{sec:maestro-preliminaries}

In this section, we introduce a running example and a few preliminary concepts that we refer to throughout this chapter.

\begin{figure*}[htbp]
	\begin{center}
	    
	\includegraphics[width=6in]{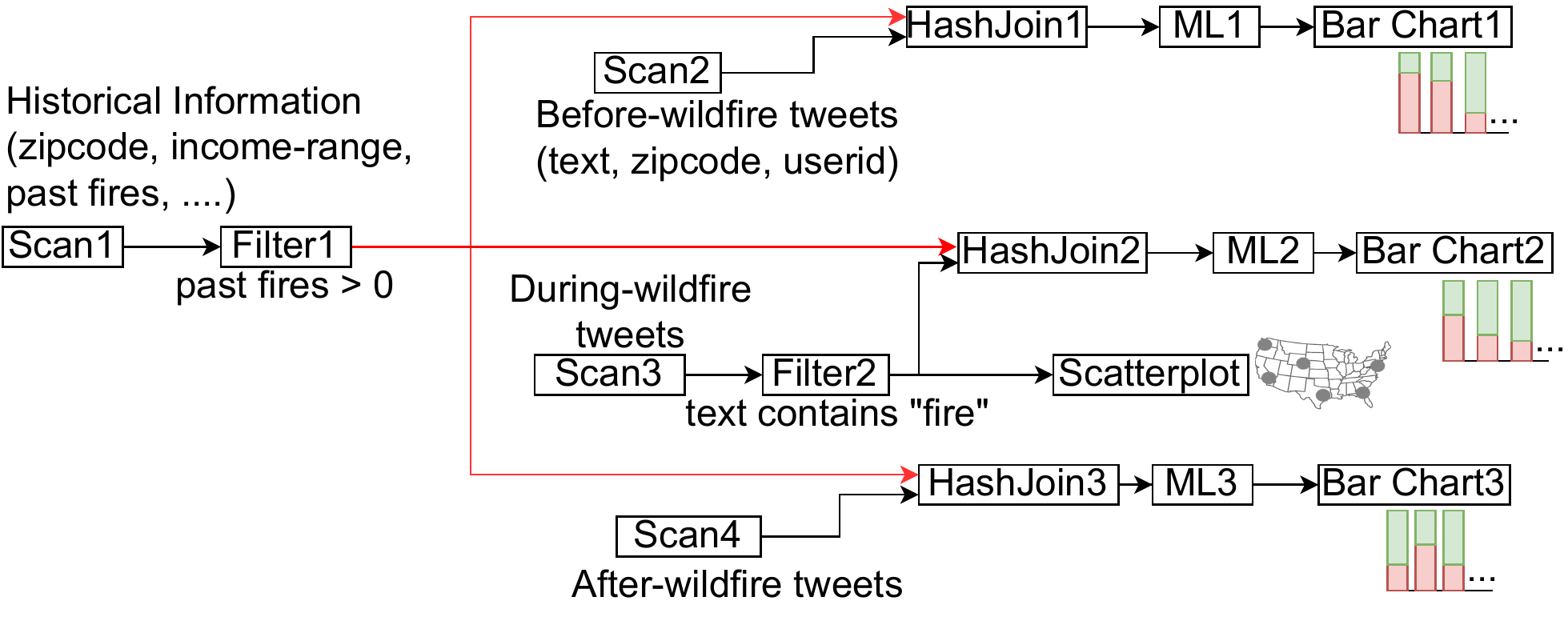} 
	\caption{\label{fig:maestro-running-example}
		\textbf{Analyzing climate change awareness in tweets before, during and after the wildfire season. The build inputs to join are shown in red. All joins are on the zipcode column. The ML operators determine if the tweet is about climate change.}
	}
		\end{center}

\end{figure*}

\boldstart{Running Example.} Figure~\ref{fig:maestro-running-example}(a) shows a workflow, which will serve as our running example, where the user wants to analyze the effect of wildfires on people's perception of climate change. The user wants to analyze the text in the tweets from before, during, and after the wildfire season. The historical information about the count of past fires in various zipcodes is read by {\sf Scan1} operator. The zipcodes with no history of wildfires are filtered out. The resulting historical information is joined with the tweets on the zipcode attribute. The join is a two-phase {\sf HashJoin} operator, where the historical information is the build input and the tweet data is the probe input. The output of the join operator is input into an {\sf ML} operator that classifies the tweet text as being related to climate change or not. The result is plotted as stacked bar charts where the $x$ axis is the income range of zipcode and the $y$ axis is the proportion of tweets related and unrelated to climate change. For the tweets collected during the wildfire season, only those that have the word ``fire'' are input into the join operator. The tweets during the wildfire season are also plotted on a scatterplot operator to show their spatial distribution.

As discussed in Chapter~\ref{chap:amber}, each operator physically corresponds to a set of workers that process input data and send output data to the workers of the downstream operators. There is a centralized controller that oversees the execution of the workflow. Next we define a few preliminary terms and concepts that will be used in this chapter.

\begin{definition}[Result operators]
We use the term {\em result operators} to refer to operators that produce tabular or visual results. These operators do not have any output links.
\end{definition}
In Figure~\ref{fig:maestro-running-example}, the {\sf bar chart} and the {\sf scatterplot} operators are result operators.

\begin{definition}[Blocking and pipelined links]
We call a link between two operators as {\em blocking} if the destination operator does not produce any output till it processes all the input data on that link. We call a link as {\em pipelined} if the destination operator starts producing output before it has processed all the input data on that link.
\end{definition}
For example, the build input link to a two-phase {\sf HashJoin} is a blocking link because the entire build input is needed to create a complete build hash table and no output is produced by the operator before that. The probe input link to a two-phase {\sf HashJoin} is a pipelined link because the operator may start producing output before it has processed the entire probe input. {\sf Sort} and {\sf GroupBy} operators are examples of operators that have just a single blocking input and no pipelined inputs.

\begin{figure*}[htbp]
	\begin{center}
	    
	\includegraphics[width=3in]{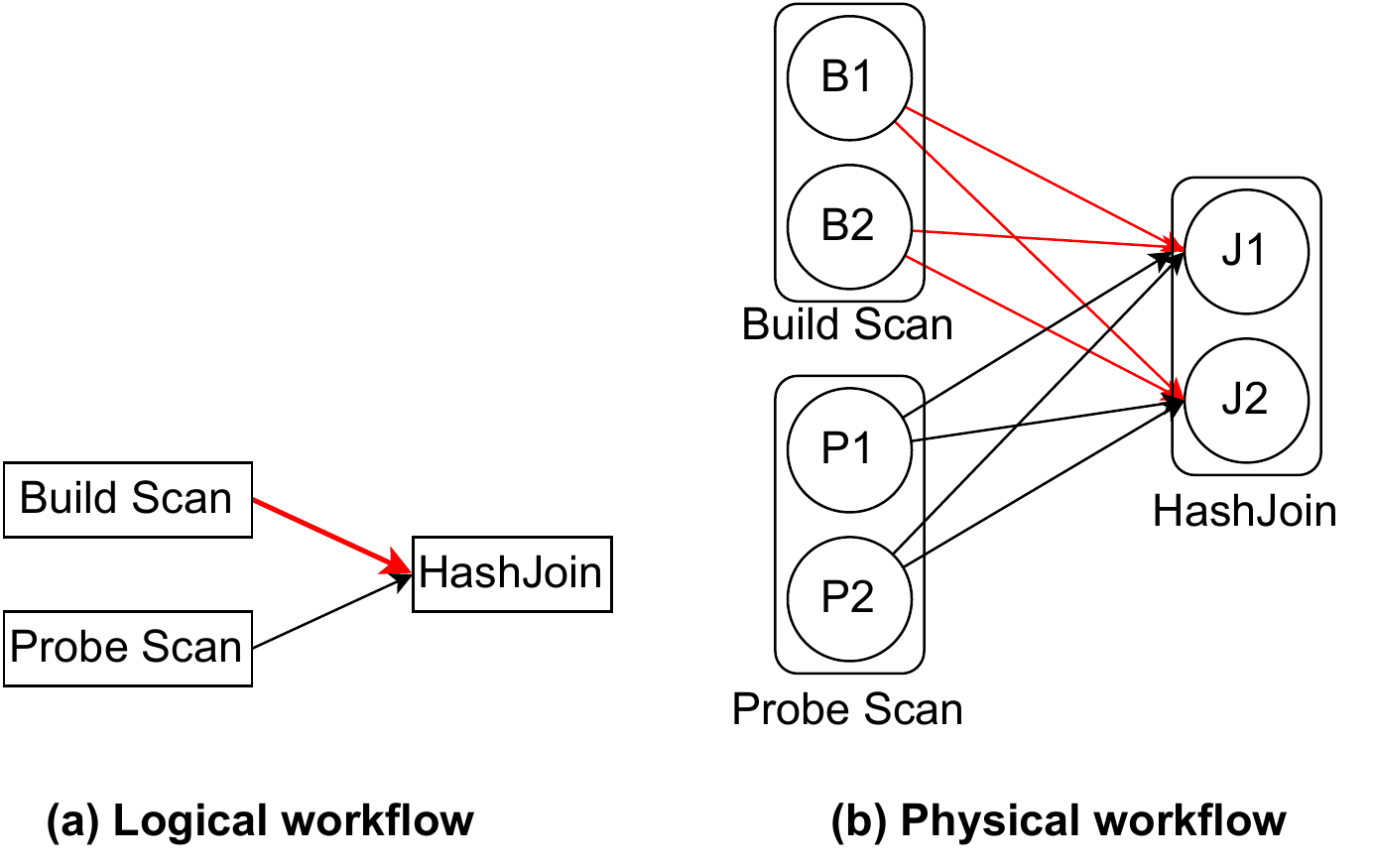} 
	\caption{\label{fig:hash-join-impl}
		\textbf{Implementation of {\sf HashJoin} assumed in this chapter. The {\sf HashJoin} operator is implemented as a set of workers and each worker performs both build and probe phases of the operator.}
	}
		\end{center}

\end{figure*}

\boldstart{Implementation of a {\sf HashJoin} operator.} In this chapter, we assume the specific implementation of the {\sf HashJoin} in \amberfrmname. The {\sf HashJoin} operator is implemented as a collection of workers and each worker performs both build and probe phases of the operator. For example, in Figure~\ref{fig:hash-join-impl}, $J1$ and $J2$ process the build input from $B1$ and $B2$ in the build phase and the probe input from $P1$ and $P2$ in the probe phase. Although this chapter uses {\sf HashJoin} in most of its examples, the results can be generalized to other operators that expect certain inputs to be processed before other inputs. For example, the ideas in this chapter can also be used to schedule an {\sf ML} operator that takes two models as two inputs before processing the test data that arrives on a third input.

\section{Overview of \schfrmname scheduler}
\label{ssec:working-of-scheduler}

\begin{figure}[htbp]
 \begin{center}
	\includegraphics[width=4in]{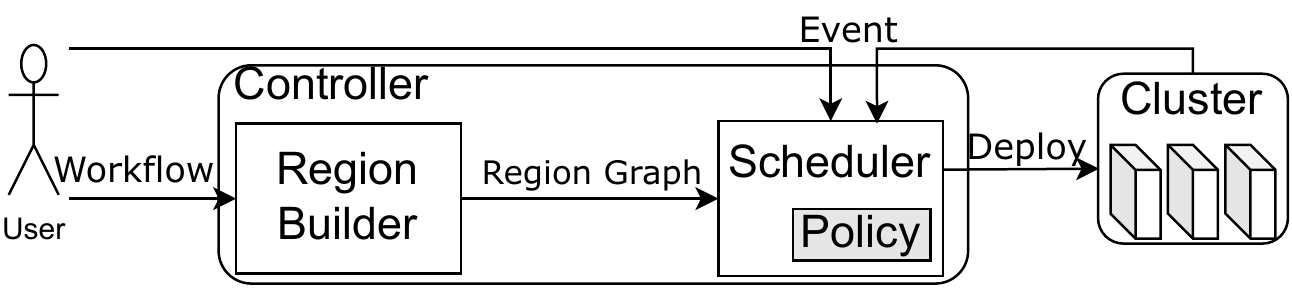}
	\caption{\label{fig:maestro-architecture}
		\textbf{Scheduling in \schfrmname}
	}
\end{center}
\end{figure}

A workflow DAG input into the engine needs to be scheduled for execution by the scheduler. This section gives a high level overview of the working of the \schfrmname scheduler. The \schfrmname scheduler is a module inside the controller of the workflow as shown in Figure~\ref{fig:maestro-architecture}. 
 
 The user submits a workflow to the data processing engine through an interface. The {\em region builder} compiles the workflow DAG to create a region graph. A region is a sub-DAG of the workflow and the region graph is a graph capturing the order in which the regions of a workflow should be executed. We discuss the concepts of region and region graph in detail in Section~\ref{sec:schedule-a-workflow}. The region graph is passed to the scheduler. The scheduler executes the regions according to the configured scheduling policy.


The scheduler receives certain {\em events} from the user or the workers executing the operators. Upon the receipt of an event, the scheduler makes scheduling decisions based on the policy. For example, when the user sends a command to start the workflow, the scheduler receives an event to begin the execution of the workflow. During the execution of a workflow, the scheduler receives various events about the execution status of the regions. An example of an event received during execution is the completion of a worker of an operator. The scheduler uses these events to decide on the scheduling of other regions.

\section{Building an acyclic region graph in \schfrmname}
\label{sec:schedule-a-workflow}

In this section, we describe how an input workflow is scheduled in \schfrmname. First, we define the concepts of region, dependencies between regions, and region graph (Section~\ref{ssec:defining-regions-and-dependencies}). In order to schedule a region graph, it should be acyclic. We show that in certain cases workflows may produce a cyclic region graph that cannot be scheduled. We also discuss the solutions to modify such workflows so that the region graph of the modified workflow is acyclic (Section~\ref{ssec:problem-cyclic-region-graph}). Lastly, we discuss the algorithm that takes in a workflow DAG and produces an acyclic region graph (Section~\ref{ssec:creating-region-graph}).


\subsection{Regions and their dependencies}
\label{ssec:defining-regions-and-dependencies}

\schfrmname divides a workflow DAG into sub-DAGs called ``regions'', each of which can be separately scheduled. We define the concepts of source operators, regions, dependencies between regions, region graph and other related terms below.

\begin{definition}[Source operator]
{\em Source operators} are the operators in a workflow that have no pipelined input links. In other words, the operator should have either no input links at all or only blocking input links. 
\end{definition}
The operators with no inputs are the scan operators that produce data in the workflow. Examples of operators with only blocking inputs are {\sf GroupBy} and {\sf Sort}. We use the concept of source operators to define regions as we will see soon.

\begin{figure*}[htbp]
	\begin{center}
	\includegraphics[width=4in]{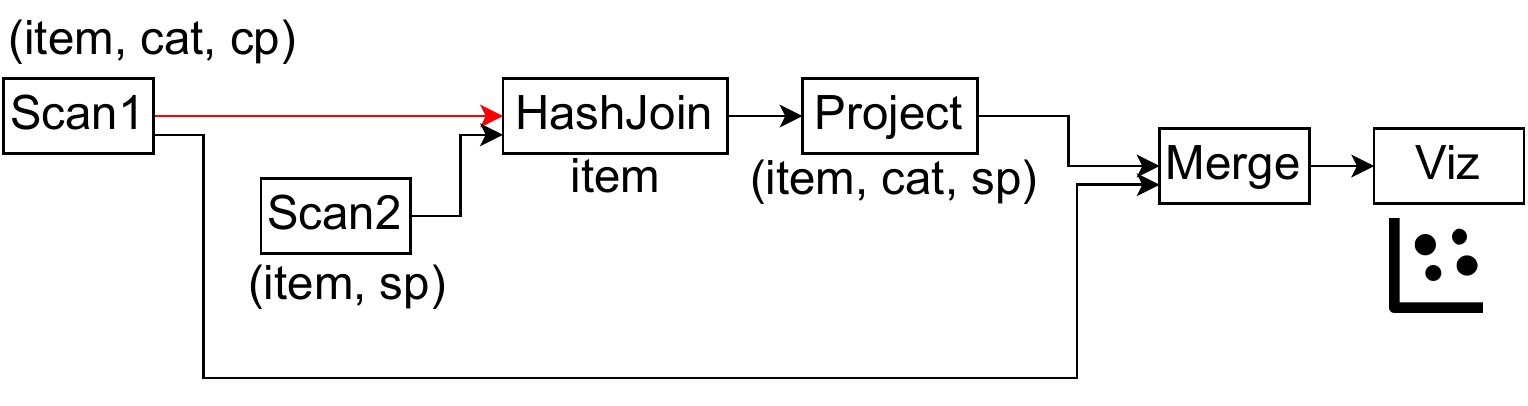} 
	\caption{\label{fig:finer-regions-need}
		\textbf{Workflow to understand regions. The blocking edges are shown in red. The {\sf HashJoin} requires the build input to be processed before the probe input arrives.}
	}
		\end{center}
\end{figure*}

Let us consider the following example. Figure~\ref{fig:finer-regions-need} shows a workflow to create a scatterplot of the selling price ({\em sp}) and the cost price ({\em cp}) of items belonging to different categories and compare the profit margins for the items. The probe input ({\sf Scan2}) containing the selling prices of items is joined with the build input ({\sf Scan1}) containing the cost price of the items on the {\em item} attribute and the output is projected to have the {\em item}, {\em category}, and {\em sp} attributes. The output from {\sf Scan1} and {\sf Project} are merged in the {\sf Merge} operator, which adds an extra column specifying if the tuple is from {\sf Scan1} or {\sf Project}. The merged data is visualized on a scatter plot.

A way to schedule the workflow without causing any exception is to first send a {\em start} message to the {\sf Scan1} operator. The data from {\sf Scan1} goes into {\sf HashJoin} as the build input and also to the {\sf Merge} operator that processes it and passes the output to the {\sf Viz} operator. After {\sf Scan1} has produced data completely and the data has been processed by the {\sf HashJoin} operator, a {\em start} message is sent to the {\sf Scan2} operator that provides the probe input to {\sf HashJoin}. The output of {\sf HashJoin} then goes to the {\sf Viz} operator through the {\sf Project} and {\sf Merge} operators. Thus, we define a region as the following.

\begin{definition}[Region]
A region is a sub-DAG of the workflow that starts at a source operator and contains all the operators reachable from the source operator using only pipelined edges.
\end{definition}

\begin{figure*}[htbp]
	\begin{center}
	\includegraphics[width=5in]{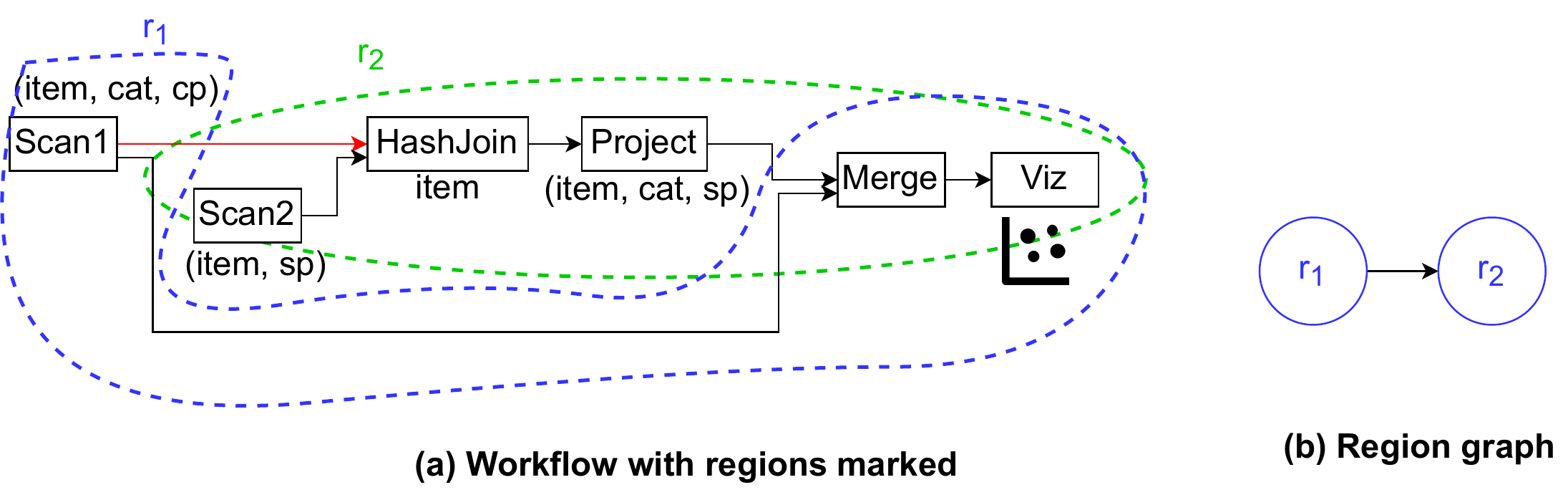} 
	\caption{\label{fig:finer-regions-example}
		\textbf{Regions in the workflow in Figure~\ref{fig:finer-regions-need}.}
	}
		\end{center}
\end{figure*}

The regions for the workflow in Figure~\ref{fig:finer-regions-need} are shown in Figure~\ref{fig:finer-regions-example}(a). Figure~\ref{fig:finer-regions-example}(b) shows the region graph for the workflow. We will discuss the concept of a region graph soon. When a {\sf Scan} operator in the workflow shown in the figure receives a command from the controller to start producing tuples, its output tuples propagate to all the operators that are reachable from the {\sf Scan} operator by pipelined links, and these reachable operators execute simultaneously with the {\sf Scan} operator.

Note that the {\sf HashJoin} operator in the Figure~\ref{fig:finer-regions-example} is active when region $r_1$ is executing so that it can receive and process the output of {\sf Scan1}. Systems such as Hyracks include {\sf HashJoin} in both $r_1$ and $r_2$~\cite{conf/icde/BorkarCGOV11}. In this chapter, we do not consider {\sf HashJoin} to be a part of $r_1$ for the purpose of simplicity of explanation. This way of representing regions is also adopted by systems such as Flink where operators do not belong to the region of their upstream operators producing blocking input~\cite{misc/flink-pipelined-region-scheduling}.

\begin{definition}[Execution of a region]
The {\em execution} of a region refers to the deployment of the operators of the region on the cluster and the subsequent processing of input data by the operators.
\end{definition}
A region is executed by the controller by sending a start message to its source operators. If a region has not been executed yet, it is called an {\em unscheduled region}.

\begin{definition}[Completion of a link and region]
A link between operators is said to be {\em completed} if all the data on the link has been received by the destination operator. A region is said to be {\em completed} if all its operators have processed all their input data and produced all their output data.
\end{definition}

\begin{definition}[Dependency between regions]
A region $Y$ is said to be {\em dependent} on another region $X$ if there exists at least one blocking link $l$ from an operator in $X$ to an operator in $Y$ and all the data on $l$ should be completely received before the operators in $Y$ start executing. 
\end{definition}
In Figure~\ref{fig:finer-regions-example}(a), regions $r_2$ depends on $r_1$ because $r_1$ produces the build input for the {\sf HashJoin} operator in the regions $r_2$.

\begin{definition}[Ready region]
Let a region $r$ depend on a set of regions represented by $D$. The region $r$ is said to be {\em ready} when all the regions in $D$ have completed and all blocking links to $r$ from the regions in $D$ have also completed.  When a region is ready, it can be executed.
\end{definition}

When a region has to be scheduled, all the regions that it depends on should have already  completed. Thus, the scheduler has to follow an order when executing the regions of a workflow. This order is captured in the region graph for a workflow, which we define below.

\begin{definition}[Region graph]
The {\em region graph} for a workflow is a graph with the regions of the workflow as its vertices and the dependencies between the regions as its edges. An edge from a region $X$ to a region $Y$ in the region graph means that $Y$ is dependent on $X$.
\end{definition}

\begin{figure*}[htbp]
	\begin{center}
	    
	\includegraphics[width=6in]{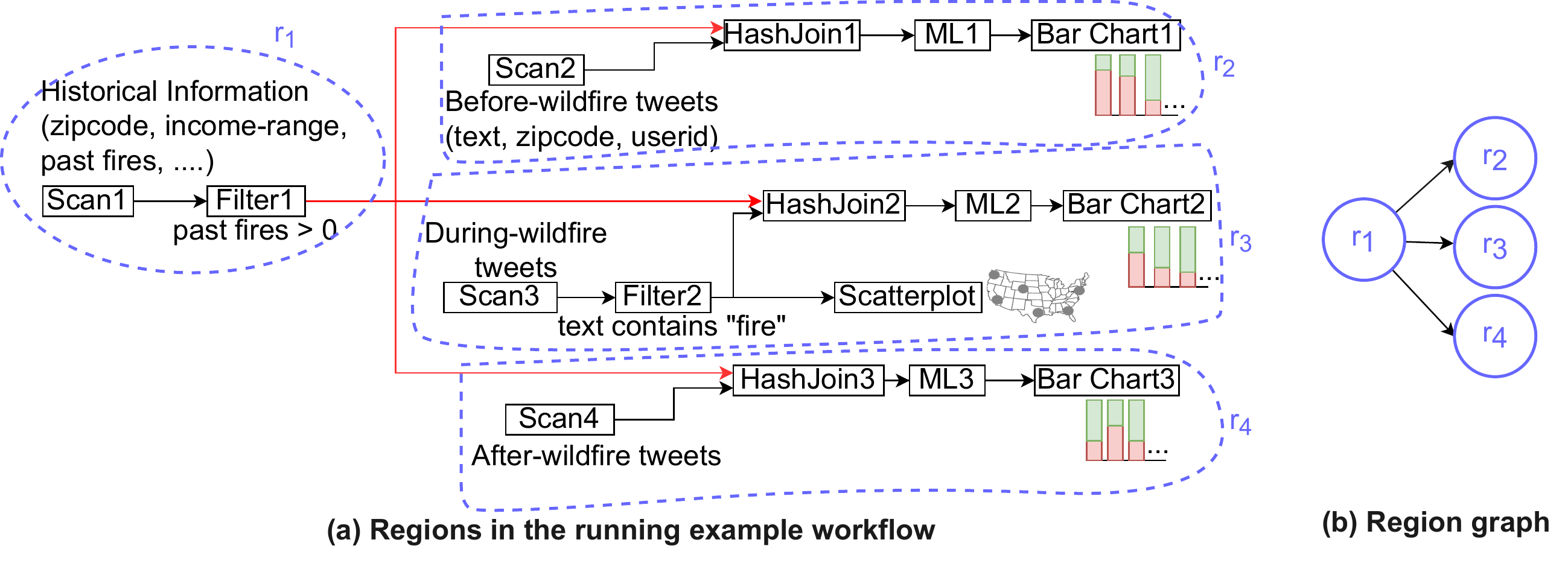} 
	\caption{\label{fig:maestro-running-example-with-regions}
		\textbf{Regions and region graph from the running example workflow. Blocking input links are in red.}
	}
		\end{center}

\end{figure*}

The region graph for the workflow in Figure~\ref{fig:finer-regions-example}(a) is shown in Figure~\ref{fig:finer-regions-example}(b). We also show the regions in the workflow in Figure~\ref{fig:maestro-running-example} and the corresponding region graph in Figure~\ref{fig:maestro-running-example-with-regions}.

\subsection{Avoiding cycles in region graphs}
\label{ssec:problem-cyclic-region-graph}

The regions must be scheduled in a topological order based on the region graph, in order to ensure that a region is executed only when all the regions it depends on have been executed. In order to obtain a topological order, the region graph should not have any cycles, i.e., it should be a DAG. As we show soon, a workflow may yield a region graph with cycles. Such a workflow has to be modified in a way that the modified workflow yields an acyclic region graph. Next we show an example of a workflow that yields a cyclic region graph. We also discuss ways to modify the workflow so that the modified workflow has an acyclic region graph.

\begin{figure*}[htbp]
	\begin{center}
	\includegraphics[width=4.8in]{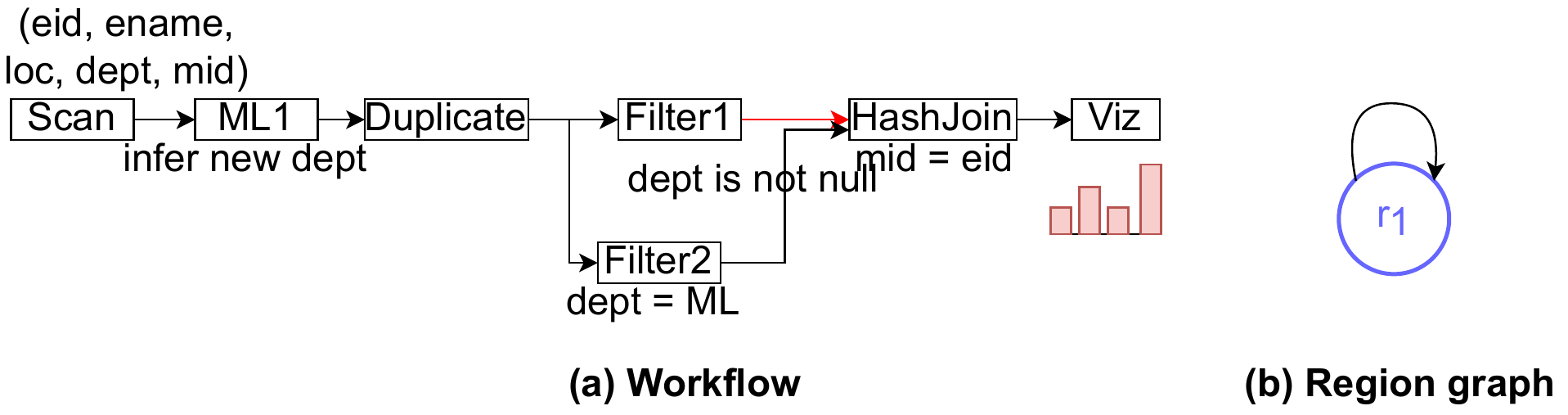} 
	\caption{\label{fig:cyclic-region-simple-workflow}
		\textbf{Workflow with cyclic region graph. When there is a cycle in the region graph, there is no feasible schedule of regions.}
	}
		\end{center}
\end{figure*}

\boldstart{Example workflow with cyclic region graph.} Let us see an example of a workflow that yields a cyclic region graph. Figure~\ref{fig:cyclic-region-simple-workflow} shows a workflow DAG and its corresponding region graph. The workflow tries to find a new department for the employees in the machine learning department and their managers. It visualizes the count of employees and managers by the new departments they have been allotted to. There is only one region in the workflow whose source is the {\sf Scan} operator. The {\sf HashJoin} expects to first completely receive and process its build input from {\sf Filter1}. Then, it expects to receive the probe input from {\sf Filter2}. Since both {\sf Filter1} and {\sf Filter2} belong to the same region $r_1$, the region graph has a cycle, as shown in the figure. Such a cyclic region graph cannot be scheduled.

\begin{figure*}[htbp]
	\begin{center}
	\includegraphics[width=\linewidth]{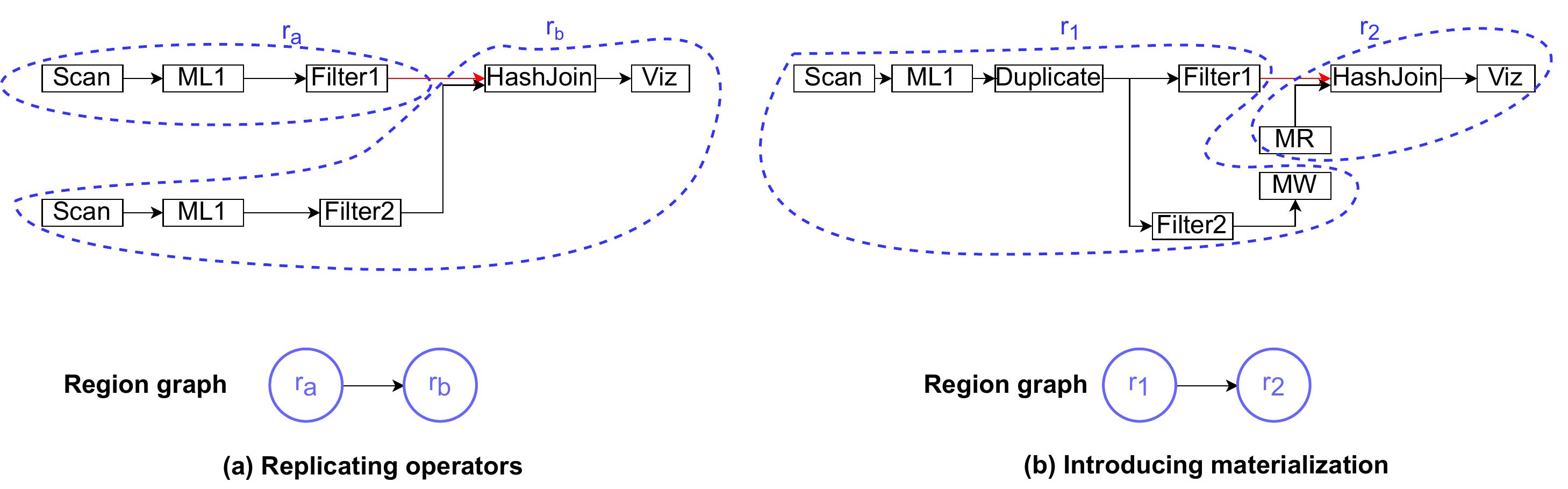} 
	\caption{\label{fig:cyclic-region-simple-workflow-solving}
		\textbf{Modifying workflow in Figure~\ref{fig:cyclic-region-simple-workflow} to create an acyclic region graph.}
	}
		\end{center}
\end{figure*}

\boldstart{Modifying the workflow to get an acylic region graph.} In order to enforce the requirement of the probe input of {\sf HashJoin} arriving after the build input, the two inputs need to be produced by different regions. The region producing the build input will need to be scheduled first, followed by the region producing the probe input. There are different ways to modify the workflow so that the probe input for {\sf HashJoin} arrives from a different region than the build input. One way is to remove the {\sf Duplicate} operator and replicate the operators {\sf Scan1} and {\sf ML1} to supply the data to {\sf Filter2} as shown in Figure~\ref{fig:cyclic-region-simple-workflow-solving}(a). This creates two regions $r_1$ and $r_2$, and $r_2$ has a dependency on $r_1$. This way incurs the overhead of repeated computation. For example, in Figure~\ref{fig:cyclic-region-simple-workflow-solving}(a), the input data is processed by the same ML operator twice. 

Another way to create a new region is to introduce materialization. Materialization is introduced on a link by adding a materialization writer and a materialization reader operator to the link. The materialization writer writes all the data produced by the source operator of the link to a memory or disk. The materialization reader reads the data from the storage and outputs it to the destination operator of the link. In Figure~\ref{fig:cyclic-region-simple-workflow-solving}(b), materialization writer and reader operators have been put between the operators {\sf Filter2} and {\sf HashJoin}. The introduction of materialization also creates two regions $r_1$ and $r_2$, where $r_2$ is dependent on $r_1$. The {\sf Scan} operator is the source of $r_1$ and the materialization reader operator ({\sf MR}) is the source of $r_2$. We focus on the case where computational resources are more expensive than storage. We will follow the materialization way to modify the workflow DAG in case of cyclic region graphs.


\subsection{Creating an acyclic region graph}
\label{ssec:creating-region-graph}

In Section~\ref{ssec:problem-cyclic-region-graph}, we saw the problem of cyclic region graphs and discussed ways to modify the workflow to yield an acyclic region graph. In this section, we discuss an algorithm (Algorithm~\ref{alg:region-graph-creation}) that takes a workflow DAG $W$ as input and yields an acyclic region graph $\mathcal{G}$. If needed, the algorithm may modify the workflow DAG $W$ by introducing materialization operators to yield the acyclic region graph. We use the workflow in Figure~\ref{fig:cyclic-region-simple-workflow}(a) to explain the various steps of the algorithm.

\begin{figure*}[htbp]
	\begin{center}
	\includegraphics[width=4in]{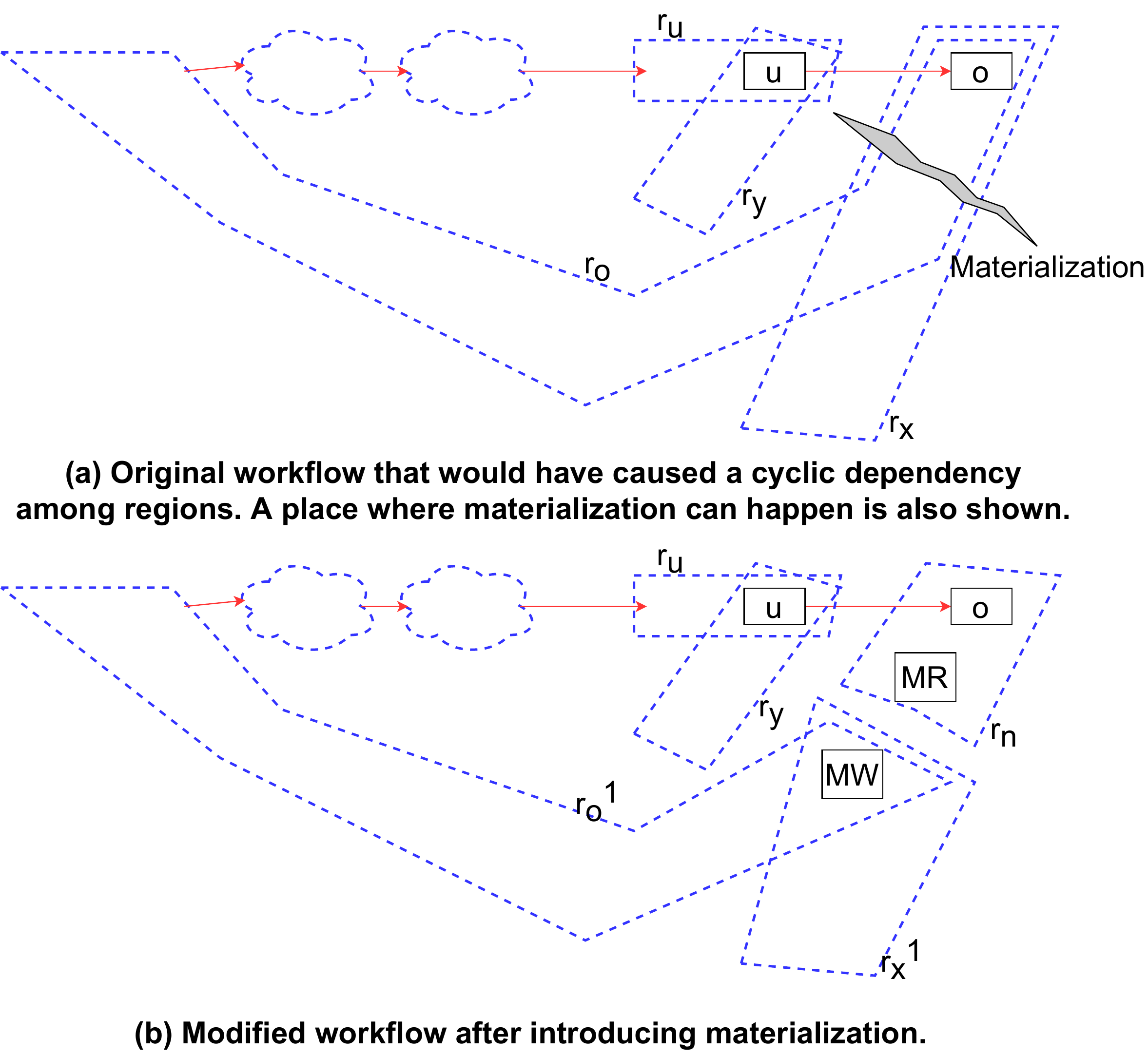} 
	\caption{\label{fig:algorithm-to-produce-region-graph}
		\textbf{Diagram to help explain the intuition behind lines~\ref{algline:blocking-input-o} to \ref{algline:blocking-input-o-end} in Algorithm~\ref{alg:region-graph-creation}. Regions are marked by dotted blue shapes.}
	}
		\end{center}
\end{figure*}

\begin{algorithm}[htbp]
\caption{\hl{Acyclic region graph creation.} \label{alg:region-graph-creation}}
        \KwIn{$W \gets$ A workflow DAG}
        \KwOut{$\mathcal{G}$: A region graph}
     $\mathcal{S} \gets$ Set of source operators $W$ \\ \label{algline:all-sources-in-workflow}
     Initialize $\mathcal{G}$ to an empty graph \\ \label{algline:empty-g}
     Initialize $\mathcal{R}$ to an empty set of regions \\ \label{algline:empty-r}
     $O \gets$ A topological order of the operators in $W$ \\ \label{algline:topological-order}
     \ForEach{operator o in O} {
        $s_o \gets$ set of sources in $\mathcal{S}$ from which $o$ is reachable using pipelined links \\ \label{algline:sources-of-o}
        \ForEach{source $s$ in $s_o$}{
            \If{\textbf{not}(there exists a region in $\mathcal{R}$ with $s$ as the source)}{
                Add a new region $r$ in $\mathcal{R}$ with $s$ as the source \\ \label{algline:add-s-to-R}
                Add a vertex in $\mathcal{G}$ corresponding to $r$ \\ \label{algline:add-region-vertex}
            }
            Add operator $o$ to the region in $\mathcal{R}$ that has $s$ as source \\ \label{algline:add-o-to-r}
        }
        \If{blocking input link b exists to o}{ \label{algline:blocking-input-o}
            $u \gets$ Upstream operator of link $b$ \\
            \While{\textbf{true}}{
                $SR_u \gets$ Set of regions that $u$ belongs to \\ \label{algline:blocking-input-o-region}
                $SR_o \gets$ Set of regions that $o$ belongs to \\ \label{algline:o-region}
                
                \If{($\exists r_u$ in $SR_u$ and $\exists r_o$ in $SR_o$ such that adding edge $r_u \rightarrow r_o$ causes a cycle in $\mathcal{G}$)}{ \label{algline:check-if-cycle}
                    call Add-Materialization($W$, $\mathcal{R}$, $\mathcal{G}$, $r_u$, $r_o$, $o$) \\ \label{algline:add-mat-ro} \tcp{This subroutine is discussed in Algorithm~\ref{alg:adding-mat-cylic-region-graph}. It adds materialization to region $r_o$. The materialization affects the structure of region $r_o$ and possible other regions too. This subroutine updates $\mathcal{R}$ and $\mathcal{G}$.}
                }
                \Else{
                    \ForEach{$r_u \in SR_u$}{
                        \ForEach{$r_o \in SR_o$}{
                            Add edge $r_u \rightarrow r_o$ to $\mathcal{G}$ \\
                        }
                    }
                    \textbf{break} \\ \label{algline:break}
                } \label{algline:blocking-input-o-end}
                
            }
        }
     } \label{algline:added-edges}
     return $\mathcal{G}$
\end{algorithm}

Let $\mathcal{S}$ be the set of all the source operators in the workflow (line~\ref{algline:all-sources-in-workflow}). For the workflow in Figure~\ref{fig:cyclic-region-simple-workflow}(a), there is only one source, which is the {\sf Scan} operator. The region graph $\mathcal{G}$ is initially empty (line~\ref{algline:empty-g}). The set of regions $\mathcal{R}$ is also initially empty (line~\ref{algline:empty-r}). Let $O$ be a topologically ordered sequence of operators in $W$ (line~\ref{algline:topological-order}). A topological order for the operators in the workflow in the figure is $\bigl[${\sf Scan}, {\sf ML1}, {\sf Duplicate}, {\sf Filter1}, {\sf Filter2}, {\sf HashJoin}, {\sf Viz}$\bigr]$. The operators in $O$ are sequentially considered and the following is done for each of them. Let the current operator in the sequence be $o$. Let $s_o$ be the set of sources from which $o$ is reachable using pipelined edges (line~\ref{algline:sources-of-o}). For each source $s$ in $s_o$, we check if the region corresponding to the source $s$ is present in $\mathcal{R}$. If it is not present, we add a region $r$ that has $s$ as the source in $\mathcal{R}$ (line~\ref{algline:add-s-to-R}). We also add a vertex representing $r$ in $\mathcal{G}$ (line~\ref{algline:add-region-vertex}). In the example workflow, the {\sf Scan} operator is considered first. Since the region that has {\sf Scan} as the source is not present in $\mathcal{R}$ and $\mathcal{G}$, it is added to both. We denote this region as $r_1$ in our example. The next operators in the order are {\sf ML1}, {\sf Duplicate}, {\sf Filter1}, and {\sf Filter2} and they are considered one by one. All these operators only have one source from which they are reachable ({\sf Scan}) and the region corresponding to that source ($r_1$) is already in $\mathcal{R}$ and $\mathcal{G}$. 

Operators such as {\sf HashJoin} have one blocking input link. On the other hand, a machine-learning operator that accepts two models on blocking inputs before processing test tuples using the models has two blocking input links. For simplicity, in the algorithm, we assume that the operator has at most one blocking input link. Let the blocking input link (if it exists) be $b$. We denote the set of regions that the upstream operator (say, $u$) of $b$ belongs to as $SR_u$ (line~\ref{algline:blocking-input-o-region}). We denote the set of regions that the operator $o$ belongs to as $SR_o$ (line~\ref{algline:o-region}). Figure~\ref{fig:algorithm-to-produce-region-graph} shows a case where $u$ belongs to two regions ($r_u$ and $r_y$) and $o$ belongs to two regions ($r_o$ and $r_x$). An edge needs to be added from each region in $SR_u$ to each region in $SR_o$. If one of these edges causes a cycle in $\mathcal{G}$, a subroutine is called to add materialization (lines~\ref{algline:check-if-cycle} to \ref{algline:break}). The details of this subroutine are presented in the next section. This subroutine adds materialization and modifies $\mathcal{R}$ and $\mathcal{G}$ to reflect the newly created regions and edges due to materialization. Note that the set of regions $SR_u$ and $SR_o$ may change after materialization, as we will show soon using Figure~\ref{fig:algorithm-to-produce-region-graph}. The Algorithm~\ref{alg:region-graph-creation} again computes the sets of regions $SR_u$ and $SR_o$ and tries to add edges between each pair of regions in these sets. This process continues till edges have been added from each region in $SR_u$ to each region in $SR_o$ without causing cycles.

In Figure~\ref{fig:algorithm-to-produce-region-graph}(a), adding an edge from the region $r_u$ to region $r_o$ causes a cycle. Thus, we need to add materialization to $r_o$. A materialization option is shown in the figure. This materialization breaks the region $r_o$ and $r_x$ as shown in Figure~\ref{fig:algorithm-to-produce-region-graph}(b) and creates a new region $r_n$ with the materialization reader as the source. Thus, the set $SR_o$ changes from \{$r_o$, $r_x$\} (before materialization) to \{$r_n$\} (after materialization). Edges $r_u \rightarrow r_n$ and $r_y \rightarrow r_n$ are added to the region graph.

In our example workflow in Figure~\ref{fig:cyclic-region-simple-workflow}(a), when the {\sf HashJoin} operator is considered, the regions of {\sf Filter1} and {\sf HashJoin} are both $r_1$. Since adding an edge from $r_1$ to $r_1$ will cause a self-loop, materialization is added on the link between {\sf Filter2} and {\sf HashJoin}. The added materialization reader becomes the source of the region $r_2$. The {\sf HashJoin} operator is now in the region $r_2$. In this case, $SR_u = \{r_1\}$ and $SR_o = \{r_2\}$. An edge is then added from $r_1$ to $r_2$. Once we have gone through all the operators in the topological order, the graph $\mathcal{G}$ is returned.

Note that Algorithm~\ref{alg:region-graph-creation} also works for operators that only have one input link which is blocking. Examples of such operators are {\sf GroupBy} and {\sf Sort}. For such an operator (say $o$), the algorithm adds an edge from the region containing the upstream operator of $o$ to the region containing $o$.

\section{Choosing a materialization option}
\label{sec:materialization-choices}

The Algorithm~\ref{alg:region-graph-creation} in Section~\ref{sec:schedule-a-workflow} may add materialization to the input workflow in order to produce an acyclic region graph. In this section, we enumerate the different options where the materialization operators  can be placed in the workflow (Section~\ref{ssec:enumerate-materialization-choices}) and discuss how it affects result awareness (Section~\ref{ssec:mat-choice-result-aware}).

\subsection{Enumerating the materialization choices}
\label{ssec:enumerate-materialization-choices}

In Figure~\ref{fig:cyclic-region-simple-workflow-solving}(b), the materialization operators were placed between the {\sf Filter2} and the {\sf HashJoin} operators. However, another choice is to place them between the {\sf ML1} and {\sf Filter2} operators. Both choices create two regions and result in the same region graph as shown in Figure~\ref{fig:cyclic-region-simple-workflow-solving}(b). In general, there can be multiple choices where materialization operators can be placed in a workflow. We are going to describe the process to enumerate all those choices.

We first give an intuitive explanation of how to enumerate the different materialization choices by giving an example. Then, we give a formal approach for enumerating the choices.

\begin{figure*}[htbp]
	\begin{center}
	\includegraphics[width=\linewidth]{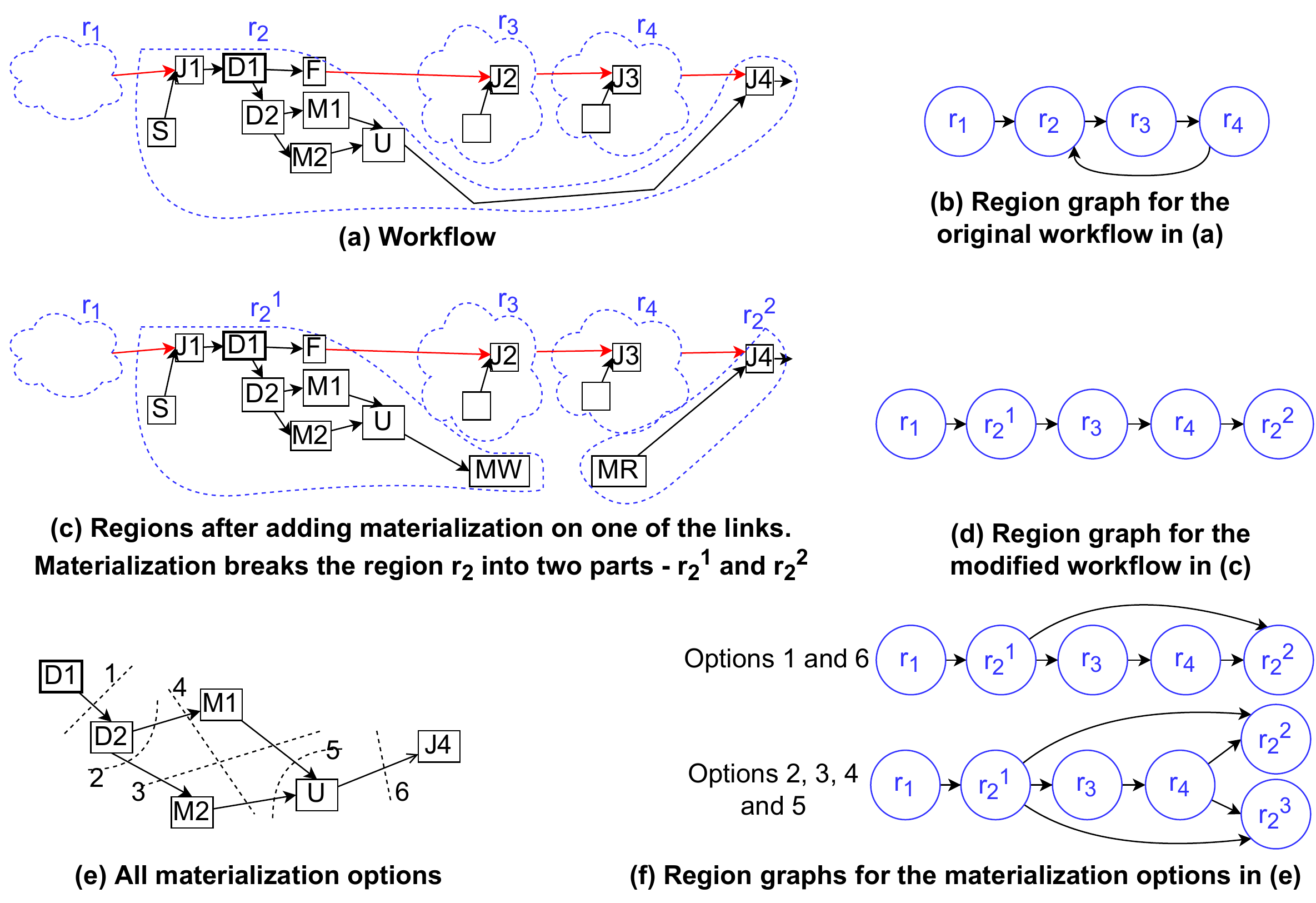} 
	\caption{\label{fig:cyclic-region-mat-choices}
		\textbf{Enumerating all the materialization choices. {\sf J1}, {\sf J2}, {\sf J3}, and {\sf J4} are {\sf HashJoin} operators. {\sf D1} and {\sf D2} are replication operators. {\sf M1} and {\sf M2} are ML operators. {\sf S} is scan, {\sf F} is filter, and {\sf U} is a union operator.}
	}
	\end{center}
\end{figure*}

\boldstart{Intuition.} Figure~\ref{fig:cyclic-region-mat-choices}(a) shows the outline of a workflow with its regions marked. The corresponding region graph containing a cycle is shown in Figure~\ref{fig:cyclic-region-mat-choices}(b). The problem exists at the operator {\sf J4} that receives its probe input from the region $r_2$ before the build input arrives from the region $r_4$. In order to solve this problem, we use materialization to modify the workflow DAG in such a way that the probe input to {\sf J4} can be delayed. The materialization should split the region $r_2$ into two parts, thereby ``cutting off'' the flow of data from $r_2$ into the probe input of {\sf J4}.  This is similar to the network flow problem where one needs to determine the cut edges in the graph that make a target vertex unreachable from the source vertices~\cite{fulkerson1962flows}. In our case, the target vertex is the operator whose probe input has to be cut off ({\sf J4}) and the cut edges are the links where data has to be materialized. Figure~\ref{fig:cyclic-region-mat-choices}(c) shows materialization added to the link between {\sf U} and {\sf J4}. It cuts off the probe input to {\sf J4} and divides the region $r_2$ into two parts $r_2^{1}$ and $r_2^{2}$ as shown in Figure~\ref{fig:cyclic-region-mat-choices}(d).

A key consideration while adding the materialization is that the materialization should cut off the probe input to {\sf J4} from $r_2$, but the build input to {\sf J2} (in $r_3$) should still come from $r_2$. If this does not happen and the build input to {\sf J2} is produced by the newly created region, then there still will be a cycle in the region graph. Thus, we need to identify the sub-DAG of the workflow where the materialization can happen such that it cuts off the probe input to {\sf J4} from $r_2$ but does not impact the build input to {\sf J2} from $r_2$. This sub-DAG for the workflow in Figure~\ref{fig:cyclic-region-mat-choices}(a) is shown in Figure~\ref{fig:cyclic-region-mat-choices}(e). The figure also shows $6$ different sets of cuts, each of which disconnects {\sf D1} and {\sf J4}. These cuts form the $6$ materialization choices. Each cut includes a set of edges where the data has to be materialized. The choices $1$ and $6$ consist of one edge only, which means that they require a single materialization writer and reader operators. When one of these two options is chosen, the region $r_2$ is broken into $r_2^{1}$ and $r_2^{2}$. The single materialization reader becomes the source of the region $r_2^{2}$ and the resulting region graph is shown in Figure~\ref{fig:cyclic-region-mat-choices}(f). The choices $2$, $3$, $4$, and $5$ consist of two edges each, which means that they require two materialization writer and reader operators, one pair on each edge. When one of these four options is chosen, the region $r_2$ is broken into three regions $r_2^{1}$, $r_2^{2}$ and $r_2^{3}$. The two materialization readers become the source of the regions $r_2^{2}$ and $r_2^{3}$, and the resulting region graph is shown in Figure~\ref{fig:cyclic-region-mat-choices}(f).

\begin{figure*}[htbp]
	\begin{center}
	\includegraphics[width=\linewidth]{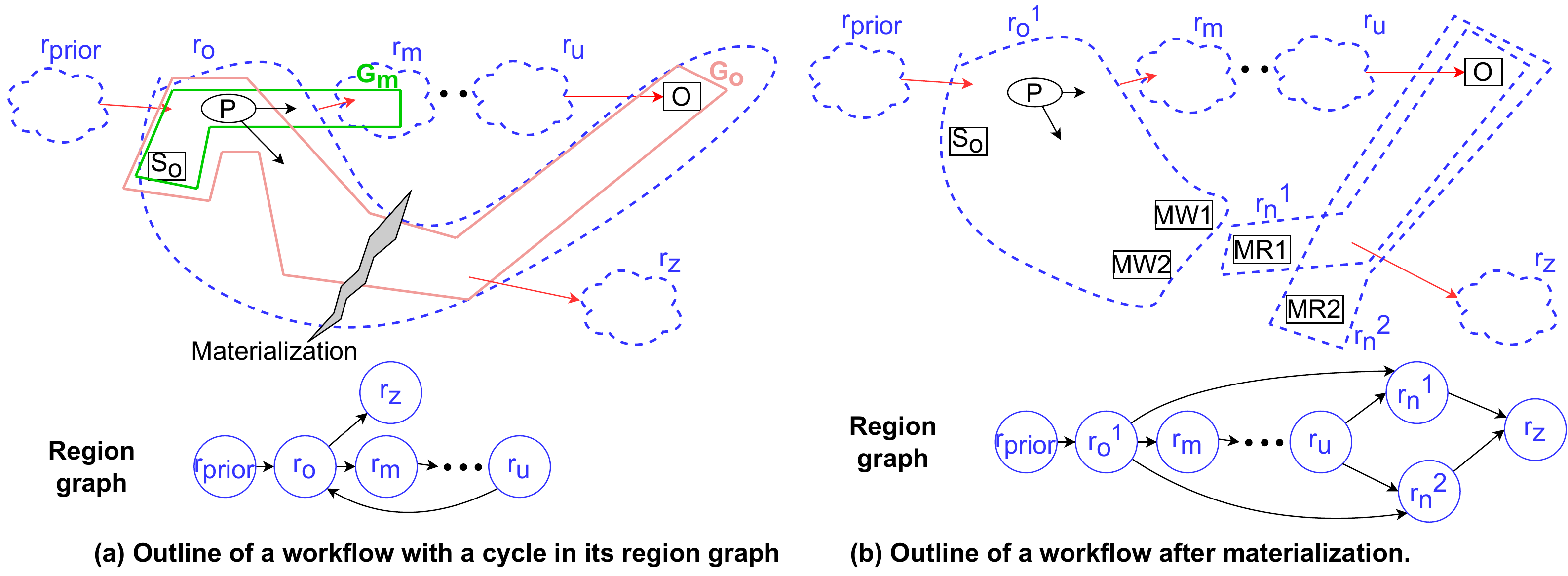} 
	\caption{\label{fig:cyclic-region-mat-choices-formal-approach-outline}
		\textbf{Understanding the formal approach to find the sub-DAG where materialization is to be added.}
	}
	\end{center}
\end{figure*}

\boldstart{Formal approach.} We present an algorithm to describe how materialization is added to the workflow DAG in the line~\ref{algline:add-mat-ro} of Algorithm~\ref{alg:region-graph-creation}. Consider that $W$, $R$, and $\mathcal{G}$ are the workflow DAG, set of regions and the region graph, respectively, before that line and $O$ is the operator that is responsible for an edge from a region $r_u$ to $r_o$ that would result in a cycle in the region graph $\mathcal{G}$.  Algorithm~\ref{alg:adding-mat-cylic-region-graph} adds materialization to $W$ to obtain a modified workflow DAG $W'$ and prevents the creation of a cycle in $\mathcal{G}$. Figure~\ref{fig:cyclic-region-mat-choices-formal-approach-outline}(a)
shows an example outline of the workflow $W$ with region $r_o$ and $r_u$ marked. Let $r_m$ be the region after $r_o$ in the cycle created by $O$ in the region graph of $W$ (line~\ref{algline:formal-find-rm}). The operators in the region $r_m$ receives blocking input from the operators in $r_o$. As explained above, the materialization should cut off the input to $O$ from $r_o$ but not impact the blocking input to $r_m$ from $r_o$.

\begin{algorithm}[htbp]
\caption{\hl{Adding materialization to a workflow.} \label{alg:adding-mat-cylic-region-graph}}
        \KwIn{$W, \mathcal{R}, \mathcal{G} \gets$ Workflow DAG, set of regions and region graph before line~\ref{algline:add-mat-ro} in Algorithm~\ref{alg:region-graph-creation}}
        \KwIn{$O \gets$ Current operator at line~\ref{algline:add-mat-ro} in Algorithm~\ref{alg:region-graph-creation} due to which Algorithm~\ref{alg:region-graph-creation} needs to add an edge from a region $r_u$ to a region $r_o$, that results in a cycle in $\mathcal{G}$}
        \KwOut{$W'$, $\mathcal{R'}$, $\mathcal{G'}$: Modified workflow DAG, set of regions and region graph after adding materialization to $W$}
    
    $r_m \gets$ The region that is after $r_o$ in the cycle in $\mathcal{G}$ \\ \label{algline:formal-find-rm}
    $S_o \gets$ The source of $r_o$ \\ \label{algline:formal-find-so}
    $G_o \gets$ The sub-DAG supplying the pipelined input from $S_o$ to $O$ \\ \label{algline:formal-find-go}
    $G_m \gets$ The sub-DAG supplying the blocking input from $S_o$ to the operators in region $r_m$ \\ \label{algline:formal-find-gm}
    $G_f \gets G_o - G_m$ \tcp*[h]{The sub-DAG where materialization can be done without affecting input to $r_m$} \\ \label{algline:formal-find-gf}
    $P \gets $ The set of operators in $G_m$ that supply the pipelined input to $O$ \\ \label{algline:formal-find-p}
    $E \gets$ A cut-edge set in $G_f$ that makes $O$ unreachable from all operators in $P$ \\ \label{algline:formal-find-e}
    Modify $W$ by adding materialization operators to all the edges in $E$ \\ \label{algline:formal-find-wm}
    $SR_b \gets$ The set of regions that will be broken due to materialization \\
    For each region in $SR_b$, remove it from  $\mathcal{R}$ and the corresponding vertex from $\mathcal{G}$ \\ \label{algline:remove-srb-vertices}
    $SS_b \gets $ The set of sources of the regions in $SR_b$ \\
    $SS_m \gets $ The set of materialization readers introduced \\
    \ForEach{source $s \in SS_b \cup SS_m$}{ \label{algline:add-srw-s}
        Add a region $r$ to $\mathcal{R}$ representing the region that has $s$ as the source and containing all the operators reachable from $s$ using pipelined edges \\
        Add a vertex to $\mathcal{G}$ representing the region $r$ \\ 
    } \label{algline:add-srw-e} 
    \ForEach{materialization reader $t$ in $SS_m$}{ \label{algline:add-srr-s}
        Add an edge in $\mathcal{G}$ from $r_u$ to the region that has $t$ as the source  \\ \label{algline:add-edge-ru-to-t}
    } \label{algline:add-srr-e}
    $SR_w \gets$ the new regions that have the sources of the regions in $SR_b$ as their sources \\
    $SR_r \gets$ the new regions that have the materialization readers as their sources \\
    \ForEach{region $r$ in $SR_w$}{ \label{algline:srw-srr-s}
        Add edges from $r$ to the regions in $SR_r$ that read a materialization output of $r$ \\
    } \label{algline:srw-srr-e}
    call Add-edges-to-new-regions-in-region-graph($SR_w$, $SR_r$, $\mathcal{G}$, $W$) \\ 
    return modified $W$, $\mathcal{R}$, and $\mathcal{G}$ \\ \label{algline:add-edge-subroutine}

\end{algorithm}

Let $S_o$ be the source of the region $r_o$ (line~\ref{algline:formal-find-so}). Let the sub-DAG that supplies pipelined input to $O$ from $S_o$ be called $G_o$ (line~\ref{algline:formal-find-go}), as shown in the figure. Let the sub-DAG that supplies blocking input from $S_o$ to $r_m$ be called $G_m$ (line~\ref{algline:formal-find-gm}), as shown in the figure. Let $G_f = G_o - G_m$. On a high level, the graph $G_f$ is the part of $r_o$ that produces pipelined input for $O$, but does not produce the blocking input for $r_m$. The input to $G_f$ comes from a set of operators in $G_m$ that are represented by $P$ in the figure. Considering the sub-DAG $G_f$, we need to identify all possible cut-edges sets~\cite{fulkerson1962flows} that make the operator $O$ unreachable from the operators in $P$. Each element of the set is a materialization choice and contains a collection of edges where materialization can be added. We choose one materialization option and apply it to $W$ (line~\ref{algline:formal-find-wm}).

Let us assume that the option shown in Figure~\ref{fig:cyclic-region-mat-choices-formal-approach-outline}(a) is chosen for materialization and use it for further discussion. Let the set of regions affected by the materialization be $SR_b$. In the figure, $r_o$ is the only region affected by materialization. The vertices and regions corresponding to these regions are removed from  $\mathcal{G}$ and $\mathcal{R}$, respectively (line~\ref{algline:remove-srb-vertices}) because materialization breaks these regions into multiple smaller regions. Then, we begin the process of adding the new regions to $\mathcal{G}$ and $\mathcal{R}$. For the source of each region in $SR_b$, we add a new region to $\mathcal{G}$ and $\mathcal{R}$, that contains all the operators reachable from the source using pipelined edges (lines~\ref{algline:add-srw-s}-\ref{algline:add-srw-e}). These are the regions that contain the materialization writer operators and the set of these regions is denoted by $SR_w$. In the figure, $SR_w$ only contains $r_o^1$. Similarly, for each materialization reader introduced, we add a corresponding region to $\mathcal{R}$ and $\mathcal{G}$. We also add an edge in $\mathcal{G}$ from $r_u$ to the newly added regions with materialization readers as the sources (lines~\ref{algline:add-srr-s}-\ref{algline:add-srr-e}). In the figure, two materialization readers are introduced that create two new regions, namely $r_n^1$ and $r_n^2$. These regions now contain the operator $O$ that receives the blocking input from $r_u$. Thus, edges are added from $r_u$ to the two new regions in the region graph. Let $SR_r$ be the set of new regions that have the materialization readers as the sources. Next, we iterate through each region (say $r_w$) in $SR_w$ and add an edge from $r_w$ to the regions in $SR_r$ that consume the materialized output of $r_w$ (lines~\ref{algline:srw-srr-s}-\ref{algline:srw-srr-e}). In the figure, this leads to an edge from $r_o^1$ to $r_n^1$ and $r_n^2$.

When we removed the regions in $SR_b$ from $\mathcal{G}$, we also removed the corresponding edges of those regions. Thus, the edges from the regions in $SR_w$ and $SR_r$ to the other regions in $\mathcal{G}$ have to be found and added (line~\ref{algline:add-edge-subroutine}). For example, in the figure, the region $r_o$ used to produce the blocking input to the region $r_z$. After materialization, the regions $r_n^1$ and $r_n^2$ produce the blocking input to $r_z$. Thus, the edges $r_n^1 \rightarrow r_z$ and $r_n^2 \rightarrow r_z$ need to be added to $\mathcal{G}$. The computation of these edges can be done from scratch by iterating over all the operators in the workflow processed till now according to the topological sequence. Another way is to compute these edges incremental by remembering the edges to the regions in $SR_b$ before materialization and adjusting only those edges. At the end, the modified $W$, $\mathcal{R}$, and $\mathcal{G}$ are returned.

\subsection{Result-aware materialization choice selection}
\label{ssec:mat-choice-result-aware}

We saw how to enumerate different materialization choices in Section~\ref{ssec:enumerate-materialization-choices}. In this section, we discuss how to consider result awareness while choosing a materialization option. We assume that there is a single cycle in the region graph and there is a single result operator in the workflow. We first define first response time, which is used as a metric for result awareness in this chapter. Then, we discuss how to choose a materialization option that gives a good first response time.

\subsection{First response time}
\label{ssec:first-response-time}

In exploratory analysis, the users want to quickly observe some initial output from the result operators. This allows the users to find issues in the analysis early without waiting for the entire workflow to finish executing~\cite{journals/pvldb/ShangZBEKMRK21}. Consider the following scenario. In the running example, when $r_3$ runs, the user may observe that the scatterplot operator contains many tweets related to other types of fires different from wildfire. Since the goal of the analysis is to determine whether wildfires lead to climate change awareness, the user may decide to add more keywords to the {\sf Filter2} operator to make the tweets more relevant to wildfire.

Considering a workflow with a single result operator, we define the {\em first response time} as follows. It is the time required from the start of execution of a workflow to the time when the result operator has produced a specific number of tuples (say, $d$). We denote the first response time by $\tau(d)$, where $d$ is the number of tuples that the result operators should produce. For simplicity, we consider $d=1$ in this chapter.

\begin{figure*}[htbp]
	\begin{center}
	\includegraphics[width=4in]{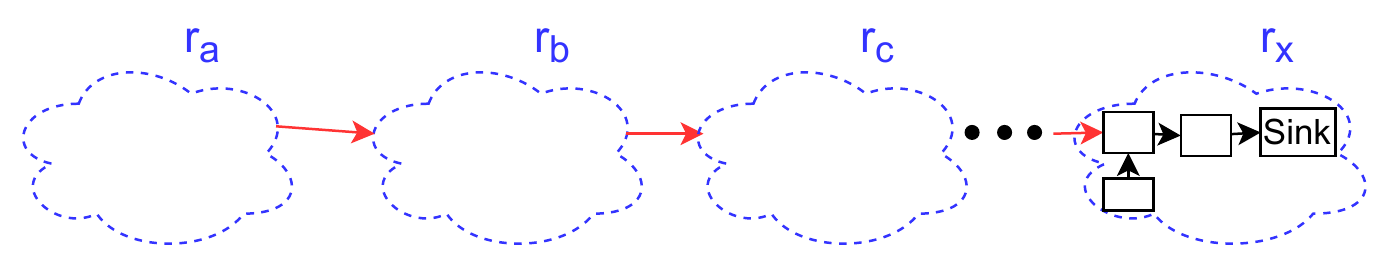} 
	\caption{\label{fig:first-response-time-def}
		\textbf{The outline of a workflow showing regions. The sink operator in the last region. For first response time, we consider the time to completely execute all regions except the region containing the sink. For the region containing the sink, the time to produce only a single tuple is considered in the calculation of the first response time.}
	}
	\end{center}
\end{figure*}

\boldstart{First response time for a sequence of regions.} Let $r_a \rightarrow r_b \rightarrow r_c \rightarrow \cdots \rightarrow r_x $ be a sequence of regions (e.g., a topological order of regions from an acyclic region graph), where the result operator is present in $r_x$ (Figure~\ref{fig:first-response-time-def}). Assume each region in the sequence is completed before the next region begins execution. Let $t(r_i)$ represent the time to execute a region $r_i$ completely and $t(r_i(1))$ represent the time to produce a single tuple from the result operator in $r_i$. Then, the first response time for the sequence of regions is,
$$
\tau(1) = t(r_a) + t(r_b) + t(r_c) + \cdots + t(r_x(1)).
$$

\subsection{Choosing a materialization option}
\label{ssec:materialization-option-choice}

We now use the above description of first response time to discuss how to choose a materialization option. The goal is to choose an option that would reduce the first response time. 

\begin{figure*}[htbp]
	\begin{center}
	\includegraphics[width=4in]{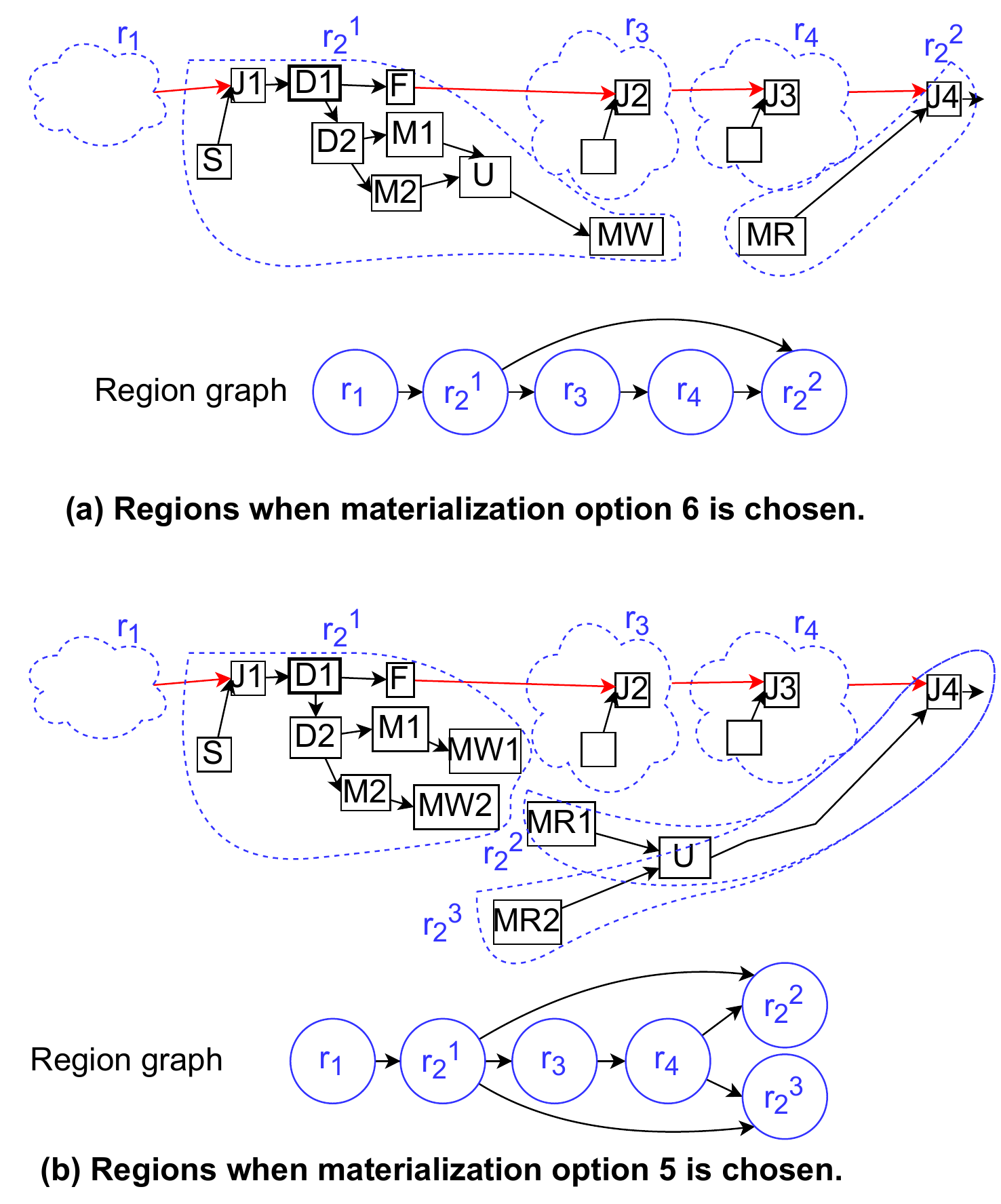} 
	\caption{\label{fig:first-response-time-def-example}
		\textbf{The outline of modified workflows showing regions when materialization is added to the workflow in Figure~\ref{fig:cyclic-region-mat-choices}(a). When there are more than one regions containing the sink, the first response time includes the minimum time among those regions to produce a single tuple.}
	}
	\end{center}
\end{figure*}

\boldstart{Explanation using example.} Let us consider the example region graph with a cycle shown in Figure~\ref{fig:cyclic-region-mat-choices}. When materialization options $1$ or $6$ are chosen, two new regions $r_2^{1}$ and $r_2^{2}$ are created out of $r_2$, and $r_2^{2}$ has the result operator (Figure~\ref{fig:first-response-time-def-example}(a)). In this case, the first response time is:
$$
\tau(1) = t(r_1) + t(r_2^{1}) + t(r_3) + t(r_4) + t(r_2^{2}(1)). 
$$
Note that the structure of $r_2^{1}$ and $r_2^{2}$ depend on which option is chosen. Thus, $t(r_2^{1})$ and $t(r_2^{2}(1))$ depend on which option is chosen. The time for the execution of other regions is not affected by the materialization choice. When materialization options $2$, $3$, $4$ or $5$ are chosen, three new regions $r_2^{1}$, $r_2^{2}$ and $r_2^{3}$ are created out of $r_2$, and both 
$r_2^{2}$ and $r_2^{3}$ have the result operators (Figure~\ref{fig:first-response-time-def-example}(b)). In this case, the first response time is:
$$
\tau(1) = t(r_1) + t(r_2^{1}) + t(r_3) + t(r_4) + min (t(r_2^{2}(1)), t(r_2^{3}(1))).
$$
Note again that the materialization option affects the structure of $r_2^{1}$, $r_2^{2}$ and $r_2^{3}$ and consequently the times $t(r_2^{1})$, $t(r_2^{2}(1))$ and $t(r_2^{3}(1))$. The execution times of other regions are unaffected. 

Query time estimation can be done to estimate $\tau(1)$ resulting from a materialization option. For a region $r$, $t(r)$ can be predicted by estimating the total query execution time~\cite{conf/icac/GuptaMD08,conf/icde/WuCZTHN13} for the query represented by the operator DAG in $r$. Similarly, $t(r)(1)$ can be predicted by estimating the time to get the first few results~\cite{conf/cikm/BayardoM96}. We choose the materialization option among the $6$ choices that has the least predicted $\tau(1)$.

\begin{figure*}[htbp]
	\begin{center}
	\includegraphics[width=4in]{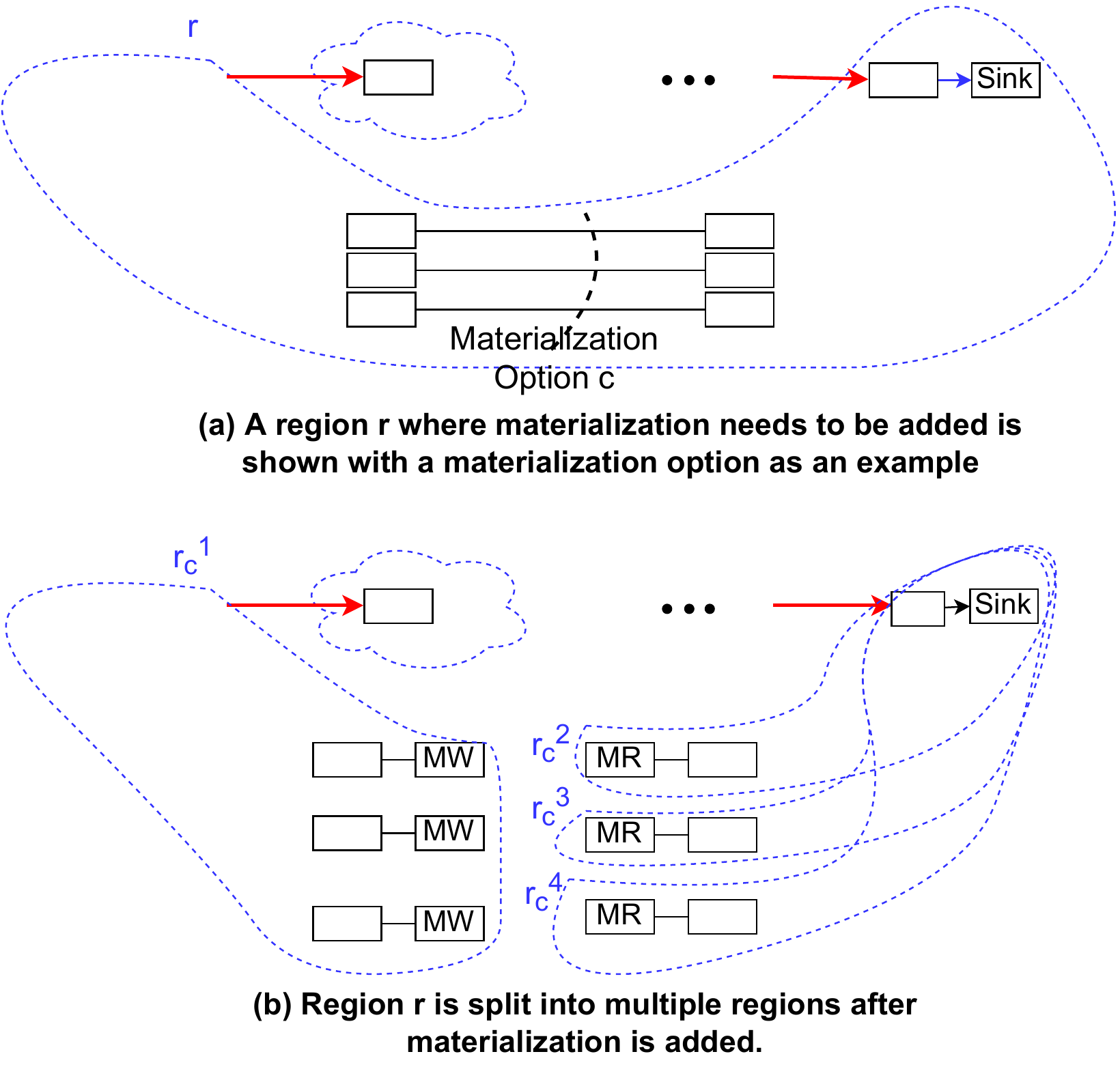} 
	\caption{\label{fig:first-response-time-def-general}
		\textbf{Example to help understand the generalization of first response time when a materialization choice $c$ is chosen in a region $r$. The region $r_c^1$ has to be fully executed, and any of the regions $r_c^2$, $r_c^3$, $\cdots$ has to produce a single tuple to get the first tuple out of the sink.}
	}
	\end{center}
\end{figure*}

\boldstart{Generalization.} In general, a materialization choice $c$ breaks a region $r$ into two sets of regions: 1) a single region $r_c^1$ that has the source of $r$ as its source and 2) a set of regions $\{r_c^2, r_c^3, \cdots\}$  that have the materialization readers as their sources. Let $\tau_c(1)$ be the first response time when the option $c$ is chosen. There are two cases to be considered when defining $\tau_c(1)$, depending on whether the result operator was in region $r$ before materialization or not. If the result operator was in region $r$, then the result operator will be part of all the regions in $\{r_c^2, r_c^3, \cdots \}$. Thus, these regions need to execute only partially for the first tuple to be produced from the result operator. On the other hand, if the result operator was not in $r$, then it will not be in any of the newly created regions. Thus, all the newly created regions will have to execute completely before the region with the result operator starts executing. Thus, we define $\tau_c(1)$ as the following:
$$ 
\tau_c(1) = \begin{cases} 
      T + t(r_c^1) + min(t(r_c^2(1)), t(r_c^3(1)), \cdots), & \text{ result operator} \in r \\
      T + t(r_c^1) + t(r_c^2) + t(r_c^3) + \cdots,  & \text{ result operator} \notin r, 
      \end{cases}
$$
where $T$ is a constant that denotes the time spent on regions other than $r$, which are not affected by materialization. The chosen materialization option is the one that has the smallest estimated $\tau_c(1)$.

\section{Experiments}
\label{sec:maestro-exp}

We performed experiments to show the presence of materialization choices in real workflows. We also performed experiments to showcase the effect of materialization choice on the first response time and the size of materialized data.

\subsection{Materialization choices in workflows}
\label{ssec:exp-analyze-workflows}

\begin{figure*}[htbp]
	\begin{center}
	\includegraphics[width=4.8in]{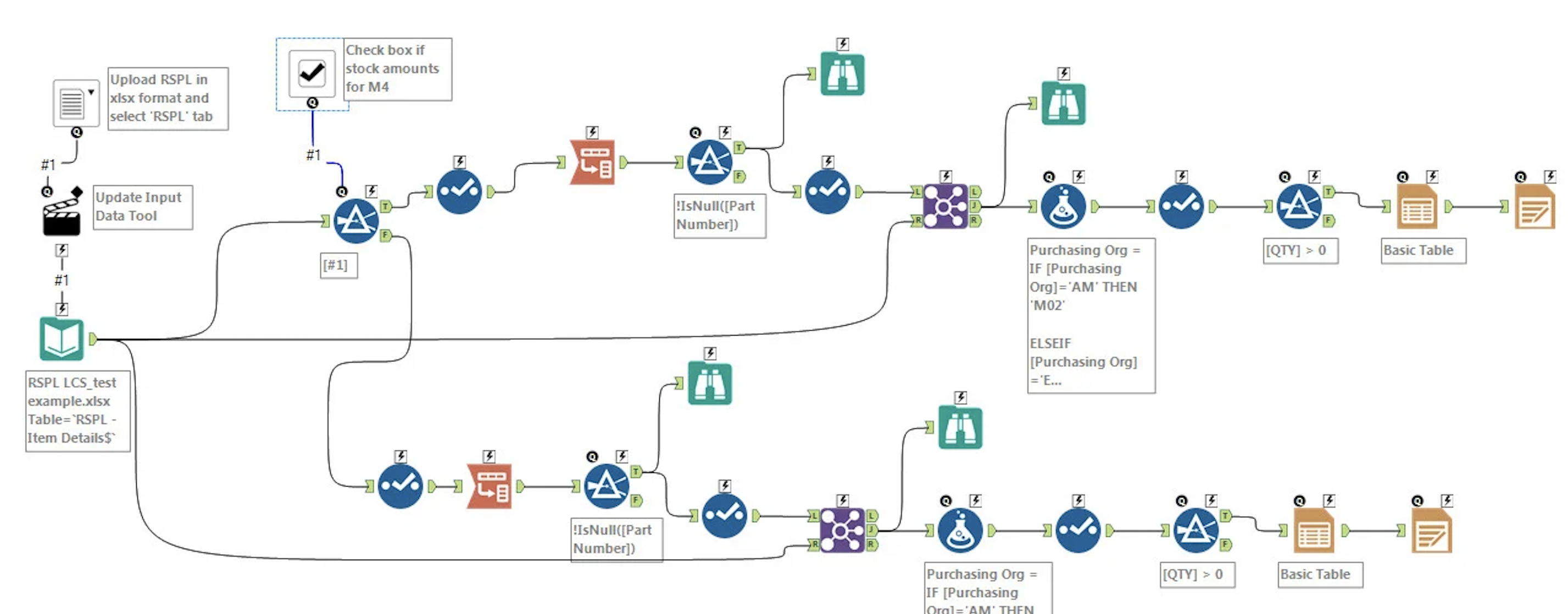} 
	\caption{\label{fig:alteryx-exp-image}
		\textbf{An Alteryx workflow.}
	}
		\end{center}
\end{figure*}

\begin{figure*}[htbp]
	\begin{center}
	\includegraphics[width=4.8in]{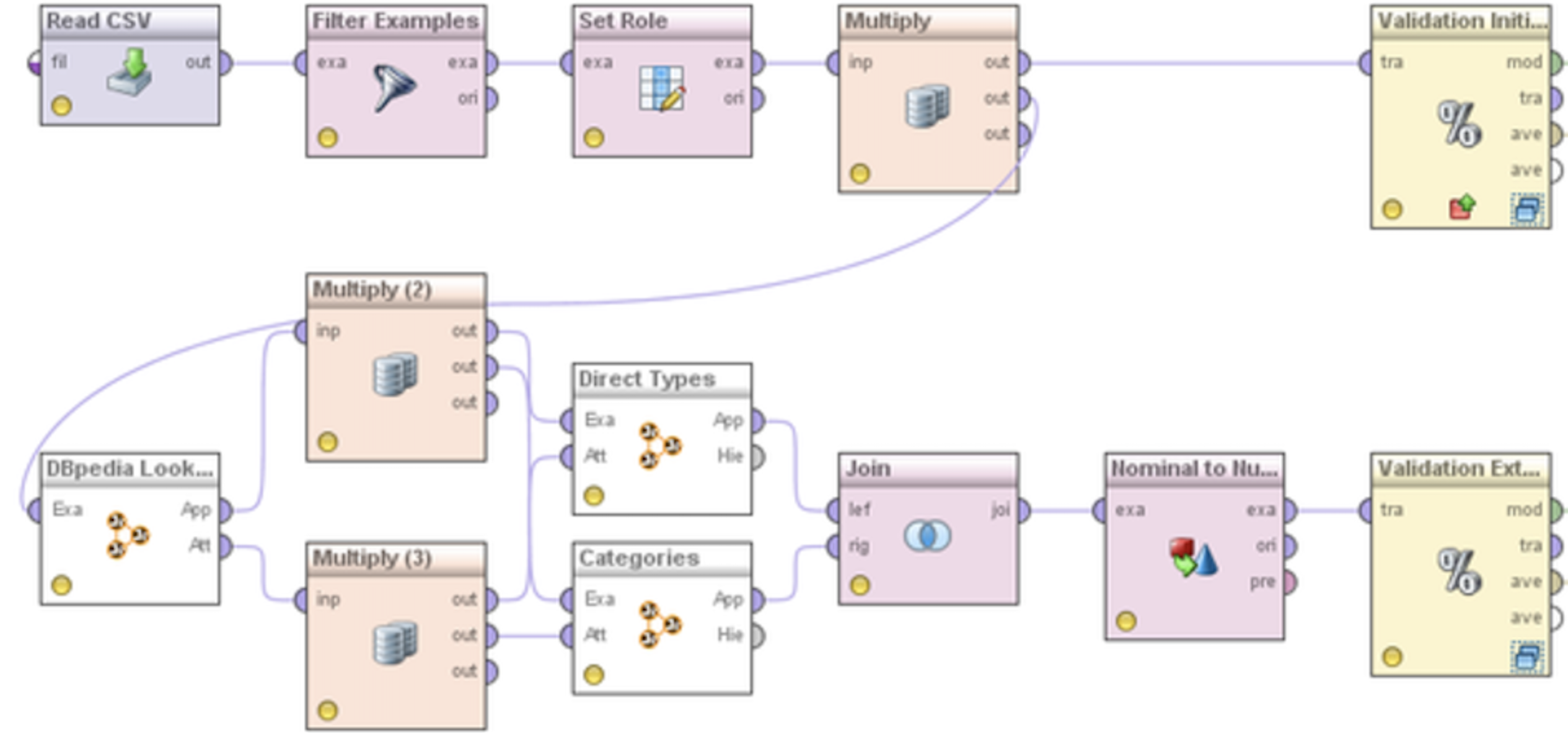} 
	\caption{\label{fig:rapidminer-exp-image}
		\textbf{A RapidMiner workflow.}
	}
		\end{center}
\end{figure*}

\begin{figure*}[htbp]
	\begin{center}
	\includegraphics[width=5.7in]{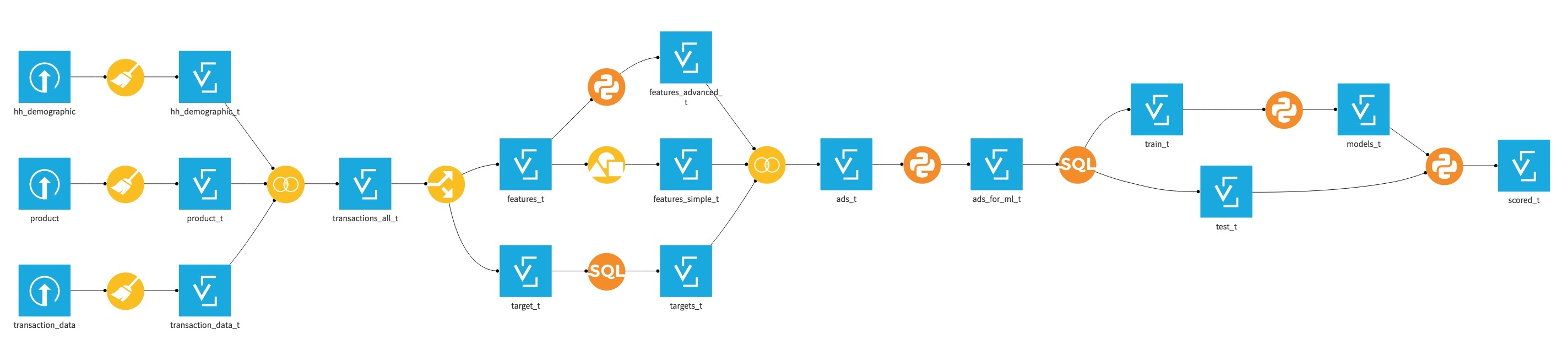} 
	\caption{\label{fig:dataiku-exp-image}
		\textbf{A Dataiku workflow.}
	}
		\end{center}
\end{figure*}

\begin{figure*}[htbp]
	\begin{center}
	\includegraphics[width=5in]{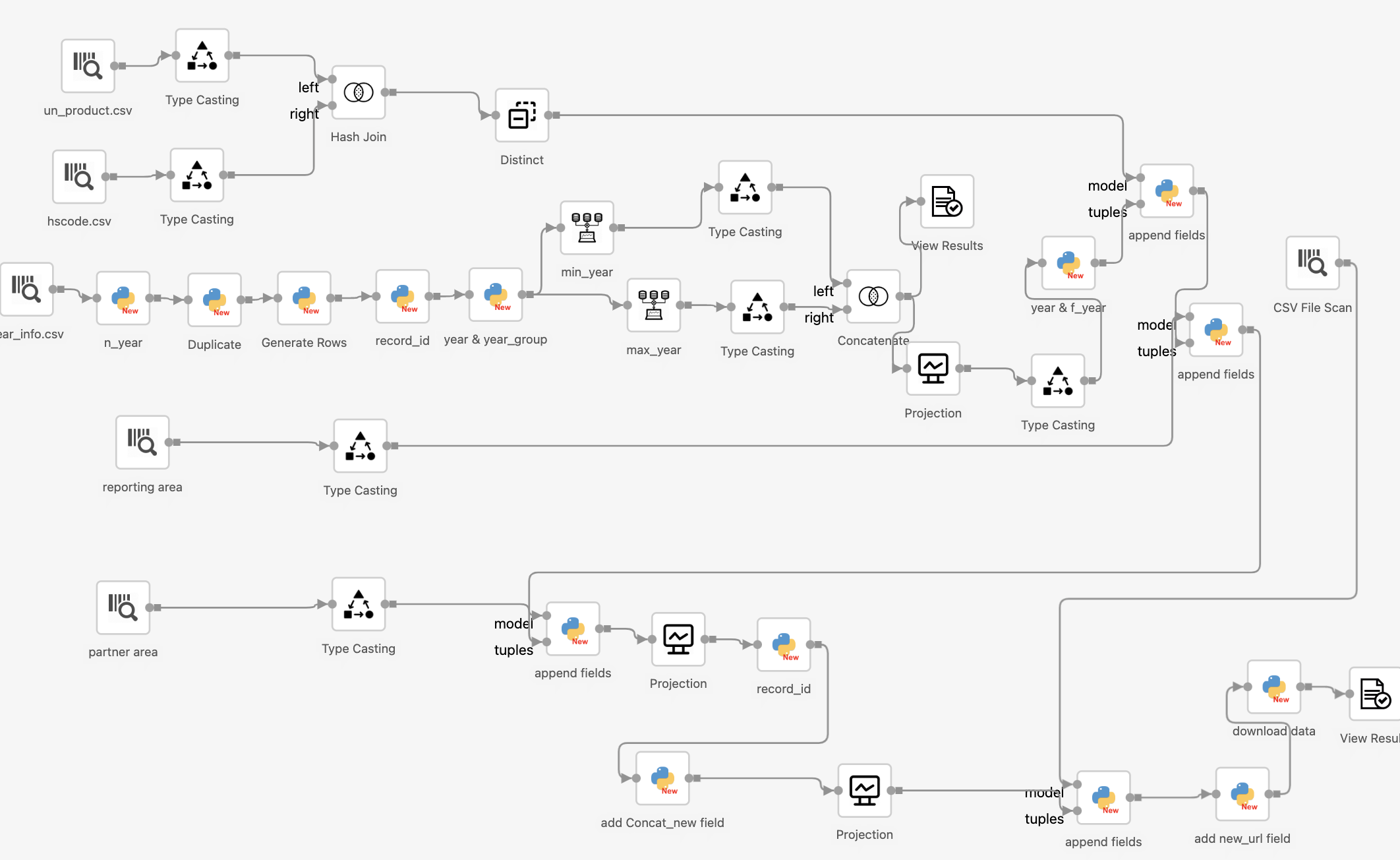} 
	\caption{\label{fig:texera-exp-image}
		\textbf{A Texera workflow.}
	}
		\end{center}
\end{figure*}

We analyzed $21$ workflows to see if they needed any materialization. For the workflows that needed materialization, we also evaluated the number of materialization choices. We did the analysis on publicly available workflows from three commercial workflow-based data processing systems, namely Alteryx, RapidMiner, and Dataiku, as well as workflows in our Texera system. Sample workflows from the systems are shown in Figures~\ref{fig:alteryx-exp-image}, \ref{fig:rapidminer-exp-image}, \ref{fig:dataiku-exp-image}, and \ref{fig:texera-exp-image}. For each workflow, we collected results about the need for materialization and the number of materialization choices.  The results are shown in Table~\ref{table:exp-analyze-workflows}.

\begin{table}[htbp]
\centering
\small{
\begin{tabular}{|C{2.3cm}|C{3cm}|C{3cm}|C{4cm}|}
\hline
{\bf System} & {\bf \# of operators} & {\bf Materialization needed?} & \textbf{\# of materialization choices} \\ \hline
\multirow{7}{*}{Alteryx} & 9 & Y & 3 \\ \cline{2-4}
& 10 & Y & 3 \\ \cline{2-4}
& 14 & Y & 4 \\ \cline{2-4}
& 19 & Y & 1 \\ \cline{2-4}
& 26 & Y & 12 \\ \cline{2-4}
& 34 & Y & 1 \\ \cline{2-4}
& 102 & Y & 1 \\ \hline \hline
\multirow{6}{*}{RapidMiner} & 5 & N & 0 \\ \cline{2-4}
& 9 & Y & 3 \\ \cline{2-4}
& 10 & Y & 2 \\ \cline{2-4}
& 10 & Y & 3 \\ \cline{2-4}
& 13 & Y & 2 \\ \cline{2-4}
& 19 & Y & 4 \\ \hline \hline
\multirow{3}{*}{Dataiku} & 31 & Y & 4 \\ \cline{2-4}
& 32 & Y & 1 \\\cline{2-4}
& 45 & N & 0 \\ \hline \hline
\multirow{5}{*}{Texera} & 8 & Y & 1 \\\cline{2-4}
& 8 & Y & 1 \\ \cline{2-4}
& 9 & N & 0 \\ \cline{2-4} 
& 10 & N & 0 \\ \cline{2-4}
& 30 & Y & 3 \\ \hline
\end{tabular}
}
\textit{}
\caption{Analyzing the workflows from various workflow processing systems.}
\label{table:exp-analyze-workflows}
\end{table}

We found a number of workflows from various systems that needed materialization. Upon examination, many of these workflows offered more than one materialization choice. The highest number of choices was found in an Alteryx workflow that offered $12$ choices. 
The analysis shows that there are many publicly available workflows that need materialization. 

\subsection{Effect of materialization choice on first response time}
\label{ssec:exp-mat-choice-first-response-time}

We took two workflows of similar structure and showed the effect of materialization choice on the first response time of the workflows. 

\begin{figure*}[htbp]
	\begin{center}
	\includegraphics[width=4in]{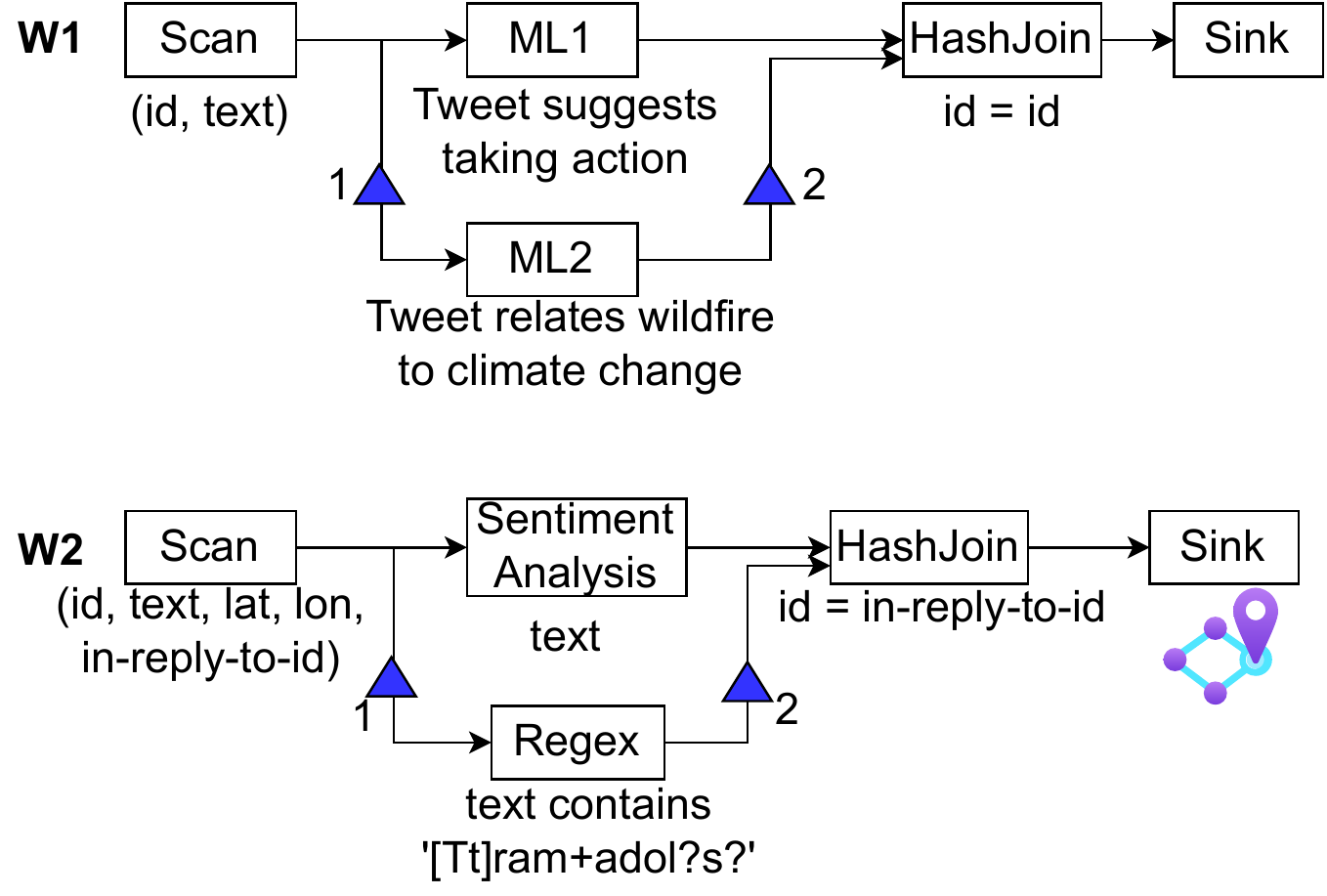} 
	\caption{\label{fig:thesis-exp-workflows}
		\textbf{Workflows used in the experiment to analyze the effect of materialization choices on the first response time. The blue triangles represent the materialization choices in the workflows.}
	}
	\end{center}
\end{figure*}

\boldstart{Data and workflows.} We used two workflows as shown in the Figure~\ref{fig:thesis-exp-workflows}. Each workflow used two different sets of tweets as input datasets. Workflow $W1$ used $5$ datasets ranging from $0.3$K tweets to $0.7$K tweets. Workflow $W2$ used $5$ datasets ranging from $1.6$K tweets to $7.6$K tweets. The datasets used were small because both the workflows contained expensive machine-learning based operators that took considerable time to process each tweet. Workflow $W1$ processes the tweets using two machine-learning based operators that add labels to the tweets specifying whether the text of a tweet is related to climate change and whether the text suggests action to be taken regarding climate change. The outputs of the two operators are then joined on the {\em id} column to create a single tuple containing both labels for a tweet. Workflow $W2$ finds the tweets related to the opioid tramadol. These tweets are then joined with their reply-tweets using the {\em in-reply-to-id} attribute. Each tuple output of the {\sf HashJoin} operator contains a tweet about tramadol and one of its reply tweets. The sentiment label of the reply-tweet is also present that has been added by the {\sf Sentiment Analysis} operator. The output of {\sf HashJoin} is plotted on a spatial graph with edges between a tweet and its reply. The reply-tweet location is colored according to its sentiment. 

\boldstart{Experimental Setting.} The experiments were carried out on GCP on a single {\em n2-highcpu-32} machine containing $32$ CPU cores, $32$ GB RAM and $100$ GB HDD disk running Ubuntu 18.04.5 LTS. All the operators were given $2$ workers except the machine-learning based UDF operators in $W1$ which were given $4$ workers because they were expensive.

\begin{figure*}[htbp]
	\begin{center}
	\includegraphics[width=3in]{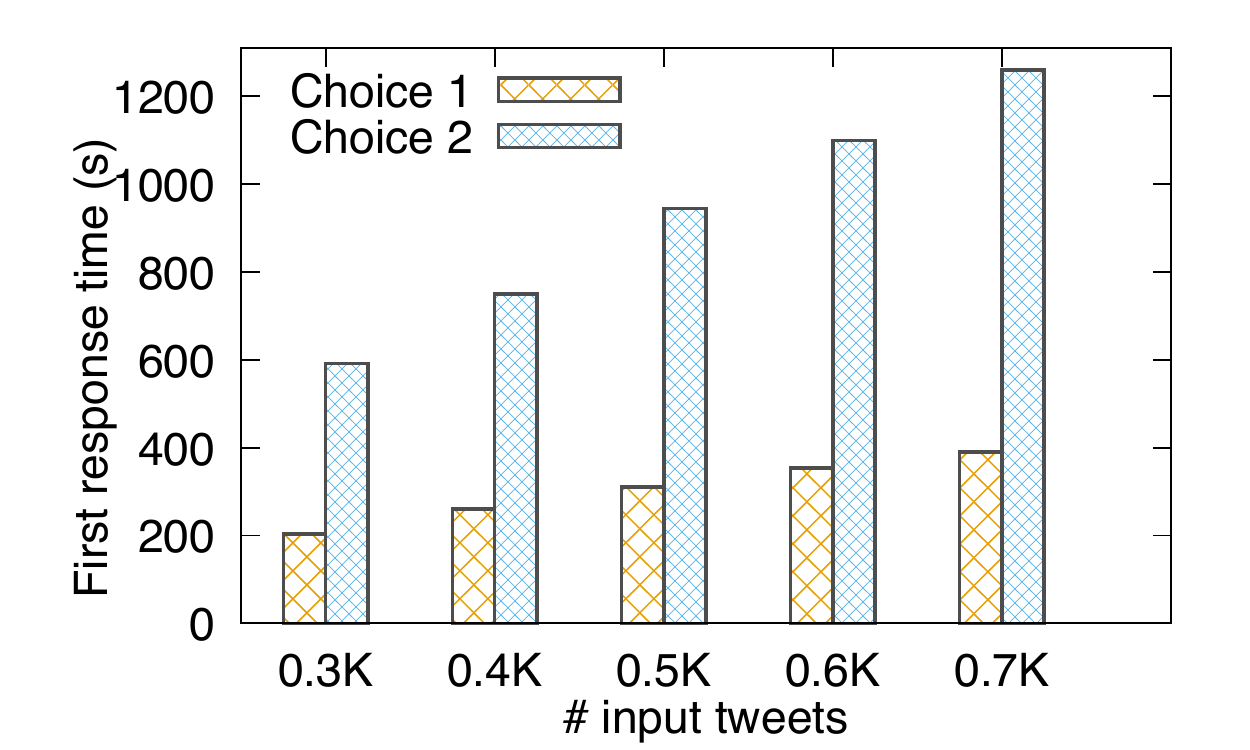} 
	\caption{\label{fig:w1-first-response-time}
		\textbf{First response time for different input data sizes in $W1$.}
	}
	\end{center}
\end{figure*}

\boldstart{Materialization choices of W1}. The workflow $W1$ had two materialization choices shown in the Figure~\ref{fig:thesis-exp-workflows}. The first response times for the two choices for various input data sizes are shown in Figure~\ref{fig:w1-first-response-time}. The first-response time when choice $1$ was used for the $0.3$K tweets dataset was $203$ seconds. The first-response time when choice $2$ was used for the $0.3$K tweets dataset was $592$ seconds. The first response times gradually increased as the input data size increased. For the $0.7$K tweets dataset, the first response times for the choices $1$ and $2$ were $390$ seconds and $1260$ seconds, respectively. The figure shows that the choice $1$ was consistently better than choice $2$. The reason is that when choice $2$ was used, the two expensive machine-learning based operators belonged to the same region and executed simultaneously. This led to competition between them for resources and slowed down computation.

\begin{figure*}[htbp]
	\begin{center}
	\includegraphics[width=3in]{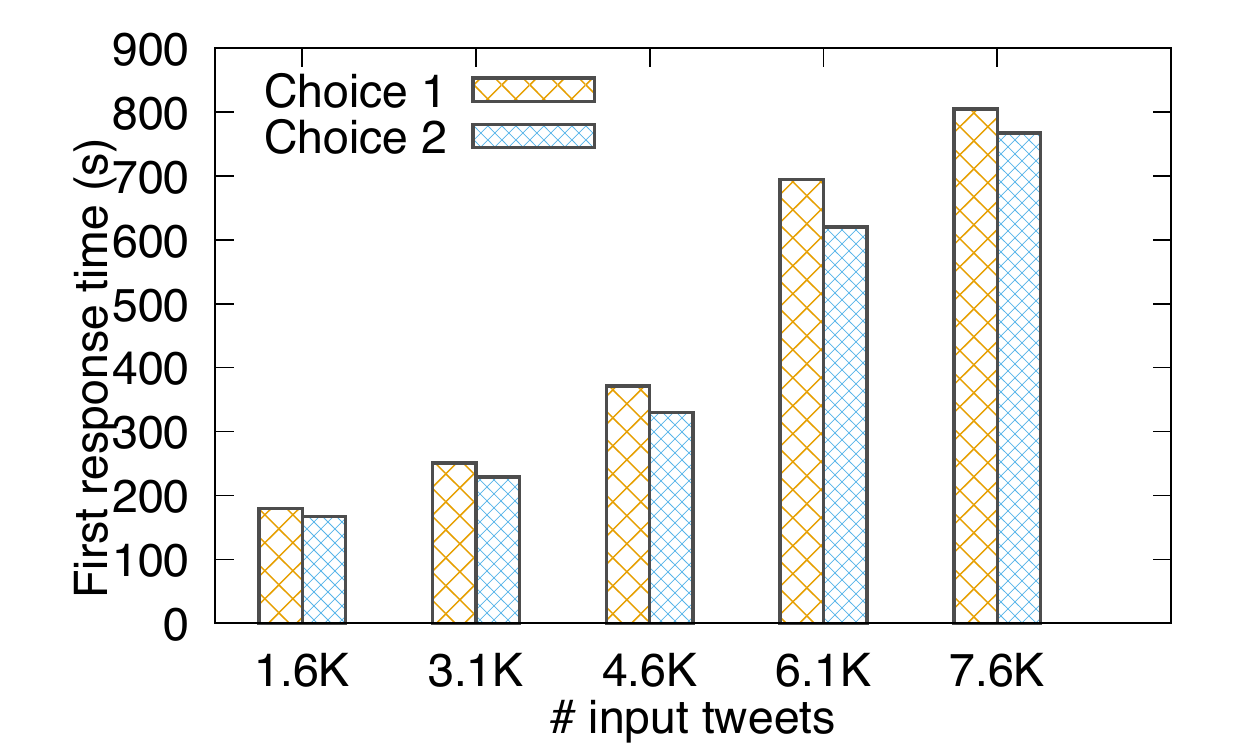} 
	\caption{\label{fig:w2-first-response-time}
		\textbf{First response time for different input data sizes in $W2$.}
	}
	\end{center}
\end{figure*}

\boldstart{Materialization choices of W2}. The workflow $W2$ also had two materialization choices. The first response times for the two choices for various input data sizes are shown in Figure~\ref{fig:w2-first-response-time}. The selectivity of the {\sf Regex} operator was low because only a few tweets contained tramadol. The first-response time when choice $1$ was used for the $1.6$K tweets dataset was $180$ seconds. The first-response time when choice $2$ was used for the $1.6$K tweets dataset was $167$ seconds. The first response times gradually increased as the input data size increased. For the $7.6$K tweets dataset, the first response times for the choices $1$ and $2$ were $805$ and $767$ seconds, respectively. The figure shows that the choice $2$ was consistently better than choice $1$. The reason for the difference in first-response time is explained next. When choice $1$ was used, the {\sf Regex} operator belonged to a region that started execution after the sentiment analysis processed the entire data. Since the {\sf Regex} operator had low selectivity, it produced the first output after a considerable time. When choice $2$ was used, {\sf Regex} and {\sf Sentiment Analysis} operators belonged to the same region and data was processed by the {\sf Regex} operator alongside {\sf Sentiment Analysis} without adding much overhead. When the next region began execution, the first tuple was immediately output from {\sf HashJoin}.

Thus, we got lower first-response time when using choice $1$ for $W1$ and choice $2$ for $W2$. This experiment shows that the materialization choices do play a role in the first-response time.

\subsection{Effect of materialization choice on materialized data size}
\label{ssec:exp-mat-choice-mat-size}

We took the two workflows from the previous experiment and performed experiments to compare the materialized data size for the various materialization options. The experiment setting was the same as the previous experiment.

\begin{figure*}[htbp]
	\begin{center}
	\includegraphics[width=3in]{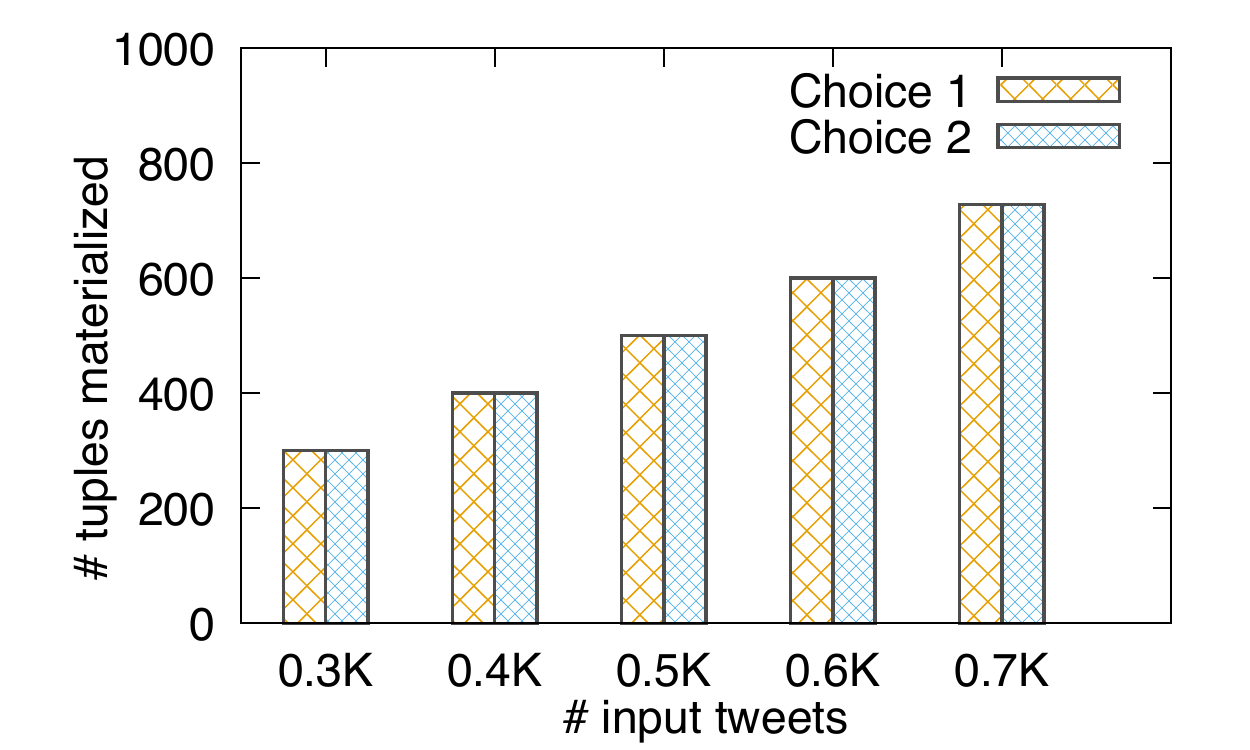} 
	\caption{\label{fig:w1-mat-size}
		\textbf{Materialization size for different input sizes in $W1$.}
	}
	\end{center}
\end{figure*}

\begin{figure*}[htbp]
	\begin{center}
	\includegraphics[width=3in]{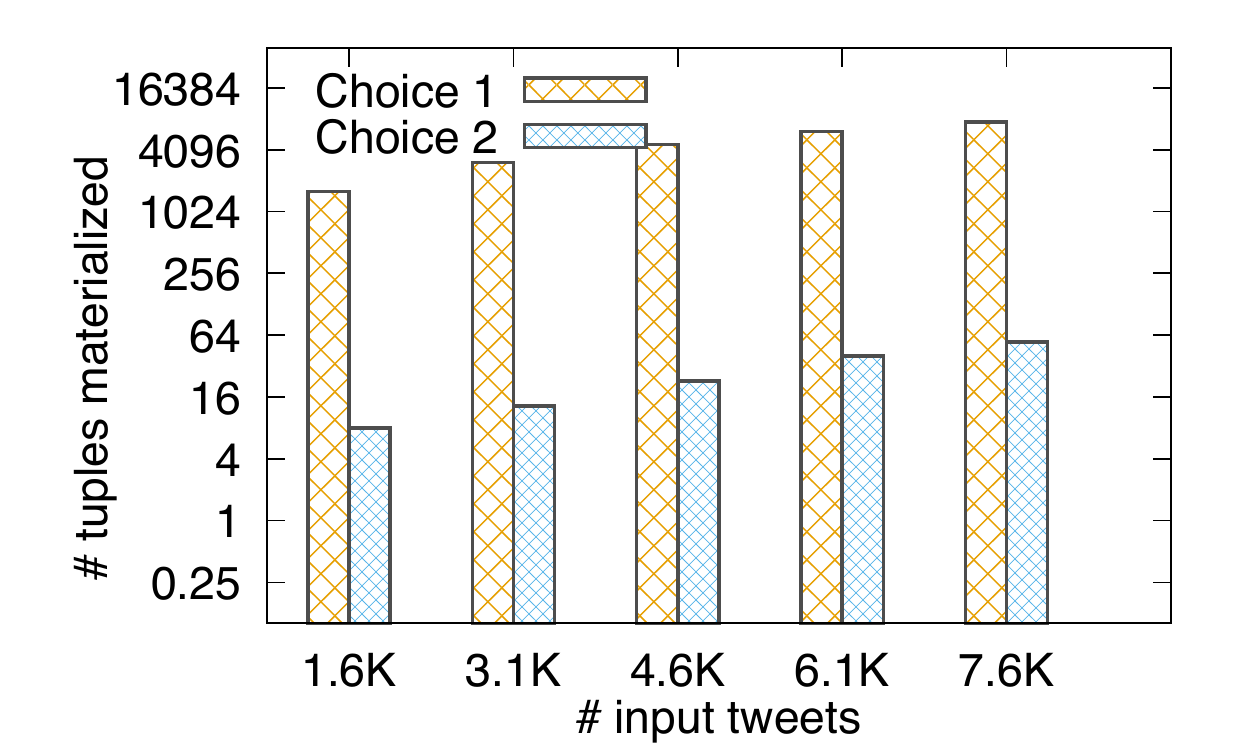} 
	\caption{\label{fig:w2-mat-size}
		\textbf{Materialization size for different input sizes in $W2$.}
	}
	\end{center}
\end{figure*}

The materialization sizes for the two choices in $W_1$ for various input data sizes are shown in Figure~\ref{fig:w1-mat-size}. In the case of $W_1$, the {\sf ML2} operator produced a single tuple for every input tuple. Thus, both choices $1$ and $2$ resulted in the same materialization size consisting of all input tuples. In the case of $W_2$ (Figure~\ref{fig:w2-mat-size}), the {\sf Regex} operator was very selective. Thus, the choice $2$ materialized few tuples, whereas the choice $1$ materialized all the tuples coming out of {\sf Scan}.

A good heuristic while choosing the materialization choices can be to add materialization after operators that filter out data so that the size of materialized data is reduced.
\section{Conclusion}
\label{sec:maestro-conclusion}

In this chapter, we presented a scheduling framework called \schfrmname. We showed how it divides a workflow into regions that can be separately scheduled. We described an algorithm to create a region graph that encapsulates the region dependencies. We showed that the the region graph can only be scheduled when it is acyclic and discussed examples where the region graph has cycles. We discussed how to use materialization to create an acyclic region graph from a cyclic region graph. We showed how to enumerate various choices for materialization. We introduced first response time as a metric of result awareness and discussed how the materialization choices differ from the perspective of result awareness. We presented experiments that showed the presence of materialization choices in real workflows and showcased the effect of materialization choice on the first-response time and materialized data size.

\chapter{Conclusions and Future Work}
\label{chap:conclusion-future-work}

In this section, we present the conclusions of the three works presented in this thesis and motivate the future work.

\section{Conclusions}
\label{sec:thesis-conclusion}

In this thesis, we presented three works that make data analytics more interactive, adaptive, and result aware.

In Chapter~\ref{chap:amber}, we presented a system called \amberfrmname that supports powerful and responsive debugging during the execution of a dataflow. It serves as the backend engine for the Texera service being developed at UC Irvine. We presented its overall system architecture based on the actor model and described the whole lifecycle of the execution of a workflow, including how control messages are sent to actors, how to expedite the processing of these control messages, and how to pause and resume the computation of each actor. We studied how to support conditional breakpoints, and presented solutions for enforcing local conditional breakpoints and global conditional breakpoints. We developed a technique to support fault tolerance in \amberfrmname, which is challenging due to the presence of control messages.  We implemented \amberfrmname on top of Orleans, and presented an extensive experimental evaluation to show its high usability and performance comparable to Spark. 

In Chapter~\ref{chap:reshape}, we presented a framework called \frmname that adaptively handles partitioning skew in the exploratory data analysis setting. We presented different approaches for load transfer and analyzed their impact on the results shown to the user. We presented an analysis about the effect of the skew-detection threshold on mitigation and used it to adaptively adjust the threshold. We generalized \frmname to other operators such as {\sf HashJoin}, {\sf Sort}, and {\sf GroupBy}, and broader execution settings. We implemented \frmname on top of two big data engines - Amber and Flink - and presented an experimental evaluation.

In Chapter~\ref{chap:maestro}, we presented a scheduling framework called \schfrmname. We showed how it divides a workflow into regions that can be separately scheduled. We described an algorithm to create a region graph that are based on the region dependencies. Since the region graph can only be scheduled when it is acyclic, we need to avoid cycles in the region graph. We discussed how to use materialization to create an acyclic region graph. We showed how to enumerate various choices for materialization. We introduced first response time as a metric of result awareness and discussed how the materialization choices differ from the perspective of result awareness. We presented experiments that showed the presence of materialization choices in real workflows and showcased the effect of materialization choice on the first-response time and materialized data size.

\section{Future Work}
\label{sec:thesis-future-work}

\boldstart{Amber.} The determination of number of workers per operator and the worker placement in  \amberfrmname is static. This can be improved to allow adaptation in the number of workers and their placement depending on the complexity of the operators and the data traffic pattern among machines. The global breakpoint detection in \amberfrmname is not deterministic, in the sense that two executions may result in different numbers of tuples being processed by the workers before the breakpoint is hit. It would be interesting to devise a technique which leads to deterministic breakpoint detection and analyze the overhead of this technique on the total processing time.

\boldstart{Reshape.} The current framework of \frmname focuses on skew handling among the workers of an operator. There can be multiple skewed operators in a workflow. It would be interesting to see how \frmname can be extended to consider the skew in multiple operators and make global decisions. It would also be interesting to extend \frmname so that a single worker can serve as the helper for multiple skewed workers. Another future work is using both {\em split by key} and {\em split by records} among the workers of an operator, i.e., load transfer for a few keys happens using {\em split by key} and for other keys using {\em split by records}. The chapter does not consider the existence of windows on the input data. When extending \frmname to work on infinite streams, it is important to consider how the existence of windows will affect the decision of \frmname.

\boldstart{Maestro.} The current definition of first-response time in \schfrmname and the process of choosing a materialization option assumes a single sink. These concepts have to be extended to the case of multiple sinks. The algorithm to run an exhaustive search over the multiple materialization choices may be expensive. We need a way, maybe using heuristics, to prune the search space. 


\clearpage
\phantomsection

\bibliographystyle{abbrv}
\bibliography{references,thesis}

\begin{thebibliography}{100}

\bibitem{Cloudberry}
{Cloudberry - Big Data Visualization.}
\newblock \url{http://cloudberry.ics.uci.edu/}, 2018.

\bibitem{conf/sigmod/AbdelhamidMDA20}
A.~S. Abdelhamid, A.~R. Mahmood, A.~Daghistani, and W.~G. Aref.
\newblock Prompt: Dynamic data-partitioning for distributed micro-batch stream
  processing systems.
\newblock In D.~Maier, R.~Pottinger, A.~Doan, W.~Tan, A.~Alawini, and H.~Q.
  Ngo, editors, {\em Proceedings of the 2020 International Conference on
  Management of Data, {SIGMOD} Conference 2020, online conference [Portland,
  OR, USA], June 14-19, 2020}, pages 2455--2469. {ACM}, 2020.

\bibitem{agha1985actors}
G.~A. Agha.
\newblock Actors: A model of concurrent computation in distributed systems.
\newblock Technical report, Massachusetts Inst Of Tech Cambridge Artificial
  Intelligence Lab, 1985.

\bibitem{Akka}
Akka Website, \url{https://akka.io/}.

\bibitem{journals/corr/AlsubaieeAABBBCCCFGGHKLLOOPTVWW14}
S.~Alsubaiee, Y.~Altowim, H.~Altwaijry, A.~Behm, V.~R. Borkar, Y.~Bu, M.~J.
  Carey, I.~Cetindil, M.~Cheelangi, K.~Faraaz, E.~Gabrielova, R.~Grover,
  Z.~Heilbron, Y.~Kim, C.~Li, G.~Li, J.~M. Ok, N.~Onose, P.~Pirzadeh, V.~J.
  Tsotras, R.~Vernica, J.~Wen, and T.~Westmann.
\newblock Asterixdb: {A} scalable, open source {BDMS}.
\newblock {\em CoRR}, abs/1407.0454, 2014.

\bibitem{alteryx}
Alteryx Website, \url{https://www.alteryx.com/}.

\bibitem{conf/debs/AnielloBQ13}
L.~Aniello, R.~Baldoni, and L.~Querzoni.
\newblock Adaptive online scheduling in storm.
\newblock In S.~Chakravarthy, S.~D. Urban, P.~R. Pietzuch, and E.~A.
  Rundensteiner, editors, {\em The 7th {ACM} International Conference on
  Distributed Event-Based Systems, {DEBS} '13, Arlington, TX, {USA} - June 29 -
  July 03, 2013}, pages 207--218. {ACM}, 2013.

\bibitem{journals/pvldb/AntonopoulosKTU17}
P.~Antonopoulos, H.~Kodavalla, A.~Tran, N.~Upreti, C.~Shah, and M.~Sztajno.
\newblock Resumable online index rebuild in {SQL} server.
\newblock {\em {PVLDB}}, 10(12):1742--1753, 2017.

\bibitem{asterix:website}
Apache AsterixDB, http://asterixdb.apache.org.

\bibitem{misc/flink}
Apache Flink http://flink.apache.org.

\bibitem{misc/hadoopmapreduce}
Apache Hadoop MapReduce,
  https://hadoop.apache.org/docs/stable/hadoop-mapreduce-client/hadoop-mapreduce-client-core/MapReduceTutorial.html.

\bibitem{Samza:website}
Apache samza.
\newblock \url{http://samza.apache.org/}.

\bibitem{misc/spark}
Apache Spark http://spark.apache.org.

\bibitem{apacheStorm}
Apache Storm, \url{http://storm.apache.org/}.

\bibitem{conf/sigmod/ArmbrustDTYZX0S18}
M.~Armbrust, T.~Das, J.~Torres, B.~Yavuz, S.~Zhu, R.~Xin, A.~Ghodsi, I.~Stoica,
  and M.~Zaharia.
\newblock Structured streaming: {A} declarative {API} for real-time
  applications in apache spark.
\newblock In G.~Das, C.~M. Jermaine, and P.~A. Bernstein, editors, {\em
  Proceedings of the 2018 International Conference on Management of Data,
  {SIGMOD} Conference 2018, Houston, TX, USA, June 10-15, 2018}, pages
  601--613. {ACM}, 2018.

\bibitem{journals/ml/AuerCF02}
P.~Auer, N.~Cesa{-}Bianchi, and P.~Fischer.
\newblock Finite-time analysis of the multiarmed bandit problem.
\newblock {\em Mach. Learn.}, 47(2-3):235--256, 2002.

\bibitem{misc/backpressure}
Handling Backpressure
  https://medium.com/@jayphelps/backpressure-explained-the-flow-of-data-through-software-2350b3e77ce7.

\bibitem{conf/chi/BattleFDBCG16}
L.~Battle, D.~Fisher, R.~DeLine, M.~Barnett, B.~Chandramouli, and J.~Goldstein.
\newblock Making sense of temporal queries with interactive visualization.
\newblock In {\em Proceedings of the 2016 {CHI} Conference on Human Factors in
  Computing Systems, San Jose, CA, USA, May 7-12, 2016}, pages 5433--5443,
  2016.

\bibitem{journals/pvldb/BegoliACHKKMS21}
E.~Begoli, T.~Akidau, S.~Chernyak, F.~Hueske, K.~Knight, K.~Knowles, D.~Mills,
  and D.~Sotolongo.
\newblock Watermarks in stream processing systems: Semantics and comparative
  analysis of apache flink and google cloud dataflow.
\newblock {\em Proc. {VLDB} Endow.}, 14(12):3135--3147, 2021.

\bibitem{conf/sigmod/BehmPAACDGHJKLL22}
A.~Behm, S.~Palkar, U.~Agarwal, T.~Armstrong, D.~Cashman, A.~Dave,
  T.~Greenstein, S.~Hovsepian, R.~Johnson, A.~S. Krishnan, P.~Leventis,
  A.~Luszczak, P.~Menon, M.~Mokhtar, G.~Pang, S.~Paranjpye, G.~Rahn, B.~Samwel,
  T.~van Bussel, H.~V. Hovell, M.~Xue, R.~Xin, and M.~Zaharia.
\newblock Photon: {A} fast query engine for lakehouse systems.
\newblock In Z.~Ives, A.~Bonifati, and A.~E. Abbadi, editors, {\em {SIGMOD}
  '22: International Conference on Management of Data, Philadelphia, PA, USA,
  June 12 - 17, 2022}, pages 2326--2339. {ACM}, 2022.

\bibitem{conf/sigir/BeitzelJCGF04}
S.~M. Beitzel, E.~C. Jensen, A.~Chowdhury, D.~A. Grossman, and O.~Frieder.
\newblock Hourly analysis of a very large topically categorized web query log.
\newblock In M.~Sanderson, K.~J{\"{a}}rvelin, J.~Allan, and P.~Bruza, editors,
  {\em {SIGIR} 2004: Proceedings of the 27th Annual International {ACM} {SIGIR}
  Conference on Research and Development in Information Retrieval, Sheffield,
  UK, July 25-29, 2004}, pages 321--328. {ACM}, 2004.

\bibitem{journals/csur/BenoitCRS13}
A.~Benoit, {\"{U}}.~V. {\c{C}}ataly{\"{u}}rek, Y.~Robert, and E.~Saule.
\newblock A survey of pipelined workflow scheduling: Models and algorithms.
\newblock {\em {ACM} Comput. Surv.}, 45(4):50:1--50:36, 2013.

\bibitem{conf/cidr/BernsteinDKM17}
P.~A. Bernstein, M.~Dashti, T.~Kiefer, and D.~Maier.
\newblock Indexing in an actor-oriented database.
\newblock In {\em {CIDR} 2017, 8th Biennial Conference on Innovative Data
  Systems Research, Chaminade, CA, USA, January 8-11, 2017, Online
  Proceedings}, 2017.

\bibitem{journals/cacm/BeschastnikhWBE16}
I.~Beschastnikh, P.~Wang, Y.~Brun, and M.~D. Ernst.
\newblock Debugging distributed systems.
\newblock {\em Commun. {ACM}}, 59(8):32--37, 2016.

\bibitem{conf/eurosys/BindschaedlerMS18}
L.~Bindschaedler, J.~Malicevic, N.~Schiper, A.~Goel, and W.~Zwaenepoel.
\newblock Rock you like a hurricane: taming skew in large scale analytics.
\newblock In R.~Oliveira, P.~Felber, and Y.~C. Hu, editors, {\em Proceedings of
  the Thirteenth EuroSys Conference, EuroSys 2018, Porto, Portugal, April
  23-26, 2018}, pages 20:1--20:15. {ACM}, 2018.

\bibitem{phd/us/Borkar16}
V.~R. Borkar.
\newblock {\em An Efficient Foundation for Big Data Processing on Modern
  Clusters}.
\newblock PhD thesis, University of California, Irvine, {USA}, 2016.

\bibitem{conf/icde/BorkarCGOV11}
V.~R. Borkar, M.~J. Carey, R.~Grover, N.~Onose, and R.~Vernica.
\newblock Hyracks: A flexible and extensible foundation for data-intensive
  computing.
\newblock In {\em International Conference on Data Engineering}, pages
  1151--1162, 2011.

\bibitem{caf}
CAF Website, \url{https://actor-framework.org/}.

\bibitem{journals/pvldb/CarboneEFHRT17}
P.~Carbone, S.~Ewen, G.~F{\'{o}}ra, S.~Haridi, S.~Richter, and K.~Tzoumas.
\newblock State management in apache flink{\textregistered}: Consistent
  stateful distributed stream processing.
\newblock {\em Proc. {VLDB} Endow.}, 10(12):1718--1729, 2017.

\bibitem{journals/corr/CarboneFEHT15}
P.~Carbone, G.~F{\'{o}}ra, S.~Ewen, S.~Haridi, and K.~Tzoumas.
\newblock Lightweight asynchronous snapshots for distributed dataflows.
\newblock {\em CoRR}, abs/1506.08603, 2015.

\bibitem{conf/sigmod/CarboneFKK20}
P.~Carbone, M.~Fragkoulis, V.~Kalavri, and A.~Katsifodimos.
\newblock Beyond analytics: The evolution of stream processing systems.
\newblock In D.~Maier, R.~Pottinger, A.~Doan, W.~Tan, A.~Alawini, and H.~Q.
  Ngo, editors, {\em Proceedings of the 2020 International Conference on
  Management of Data, {SIGMOD} Conference 2020, online conference [Portland,
  OR, USA], June 14-19, 2020}, pages 2651--2658. {ACM}, 2020.

\bibitem{conf/sigmod/ChandramouliBBY07}
B.~Chandramouli, C.~N. Bond, S.~Babu, and J.~Yang.
\newblock Query suspend and resume.
\newblock In {\em Proceedings of the {ACM} {SIGMOD} International Conference on
  Management of Data, Beijing, China, June 12-14, 2007}, pages 557--568, 2007.

\bibitem{journals/tocs/ChandyL85}
K.~M. Chandy and L.~Lamport.
\newblock Distributed snapshots: Determining global states of distributed
  systems.
\newblock {\em {ACM} Trans. Comput. Syst.}, 3(1):63--75, 1985.

\bibitem{conf/sigmod/ChaudhuriMN98}
S.~Chaudhuri, R.~Motwani, and V.~R. Narasayya.
\newblock Random sampling for histogram construction: How much is enough?
\newblock In L.~M. Haas and A.~Tiwary, editors, {\em {SIGMOD} 1998, Proceedings
  {ACM} {SIGMOD} International Conference on Management of Data, June 2-4,
  1998, Seattle, Washington, {USA}}, pages 436--447. {ACM} Press, 1998.

\bibitem{journals/tpds/ChenYX15}
Q.~Chen, J.~Yao, and Z.~Xiao.
\newblock {LIBRA:} lightweight data skew mitigation in mapreduce.
\newblock {\em {IEEE} Trans. Parallel Distributed Syst.}, 26(9):2520--2533,
  2015.

\bibitem{conf/infocom/ChengCZGM17}
D.~Cheng, Y.~Chen, X.~Zhou, D.~Gmach, and D.~S. Milojicic.
\newblock Adaptive scheduling of parallel jobs in spark streaming.
\newblock In {\em 2017 {IEEE} Conference on Computer Communications, {INFOCOM}
  2017, Atlanta, GA, USA, May 1-4, 2017}, pages 1--9. {IEEE}, 2017.

\bibitem{dao2009live}
D.~Dao, J.~Albrecht, C.~Killian, and A.~Vahdat.
\newblock Live debugging of distributed systems.
\newblock In {\em International Conference on Compiler Construction}, pages
  94--108. Springer, 2009.

\bibitem{conf/osdi/DeanG04}
J.~Dean and S.~Ghemawat.
\newblock Mapreduce: Simplified data processing on large clusters.
\newblock In {\em OSDI}, pages 137--150, 2004.

\bibitem{journals/tkde/DeWittGSBHR90}
D.~J. DeWitt, S.~Ghandeharizadeh, D.~A. Schneider, A.~Bricker, H.-I. Hsiao, and
  R.~Rasmussen.
\newblock The {Gamma} database machine project.
\newblock {\em IEEE Trans. Knowl. Data Eng.}, 2(1):44--62, 1990.

\bibitem{journals/cacm/DeWittG92}
D.~J. DeWitt and J.~Gray.
\newblock Parallel database systems: The future of high performance database
  systems.
\newblock {\em Commun. ACM}, 35(6):85--98, 1992.

\bibitem{conf/vldb/DeWittNSS92}
D.~J. DeWitt, J.~F. Naughton, D.~A. Schneider, and S.~Seshadri.
\newblock Practical skew handling in parallel joins.
\newblock In L.~Yuan, editor, {\em 18th International Conference on Very Large
  Data Bases, August 23-27, 1992, Vancouver, Canada, Proceedings}, pages
  27--40. Morgan Kaufmann, 1992.

\bibitem{conf/sc/DiKC12}
S.~Di, D.~Kondo, and W.~Cirne.
\newblock Host load prediction in a google compute cloud with a bayesian model.
\newblock In J.~K. Hollingsworth, editor, {\em {SC} Conference on High
  Performance Computing Networking, Storage and Analysis, {SC} '12, Salt Lake
  City, UT, {USA} - November 11 - 15, 2012}, page~21. {IEEE/ACM}, 2012.

\bibitem{journals/pvldb/DingCGN21}
B.~Ding, S.~Chaudhuri, J.~Gehrke, and V.~R. Narasayya.
\newblock {DSB:} {A} decision support benchmark for workload-driven and
  traditional database systems.
\newblock {\em Proc. {VLDB} Endow.}, 14(13):3376--3388, 2021.

\bibitem{journals/corr/DingFMWYZC15}
J.~Ding, T.~Z.~J. Fu, R.~T.~B. Ma, M.~Winslett, Y.~Yang, Z.~Zhang, and H.~Chao.
\newblock Optimal operator state migration for elastic data stream processing.
\newblock {\em CoRR}, abs/1501.03619, 2015.

\bibitem{Einblick}
Einblick, \url{https://www.einblick.ai}.

\bibitem{journals/pvldb/ElseidyEVK14}
M.~Elseidy, A.~Elguindy, A.~Vitorovic, and C.~Koch.
\newblock Scalable and adaptive online joins.
\newblock {\em Proc. {VLDB} Endow.}, 7(6):441--452, 2014.

\bibitem{conf/icde/EngOST03}
P.~Eng, B.~C. Ooi, H.~S. Sim, and K.~Tan.
\newblock Preference-driven query processing.
\newblock In U.~Dayal, K.~Ramamritham, and T.~M. Vijayaraman, editors, {\em
  Proceedings of the 19th International Conference on Data Engineering, March
  5-8, 2003, Bangalore, India}, pages 671--673. {IEEE} Computer Society, 2003.

\bibitem{conf/sigmod/FernandezMKP13}
R.~C. Fernandez, M.~Migliavacca, E.~Kalyvianaki, and P.~R. Pietzuch.
\newblock Integrating scale out and fault tolerance in stream processing using
  operator state management.
\newblock In K.~A. Ross, D.~Srivastava, and D.~Papadias, editors, {\em
  Proceedings of the {ACM} {SIGMOD} International Conference on Management of
  Data, {SIGMOD} 2013, New York, NY, USA, June 22-27, 2013}, pages 725--736.
  {ACM}, 2013.

\bibitem{journals/interactions/FisherDCD12}
D.~Fisher, R.~DeLine, M.~Czerwinski, and S.~M. Drucker.
\newblock Interactions with big data analytics.
\newblock {\em Interactions}, 19(3):50--59, 2012.

\bibitem{misc/flink-pipelined-region-scheduling}
Apache flink pipelined region scheduling.
\newblock
  \url{https://flink.apache.org/2020/12/15/pipelined-region-sheduling.html}.

\bibitem{conf/icdcs/FowlerZ90}
J.~Fowler and W.~Zwaenepoel.
\newblock Causal distributed breakpoints.
\newblock In {\em 10th International Conference on Distributed Computing
  Systems {(ICDCS} 1990), May 28 - June 1, 1990, Paris, France}, pages
  134--141, 1990.

\bibitem{fulkerson1962flows}
D.~R. Fulkerson and L.~R. Ford.
\newblock {\em Flows in networks}.
\newblock Princeton University Press Princeton, 1962.

\bibitem{conf/isorc/GarraghanOTX15}
P.~Garraghan, X.~Ouyang, P.~Townend, and J.~Xu.
\newblock Timely long tail identification through agent based monitoring and
  analytics.
\newblock In {\em {IEEE} 18th International Symposium on Real-Time Distributed
  Computing, {ISORC} 2015, Auckland, New Zealand, 13-17 April, 2015}, pages
  19--26. {IEEE} Computer Society, 2015.

\bibitem{journals/csur/Graefe93}
G.~Graefe.
\newblock Query evaluation techniques for large databases.
\newblock {\em ACM Comput. Surv.}, 25(2):73--170, 1993.

\bibitem{conf/icde/GuflerARK12}
B.~Gufler, N.~Augsten, A.~Reiser, and A.~Kemper.
\newblock Load balancing in mapreduce based on scalable cardinality estimates.
\newblock In A.~Kementsietsidis and M.~A.~V. Salles, editors, {\em {IEEE} 28th
  International Conference on Data Engineering {(ICDE} 2012), Washington, DC,
  {USA} (Arlington, Virginia), 1-5 April, 2012}, pages 522--533. {IEEE}
  Computer Society, 2012.

\bibitem{conf/sigmod/GulzarICK17}
M.~A. Gulzar, M.~Interlandi, T.~Condie, and M.~Kim.
\newblock Debugging big data analytics in spark with \emph{BigDebug}.
\newblock In {\em Proceedings of the 2017 {ACM} International Conference on
  Management of Data, {SIGMOD} Conference 2017, Chicago, IL, USA, May 14-19,
  2017}, pages 1627--1630, 2017.

\bibitem{conf/cloud/GulzarIHLCK17}
M.~A. Gulzar, M.~Interlandi, X.~Han, M.~Li, T.~Condie, and M.~Kim.
\newblock Automated debugging in data-intensive scalable computing.
\newblock In {\em Proceedings of the 2017 Symposium on Cloud Computing, SoCC
  2017, Santa Clara, CA, USA, September 24-27, 2017}, pages 520--534, 2017.

\bibitem{conf/icac/GuptaMD08}
C.~Gupta, A.~Mehta, and U.~Dayal.
\newblock {PQR:} predicting query execution times for autonomous workload
  management.
\newblock In J.~Strassner, S.~A. Dobson, J.~A.~B. Fortes, and K.~K. Goswami,
  editors, {\em 2008 International Conference on Autonomic Computing, {ICAC}
  2008, June 2-6, 2008, Chicago, Illinois, {USA}}, pages 13--22. {IEEE}
  Computer Society, 2008.

\bibitem{haban1988global}
D.~Haban and W.~Weigel.
\newblock Global events and global breakpoints in distributed systems.
\newblock In {\em [1988] Proceedings of the Twenty-First Annual Hawaii
  International Conference on System Sciences. Volume II: Software track},
  volume~2, pages 166--175. IEEE, 1988.

\bibitem{misc/hadoop-mapreduce-configs}
Apache hadoop mapreduce configuration parameters.
\newblock
  \url{https://hadoop.apache.org/docs/stable/hadoop-mapreduce-client/hadoop-mapreduce-client-core/mapred-default.xml}.

\bibitem{Halo}
Halo Website, \url{https://www.halowaypoint.com/en-us}.

\bibitem{conf/IEEEscc/HerathLMGSVMP12}
C.~Herath, F.~Liu, S.~Marru, L.~Gunathilake, M.~Sosonkina, J.~P. Vary,
  P.~Maris, and M.~E. Pierce.
\newblock Web service andworkflow abstractions to large scale nuclear physics
  calculations.
\newblock In {\em 2012 {IEEE} Ninth International Conference on Services
  Computing, Honolulu, HI, USA, June 24-29, 2012}, pages 703--710, 2012.

\bibitem{conf/ijcai/HewittBS73}
C.~Hewitt, P.~B. Bishop, and R.~Steiger.
\newblock A universal modular {ACTOR} formalism for artificial intelligence.
\newblock In {\em Proceedings of the 3rd International Joint Conference on
  Artificial Intelligence. Standford, CA, USA, August 20-23, 1973}, pages
  235--245, 1973.

\bibitem{conf/nsdi/HindmanKZGJKSS10}
B.~Hindman, A.~Konwinski, M.~Zaharia, A.~Ghodsi, A.~D. Joseph, R.~H. Katz,
  S.~Shenker, and I.~Stoica.
\newblock Mesos: {A} platform for fine-grained resource sharing in the data
  center.
\newblock In D.~G. Andersen and S.~Ratnasamy, editors, {\em Proceedings of the
  8th {USENIX} Symposium on Networked Systems Design and Implementation, {NSDI}
  2011, Boston, MA, USA, March 30 - April 1, 2011}. {USENIX} Association, 2011.

\bibitem{journals/pvldb/HoffmannLMKLR19}
M.~Hoffmann, A.~Lattuada, F.~McSherry, V.~Kalavri, J.~Liagouris, and T.~Roscoe.
\newblock Megaphone: Latency-conscious state migration for distributed
  streaming dataflows.
\newblock {\em Proc. {VLDB} Endow.}, 12(9):1002--1015, 2019.

\bibitem{conf/osdi/HsiehABVBPGM18}
K.~Hsieh, G.~Ananthanarayanan, P.~Bod{\'{\i}}k, S.~Venkataraman, P.~Bahl,
  M.~Philipose, P.~B. Gibbons, and O.~Mutlu.
\newblock Focus: Querying large video datasets with low latency and low cost.
\newblock In A.~C. Arpaci{-}Dusseau and G.~Voelker, editors, {\em 13th {USENIX}
  Symposium on Operating Systems Design and Implementation, {OSDI} 2018,
  Carlsbad, CA, USA, October 8-10, 2018}, pages 269--286. {USENIX} Association,
  2018.

\bibitem{journals/pvldb/InterlandiSTGYK15}
M.~Interlandi, K.~Shah, S.~D. Tetali, M.~A. Gulzar, S.~Yoo, M.~Kim, T.~D.
  Millstein, and T.~Condie.
\newblock Titian: Data provenance support in spark.
\newblock {\em {PVLDB}}, 9(3):216--227, 2015.

\bibitem{conf/cikm/BayardoM96}
R.~J.~B. Jr. and D.~P. Miranker.
\newblock Processing queries for first few answers.
\newblock In {\em {CIKM} '96, Proceedings of the Fifth International Conference
  on Information and Knowledge Management, November 12 - 16, 1996, Rockville,
  Maryland, {USA}}, pages 45--52. {ACM}, 1996.

\bibitem{conf/IEEEcloud/Kim0QH16}
I.~K. Kim, W.~Wang, Y.~Qi, and M.~Humphrey.
\newblock Empirical evaluation of workload forecasting techniques for
  predictive cloud resource scaling.
\newblock In {\em 9th {IEEE} International Conference on Cloud Computing,
  {CLOUD} 2016, San Francisco, CA, USA, June 27 - July 2, 2016}, pages 1--10.
  {IEEE} Computer Society, 2016.

\bibitem{knime}
Knime Website, \url{https://www.knime.com/}.

\bibitem{journals/csur/Kossmann00}
D.~Kossmann.
\newblock The state of the art in distributed query processing.
\newblock {\em ACM Comput. Surv.}, 32(4):422--469, 2000.

\bibitem{conf/wsdm/KulkarniTSD11}
A.~Kulkarni, J.~Teevan, K.~M. Svore, and S.~T. Dumais.
\newblock Understanding temporal query dynamics.
\newblock In I.~King, W.~Nejdl, and H.~Li, editors, {\em Proceedings of the
  Forth International Conference on Web Search and Web Data Mining, {WSDM}
  2011, Hong Kong, China, February 9-12, 2011}, pages 167--176. {ACM}, 2011.

\bibitem{journals/pvldb/KumarWNL20}
A.~Kumar, Z.~Wang, S.~Ni, and C.~Li.
\newblock Amber: {A} debuggable dataflow system based on the actor model.
\newblock {\em Proc. {VLDB} Endow.}, 13(5):740--753, 2020.

\bibitem{conf/sigmod/KwonBHR12}
Y.~Kwon, M.~Balazinska, B.~Howe, and J.~A. Rolia.
\newblock Skewtune: mitigating skew in mapreduce applications.
\newblock In K.~S. Candan, Y.~Chen, R.~T. Snodgrass, L.~Gravano, and A.~Fuxman,
  editors, {\em Proceedings of the {ACM} {SIGMOD} International Conference on
  Management of Data, {SIGMOD} 2012, Scottsdale, AZ, USA, May 20-24, 2012},
  pages 25--36. {ACM}, 2012.

\bibitem{conf/vldb/Lawrence05}
R.~Lawrence.
\newblock Early hash join: {A} configurable algorithm for the efficient and
  early production of join results.
\newblock In K.~B{\"{o}}hm, C.~S. Jensen, L.~M. Haas, M.~L. Kersten, P.~Larson,
  and B.~C. Ooi, editors, {\em Proceedings of the 31st International Conference
  on Very Large Data Bases, Trondheim, Norway, August 30 - September 2, 2005},
  pages 841--852. {ACM}, 2005.

\bibitem{journals/pvldb/LiuWNAHKL22}
X.~Liu, Z.~Wang, S.~Ni, S.~Alsudais, Y.~Huang, A.~Kumar, and C.~Li.
\newblock Demonstration of collaborative and interactive workflow-based data
  analytics in texera.
\newblock {\em Proc. {VLDB} Endow.}, 15(12):3738--3741, 2022.

\bibitem{ludascher2006scientific}
B.~Lud{\"a}scher, I.~Altintas, C.~Berkley, D.~Higgins, E.~Jaeger, M.~Jones,
  E.~A. Lee, J.~Tao, and Y.~Zhao.
\newblock Scientific workflow management and the kepler system.
\newblock {\em Concurrency and Computation: Practice and Experience},
  18(10):1039--1065, 2006.

\bibitem{journals/pvldb/MaiZPXSVCKMKDR18}
L.~Mai, K.~Zeng, R.~Potharaju, L.~Xu, S.~Suh, S.~Venkataraman, P.~Costa,
  T.~Kim, S.~Muthukrishnan, V.~Kuppa, S.~Dhulipalla, and S.~Rao.
\newblock Chi: {A} scalable and programmable control plane for distributed
  stream processing systems.
\newblock {\em Proc. {VLDB} Endow.}, 11(10):1303--1316, 2018.

\bibitem{conf/sigcomm/MaoSVMA19}
H.~Mao, M.~Schwarzkopf, S.~B. Venkatakrishnan, Z.~Meng, and M.~Alizadeh.
\newblock Learning scheduling algorithms for data processing clusters.
\newblock In J.~Wu and W.~Hall, editors, {\em Proceedings of the {ACM} Special
  Interest Group on Data Communication, {SIGCOMM} 2019, Beijing, China, August
  19-23, 2019}, pages 270--288. {ACM}, 2019.

\bibitem{conf/sc/MarruGHTPMSGCGSDPW11}
S.~Marru, L.~Gunathilake, C.~Herath, P.~Tangchaisin, M.~E. Pierce, C.~Mattmann,
  R.~Singh, T.~Gunarathne, E.~Chinthaka, R.~Gardler, A.~Slominski, A.~Douma,
  S.~Perera, and S.~Weerawarana.
\newblock Apache airavata: a framework for distributed applications and
  computational workflows.
\newblock In {\em Proceedings of the 2011 {ACM} {SC} Workshop on Gateway
  Computing Environments, {GCE} 2011, Seattle, WA, USA, November 18, 2011},
  pages 21--28, 2011.

\bibitem{conf/icdt/MetwallyAA05}
A.~Metwally, D.~Agrawal, and A.~E. Abbadi.
\newblock Efficient computation of frequent and top-k elements in data streams.
\newblock In T.~Eiter and L.~Libkin, editors, {\em Database Theory - {ICDT}
  2005, 10th International Conference, Edinburgh, UK, January 5-7, 2005,
  Proceedings}, volume 3363 of {\em Lecture Notes in Computer Science}, pages
  398--412. Springer, 2005.

\bibitem{conf/icdcs/MillerC88}
B.~P. Miller and J.~Choi.
\newblock Breakpoints and halting in distributed programs.
\newblock In {\em Proceedings of the 8th International Conference on
  Distributed Computing Systems, San Jose, California, USA, June 13-17, 1988},
  pages 316--323, 1988.

\bibitem{missier2010taverna}
P.~Missier, S.~Soiland-Reyes, S.~Owen, W.~Tan, A.~Nenadic, I.~Dunlop,
  A.~Williams, T.~Oinn, and C.~Goble.
\newblock Taverna, reloaded.
\newblock In {\em International conference on scientific and statistical
  database management}, pages 471--481. Springer, 2010.

\bibitem{conf/sigmod/MonteZRM20}
B.~D. Monte, S.~Zeuch, T.~Rabl, and V.~Markl.
\newblock Rhino: Efficient management of very large distributed state for
  stream processing engines.
\newblock In D.~Maier, R.~Pottinger, A.~Doan, W.~Tan, A.~Alawini, and H.~Q.
  Ngo, editors, {\em Proceedings of the 2020 International Conference on
  Management of Data, {SIGMOD} Conference 2020, online conference [Portland,
  OR, USA], June 14-19, 2020}, pages 2471--2486. {ACM}, 2020.

\bibitem{journals/cgf/MoritzHHH15}
D.~Moritz, D.~Halperin, B.~Howe, and J.~Heer.
\newblock Perfopticon: Visual query analysis for distributed databases.
\newblock {\em Comput. Graph. Forum}, 34(3):71--80, 2015.

\bibitem{conf/icde/NasirMGKS15}
M.~A.~U. Nasir, G.~D.~F. Morales, D.~Garc{\'{\i}}a{-}Soriano, N.~Kourtellis,
  and M.~Serafini.
\newblock The power of both choices: Practical load balancing for distributed
  stream processing engines.
\newblock In J.~Gehrke, W.~Lehner, K.~Shim, S.~K. Cha, and G.~M. Lohman,
  editors, {\em 31st {IEEE} International Conference on Data Engineering,
  {ICDE} 2015, Seoul, South Korea, April 13-17, 2015}, pages 137--148. {IEEE}
  Computer Society, 2015.

\bibitem{conf/icde/NasirMKS16}
M.~A.~U. Nasir, G.~D.~F. Morales, N.~Kourtellis, and M.~Serafini.
\newblock When two choices are not enough: Balancing at scale in distributed
  stream processing.
\newblock In {\em 32nd {IEEE} International Conference on Data Engineering,
  {ICDE} 2016, Helsinki, Finland, May 16-20, 2016}, pages 589--600. {IEEE}
  Computer Society, 2016.

\bibitem{conf/icdm/NeumeyerRNK10}
L.~Neumeyer, B.~Robbins, A.~Nair, and A.~Kesari.
\newblock {S4:} distributed stream computing platform.
\newblock In {\em {ICDMW} 2010, The 10th {IEEE} International Conference on
  Data Mining Workshops, Sydney, Australia, 13 December 2010}, pages 170--177,
  2010.

\bibitem{misc/nytimes-80percent-data-wrangling}
For big-data scientists, 'janitor work' is key hurdle to insights.
\newblock
  \url{https://www.nytimes.com/2014/08/18/technology/for-big-data-scientists-hurdle-to-insights-is-janitor-work.html}.

\bibitem{conf/sigmod/OlstonR11}
C.~Olston and B.~Reed.
\newblock Inspector gadget: a framework for custom monitoring and debugging of
  distributed dataflows.
\newblock In {\em Proceedings of the {ACM} {SIGMOD} International Conference on
  Management of Data, {SIGMOD} 2011, Athens, Greece, June 12-16, 2011}, pages
  1221--1224, 2011.

\bibitem{orleans}
Orleans Website, \url{https://dotnet.github.io/orleans/}.

\bibitem{books/sp/OzsuV20}
M.~T. {\"{O}}zsu and P.~Valduriez.
\newblock {\em Principles of Distributed Database Systems, 4th Edition}.
\newblock Springer, 2020.

\bibitem{conf/middleware/PengHHFC15}
B.~Peng, M.~Hosseini, Z.~Hong, R.~Farivar, and R.~H. Campbell.
\newblock R-storm: Resource-aware scheduling in storm.
\newblock In R.~Lea, S.~Gopalakrishnan, E.~Tilevich, A.~L. Murphy, and
  M.~Blackstock, editors, {\em Proceedings of the 16th Annual Middleware
  Conference, Vancouver, BC, Canada, December 07 - 11, 2015}, pages 149--161.
  {ACM}, 2015.

\bibitem{conf/icde/PopescuEBBA12}
A.~D. Popescu, V.~Ercegovac, A.~Balmin, M.~Branco, and A.~Ailamaki.
\newblock Same queries, different data: Can we predict runtime performance?
\newblock In A.~Kementsietsidis and M.~A.~V. Salles, editors, {\em Workshops
  Proceedings of the {IEEE} 28th International Conference on Data Engineering,
  {ICDE} 2012, Arlington, VA, USA, April 1-5, 2012}, pages 275--280. {IEEE}
  Computer Society, 2012.

\bibitem{misc/prediction-interval}
Prediction interval, \url{https://otexts.com/fpp2/prediction-intervals.html}.

\bibitem{ProtoActor}
ProtoActor Website, \url{http://proto.actor}.

\bibitem{ptolemyII}
Ptolemy II Website,
  \url{https://ptolemy.berkeley.edu/ptolemyII/ptII11.0/index.htm}.

\bibitem{conf/cloud/RamakrishnanSU12}
S.~R. Ramakrishnan, G.~Swart, and A.~Urmanov.
\newblock Balancing reducer skew in mapreduce workloads using progressive
  sampling.
\newblock In M.~J. Carey and S.~Hand, editors, {\em {ACM} Symposium on Cloud
  Computing, {SOCC} '12, San Jose, CA, USA, October 14-17, 2012}, page~16.
  {ACM}, 2012.

\bibitem{rapidMiner}
RapidMiner Website, \url{https://rapidminer.com/}.

\bibitem{conf/icde/RodigerIK016}
W.~R{\"{o}}diger, S.~Idicula, A.~Kemper, and T.~Neumann.
\newblock Flow-join: Adaptive skew handling for distributed joins over
  high-speed networks.
\newblock In {\em 32nd {IEEE} International Conference on Data Engineering,
  {ICDE} 2016, Helsinki, Finland, May 16-20, 2016}, pages 1194--1205. {IEEE}
  Computer Society, 2016.

\bibitem{conf/sigmod/SabekUK22}
I.~Sabek, T.~S. Ukyab, and T.~Kraska.
\newblock Lsched: {A} workload-aware learned query scheduler for analytical
  database systems.
\newblock In Z.~Ives, A.~Bonifati, and A.~E. Abbadi, editors, {\em {SIGMOD}
  '22: International Conference on Management of Data, Philadelphia, PA, USA,
  June 12 - 17, 2022}, pages 1228--1242. {ACM}, 2022.

\bibitem{sentimentAnalysis}
Sentiment Analysis operator for .Net,
  \url{https://github.com/arafattehsin/CognitiveRocket/tree/master/CognitiveLibrary/SentimentAnalyzer}.

\bibitem{conf/icde/ShahHCF03}
M.~A. Shah, J.~M. Hellerstein, S.~Chandrasekaran, and M.~J. Franklin.
\newblock Flux: An adaptive partitioning operator for continuous query systems.
\newblock In U.~Dayal, K.~Ramamritham, and T.~M. Vijayaraman, editors, {\em
  Proceedings of the 19th International Conference on Data Engineering, March
  5-8, 2003, Bangalore, India}, pages 25--36. {IEEE} Computer Society, 2003.

\bibitem{journals/pvldb/ShangZBEKMRK21}
Z.~Shang, E.~Zgraggen, B.~Buratti, P.~Eichmann, N.~Karimeddiny, C.~Meyer,
  W.~Runnels, and T.~Kraska.
\newblock Davos: {A} system for interactive data-driven decision making.
\newblock {\em Proc. {VLDB} Endow.}, 14(12):2893--2905, 2021.

\bibitem{conf/cloud/ShenSGW11}
Z.~Shen, S.~Subbiah, X.~Gu, and J.~Wilkes.
\newblock Cloudscale: elastic resource scaling for multi-tenant cloud systems.
\newblock In J.~S. Chase and A.~E. Abbadi, editors, {\em {ACM} Symposium on
  Cloud Computing in conjunction with {SOSP} 2011, {SOCC} '11, Cascais,
  Portugal, October 26-28, 2011}, page~5. {ACM}, 2011.

\bibitem{misc/spark-job-scheduling}
Apache spark job scheduling.
\newblock \url{https://spark.apache.org/docs/latest/job-scheduling.html}.

\bibitem{spark-dataframe-optimization}
Spark DataFrame API optimizations in Projet Tungsten,
  \url{https://databricks.com/blog/2015/04/28/project-tungsten-bringing-spark-closer-to-bare-metal.html}.

\bibitem{spark-rdd-vs-dataframe}
Differences between Spark RDD API and Dataframe API,
  \url{https://databricks.com/blog/2016/07/14/a-tale-of-three-apache-spark-apis-rdds-dataframes-and-datasets.html}.

\bibitem{SparkAQE:website}
{Adaptive Query Execution in Spark}.
\newblock
  \texttt{https://spark.apache.org/docs/latest/sql-performance-tuning.html\#adaptive-query-execution}.

\bibitem{spark-stage}
Spark docs Stage explanation,
  \url{https://spark.apache.org/docs/1.2.2/api/java/org/apache/spark/scheduler/Stage.html}.

\bibitem{misc/statistical-forecasting}
Statistical forecasting, \url{https://people.duke.edu/~rnau/411home.htm}.

\bibitem{Tapad}
Tapad Website, \url{https://www.tapad.com}.

\bibitem{taxi-data}
NYC TLC Trip Record Data,
  \url{https://www1.nyc.gov/site/tlc/about/tlc-trip-record-data.page}.

\bibitem{texera}
Texera Website, \url{https://github.com/Texera/texera}.

\bibitem{TexeraWebsite}
Texera.
\newblock {Texera Website}.
\newblock \url{https://github.com/Texera/texera/}, 2021.

\bibitem{TPC-H}
TPC-H Website, \url{http://www.tpc.org/tpch/}.

\bibitem{misc/TPC-H}
TPC-H Website, \url{http://www.tpc.org/tpch/}.

\bibitem{journals/debu/UrhanF00}
T.~Urhan and M.~J. Franklin.
\newblock Xjoin: {A} reactively-scheduled pipelined join operator.
\newblock {\em {IEEE} Data Eng. Bull.}, 23(2):27--33, 2000.

\bibitem{conf/vldb/UrhanF01}
T.~Urhan and M.~J. Franklin.
\newblock Dynamic pipeline scheduling for improving interactive query
  performance.
\newblock In P.~M.~G. Apers, P.~Atzeni, S.~Ceri, S.~Paraboschi,
  K.~Ramamohanarao, and R.~T. Snodgrass, editors, {\em {VLDB} 2001, Proceedings
  of 27th International Conference on Very Large Data Bases, September 11-14,
  2001, Roma, Italy}, pages 501--510. Morgan Kaufmann, 2001.

\bibitem{conf/sigmod/VartakSLVHMZ16}
M.~Vartak, H.~Subramanyam, W.~Lee, S.~Viswanathan, S.~Husnoo, S.~Madden, and
  M.~Zaharia.
\newblock Modeldb: a system for machine learning model management.
\newblock In {\em HILDA@SIGMOD'16}.

\bibitem{conf/cloud/VavilapalliMDAKEGLSSSCORRB13}
V.~K. Vavilapalli, A.~C. Murthy, C.~Douglas, S.~Agarwal, M.~Konar, R.~Evans,
  T.~Graves, J.~Lowe, H.~Shah, S.~Seth, B.~Saha, C.~Curino, O.~O'Malley,
  S.~Radia, B.~Reed, and E.~Baldeschwieler.
\newblock Apache hadoop {YARN:} yet another resource negotiator.
\newblock In G.~M. Lohman, editor, {\em {ACM} Symposium on Cloud Computing,
  {SOCC} '13, Santa Clara, CA, USA, October 1-3, 2013}, pages 5:1--5:16. {ACM},
  2013.

\bibitem{conf/eurosys/VermaPKOTW15}
A.~Verma, L.~Pedrosa, M.~Korupolu, D.~Oppenheimer, E.~Tune, and J.~Wilkes.
\newblock Large-scale cluster management at google with borg.
\newblock In L.~R{\'{e}}veill{\`{e}}re, T.~Harris, and M.~Herlihy, editors,
  {\em Proceedings of the Tenth European Conference on Computer Systems,
  EuroSys 2015, Bordeaux, France, April 21-24, 2015}, pages 18:1--18:17. {ACM},
  2015.

\bibitem{conf/icde/VitorovicE016}
A.~Vitorovic, M.~Elseidy, and C.~Koch.
\newblock Load balancing and skew resilience for parallel joins.
\newblock In {\em 32nd {IEEE} International Conference on Data Engineering,
  {ICDE} 2016, Helsinki, Finland, May 16-20, 2016}, pages 313--324. {IEEE}
  Computer Society, 2016.

\bibitem{journals/pvldb/WangKNL20}
Z.~Wang, A.~Kumar, S.~Ni, and C.~Li.
\newblock Demonstration of interactive runtime debugging of distributed
  dataflows in texera.
\newblock {\em Proc. {VLDB} Endow.}, 13(12):2953--2956, 2020.

\bibitem{conf/pdis/WilschutA91}
A.~N. Wilschut and P.~M.~G. Apers.
\newblock Dataflow query execution in a parallel main-memory environment.
\newblock In {\em Proceedings of the First International Conference on Parallel
  and Distributed Information Systems {(PDIS} 1991), Fontainebleu Hilton
  Resort, Miami Beach, Florida, USA, December 4-6, 1991}, pages 68--77. {IEEE}
  Computer Society, 1991.

\bibitem{conf/icde/WuCZTHN13}
W.~Wu, Y.~Chi, S.~Zhu, J.~Tatemura, H.~Hacig{\"{u}}m{\"{u}}s, and J.~F.
  Naughton.
\newblock Predicting query execution time: Are optimizer cost models really
  unusable?
\newblock In C.~S. Jensen, C.~M. Jermaine, and X.~Zhou, editors, {\em 29th
  {IEEE} International Conference on Data Engineering, {ICDE} 2013, Brisbane,
  Australia, April 8-12, 2013}, pages 1081--1092. {IEEE} Computer Society,
  2013.

\bibitem{journals/tpds/XiaoSC13}
Z.~Xiao, W.~Song, and Q.~Chen.
\newblock Dynamic resource allocation using virtual machines for cloud
  computing environment.
\newblock {\em {IEEE} Trans. Parallel Distributed Syst.}, 24(6):1107--1117,
  2013.

\bibitem{conf/sigmod/XuKAR22}
Z.~Xu, G.~T. Kakkar, J.~Arulraj, and U.~Ramachandran.
\newblock {EVA:} {A} symbolic approach to accelerating exploratory video
  analytics with materialized views.
\newblock In Z.~Ives, A.~Bonifati, and A.~E. Abbadi, editors, {\em {SIGMOD}
  '22: International Conference on Management of Data, Philadelphia, PA, USA,
  June 12 - 17, 2022}, pages 602--616. {ACM}, 2022.

\bibitem{conf/bigdataconf/YanXM13}
W.~Yan, Y.~Xue, and B.~A. Malin.
\newblock Scalable and robust key group size estimation for reducer load
  balancing in mapreduce.
\newblock In X.~Hu, T.~Y. Lin, V.~V. Raghavan, B.~W. Wah, R.~Baeza{-}Yates,
  G.~C. Fox, C.~Shahabi, M.~Smith, Q.~Yang, R.~Ghani, W.~Fan, R.~Lempel, and
  R.~Nambiar, editors, {\em Proceedings of the 2013 {IEEE} International
  Conference on Big Data, 6-9 October 2013, Santa Clara, CA, {USA}}, pages
  156--162. {IEEE} Computer Society, 2013.

\bibitem{YouScan}
YouScan Website, \url{https://youscan.io/en/}.

\bibitem{conf/networking/YunLWRK20}
D.~Yun, W.~Liu, C.~Q. Wu, N.~S.~V. Rao, and R.~Kettimuthu.
\newblock Performance prediction of big data transfer through experimental
  analysis and machine learning.
\newblock In {\em 2020 {IFIP} Networking Conference, Networking 2020, Paris,
  France, June 22-26, 2020}, pages 181--189. {IEEE}, 2020.

\bibitem{journals/tse/ZellerH02}
A.~Zeller and R.~Hildebrandt.
\newblock Simplifying and isolating failure-inducing input.
\newblock {\em {IEEE} Trans. Software Eng.}, 28(2):183--200, 2002.

\end{thebibliography}

\captionsetup[figure]{list=no}
\captionsetup[table]{list=no}


\end{document}